\documentclass[]{aa}
\usepackage{epsfig,psfig}
\begin{document} 
\thesaurus{06     
              (03.11.1;  
               16.06.1;  
               19y.06.1;  
               19.37.1;  
               19.53.1;  
               19.63.1)} 
\title{VLA radio continuum observations of a new sample of high redshift radio galaxies }


\author{L.Pentericci \inst{1,2}, 
           W. Van Reeven \inst{2}, 
           C.L. Carilli \inst{3},
           H.J.A. R\"ottgering \inst{2}
           G.K. Miley \inst{2}
          }

\offprints{L. Pentericci: laura@mpia-hd.mpg.de}

   \institute{Max Planck Institute f\"ur Astronomie, K\"onigstuhl 17 - 69117
   Heidelberg, Germany \and
Leiden Observatory P.O.~Box 9513, NL - 2300 RA Leiden,
The Netherlands \and
NRAO P.O. Box 0, Socorro NM, 87801 USA
             }

\date{Received ; accepted}
\titlerunning{VLA observations of HzRGs}
\authorrunning{Pentericci et al.} 
\maketitle

\begin{abstract}
We present new deep multi-frequency radio-polarimetric images of a sample of
high redshift radio galaxies (HzRGs), having redshift between 1.7 and 4.1.
The radio data at 4.7 and 8.2 GHz were taken with the Very Large Array in the A
configuration and provide a highest angular resolution of 0.2$''$. 
Maps of total intensity, 
radio spectral index, radio polarization and internal magnetic field are
presented for each source.  
\\
The morphology of most objects is that of standard FRII double radio sources,
but several contain multiple hot-spots in one or both lobes. 
Compared to similar samples of HzRGs previously imaged, 
there is a higher fraction (29 \%) of compact steep spectrum sources
(i.e. sources with a projected linear size less than 20 kpc). 
Radio cores are identified in about half of the sample and tend to have
relatively steep spectra ($\alpha \le -1$). 
\\
Polarization is detected in all but 4 sources, with  typical polarization 
at 8.2 GHz of around 10-20\%. The Faraday rotation can be measured 
in most of the radio galaxies: the observed rotation measure (RM) of 8 radio sources
exceeds 100 rad m$^{-2}$ in at least one of the lobes, with large gradients
between the two lobes. We find no dependence of  Faraday rotation with 
other
properties of the radio sources. If the origin of the
Faraday rotation is local to the sources, as we believe, 
then the intrinsic RM is more than a 1000 rad m$^{-2}$. 
Because low redshift 
radio galaxies residing at the center of clusters usually show extreme RMs,
we suggest that the high-z large RM sources also lie in  very dense 
environments. 

Finally, we find that the fraction of powerful 
radio galaxies with extreme Faraday rotation increases with redshift, as
would be expected if their average environment  tends to become denser 
with decreasing cosmic epoch. However this result has to be 
taken with caution, given the limitations of our analysis. 
   \end{abstract}

%

\section{Introduction}
High redshift radio galaxies play an important role in the study
of the early universe: thanks to their extreme luminosity at
different wavelengths it has been possible to us them as
cosmological probes already for several decades. Currently there are
more than 150 radio galaxies known
with redshift greater than 2, and recently a powerful radio source at a
redshift of 5.19 has been discovered by  \cite{bre99}, 
becoming the
most distant known AGN.
\\ High redshift radio galaxies (HzRGs) comprise a different
population to high redshift radio-quiet galaxies, e.g. Ly-dropouts: 
there is evidence that they are older and more massive, and
will evolve into brightest cluster galaxies  rather than 
L$_{*}$ ellipticals (Best et al. 1997; van Breugel et al. 1998). 
\nocite{bes97,bre98}
\\
In the past few years a number of studies have concentrated on the properties
of HzRG host galaxies, at visual and near infrared wavelengths
(e.g. Pentericci et al. 1999, van Breugel et al. 1998,
Best et al. 1997, Eales et al. 1997) \nocite{pen99,bre98,bes97,eal97} with the main goal of
studying the morphological 
evolution of these  host galaxies, understanding 
the nature of the radio-optical alignment effect and discerning  
the various components 
(stellar light, scattered light, etc.) 
that  contribute to the optical and infrared continuum emission.
\\
However,
one of the potentially most important results from recent studies
on powerful radio galaxies came from radio observations and 
with the discovery that a significant fraction ($\sim 20\%$) 
of HzRGs have extremely large Faraday rotation, of the order of
several thousands rad m$^{-2}$ (Carilli et al. 1997; Athreya et al.
1998), \nocite{car97,atr98} similar to low redshift 
powerful radio galaxies residing at
the center of X-ray clusters with extreme cooling flows (Taylor et
al. 1994).\nocite{tay94} 
This makes HzRGs potential excellent targets for finding
and studying  high redshift (proto) clusters.
\\
Therefore the main purpose of these new high resolution radio
polarimetric imaging observations was to enlarge the number of known HzRGs
 with high intrinsic Faraday rotation, 
and to provide cluster targets for future observations with facilities such 
as  the new X-ray telescopes, 
Chandra and XMM. \\ 
High resolution radio
imaging is important not only for finding high Faraday rotation radio galaxies,
but also for a number of other important issues, 
such as the identification of
the location of the active nucleus, the study of the correlation
between the optical morphology and the radio jets, or the line
emission gas and the radio jets, and the study of the evolution of
radio size structure.
\\
Throughout the paper we adopt a cosmology with $H_0=50$ km s$^{-1}$ Mpc$^{-1}$
and $q_0=0.5$. 
\section{Sample selection}
A large fraction of presently known HzRGs have been found 
by selecting ultra steep
spectrum  (USS) radio sources. Various groups have been involved
in the search using this technique (R\"ottgering et al. 1994; Chambers et al.
1996; Blundell et al. 1998). \nocite{rot94,cha96,blu98} 
At present there are about 150 radio galaxies
know with $z > 2$ (de Breuck et al. in preparation) of which
25 have  $z > 3$, 5 greater than 4 and 1 has a redshift greater than 5.
About 60$\%$ of all HzRGs have been found with the USS technique and 
this percentage increases with redshift (e.g. de Breuck et al. 1998). 
\nocite{deb98}
\\
The observations presented in this paper are an extension of 
the sample observed by \cite{car97}.  
We selected the present sample  
from the $\sim$110 radio galaxies  that were known  at the time (1997). 
Of these 110 about 50 (most of the optically bright ones) 
had already been observed at high
resolution (VLA in A configuration) at several frequencies 
(mostly by Carilli et al. 1997; few others by Carilli et al. 1994 and 
 Athreya et al. 1997). \nocite{car97,car94,atr97} From
the remaining list we selected all the radio galaxies having total
(estimated) flux density at 4.5 GHz greater than 25 $\mu$Jy, which allowed
a reasonable detection of the polarized flux, even in sources with only 
$\sim$0.5$\%$ percentage polarization. The final selection to reduce 
the sample to 27 objects was done randomly. Three
sources at $z < 2$, 0152$-$209, 1017$-$220 and
2224$-$273 (respectively at z$=$1.89, 1.77 and 1.68) from the MRC catalogue (McCarthy et al. 1996), 
were added, since they are part of
a sample of high redshift radio galaxies observed with HST/NICMOS
(Pentericci 1999)  and the  VLA was needed to match the resolution of
the near infrared images. 
Compared to the previous sample, the galaxies observed in this sample have,  
on average, slightly lower optical magnitudes.  
\\
In Table 1 we
report the radio and optical properties of our sample: for each
radio galaxy we list the radio catalogue from which it was
originally taken, the redshift (usually
determined from the Ly$\alpha$  emission line), and the optical magnitude
of the host galaxies.   
In a number of cases no optical magnitude was available, either 
because the sources were identified by observations in a different band
(e.g. K band)  or because they have not been measured.
We also report the
total flux densities at 1.4, 4.5 and 8.2 GHz (the first value is taken
from published radio catalogues listed in the table,  the others are
 determined from the  present observations), and  the total extent 
(in arcseconds) measured from the 8.2 GHz images for all radio sources.
In the last column we list the  references for the optical identification and
redshift determination.
\begin{table*}
\caption{\rm Source Parameters}
\begin{center}
\begin{tabular}{l l l c r r r r c} \hline \hline
\multicolumn{1}{c}{Source}& & &\multicolumn{1}{c}{R}&\multicolumn{1}{c}{Size}
&\multicolumn{1}{c}{$F_{1.4}$}&\multicolumn{1}{c}{$F_{4.7}$}
&\multicolumn{1}{c}{$F_{8.2}$}& \\
\multicolumn{1}{c}{(B1950)}&\multicolumn{1}{c}{z}&\multicolumn{1}{c}{Catalog}
&\multicolumn{1}{c}{(mag)}&\multicolumn{1}{c}{(arcsec)}&\multicolumn{1}{c}{(mJy)}
&\multicolumn{1}{c}{(mJy)}&\multicolumn{1}{c}{(mJy)}
&\multicolumn{1}{c}{Ref.}\\
\multicolumn{1}{c}{(1)}&\multicolumn{1}{c}{(2)}&\multicolumn{1}{c}{(3)}
&\multicolumn{1}{c}{(4)}&\multicolumn{1}{c}{(5)}&\multicolumn{1}{c}{(6)}
&\multicolumn{1}{c}{(7)}&\multicolumn{1}{c}{(8)}&\multicolumn{1}{c}{(9)}\\
\hline
0011$-$023 &2.080 &PKS &23.5  &$<0.2$ &347$^{\it a}$ &161 &95  & 1\\
0152$-$209 &1.89  &MRC &21.9  &2.2    &453$^{\it b}$ &109 &47  & 2 \\
0930+389   &2.395 &6C  &---   &4.2    &215$^{\it c}$ &73  &33  & 3\\      
1017$-$220 &1.77  &MRC &21.6  &$<0.2$ &583$^{\it d}$ &257 &148 & 2\\
J1019+053  &2.765 &MG  &23.7  &2.2    &454$^{\it a}$ &120 &59  & 4\\
1031+34    &2.1   &6C  &21.4  &41.2   &478$^{\it a}$ &128 &56  & 3\\
1039+681   &2.530 &8C  &---   &16.6   &268$^{\it a}$ &61  &25  & 5\\
1056+39    &2.171 &B2  &23.6  &14.2   &264$^{\it a}$ &71  &34  & 3\\
1132+37    &2.88  &B2  &---   &$<0.2$ &637$^{\it c}$ &227 &127 & 3\\
1134+369   &2.120 &6C  &---   &14.0   &235$^{\it a}$ &61  &26  & 3\\
1202+527   &2.73  &TX  &---   &5.3    &441$^{\it a}$ &188 &162 & 6\\
1204+401   &2.066 &B3  &23.0  &2.5    &237$^{\it c}$ &60  &28  & 7\\
J1338$-$19 &4.11  &TN  &22.0  &5.2    &--- &23  &9   & 8\\
1339+35    &2.772 &FW  &---   &13.0   &124$^{\it c}$ &21  &9   & 9\\
1357+007   &2.671 &PKS &23.7  &2.7    &296$^{\it a}$ &59  &32  & 10\\
1425$-$148 &2.355 &PKS &22.0  &11.6   &413$^{\it d}$ &120 &53  & 9\\
1558$-$003 &2.520 &TX  &23.4  &9.2    &375$^{\it a}$ &91  &38  & 10\\
1647+100   &2.509 &TX  &23.5  &23.4   &298$^{\it a}$ &44  &18  & 9\\
J1747+182  &2.281 &MG  &23.0  &7.3    &1095$^{\it a}$&329 &166 & 11\\
1908+722   &3.537 &6C  &21.4  &15.4   &259$^{\it a}$ &49  &16  & 12\\
2034+027   &2.129 &TX  &24    &3.9    &243$^{\it a}$ &61  &26  &  2\\
2048$-$272 &2.06  &MRC &$>$24 &6.7    &457$^{\it e}$ &90  &35  &  2\\
2052$-$253 &2.630 &MRC &23.8  &20.0   &219$^{\it e}$ &49  &22  &  2\\
2104$-$242 &2.49  &MRC &22.7  &23.7   &297$^{\it e}$ &58  &21  &  2\\
2211$-$251 &2.508 &MRC &23.4  &3.5    &836$^{\it d}$ &227 &112 &  2\\
2224$-$273 &1.68  &MRC &22.5  &$<0.2$ &233$^{\it e}$ &48  &19  &  2\\
2319+223   &2.554 &TX  &---   &8.9    &284$^{\it a}$ &44  &17  &  9\\ \hline
\end{tabular}\end{center}
(1) Most common name in B1950 notation (except when there is a J); (2)
redshift; (3) Radio catalogue: MRC Molonglo reference catalogue; TX Texas; PKS
Parkes; MG MIT-Greenbank;  6C/8C Cambridge; B2 Second Bologna; (4) R-band
magnitude of the host galaxies (when available); (5) Total angular extent of
the radio source; (6) Total flux density at 1.4 GHz: 
$^{\it a}$ Green Bank 1.4 GHz, White and Becker 1992;  
$^{\it b}$ NVSS, Condon et al. 1998; $^{\it c}$ FIRST, Becker et al. 1995; $^{\it d}$ extrapolated
from the 408 MHz flux of the PKSCAT90, Wright and Otrupcek (Eds) 1990;  $^{\it e}$ extrapolated
from the 408 MHz flux of the MRC Catalogue, Lange et al. 1981; 
(7) Total flux density at
4.7 GHz; (8)  Total flux density at 8.2 GHz;
(9) Reference for redshift: 1. Dunlop et al. 1989; 2. McCarthy et al. 1996;
3. Eales and Rawlings 1996; 4. Dey et al. 1995; 5. Lacy, PhD thesis; 6.Owen et
al. 1995; 7. Thompson et al. 1994; 8. Rawlings and Lacy 1996; 9. De Breuck et
al. in preparation; 10. R\"ottgering et al. 1997; 11. Eales and Rawlings 1993; 12. Dey et al. 1998.

\end{table*}
\begin{table*}
\begin{center}
\caption{\rm Core Parameters}
\begin{tabular}{l c c r r} \hline \hline
\multicolumn{1}{c}{Source} &\multicolumn{1}{c}{Core Position(J2000)} 
&\multicolumn{1}{c}{$F_{8.2}$}&\multicolumn{1}{c}{$\alpha_{4.7}^{8.2}$} &\multicolumn{1}{c}{$CF_{20}$}\\
&\multicolumn{1}{c}{RA \hskip1cm  Dec} & \multicolumn{1}{c}{(mJy beam$^{-1}$)}
& &\multicolumn{1}{c}{(\%)} \\
\multicolumn{1}{c}{(1)} &\multicolumn{1}{c}{(2)}&\multicolumn{1}{c}{(3)}
&\multicolumn{1}{c}{(4)} &\multicolumn{1}{c}{(5)}\\ \hline
0011$-$023 &00 14 25.54 $-$02 05 55.1 &88.4 &$-$0.9 &100.0\\
0152$-$209 &...                       &...   &...    &...  \\
0930+389   & 09 33 06.94 +38 41 50.8  & 0.29 &$-$0.8 &0.8 \\ 
1017$-$220 &10 19 49.02 $-$22 19 59.8 &144.7&$-$1.0 &100.0\\
1019+053J  &10 19 33.42 +05 34 34.8   &X2.06 &$-$1.0 &4.7\\
1031+34    &10 34 34.61 +33 49 27.3   &0.38  &$-$0.3 &0.6\\
1039+681   &...                       &...   &...    &$<$0.23\\
1056+39    &...                       &...   &...    &$<$0.13\\
1132+37    &11 35 05.93 +37 08 40.8   &123.9&$-$1.0 &100.0\\
1134+369   &...                       &...   &...    &$<$0.16\\
1202+527   &...                       &...   &...    &$<$0.04\\
1204+401   &12 07 06.27 +39 54 39.0   &0.66  &$-$0.5 &2.1\\
1338$-$19J &13 38 26.23 $-$19 42 33.6 &0.16  &$-$1.0 &1.9\\
1339+35    &...                       &...   &...    &$<$0.44\\
1357+007   &14 00 21.26 +00 30 20.7   &10.3 &$-$0.4 &27.0\\
1425$-$148 &14 28 41.72 $-$15 02 28.4 &0.35  &$-$0.7 &0.5\\
1558$-$003 &16 01 17.36 $-$00 28 46.3 &0.97  &$-$0.3 &1.6\\
1647+100   &...                       &...   &...    &$<$0.17\\
1747+182J  &...                       &...   &...    &$<$0.28\\
1908+722   &19 09 09.74 +72 15 15.3   &1.2  &$-$1.3 &4.8\\
2034+027   &20 36 34.78 +02 56 54.4   &1.7  &$-$1.2 &6.1\\
2048$-$272 &...                       &...   &...    &$<$0.18\\
2052$-$253 &...                       &...   &...    &$<$2.3\\
2104$-$242 &21 06 58.27 $-$24 05 09.1 &0.19  &$-$1.6 &0.7\\
2211$-$251 &...                       &...   &...    &$<$0.04\\
2224$-$273 &22 27 43.27 $-$27 05 01.7 &17.8 &$-$1.6 &100.0\\
2319+223   &...                       &...   &...    &$<$0.25\\ \hline
\end{tabular}
\end{center}
(1) Source name, (2) position of the radio core in J2000 coordinates; (3)
 core surface brightness in mJy beam$^{-1}$, measured on the 8.2 GHz map,
convolved to the resolution of the 4.7 GHz image, unless preceeded by an "X"; (4) core spectral index between 4.5 GHz and 8.2
    GHz; (5) percentage core fractions calculated at a rest frame frequency of
    20 GHz, with upper limits for the undetected cores. 
\end{table*}
\section{Observations and data reduction}
We used the VLA in its A (27 km) configuration on 1998 March 23 and 24 to make 
the observations. All sources were observed at two frequencies in the 5 GHz 
band of the VLA (4535 and 4885 MHz) and at two frequencies in the 8 GHz band 
(8085 and 8335 MHz). Bandwidths were 50 MHz for all frequencies. 
Each source was
observed for 15 minutes at 8 GHz and 8 minutes at 5 GHz.\\ 
An important limitation of
using the VLA in the A configuration is the short spacing limit. This implies a
maximum size on which we have information of about 10$''$ at 5 GHz and 
6$''$ at 8GHz. 
Although this is important for the large angular scale sources, it
will have a negligible effect on bright, small components such as the hot
spots and cores. \\
The data were gain-calibrated using 3C 286.
We used multiple scans of the calibrator 1745+173 to determine
the on-axis antenna polarization response. Two scans of 3C 286 separated in 
time by 7 hour were used to measure the absolute linear polarization position 
angles.\\
We used the Astronomical Image Processing System (AIPS) to process the data.
After calibration the data were edited and self-calibrated using standard
procedures to improve image dynamic range. The first few self-calibration
iterations involved phase self-calibration using a model derived from the same 
data. Natural weighting of the gridded visibilities was employed. 
The AIPS task IMAGR,
in which the CLEAN algorithm is implemented, was used to deconvolve the images.
The FWHM of the Gaussian restoring beams are shown in the bottom-left corners
of Fig. 6-32.
We synthesized images of the three Stokes parameters, $I$, $Q$ and $U$, and all
images were CLEANed down to the noise-levels.
The achieved noise is
25 $\mu$Jy/beam at 8 GHz and 50 $\mu$Jy/beam at 5 GHz. The resolution of the
observations is 0.23$''$  for the 8.2 GHz maps, and  0.43$''$ 
for the 4.7 GHz maps.\\
Total intensity maps were created using the combined data from the two
frequencies per band and the mean frequencies in each band. In order to make
spectral index maps we convolved the 8 GHz image with the Gaussian restoring
beam of the 5 GHz image. Since the spectrum of the sources can be approximated
by a power law, the spectral index $\alpha$ is defined as 
 as $I_{\nu} \propto{\nu}^{\alpha}$, where $I_{\nu}$ is the 
surface brightness at frequency $\nu$.
We calculated two-point spectral index values only for pixels with
surface brightness exceeding 3.5$\sigma$ (where $\sigma$ is the measured
off-source rms on an image) at both frequencies.\\
We used position angles for the polarized intensity for three frequencies
(4535, 4885 and 8200 MHz) to derive rotation measures. The rotation measures
were derived using the AIPS task RM. For the hot-spots
rotation measures were derived for the position of peak intensity. 
\section{Images and observed parameters}
For each source four images are presented (Fig. 6-32). 
They represent the
total intensity at 4.7 GHz 
({\em upper left}), the total intensity at 8.2 GHz 
({\em upper right}), the spectral index between 4.7 GHz and 8.2 GHz 
({\em lower right}) and the polarized intensity at 4.7 GHz
({\em lower left}) with overlayed vectors indicating the position angles 
of the 
magnetic field. 
The two total intensity maps are at full resolution, the
spectral index map and polarized intensity map are at the resolution of the 4.7
GHz image.
\\
For some of the sources we had to rotate the image over 90 degrees in order to
best display the source structure. The four images are placed in the same way
as with the other sources.\\
In the intensity maps, the contour levels vary with $2^{1/2}$, which implies a
change in surface brightness of a factor 2 every 2 contours including the
negative ones. The first non-negative contour is indicated in the caption to
each image and is set to 3 $\sigma$, with $\sigma$ the measured off-source rms.
The peak surface brightness in each image is also given in the caption.\\
The contour levels in the
spectral index maps are  -3,-2.8,-2.6,-2.4,-2.2,-2,-1.8,-1.6,-1.4,-1.2,-1,-0.8,
-0.6,-0.4,-0.2 and 0. The grey scale in the spectral index images also ranges
from -3 to 0.\\
In Tables 2 and 3 some of the observed properties 
of the sources are listed. All
source identifiers are for the B1950.0 coordinate system except those marked
with J.
 Table 2 lists the properties of the
radio cores, where identified. 
In column (2) we list core positions (J2000 coordinates), in column(3) the 
flux densities of the core at 8.2 GHz,
in column (4) the spectral index between 4.7 GHz and 8.2 GHz. Column (5) lists
the core fraction at a rest frame frequency of 20 GHz.\\
Table 3 lists properties of the brightest southern and northern hot spots.
Columns (2) - (5) list the properties of the southern hot spot, while columns
(6) - (10) list the properties of the northern hot spot. They are the peak
surface brightness, the spectral index between 4.7 and 8.2 GHz, the hot spot
fractional polarization at 4.7 and 8.2 GHz and the observed rotation measure at
the hot spot position respectively. The 8.2 GHz fractional polarization was
obtained in images that were convolved to the 4.7 GHz resolution.

\begin{table*}
\begin{center}
\caption{\rm Hot-Spots Parameters}
\begin{tabular}{l r r r r r r r r r r r} \hline \hline
&\multicolumn{5}{c}{Southern Hot Spot}&
&\multicolumn{5}{c}{Northern Hot Spot} \\ \cline{2-6} \cline{8-12}
&\multicolumn{1}{c}{$I_{4.5}$}
& \multicolumn{1}{c}{$\alpha_{4.7}^{8.2}$} 
&\multicolumn{1}{c}{$FP_{4.5}$}
&\multicolumn{1}{c}{$FP_{8.2}$}
&\multicolumn{1}{c}{RM} &
&\multicolumn{1}{c}{$I_{4.5}$}
&\multicolumn{1}{c}{$\alpha_{4.7}^{8.2}$}
&\multicolumn{1}{c}{$FP_{4.5}$}
&\multicolumn{1}{c}{$FP_{8.2}$}
&\multicolumn{1}{c}{RM}\\
\multicolumn{1}{c}{Source}
&\multicolumn{1}{c}{mJy/beam}
& &\multicolumn{1}{c}{(\%)}
&\multicolumn{1}{c}{(\%)}
&\multicolumn{1}{c}{(rad m$^{-2}$)}
&&\multicolumn{1}{c}{mJy/beam}
&&\multicolumn{1}{c}{(\%)}
&\multicolumn{1}{c}{(\%)}&\multicolumn{1}{c}{(rad m$^{-2}$)}\\
\multicolumn{1}{c}{(1)}&\multicolumn{1}{c}{(2)}&\multicolumn{1}{c}{(3)}&
\multicolumn{1}{c}{(4)}&\multicolumn{1}{c}{(5)}&\multicolumn{1}{c}{(6)}&&
\multicolumn{1}{c}{(7)}&\multicolumn{1}{c}{(8)}&\multicolumn{1}{c}{(9)}&
\multicolumn{1}{c}{(10)}&\multicolumn{1}{c}{(11)}\\ \hline
0011$-$023 &149.4 &$-$0.9&... &... &...   & &...  &...   &... &... &...   \\
0152$-$209 &95.9  &$-$1.4&... &0.4 &...   & &7.2&$-$1.7&0.3 &2.4 &87    \\
0930+389   &30.1  &$-$1.3&1.6 &7.4 &267   & &28.7&$-$1.3&4.0 &2.9 &29    \\
1017$-$220 &253.0 &$-$1.0&... &... &...   & &...  &...   &... &... &...   \\
1019+053J  &48.5  &$-$1.1&0.2 &6.0 &...   & &51.6&$-$1.2&1.2 &4.5 &$-$36 \\
1031+34    &48.3  &$-$1.1&6.9 &7.8 &51    & &8.7 &$-$1.5&9.3 &6.9 &19    \\
1039+681   &21.6  &$-$1.1&7.9 &15.4&$-$170& &3.5 &$-$1.4&21.1&34.7&$-$65 \\
1056+39    &9.7   &$-$1.1&13.4&17.9&2     & &42.8&$-$1.2&4.4 &11.1&$-$95 \\
1132+37    &218.6 &$-$1.0&... &... &...   & &...  &...   &... &... &...   \\
1134+369   &10.1  &$-$1.4&8.1 &15.9&140   & &29.8&$-$1.1&4.0 &6.3 &4     \\
1202+527   &6.0   &$-$1.6&5.8 &12.4&208   & &154.4&$-$0.1&3.0&0.8 &$-$100\\
1204+401   &16.6  &$-$1.4&... &... &...   & &22.3&$-$1.1&5.1 &16.2&...   \\
1338$-$19J &4.7   &$-$1.8&... &... &...   & &20.6&$-$1.6&... &1.7 &...   \\
1339+35    &2.9   &$-$2.5&... &... &...   & &12.4&$-$2.2&1.9 &3.6 &$-$160\\
1357+007   &17.3  &$-$1.2&1.8 &1.8 &$-$111& &18.0&$-$1.2&... &1.2 &...   \\
1425$-$148 &30.3  &$-$1.5&3.1 &3.4 &108   & &50.2&$-$1.4&17.4&19.7&$-$27 \\
1558$-$003 &39.9  &$-$1.3&11.0&13.0&$-$15 & &75.7&$-$1.5&10.9&5.4 &36    \\
1647+100   &8.2   &$-$1.3&1.9 &1.7 &...   & &24.6&$-$1.4&5.3 &6.3 &12    \\
1747+182J  &161.7 &$-$1.1&7.5 &16.6&73    & &92.2&$-$1.0&21.6&31.7&62    \\
1908+722   &16.0  &$-$1.7&7.3 &6.1 &34    & &7.3 &$-$1.8&13.9&16.8&91    \\
2034+027   &31.2  &$-$1.3&0.6 &4.6 &$-$45 & &9.8 &$-$1.6&9.2 &18.9&...   \\
2048$-$272 &1.4   &$-$3.1&... &... &...   & &49.7&$-$1.4&7.7 &12.3&63    \\
2052$-$253 &2.9   &$-$1.8&9.3 &26.9&...   & &11.8&$-$1.9&6.0 &16.8&$-$62 \\
2104$-$242 &2.3   &$-$1.9&6.6 &... &...   & &21.8&$-$1.5&14.9&16.1&$-$7  \\
2211$-$251 &18.9  &$-$1.5&6.3 &... &54    & &190.1&$-$1.2&0.25&... &...   \\
2224$-$273 &45.9  &$-$1.6&... &... &...   & &...  &...   &... &... &...   \\
2319+223   &32.4  &$-$1.6&8.6 &10.8&50    & &7.5 &$-$1.4&5.3 &9.7 &10    \\ \hline
\end{tabular}
\end{center}
Col (2)-(6) refer to the southern hot-spots, and (7)-(11) to the northern
one. 
(2) and (7) are the peak surface brightness; (3) and (8) the spectral index;
(4) and (9) the fractional polarization at 4.7 GHz; (5) and (10) the
fractional polarization at 8.2 GHz: (6) and (11) the rotation measure
calculated using observed frequencies with no correction for redshift.
A blank entry means that there was no detection at a 3$\sigma$ level.
\end{table*}

\section{Discussion}
\subsection{Morphology and core identification}
\begin{figure*} 
\centerline{
\psfig{figure=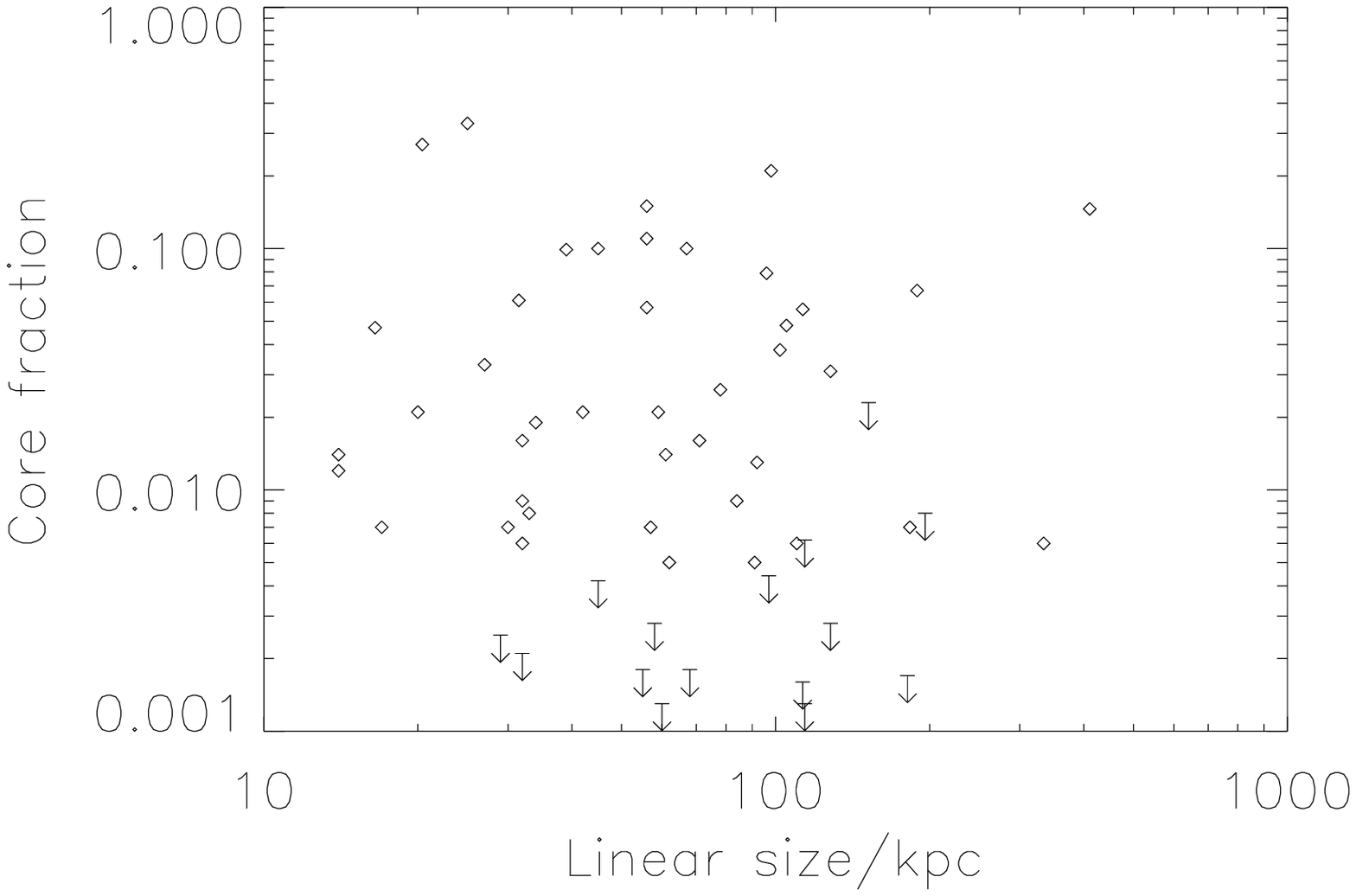,width=9cm,clip=}
\psfig{figure=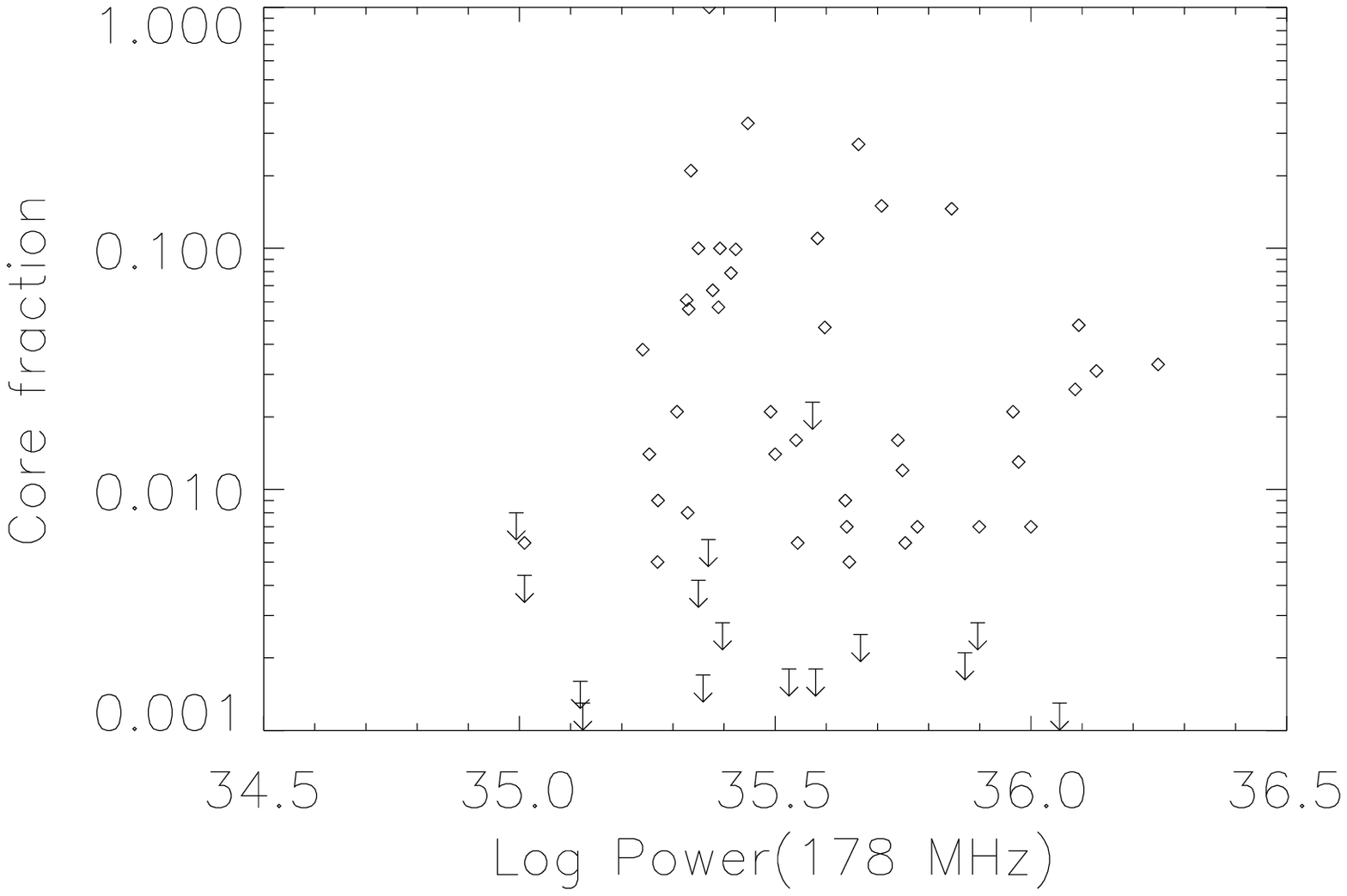,width=9cm,clip=}}
\caption{{\it Left} A log-log plot of the core fraction at 
rest frame frequency of 20 GHz  versus linear radio size in kpc (arrows
represent upper limits).{\it Right} The
core fraction plotted versus the radio luminosity at a rest-frame frequency 
of 178 MHz.}\label{f:log}
\end{figure*}

All the resolved sources have a FRII double morphology, with hot-spots at the
extremity of the radio source. 
A few sources, such as 1558$-$003, 
 have multiple hot-spots connecting the core to the outermost components,
typically on only one side. 
There are 8 small  sources that can be classified
 as compact steep spectrum (CSS) sources, having linear 
 (projected) sizes smaller than about 20 kpc (e.g. Fanti et al. 1990).
Of these, 
3 are unresolved (1017$-$220 and 1132+37 and 2224$-$273, with sizes less 
than 0.2$''$), 
one is barely resolved (0011$-$023,
which is sightly elongated in the 4.7 GHz maps)
 and another 4 (0152$-$209, 1019+053, 1204+401 and 1357+007) 
appear as small FRII radio galaxies.  
The percentage of CSS  sources  is  29\%,
significantly  higher than found in similar
 samples observed at the same frequencies and VLA array configuration: 
14\% of small sources in the sample of \cite{car97}, 
which included 38 HzRGs; 13\% in the sample of \cite{atr97},
 which included 15 HzRGs, some of which also belonged 
to the previous sample. The only difference between our sample and the
 previous ones, is the slightly lower average optical magnitude of the radio
 galaxies selected here,
but it is unclear how this should influence the average size of the sources.
Interestingly the percentage of CSS sources in this study is more similar
to that found by \cite{lon93} in a sample of steep spectrum radio
 loud quasars at z$\ge 1.5$ (27 \% of CSS sources) 
\\
In Table 2 we tabulate  the characteristics (coordinates, flux density,
 spectral index, and core fraction) of the cores.
We can identify the radio cores in  
less than half of the resolved radio sources (11 out of 24): 
the cores are identified as the unresolved 
component with the relatively flattest 
spectral index, which are not polarized. 
The percentage of identified cores is lower that 
the identification rate of  \cite{car97} (75\%). 
\\
In the
 source 1202+527 there is a component with very flat spectral index (-0.1),
but given that it is polarized and that its position would imply a highly 
asymmetric morphology, we consider it to be a hot-spot, and tabulate its
 characteristics in Table 3 instead of Table 2.
In the highest redshift 
radio source of the sample, 1338$-$19 at z=4.11, the core is
 the faint component just below the northern hot-spot, 
and appears only in the high frequency map (Fig. 14b). 
Despite its steep spectrum
 ($\alpha=-1$) and the high resulting asymmetry of the radio sources
we are confident about
 its identification given its proximity to the host galaxy 
(de Breuck et al. 1999). \nocite{deb99} 
\\
As previously mentioned, in general the radio cores are identified as 
the flat spectrum components:
however in 5 radio galaxies the  cores have very steep spectra 
(spectral index equal or steeper than -1). While it could be that 
in some cases the nucleus is not correctly identified or is blended 
with a steep spectrum component (this could be the case of 1019+681), 
most should be true unresolved cores.
\\
Steep core spectra have previously been found in many high redshift radio 
galaxies by \cite{car97} and \cite{atr97}. 
Athreya et al.  suggest that the rest frame frequencies 
at which these cores are observed (15 to 30 GHz,
depending on redshift) are higher than the turn over frequency due to 
synchrotron self-absorption. 
They suggest that the cores of radio sources exhibit 
synchrotron self-absorption turn over at 20 GHz in the rest frame of the
 emitting plasma, and that the spectra will appear steep above that
 frequency. They also explain the difference between the galaxy cores and
 quasars cores (which exhibit steepening at a much higher frequency) 
with the fact that the turn over frequency appears blue-shifted in the
 relativistically beamed quasar core and redshifted in the galaxy cores.
They predict that the size of the dominant core component should be less than
 1 mas.
\\
In Column 5 of Table 2 we list the core factions, calculated at a rest-frame
 frequency of 20 GHz: for those sources where the core is undetected we give
an upper limit, assuming that the core flux is less than 3 times the rms
 noise of the map.
Core fractions vary from less than  0.05\% to few \%, a range that is
 typical for radio galaxies. 
The only exception is 
 the CSS source 1357+007 that has a core contributing for 27\% of the total
 flux. The median core fraction is  0.6\%.
\\
According to evolutionary models  of radio sources (e.g. Kaiser et al. 1997),
\nocite{kai97} as a radio source grows older and expands, its lobe 
radio luminosity 
declines, whereas the core flux remains constant.
Therefore the prediction is that larger radio sources, 
which on average should be older, should also  have higher core fractions.
To test this model, in Fig. \ref{f:log}a we have plotted the core fractions 
of the radio sources in this sample,
 augmented by the sample of \cite{car97}, 
versus their radio sizes in kpc.
We do not detect  any increase in the average core 
fractions with increasing radio
 size: on the contrary we see the opposite
 effect, i.e. a slight decrease in the core fraction, from a median value of
 1.5 \% for sources smaller than 50 kpc, to 1.2\% for sources 
between 50 and 100,
to 0.8\% for sources larger than 100 kpc, with a large scatter 
around these median
 values at any given size 
(note that using the median value is a better estimate, expecially when
 dealing with upper limits).\\
These trends should be interpreted with caution 
since the sizes of HzRGs range  only 
from 10 to
 less than 500 kpc and there are very few large sources.
For example the only two sources larger than 400 kpc have relatively large
 core fractions.
Therefore if the effect predicted  by Kaiser et al. sets in only at rather 
large sizes (of few hundreds of kpc), 
we would not observe it given our limited
 number of large sources.
\\
The core fraction of HzRGs tend to be higher than those of matched luminosity 
3CR galaxies (Laing et al. 1983) at redshift z$\sim 1$. 
\cite{bes99} show that for these radio galaxies 
the median core fraction at
a rest frame frequency of 16 GHz is  0.2-0.3\%
and does not depend on  radio sources size
for a large range of sizes (from 10 to 1000 kpc.
 The slight difference in rest-frame frequency (we use 20 GHz instead 
of 16 GHz) should not be important. 
Therefore the core fractions at z $>$ 2 are $\sim 4$ times larger 
that at z$\sim 1$.
This could indicate that either at high redshift there
are intrinsically stronger cores, or that the beaming factor is higher at
 earlier epochs.
Alternatively, if the core fraction really depends on radio sources 
sizes as predicted by \cite{kai97}, 
the difference between the high and low redshift samples could 
be due to different average sizes of the two samples considered
A full discussion of this issue 
is beyond the scope of this paper.
\\
In Fig. \ref{f:log}b we present also a plot of core fraction versus power.
Although 
we sample only about one order of magnitude in power,
we see no  significant correlation between these two quantities.
This is in agreement with previous results at lower redshift 
(Best et al. 1999).
\\
\\
\subsection{Radio source distortion}
\begin{figure*} 
\centerline{
\psfig{figure=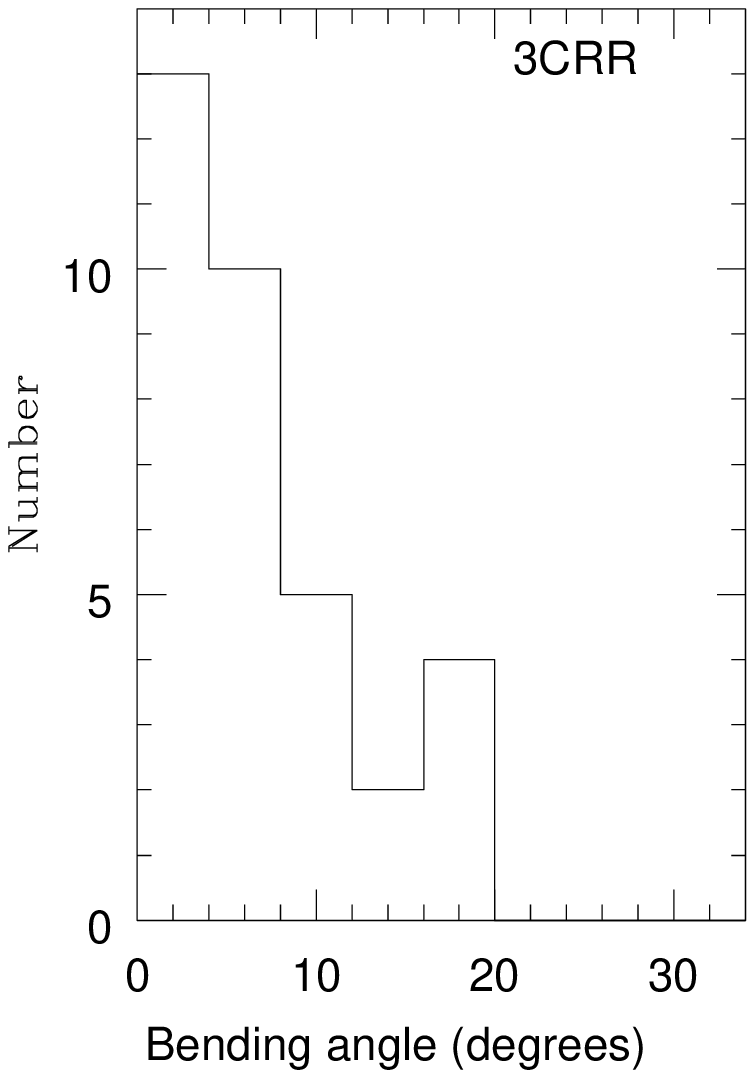,width=7cm,clip=}
\psfig{figure=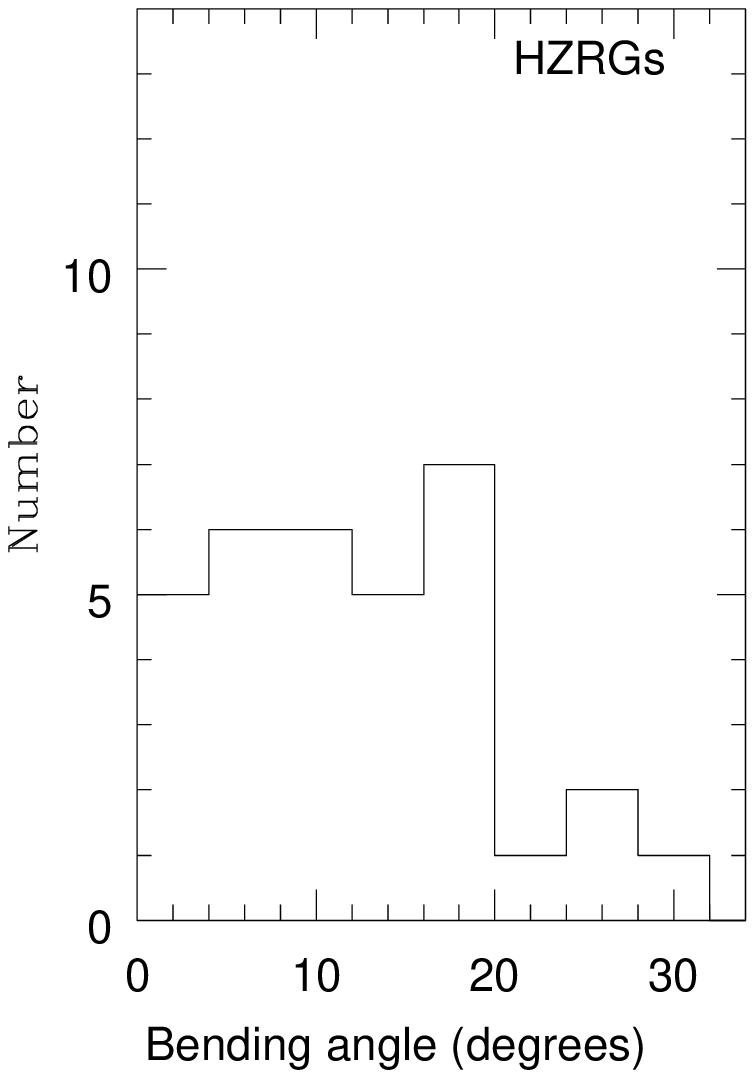,width=7cm,clip=}}
\caption{Right: the distribution of the bending angles for the sample of HzRGs
  that includes all the sources presented in this paper plus those from Carilli
  et al. (1997). Left, the analogous distribution for the  3CR sources
 (figure from Athreya 1997).}\label{f:angle}
\end{figure*}

Several sources in our sample show  "distorted" morphologies, with multiple 
hot-spots which are often not aligned (e.g. 1908+722).
Following previous authors (e.g. Barthel \& Miley 1988), \nocite{bar88} we 
measure this non-linearity 
 of a radio source 
with a "bending angle",  which is defined as 180$^{\circ}$ minus the angle
between the 
lines joining the core to opposite hot-spots on either side of the 
source.
The bending angles for objects in our sample range  
from 0$^{\circ}$ to 22$^{\circ}$.
Note that many sources such as 1908+722 and 1039+681 show multiple 
bends, therefore 
the bending angle as defined above only measures the overall bends as defined
by the outer extremities.
\\
In Fig. \ref{f:angle}a we present 
the distribution of the bending angles of our 
present sample, with the addition of data for objects in \cite{car97},
supplemented by data on a few radio sources from the literature 
(Carilli et al. 1994). \nocite{car94}
This larger sample is homogeneous, i.e. all the galaxies have been observed 
at the same  resolution  and frequencies.
The distribution of bending galaxies is basically flat 
from 0$^{\circ}$ to 20$^{\circ}$, with an
average of 12.3$^{\circ}$.
We compare this distribution with that of the 3CR 
radio sources which are
matched in luminosity and have redshifts between 0.6 and 1.6 
(Fig. \ref{f:angle}b reproduced from Athreya 1996)
The difference is striking: 
not only is the average bending angle at redshift greater than 2 more than
double that at $z \sim 1$, but also the distribution is 
different. 
\\ 
\cite{bar88} first pointed out  
the increasing distortion in the appearance of radio  
sources at high redshift, although  the angles involved for quasars are
larger, due to larger projection angles (e.g. Kaphai 1990).\nocite{kap90} 
This is 
in agreement with the predictions of unification schemes based on orientation.
The increase with redshift of asymmetries in the morphology of
powerful radio galaxies 
was also noted by \cite{mcc91} for low redshift 
3CR radio sources, comparing two samples of 
z $<$ 0.2  and 2.5$>$ z $>$ 0.7 radio  galaxies.
\\
A source can be distorted due to interaction between the radio
jets and the ambient medium:
one example is the radio galaxy 1138-262 (Pentericci et al. 1998), \nocite{pen98} which  
shows 
clear signs of interaction between the western 
jet and the emission line gas, and has large velocity gradients at the
location where the radio jet sharply bends. 
Therefore denser and clumpier environment at high redshift could 
also explain the increase of distortion with redshift.
Indeed optical and narrow band observations have shown 
that also the stars and gas distribution of HzRGs appear extremely clumpy and
asymmetric at high redshift, on a scale comparable or larger than  that 
of the radio emission
(e.g. Pentericci et al. 1999).\nocite{pen99} 
\\
The increase of ambient  density with redshift has also been invoked
to explain 
the decrease in average source size, at a fixed radio power,  
observed by many groups, although
there is still disagreement on the 
exact shape of the distance-redshift relationship (see references in Blundell
et al. 1999).
\subsection{Polarization properties}
In most sources one or more components are polarized at both frequencies;
typical polarization levels are on the order of less than 10\% at 4.5 GHz
and up to 20 \% at 8.2 GHz.
In many case there are large differences in polarization between the hot-spots,
which could be due to asymmetric properties in the environment. 
\\
In Fig. 6-32c we present the polarization maps at 4.5 GHz 
with superimposed vectors representing the direction and strength of the 
magnetic field 
(corrected for Faraday rotation).
The electric field is oriented perpendicular to these vectors.
\\
In about half the  sources the hot-spots magnetic fields 
 are oriented perpendicular to the 
jet direction, a common characteristic of the hot-spots 
of powerful radio sources
(e.g. Muxlow \& Garrington 1991). \nocite{mux91}
However there are several cases, such as 1357+007  
where the magnetic field  vectors are parallel
to the jet direction, and in many sources both parallel and perpendicular
fields are present (e.g. 2211$-$251).
\\
A possible reason is that these components are not true outer hot-spot
but are associated with jets or oblique shocks. 
The magnetic field is parallel in the jets, and perpendicular to the jet axis
in the hot-spots, while in the radio bridges it wraps around the the edges 
around the hot-spots, hence being parallel to the radio axis (Saikia \& 
 Salter 1988). \nocite{sai88} 
Therefore if the knots we observe are not real hot-spots 
but  
consist of different unresolved  structures, there can be intermediate
direction or even parallel B field.
\\
The strength of the magnetic field in each hot-spot  
can be calculated by making the standard minimum energy conditions 
(Miley 1980)\nocite{mil80}: the resulting 
magnetic fields range from 160 to 700 $\mu G$.
\subsection{Faraday rotation} 
\begin{figure*}
\centerline{
\psfig{figure=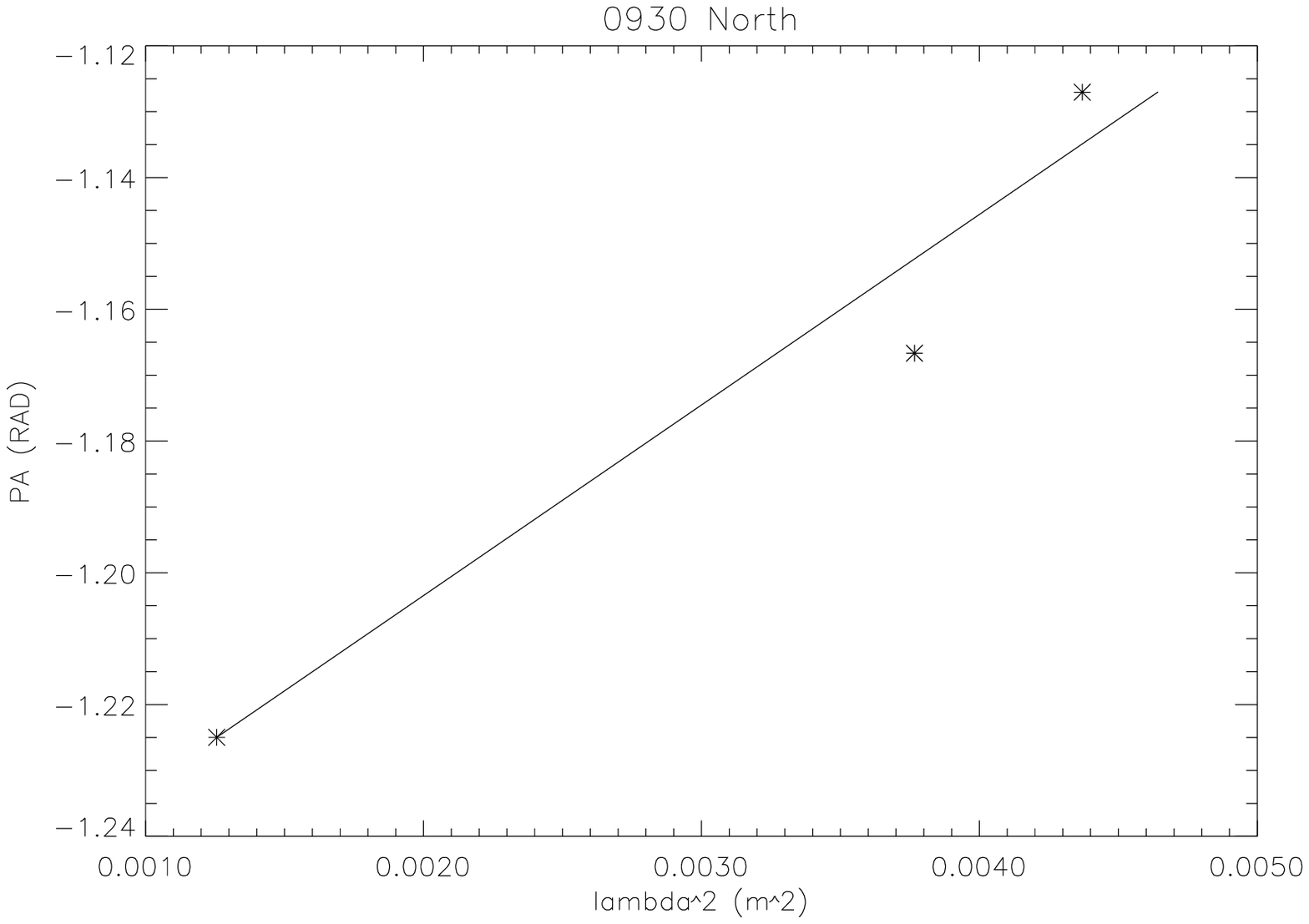,width=7cm}
\psfig{figure=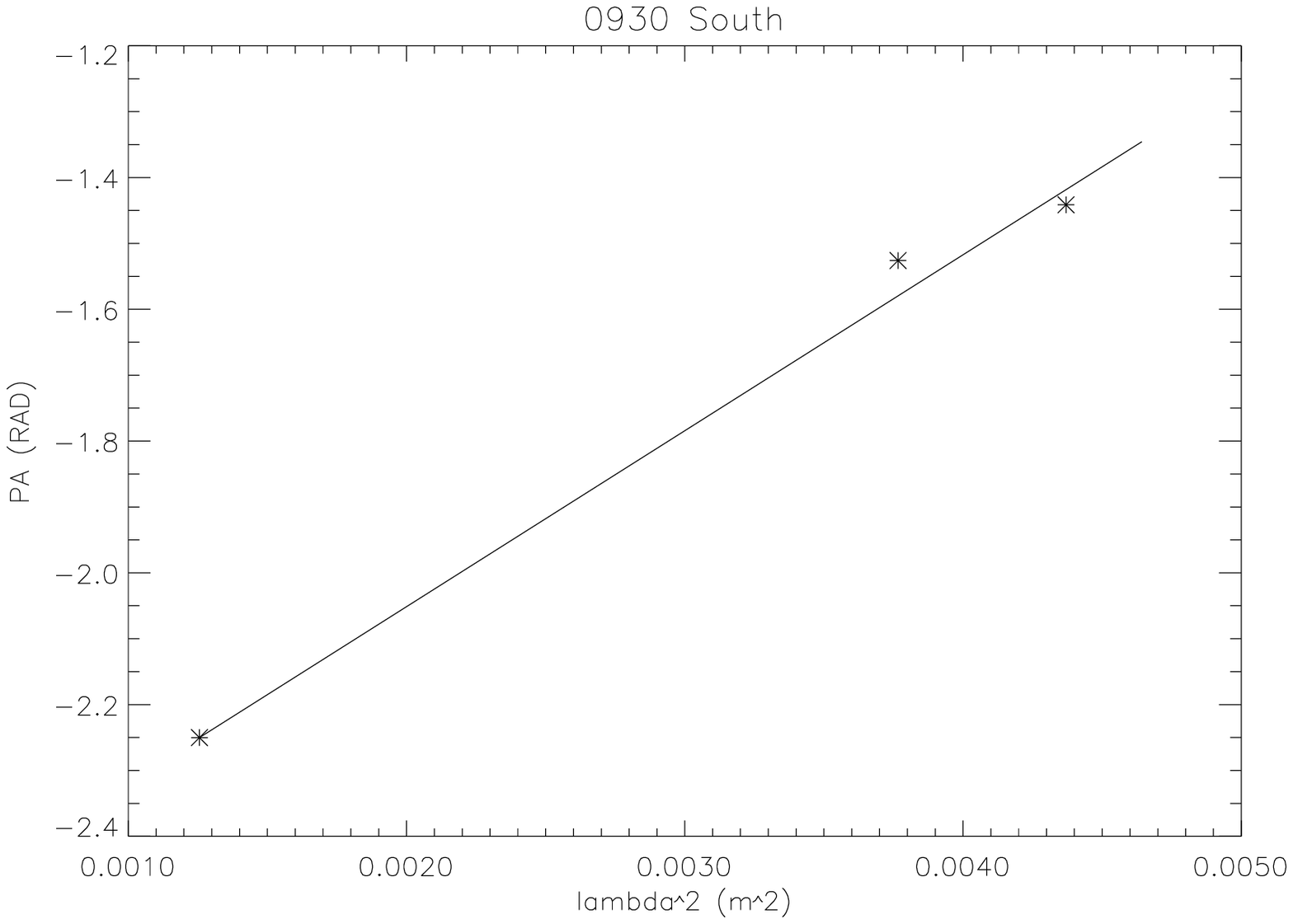,width=7cm}}
\centerline{
\psfig{figure=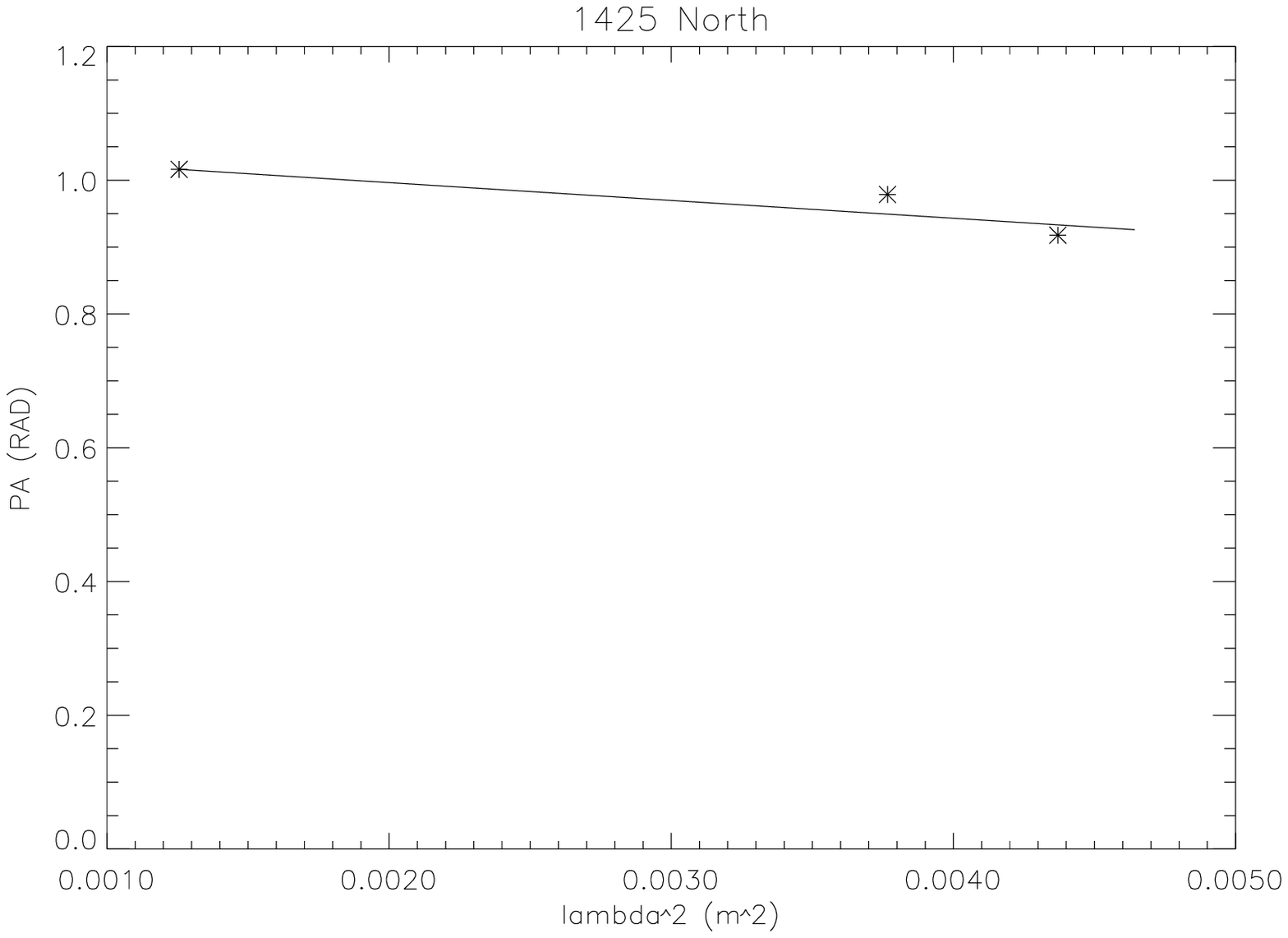,width=7cm}
\psfig{figure=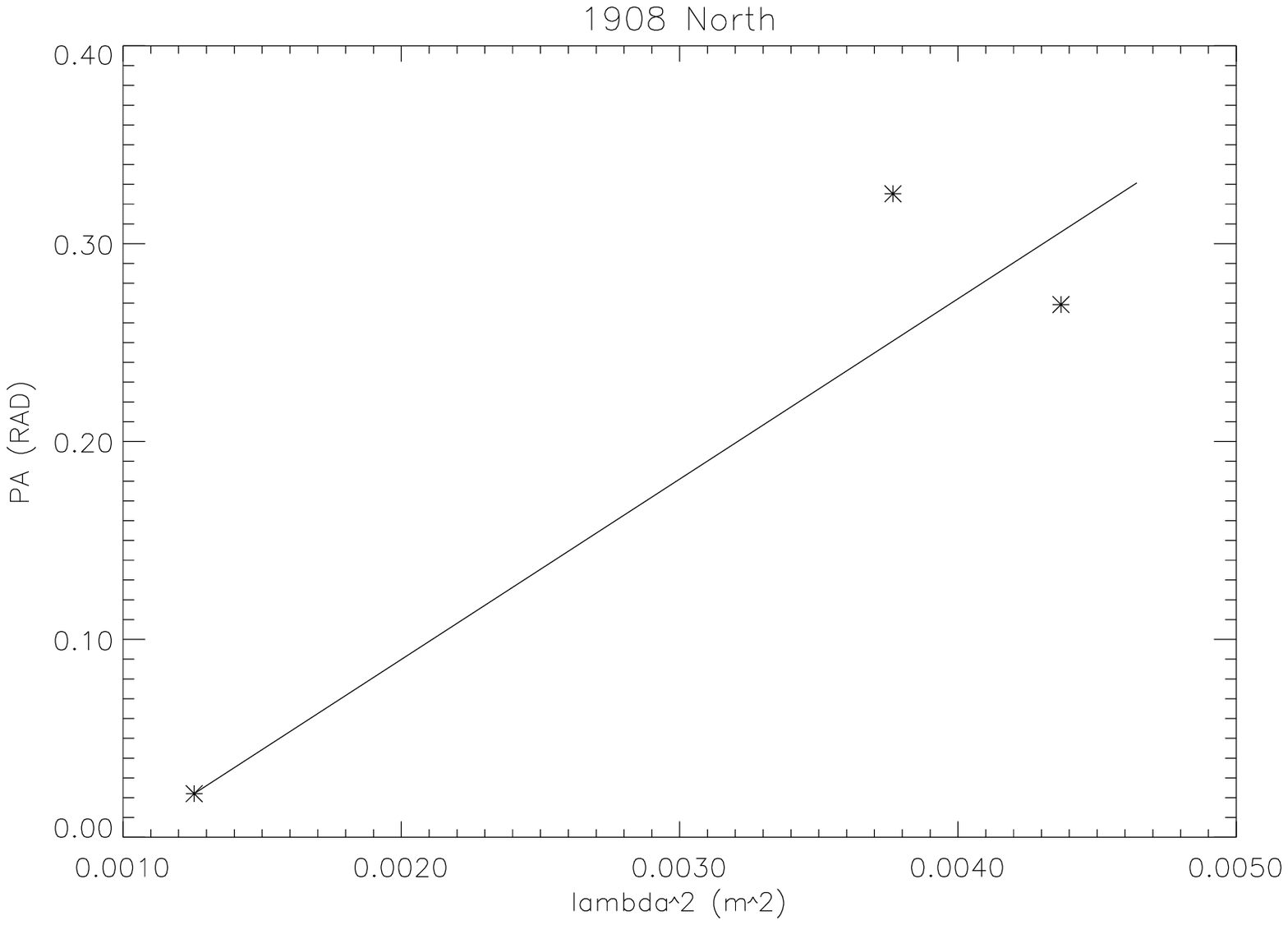,width=7cm}}
\caption{Polarization position angles versus $\lambda^2$(in cm) for the lobes
  of some HzRGs in our sample: upper panels are the components of 0930 north
  (left) and south (right). The lower panels are 1425 north (left) and 1988
  north (righ).}\label{f:rm1}
\end{figure*}

In Table 3 we list the observed values of Faraday rotation for those regions of
the radio galaxies that produce enough polarized signal to 
allow a determination of
the angle of polarization (see Section 2). 
In Fig. \ref{f:rm1} we show several plots of the polarization position angles (in
radians) versus
wavelength squared for the components of three
HzRGs. The lines represent the best linear 
fit to the data points, given by the AIPS task RM.
\\
If the Faraday screen is located at a redshift $z_F$,
then the intrinsic value of RM$_{intr}$ is related 
to the observed value RM$_{obs}$ as:
$$RM_{intr} =RM_{obs} \times(1+z_F)^2.$$
For the radio galaxies, it is most probable that the 
 Faraday screen that produces the RM is located at the same 
redshift of the radio sources.  We can exclude a Galactic origin 
for the Faraday rotation since, at latitudes 
$b > 20^{\circ }$  the contribution of the Galactic screen 
is of order of 10 rad m$^{-2}$ with RM gradients 
of $<<$10 rad m$^{-2}$ over 1$''$ (e.g. Leahy 1987)\nocite{lea87}  while
 we observe much larger gradients between the two (or more) hot-spots
of each radio source. For example in the radio galaxy 1202+527 there is a
gradient of more than 300 rad m$^{-2}$ with a sign reversal, over only 4$''$.
Contribution from intervening structures such as galaxies and clusters, which
have $\mu$G magnetic fields correlated over kpc or 10ns of kpc  scales,
or  absorption line systems, 
 can be also ruled out on the basis of small probability (e.g. 
Athreya et al. 1998). \nocite{atr98}
\\
Therefore if the RM screen is in the vicinity of the radio source, 
the values listed in
Table 3 have to be multiplied by a factor of 10 to 20 (depending on redshift),
implying RMs of the order of several 100 rad m$^{-2}$ for most sources and 
in excess of  1000 rad m$^{-2}$ for 8 sources, with a maximum of 3100 rad
m$^{-2}$ for the radio source 0930+389. 
\\
\cite{car94} were the first 
to point out the existence of very large Faraday
rotation in HzRGs. In the previous VLA observational 
study similar to this,  they found
that about 20 \% of radio galaxies had
intrinsic RM in excess with   1000 rad m$^{-2}$, while 
\cite{atr98} found high RM in 4 out of 15 radio galaxies.
Both these results are in agreement with the results from the present 
sample.
\\
At low redshift most powerful  radio galaxies
show rotation measures of only several 10s rad m$^{-2}$ which arise
in the interstellar medium (ISM) of our Galaxy.
However few  radio galaxies have a RM in excess of 1000 rad m$^{-2}$
 (Taylor et al. 1994): these are either compact (sub galactic in size) radio sources or are radio galaxies located in X-rays cooling flow clusters.
In fact \cite{tay94} found  that the strength of the cooling flow
(or alternatively the core density of the cluster)
directly correlates with the rotation measure.
An important point is that the above relation is independent of radio 
source luminosity and  morphological class: since FRI and FRII type galaxies 
have very different physical interactions, this suggests that the 
Faraday rotation is most likely a probe of cluster properties and not
 radio source properties.
The physical conclusion is that for those sources located in clusters,
the RM arises in the dense X-ray emitting gas which is substantially 
magnetized.
\subsubsection{Does Faraday rotation depend on radio source properties?}
\begin{figure*}
\centerline{
\psfig{figure=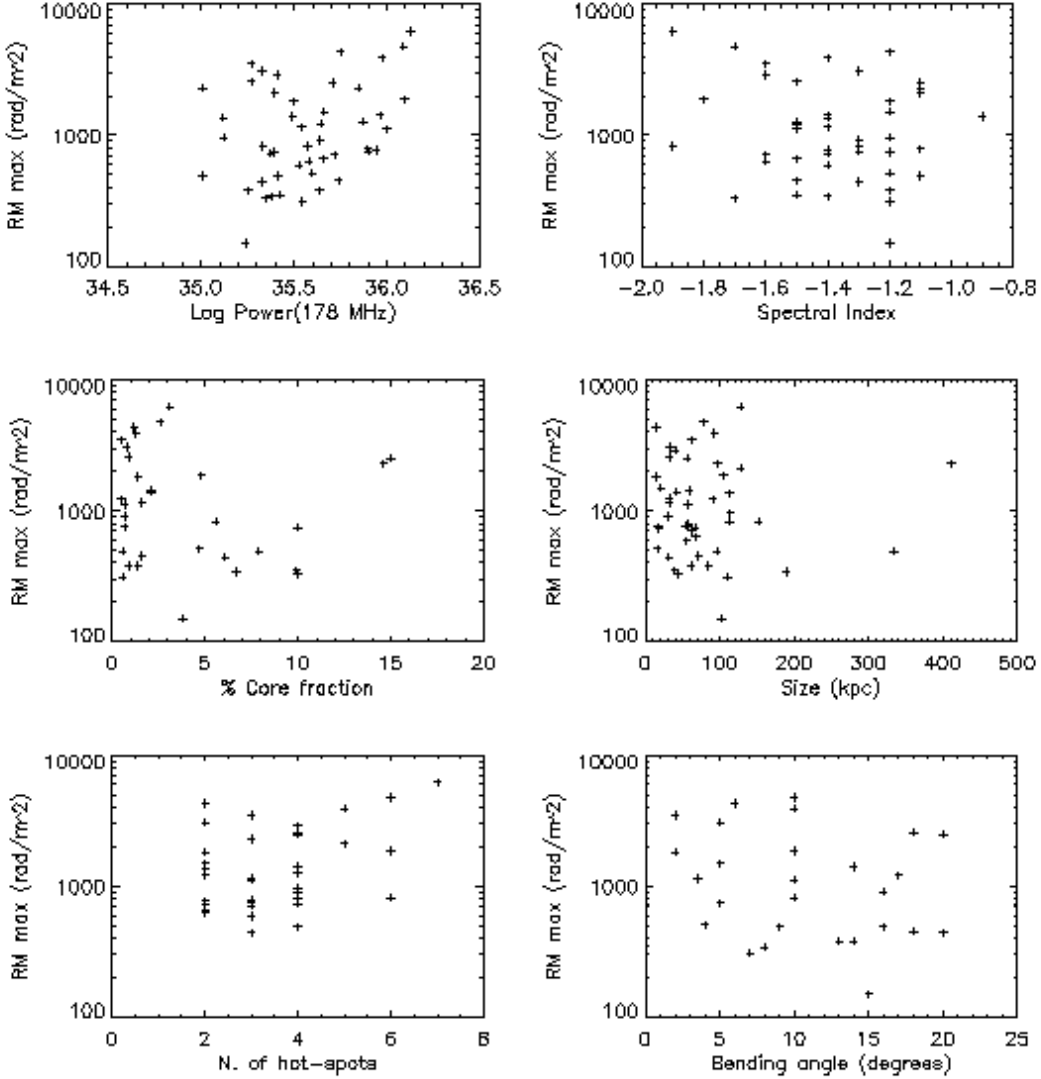,width=15cm}}
\caption{The maximum Faraday rotation observed in high redshift
radio galaxies, plotted against different  energetic and morphological
properties of the radio sources (see text).}\label{f:prop}
\end{figure*}

To determine the nature of the Faraday screen at high redshift
we first investigated whether the Faraday rotation
 is significantly correlated with the radio
source morphology or other properties. As mentioned in the previous section,
at low redshift Faraday rotation does not depend on radio source
morphology or luminosity: this might not be the case at high redshift,
where it is believed that the interactions of the radio jets with
the host galaxies and the ambient medium are much stronger, as
shown for example by the increasing distorted morphology of radio
sources discussed in section 5.2.
\\
From the combined sample of about 70 radio galaxies at $z > 2$  with 
homogeneous  high resolution
radio polarimetric VLA observations (Carilli et al. 1994 and 1997;
Athreya et al. 1998 and this paper), \nocite{car94,car97,atr97} we selected
all HzRGs with observed  RM larger than 40 rad m$^{-2}$ (lower
observed values might be effected by relatively larger errors).
The total number of selected HzRGs is 37, of
which 23 have intrinsic RM larger than 1000 rad m$^{-2}$.
\\To
characterize a radio source we used the following parameters: (a)
the monochromatic  power at a rest-frame frequency of 178 MHz; (b) the
total extent of the radio source; (c) the integrated spectral index
between 8.2 and 4.5 GHz (observed frequencies) ; (d)
the core fraction at a rest-frame frequency of 20 GHz; (e) the 
number of hot-spots defined as the number of separate emission peaks in 
the 8.2 GHz maps and (f) the bending angle.
\noindent
Assuming orientation unification models (e.g. Barthel 1989) \nocite{bar89}
the core
fraction allows us study line-of sight effects, knowing that radio
sources in general have a smaller core fraction (typically 1-2 \%) than
radio loud quasars and therefore those with higher core fractions will
be in the transition zone between radio galaxies and quasars.
The total length of the source is used to see if Faraday rotation is
related to the inner region of the galaxy, and hence is shown only by radio
sources whose size is  comparable to the galaxy, like
the CSS sources that have sizes of less than 20 kpc,
or is a larger scale effect.
Finally parameters such as the number of hot-spot and the
bending angle, give us information on the distortion of the radio
sources, which is probably related to the density of the near
environment of the object (e.g. Barthel \& Miley 1988). \nocite{bar88}
\\
In Fig. \ref{f:prop} we present the results of these investigations:  we do
not see any significant dependence of the  Faraday rotation on any of above
parameters. Therefore we conclude that, at high redshift, Faraday rotation
is independent of
radio source luminosity and morphology, and is probably not the probe of radio
sources properties but of their environment
\\
Drawing the analogy with lower redshift sources having  extreme RM, 
the conclusion is that also the HzRGs with large Faraday rotation
 might reside in dense, proto-cluster environment.
Indeed selecting the
HzRGs with highest known Faraday rotation (6600  rad m$^{-2}$) our group has
detected possibly extended X-ray emission around the radio galaxy
1138$-$262 at a redshift of 2.2 (Carilli et al. 1998).\nocite{car98}  
If confirmed (time has been
allocated at the Chandra X-ray observatory 
to re-observe the source with a much higher spatial resolution), 
this would be the highest
redshift X-ray cluster known to date.
\\
Including the new sources presented in this paper, there are now 
a total of 23 HzRGs with Faraday rotation 
exceeding 1000 rad m$^{-2}$, with redshift ranging from 2.2 to 3.8 (for a
complete list see Pentericci 1999). Of these, 3 are CSS sources 
for which the
origin of the Faraday rotation could be the local ISM, given that the radio
sources are completely embedded within the host galaxies.
The remaining 20 
are excellent targets when searching for the most distant (proto)clusters
in the early Universe.
\subsubsection{Comparison with low redshift radio galaxies}
\begin{figure}
\centerline{
\psfig{figure=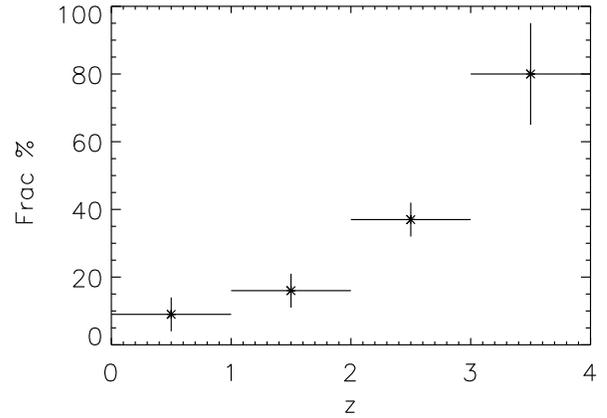,width=9.cm}}
\caption{The fraction of powerful radio galaxies with Faraday rotation 
in excess of 1000 rad m$^{-2}$ as a function of redshift; 
the horizontal error bars indicate the redshift range for each bin.}
\label{f:frac}
\end{figure}

To investigate whether the fraction of galaxies with extreme RM changes
as a function of redshift, we gathered  data from the literature for a
sample of powerful radio galaxies at low and intermediate redshift to be 
compared to our sample.
We selected all objects from the 3CR catalogue (Laing et al. 1983) 
\nocite{lai83}
having a monochromatic power at 178 MHz (rest-frame frequency) of
$log P_{178} >34.8$ (with $P_{178}$ in units of erg sec$^{-1}$ Hz$^{-1}$)
an arbitrary value chosen
in order to have relatively powerful radio sources, with still a
considerable number of low redshift sources.
We added these sources to the sample of radio galaxies at $z \ge 2$.
We then eliminated the CSS
sources from the resulting list: in  these galaxies
the radio source is generally completely embedded
within the host galaxy,
so the RM is due to the local magnetized ISM.
CSS sources generally have very high Faraday
rotation, therefore if distribution of CSS sources with redshift
is different from that of large radio sources 
(e.g. O'Dea 1998),
this will modify the final result.
The final sample contains $\sim 90$ radio galaxies
spanning a redshift range from $z=0.015$ (Cygnus A) to $z=3.8$
(4C41.17).
\\
We then searched for Faraday rotation measurements for the 3CR
galaxies: all of the values were taken from \cite{tab80} and \cite{ino95}, 
with the exception of the radio galaxy
Cygnus A for which much more detailed studies have been carried out
(Carilli and Barthel 1996). 
The results are
presented in Fig. \ref{f:frac}, where we plot the fraction 
of galaxies with RMs above 1000 rad m$^{-2}$ as a function of redshift,
for four redshift intervals.    
The fraction of galaxies with RMs above the line clearly
increases with z: $9\pm 4\%$ at $z <1$, $16\pm 5\%$ at $1< z <2$, $37 \pm 5 \%$
at $2< z <3$ and $80 \pm 15 \%$ at $z>3$ (although in the last redshift
bin there are only 6 galaxies and therefore the statistics is very
low).
\\
Some care has to be taken when considering this result, given the
following limitations
to our analysis: (i) first, most high redshift radio galaxies were found
by selecting ultra steep spectrum radio sources (radio spectral index $\alpha
<-1$, where $S_{\it \nu} \propto \nu^{\alpha}$, with
 $S_{\it \nu}$ the flux density and $\nu$ the frequency). This means that
if Faraday rotation depends on spectral index (but in the previous section we
showed that this is not the case, at least for a limited range of $\alpha$), then our
results will be biased;
(ii) the quality of the different radio
observations is not matched in  resolution: since one always
measures the average RM within the beam size, having a large beam size
implies measuring lower values of RM;  (iii) finally, and most important,
the various sets of observations have different wavelength coverage, which
determines the
highest and lowest values of RM observable. Note that the resolution effect 
(ii) would tend to decrease the
correlation found, since at higher redshift the beam size in kpc will tend 
to be larger. Furthermore, within our sample (redshift  $ z > 2$, the 
physical resolution  is nearly constant, but we still detect
 the correlation between
redshift and Faraday rotation. 
\\
Despite the above 
limitations, we regard the effect apparent in Fig. \ref{f:frac} to be  real, 
namely that the fraction of powerful radio galaxies
with  high Faraday rotation increases with redshift.
If high Faraday rotation is an indication of dense environment, 
this result is consistent with the 
fact that the  average environment of powerful radio sources
becomes denser with increasing redshift (e.g. Hill \& Lilly 1991). \nocite{hil91}
At low redshift most powerful radio sources (FRII type) reside in sparse
environment with few exceptions (e.g. Cygnus A), while at earlier epochs
more and more radio galaxies reside in dense environments (e.g. Roche et. al 1998).\nocite{roc98}
\section{Conclusions}
We have presented high resolution multi-frequency radio polarimetric
observations of a sample of 27 high redshift radio galaxies.
Maps of the sources and 
the fundamental parameters of the observations were presented.
This, together with previous samples makes now an extended data base from
which the relation between basic properties can be studied.
The main results are the following:
\begin{itemize}
\item We detect radio cores in about half of the sample.
The cores often have steep spectra ($\alpha < -1$).
The core fractions depend only weakly on radio sources size, contrary to the 
predictions of radio source evolutionary models. The median core fraction 
 is larger than that  of matched-luminosity 3CR radio galaxies 
at redshift $\sim 1$. 
\item We have shown that high redshift radio galaxies tend to be more
  distorted than at low redshift. This implies a larger density of
  the external medium in which they reside.
\item 
We have discovered 8 new radio galaxies with
very high Faraday rotation and large gradients between the different
components. Given that the Faraday rotation properties do not depend on 
radio sources parameters such as power, total size, distortion etc, 
our interpretation is that these sources reside in very high density
environments, possibly proto-clusters.
We also find that the fraction of powerful radio galaxies with extreme Faraday
rotation increases with redshift, in agreement with the change of their
average environment with cosmic epoch.

\end{itemize}

\begin{acknowledgements}
The National Radio Astronomy Observatory is operated by Associated Univ. under
contract with the National Science Foundation.
This research has made use of the NASA/IPAC Extragalactic Database (NED) 
which is operated by the Jet Propulsion Laboratory, California Institute 
of Technology, under contract with the National Aeronautics and Space 
Administration. 

\end{acknowledgements}

\begin{figure*}
\centerline{
\psfig{figure=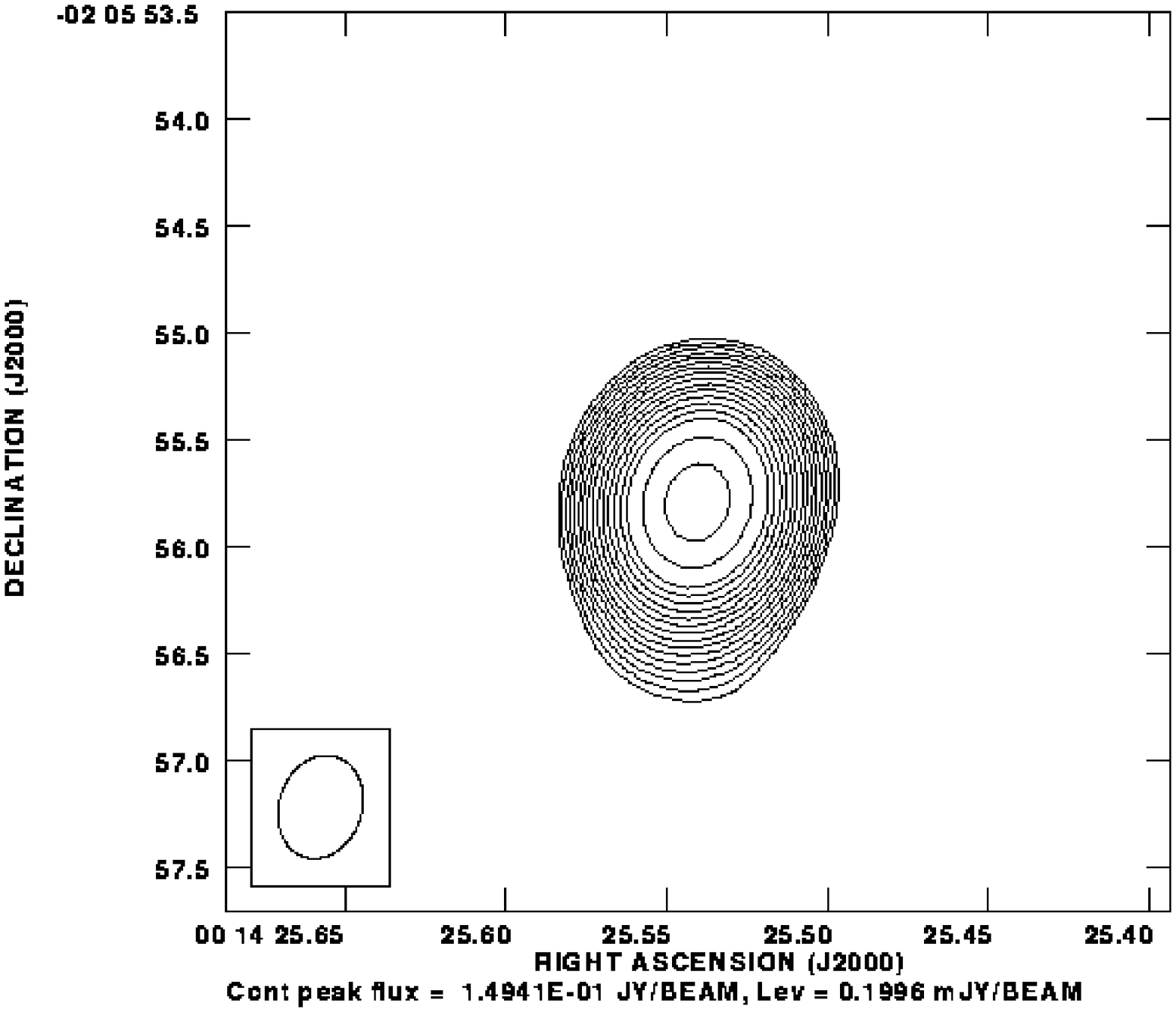,width=6.5cm,angle=-90} 
\psfig{figure=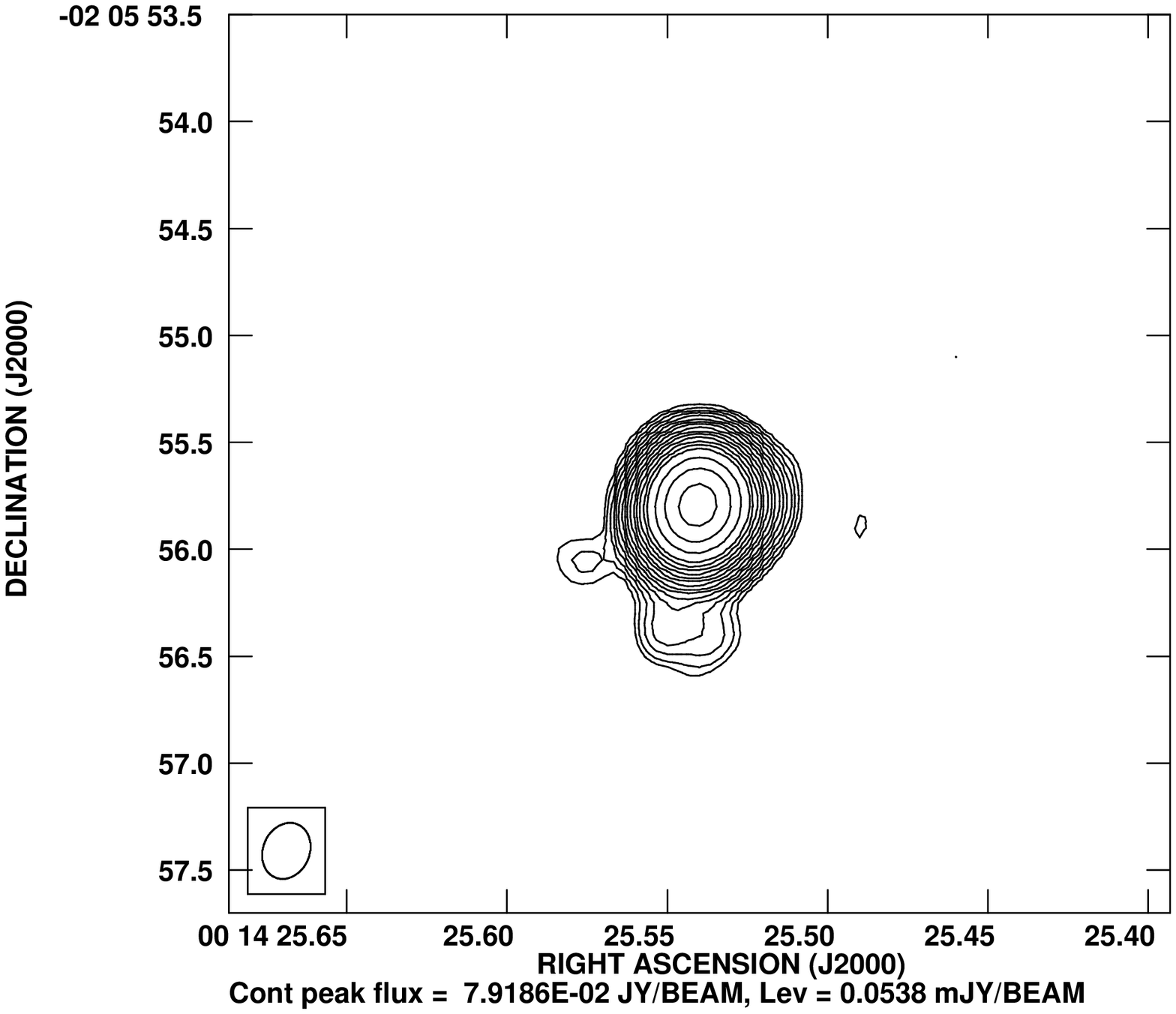,width=6.5cm,angle=-90}}\vskip-1cm\centerline{
\psfig{figure=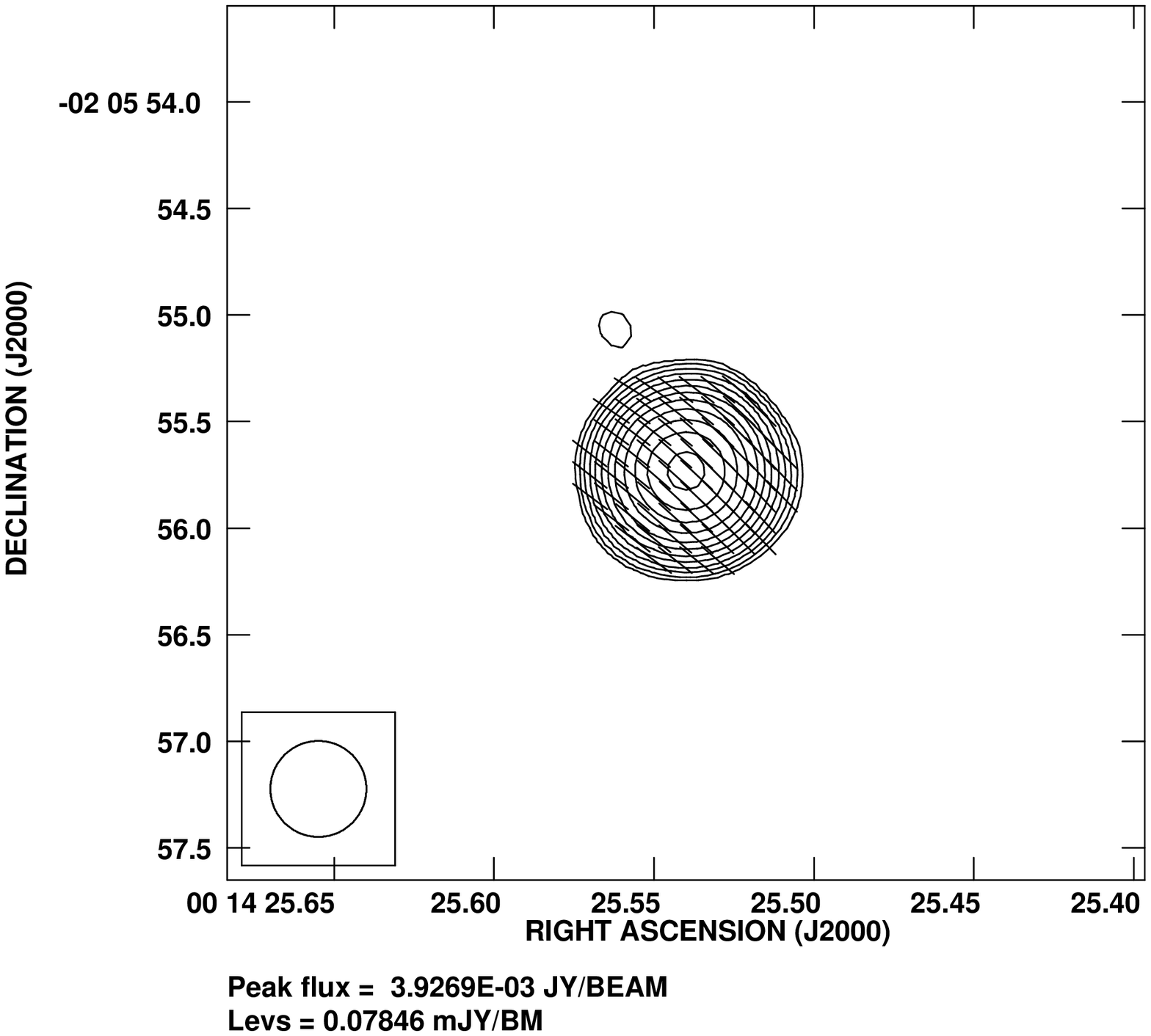,width=6.5cm}
\psfig{figure=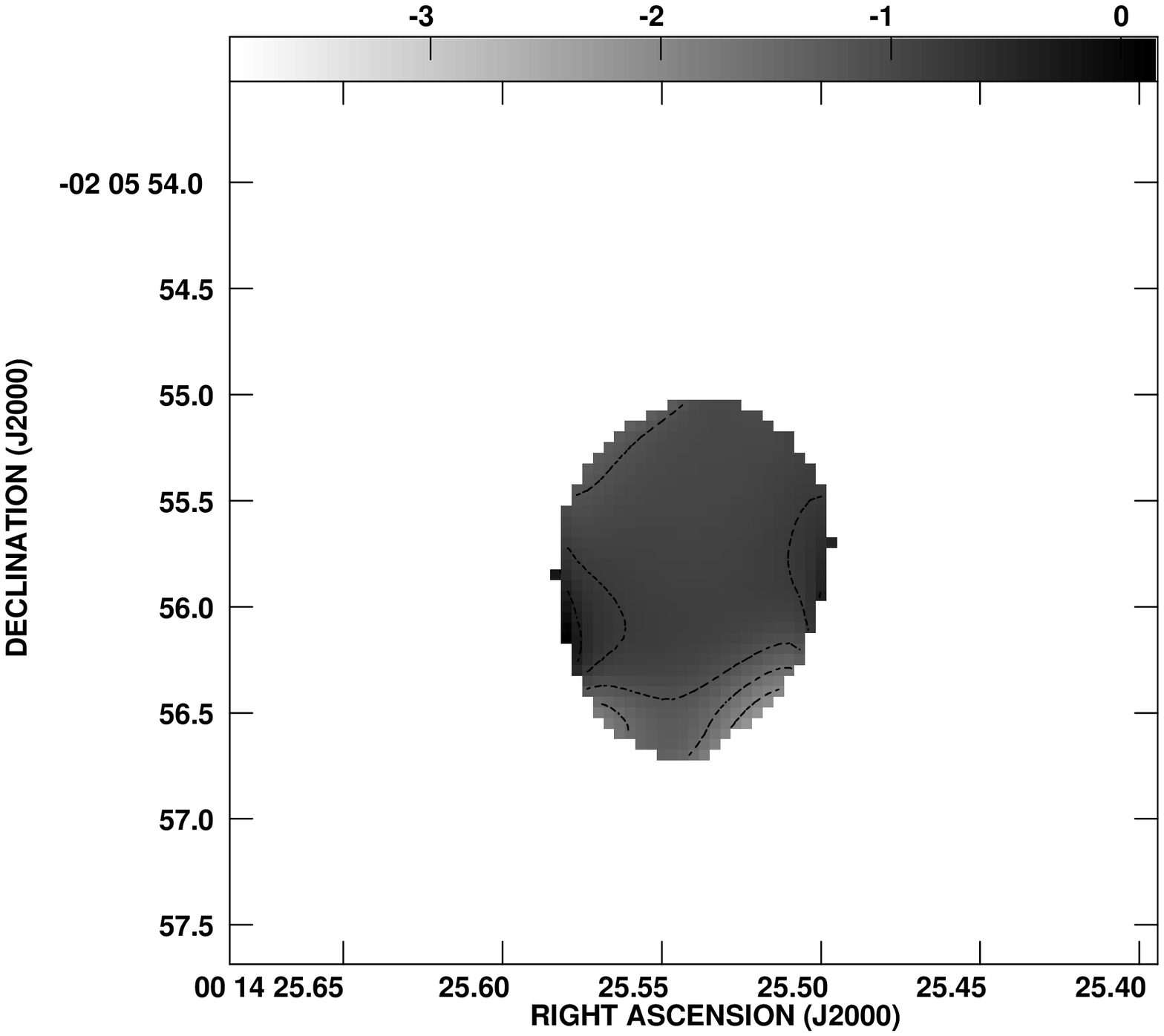,width=6.5cm}}
\caption{ Maps of the radio source 0011$-$023 at redshift z$=$2.08. 
Upper left is the 4.5 GHz total intensity map, having a FWHM of 0.5 arcsec, 
with the first contour at 0.2 mJy beam$^{-1}$ and a peak surface brightness 
of 149   mJy beam$^{-1}$. Upper right is a contour  map of total intensity at 
8.2 GHz, with the first contour at 0.054 mJy beam$^{-1}$ . 
Lower left is the map of polarized flux at 4.5 GHz, with the first contour
0.078 mJy beam$^{-1}$ and peak  surface brightness of 39 mJy beam$^{-1}$; 
superimposed are  vectors representing the strenght and direction of the
magnetic field. Lower right is a map of the spectral index calculated between 
the frequencies 4.5 GHz and 8.2 GHz.}  
\end{figure*}
\begin{figure*}
\centerline{
\psfig{figure=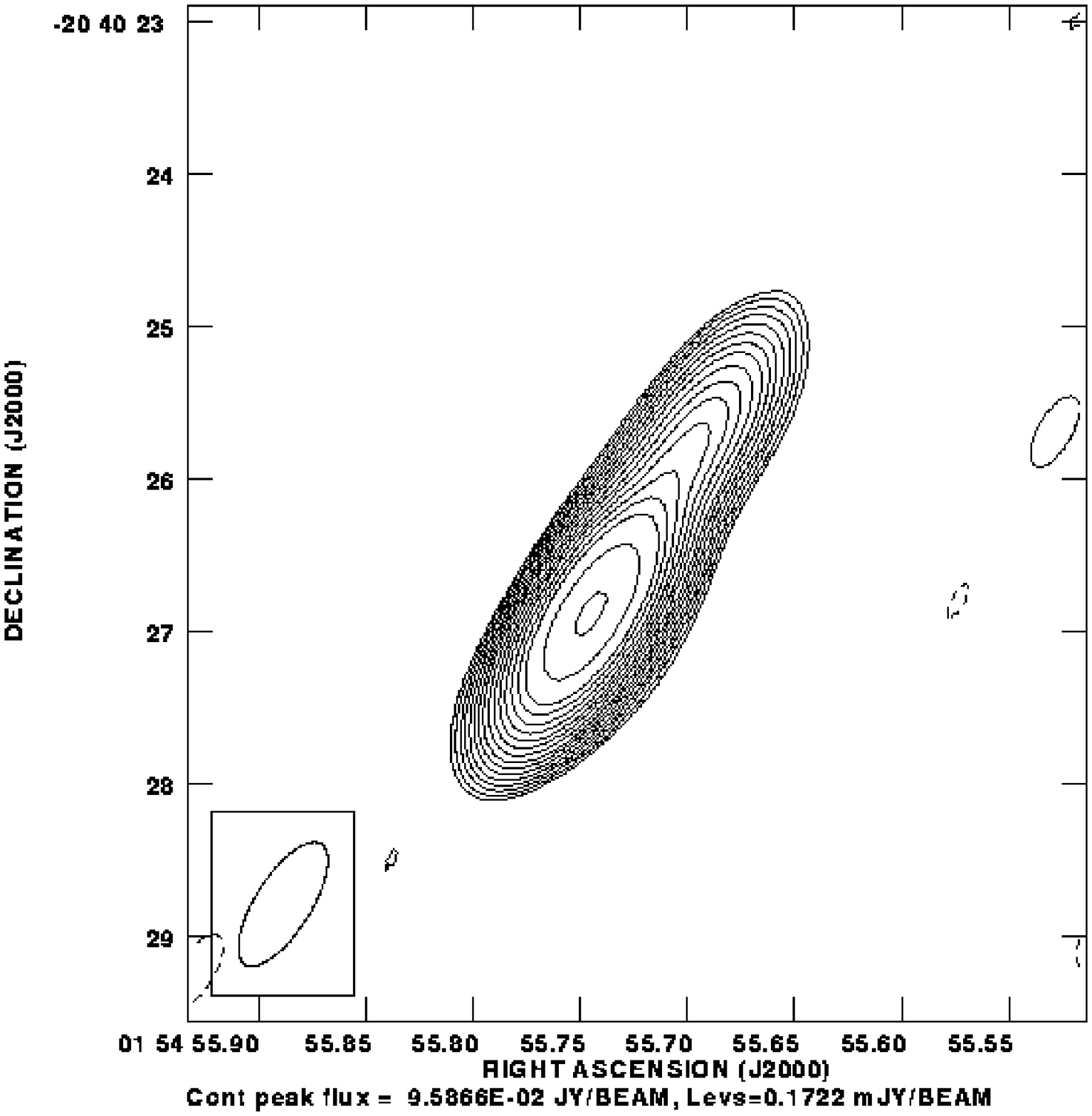,width=6.5cm}
\psfig{figure=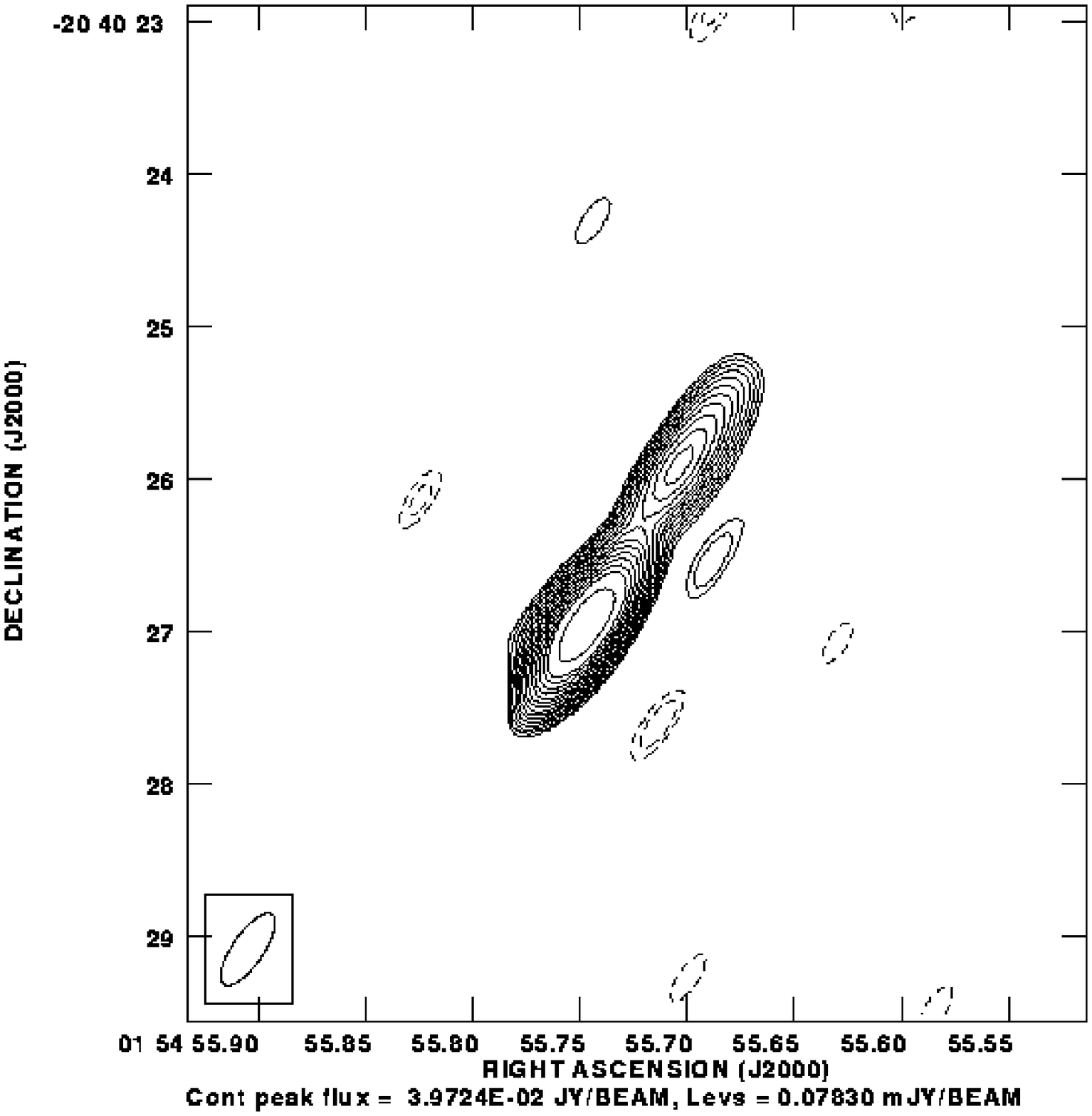,width=6.5cm}}
\vskip-1cm\centerline{
\psfig{figure=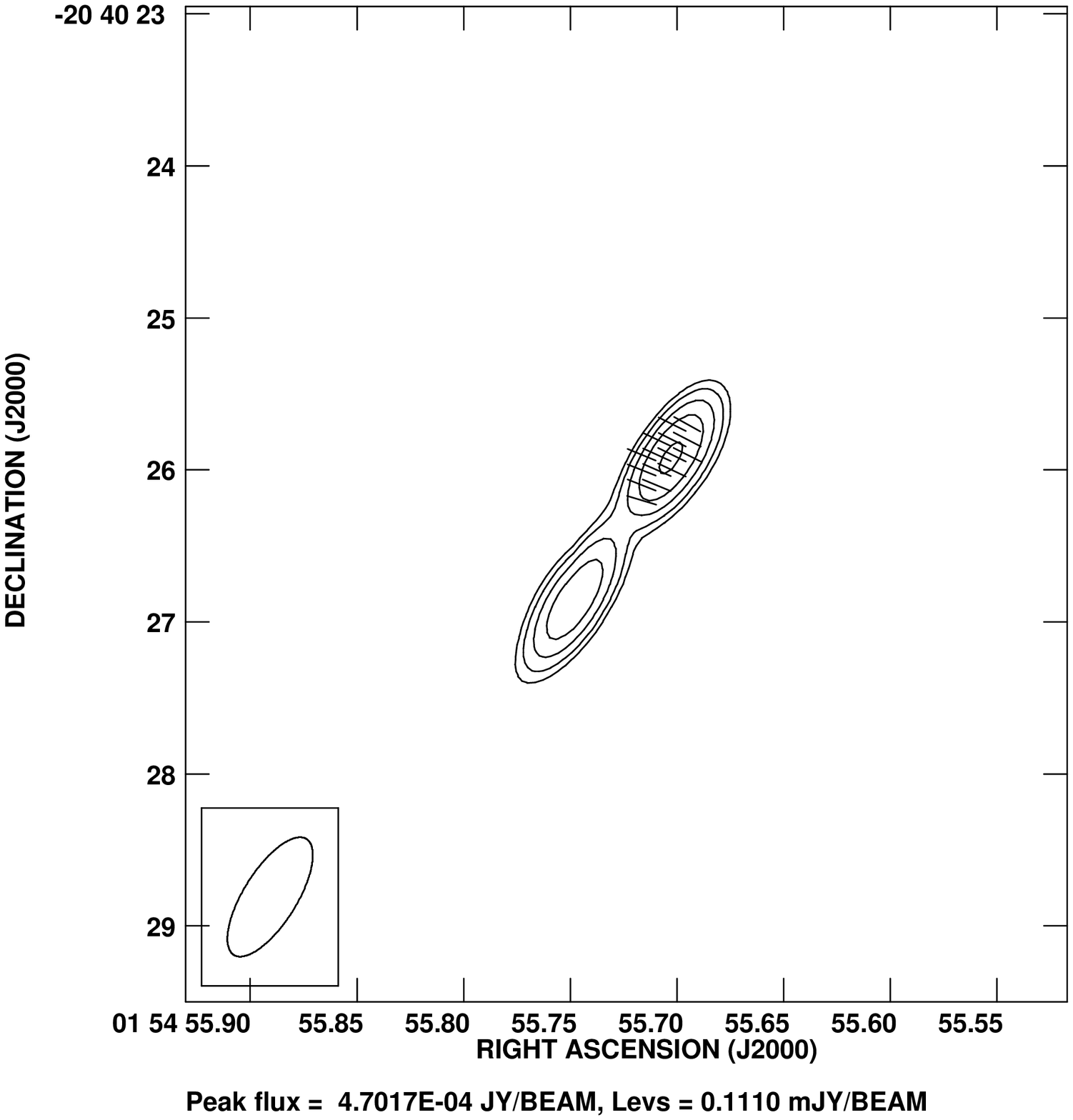,width=6.5cm}
\psfig{figure=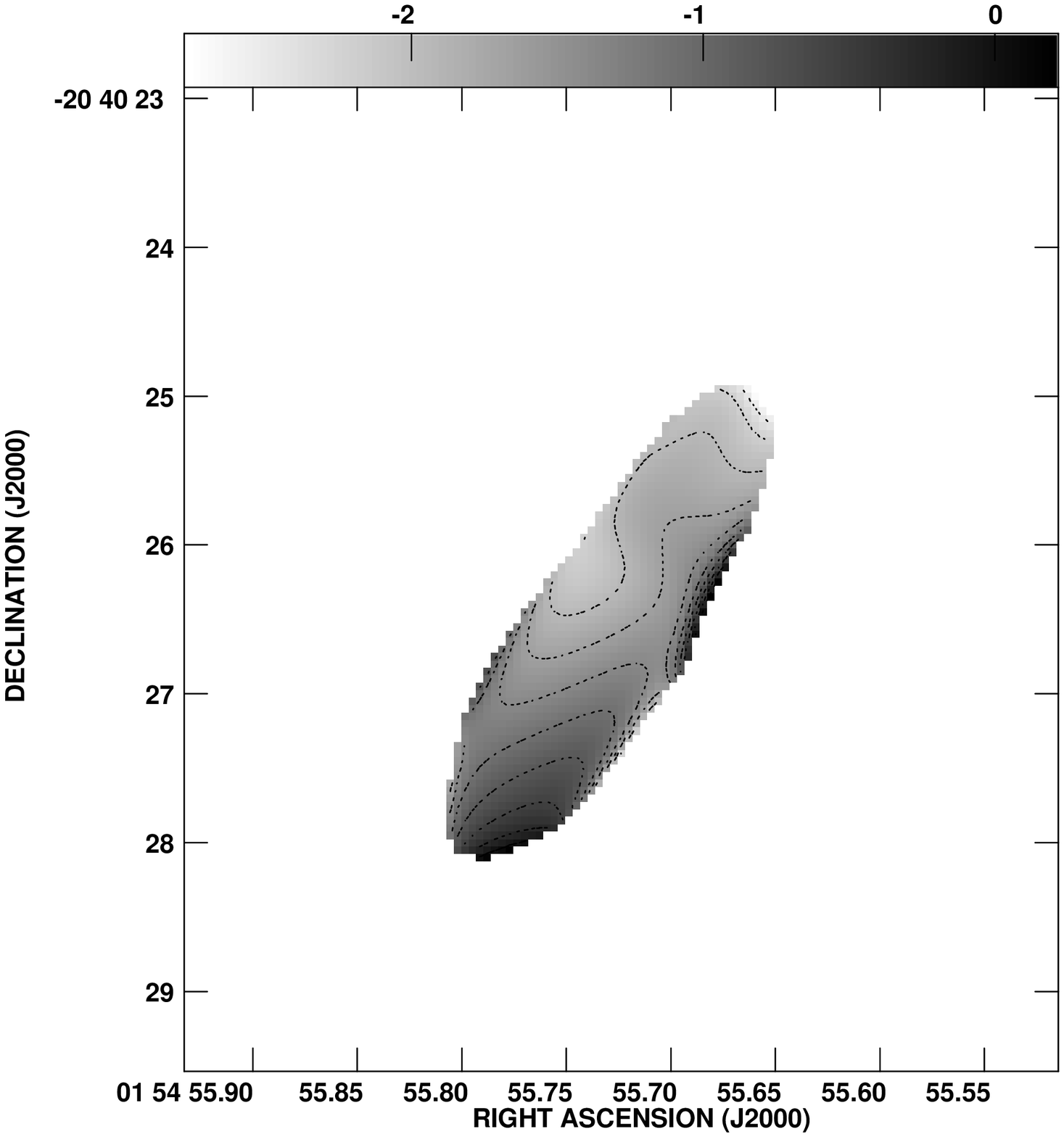,width=6.5cm}}
\caption{ Maps of the radio source 0152$-$209 at redshift z$=$1.89.
The sequence of figures is the same as in Fig. 6. The first contour level and
the  peak surface brightness  are respectively  0.172   mJy beam$^{-1}$ and 96
 mJy beam$^{-1}$ for
the 4.5 GHz map; 0.078  mJy beam$^{-1}$ and 40  mJy beam$^{-1}$ for the 8.2
GHz map; 0.11  mJy beam$^{-1}$ and 4.7  mJy beam$^{-1}$ for the 4.7 GHz
polarized intensity map.}
\end{figure*}
\clearpage
\begin{figure*}
\centerline{
\psfig{figure=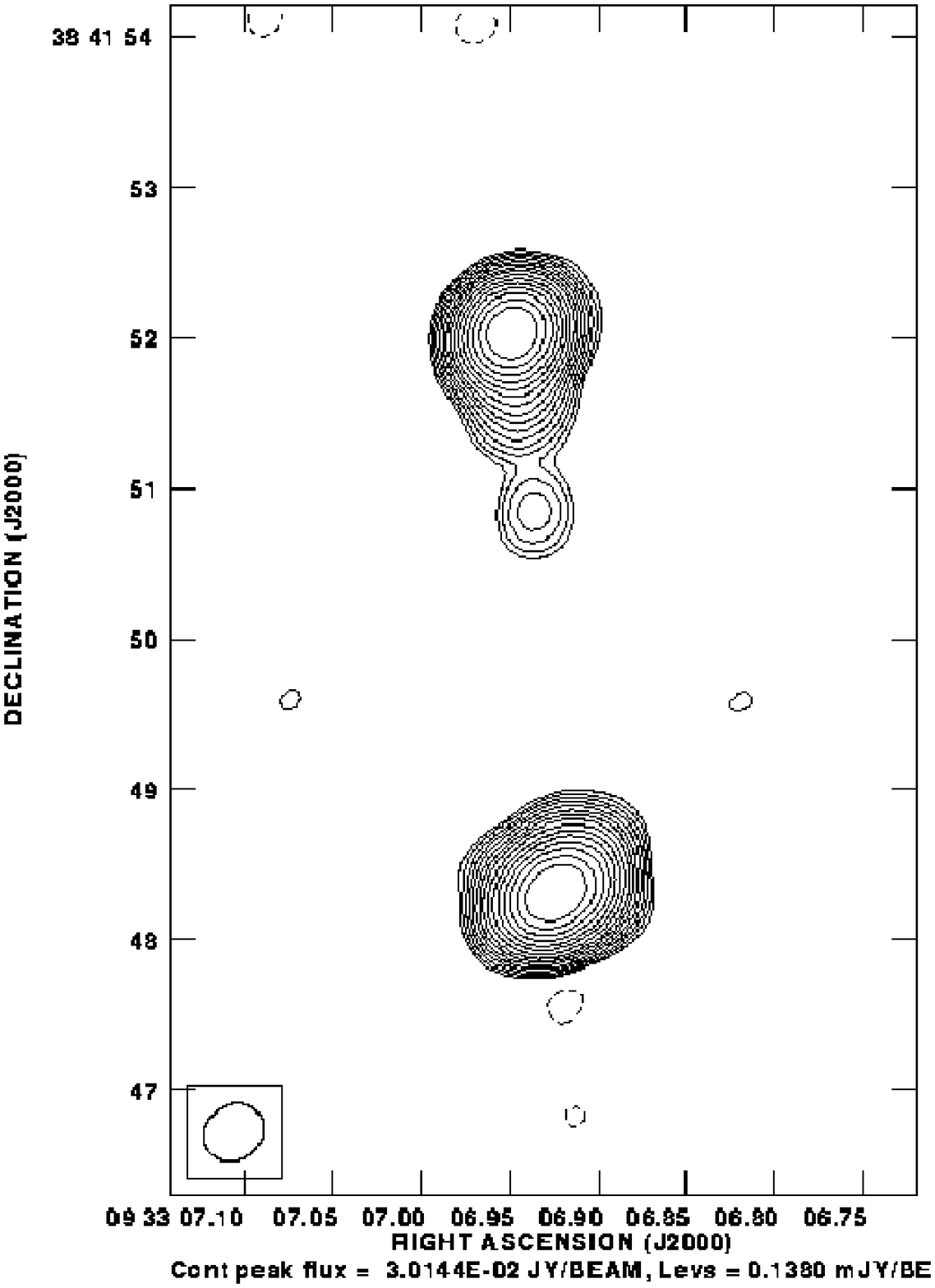,width=7.cm}
\psfig{figure=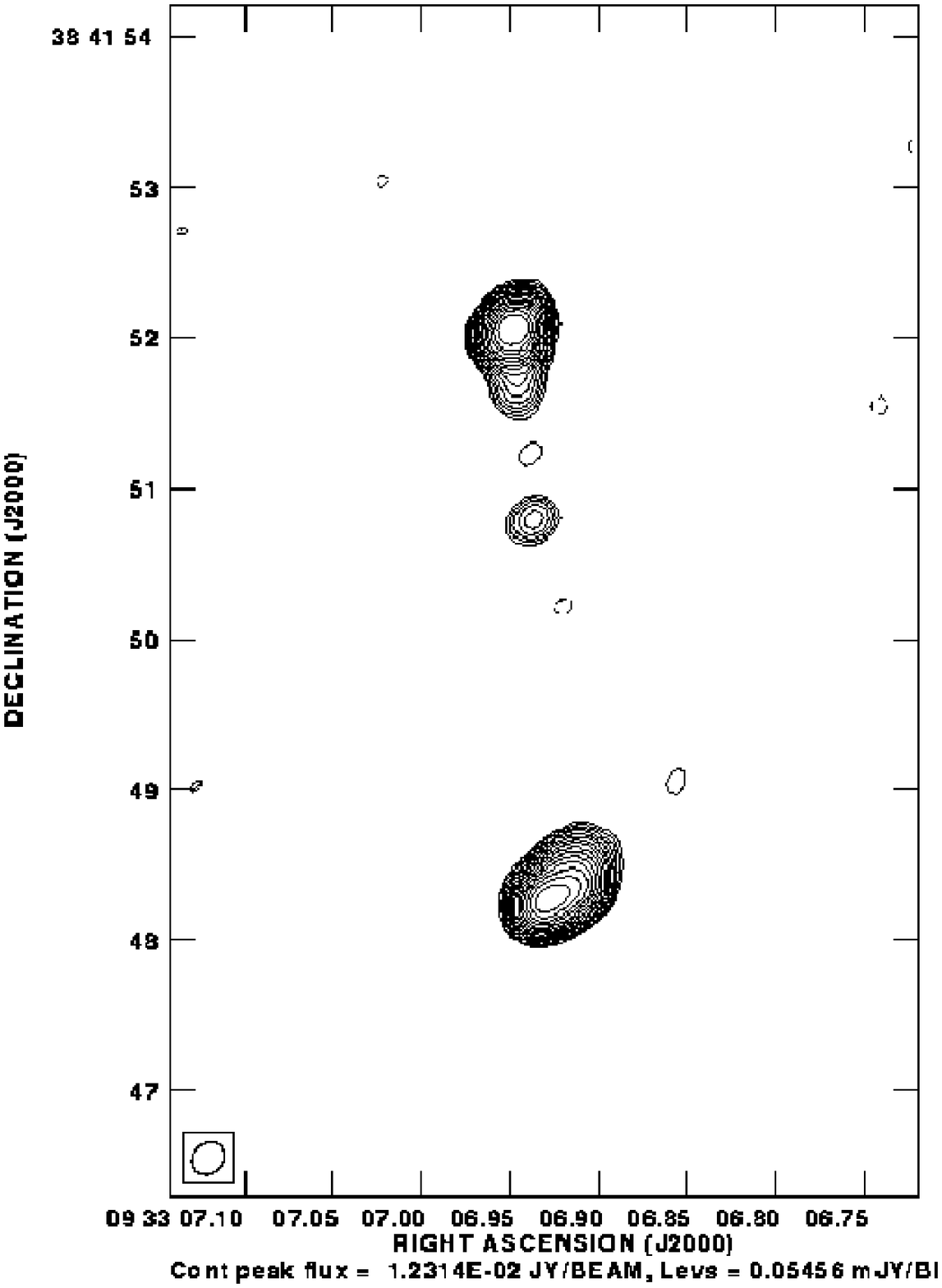,width=7.cm}}\vskip-1cm\centerline{
\psfig{figure=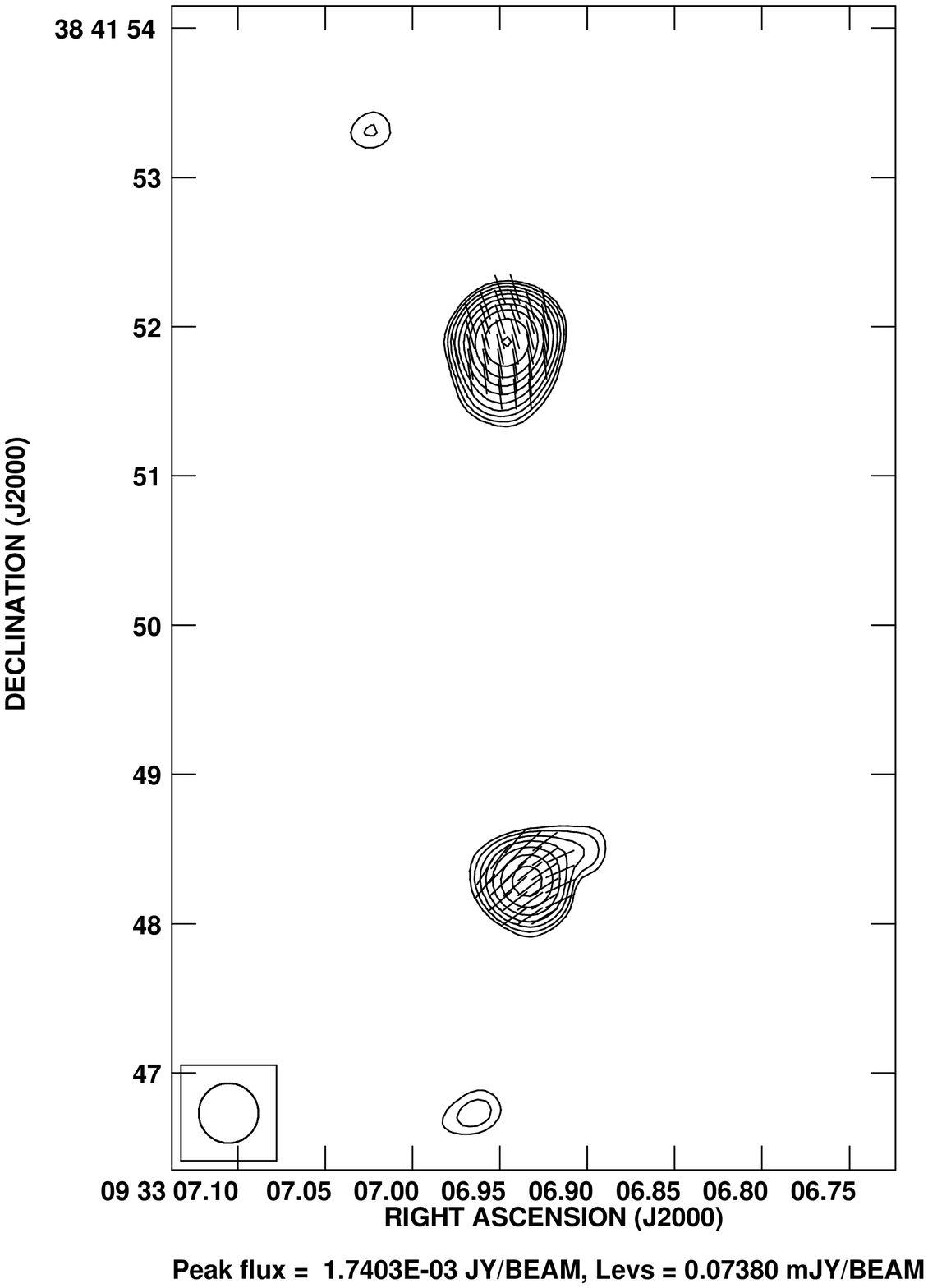,width=7.cm}
\psfig{figure=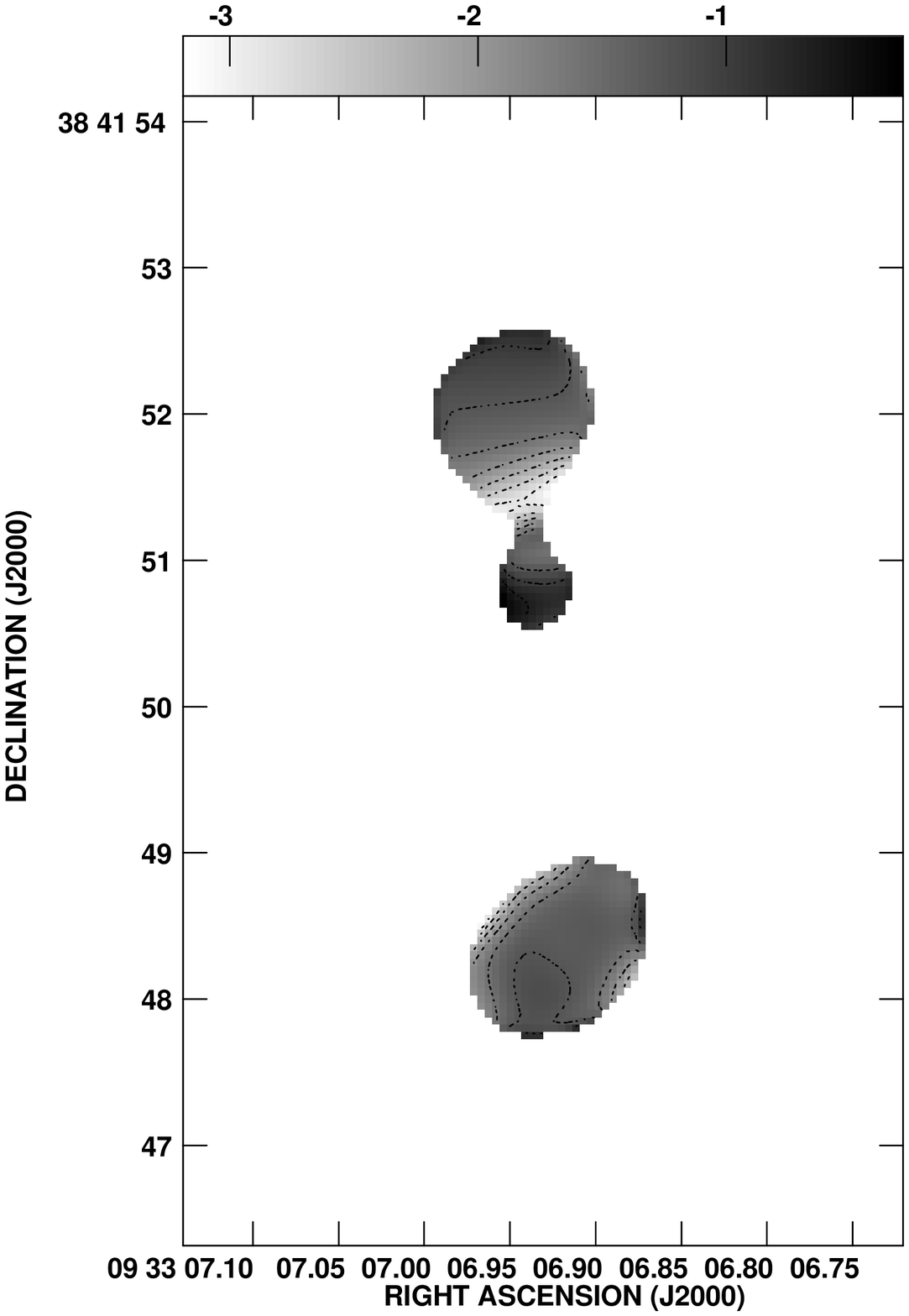,width=7.cm}} 
\caption{Maps of the radio source 0930+389 at redshift z$=$2.395.
The sequence of figures is the same as in Fig. 6. The first contour level and
the  peak surface brightness  are respectively  0.138  mJy beam$^{-1}$ and 30
 mJy beam$^{-1}$ for
the 4.5 GHz map; 0.055  mJy beam$^{-1}$ and 12  mJy beam$^{-1}$ for the 8.2
GHz map; 0.074  mJy beam$^{-1}$ and 1.7  mJy beam$^{-1}$ for the 4.7 GHz
polarized intensity map.}
\end{figure*}

\begin{figure*}
\centerline{
\psfig{figure=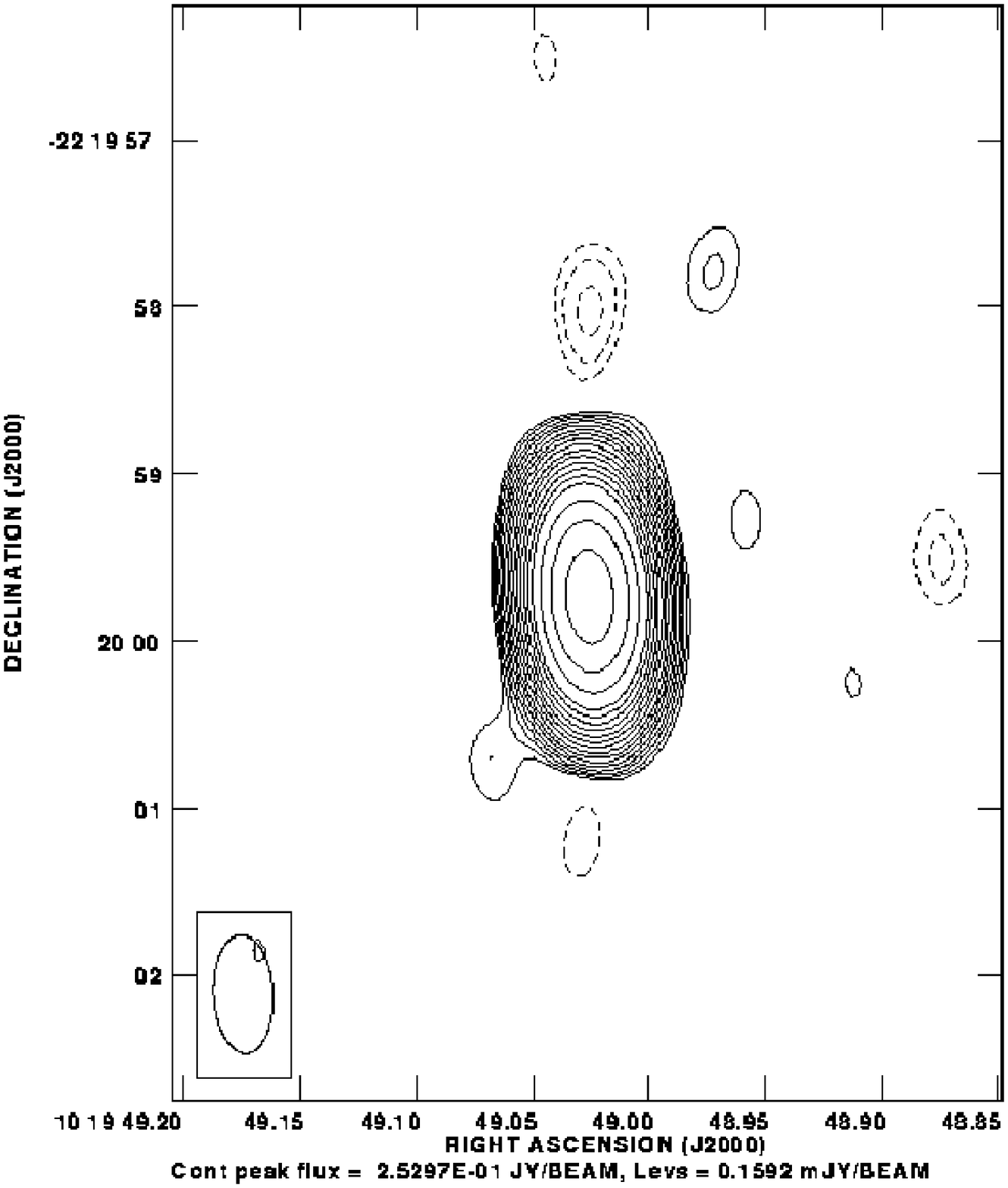,width=5.5cm}
\psfig{figure=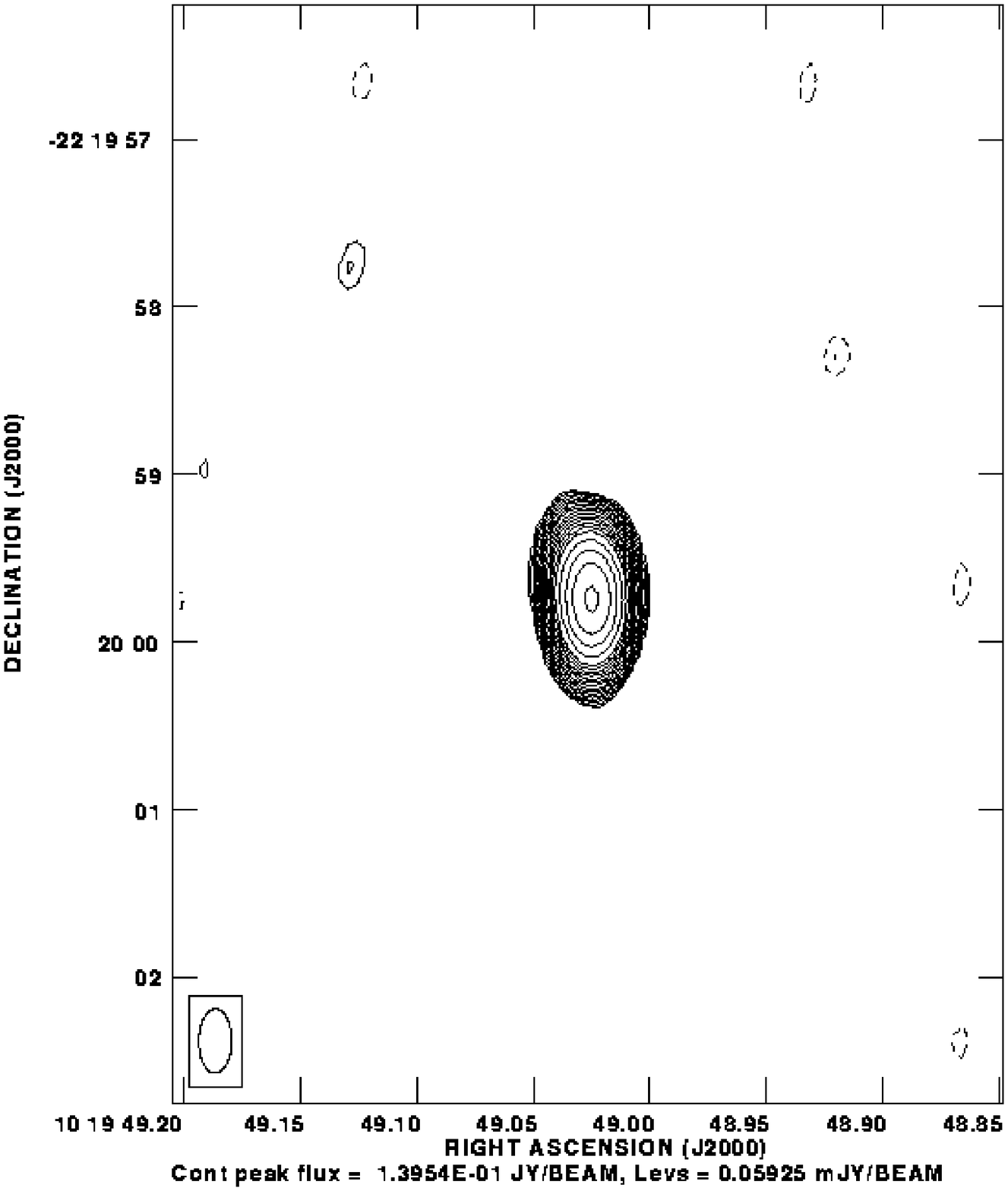,width=5.5cm}}\vskip-0.5cm\centerline{
\psfig{figure=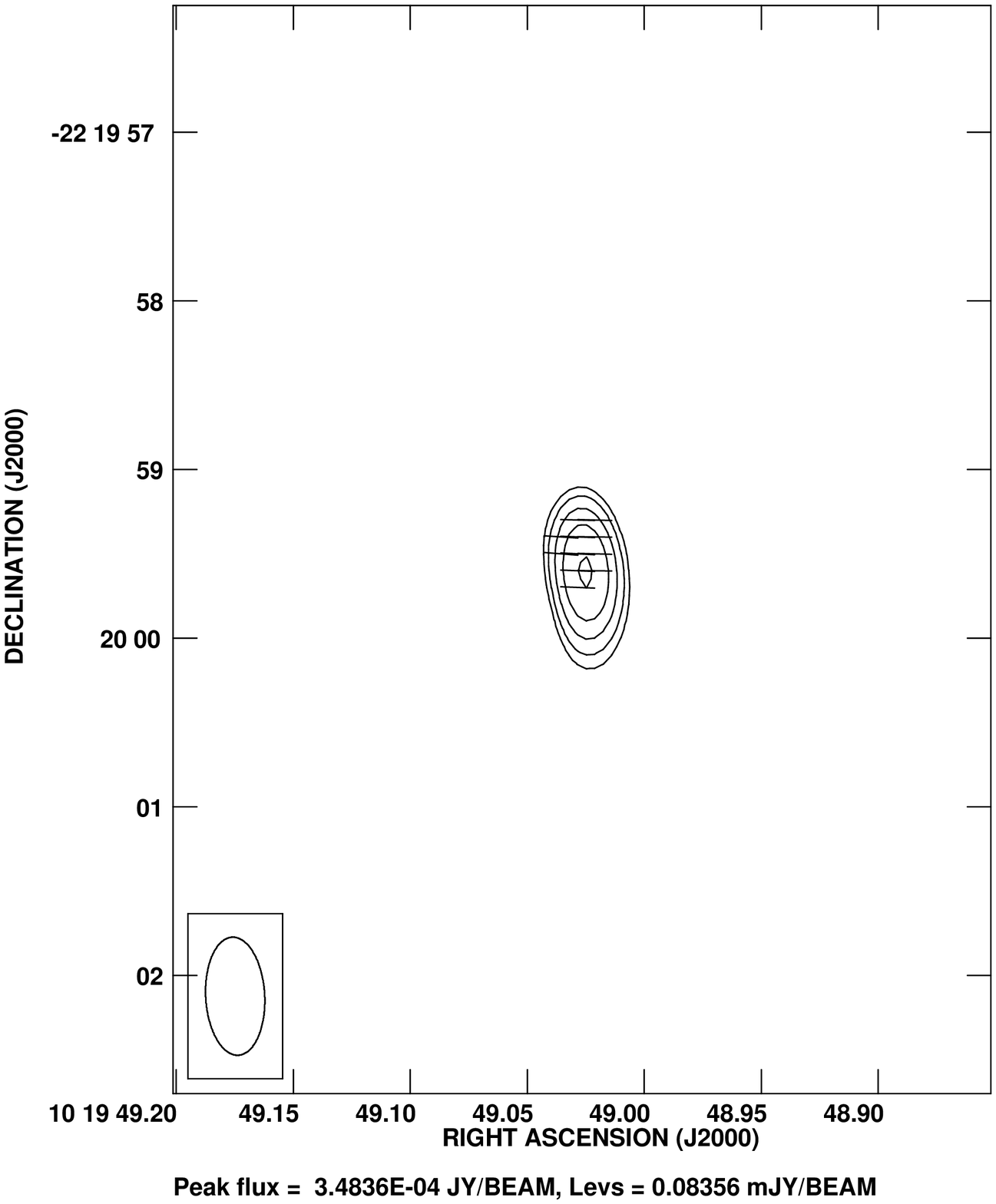,width=5.5cm}
\psfig{figure=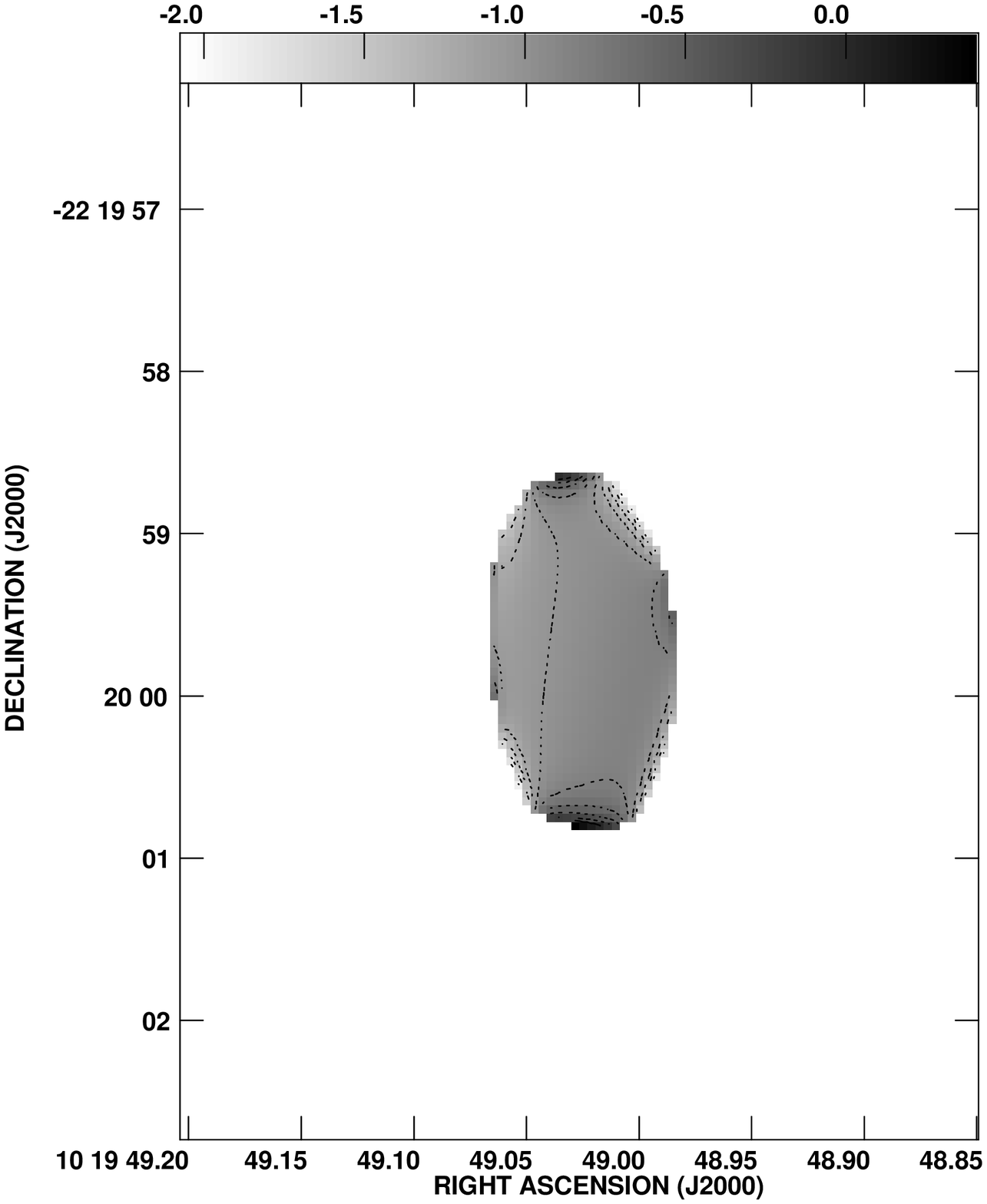,width=5.5cm}}
\caption{Maps of the radio source 1017$-$220 at redshift z$=$1.77.
The sequence of figures is the same as in Fig. 6. The first contour level and
the  peak surface brightness  are respectively  0.159  mJy beam$^{-1}$ and 253
 mJy beam$^{-1}$ for
the 4.5 GHz map; 0.059  mJy beam$^{-1}$ and 139 mJy beam$^{-1}$ for the 8.2
GHz map; 0.084  mJy beam$^{-1}$ and 0.35  mJy beam$^{-1}$ for the 4.7 GHz
polarized intensity map.}
\end{figure*}
\begin{figure*}
\centerline{
\psfig{figure=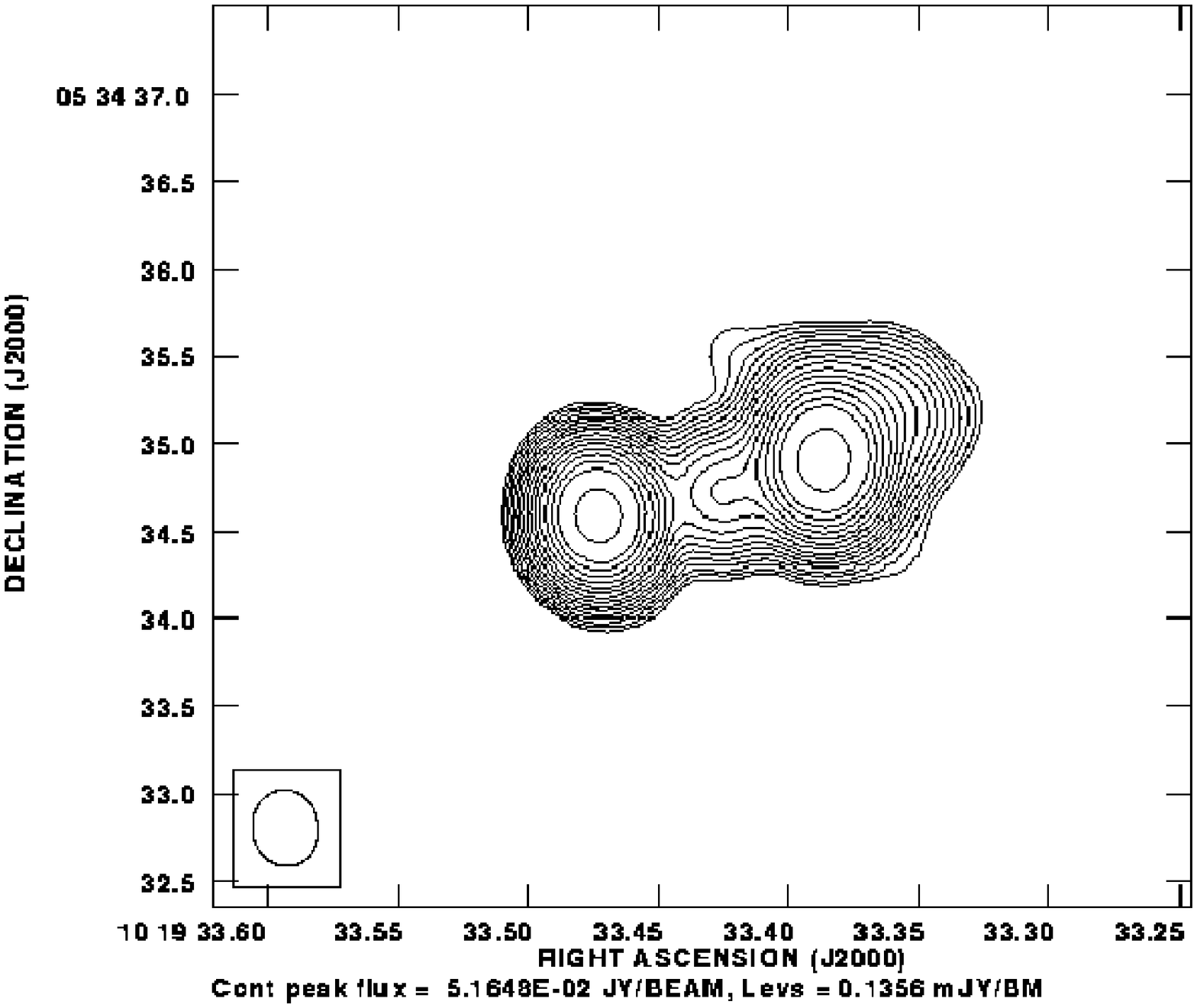,width=6.cm,angle=-90}
\psfig{figure=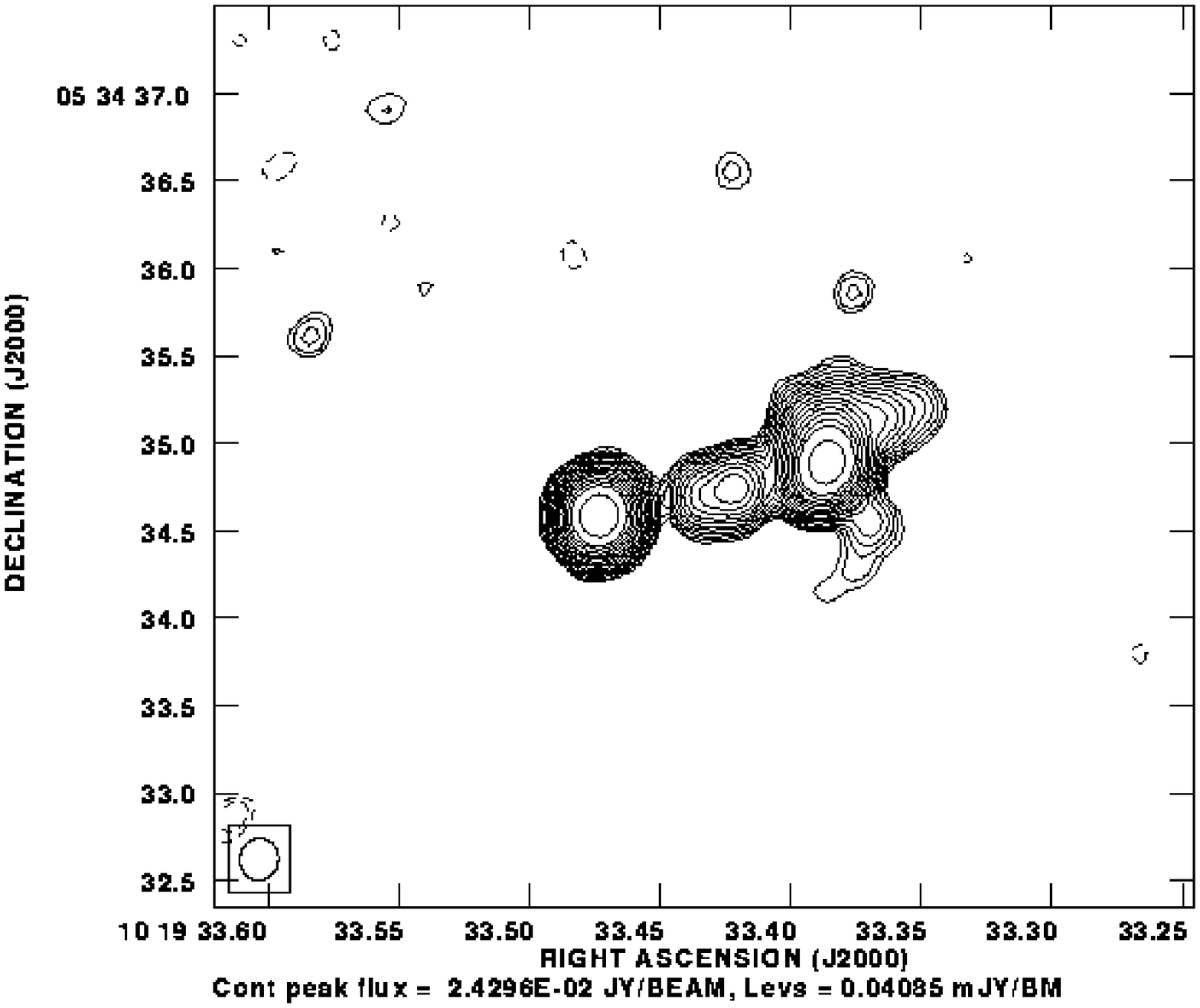,width=6.cm,angle=-90}}
\vskip-0.5cm\centerline{
\psfig{figure=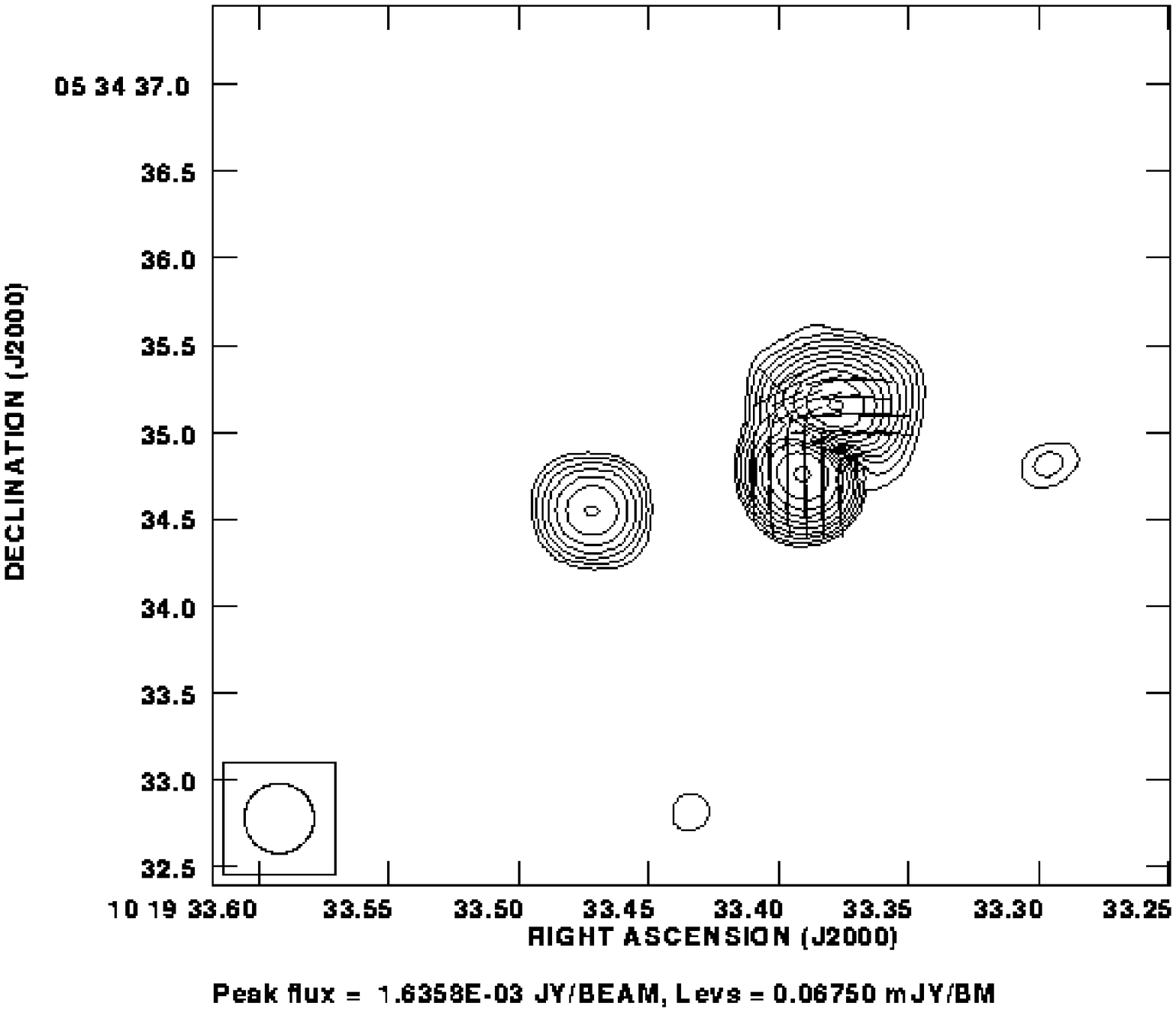,width=7.cm}
\psfig{figure=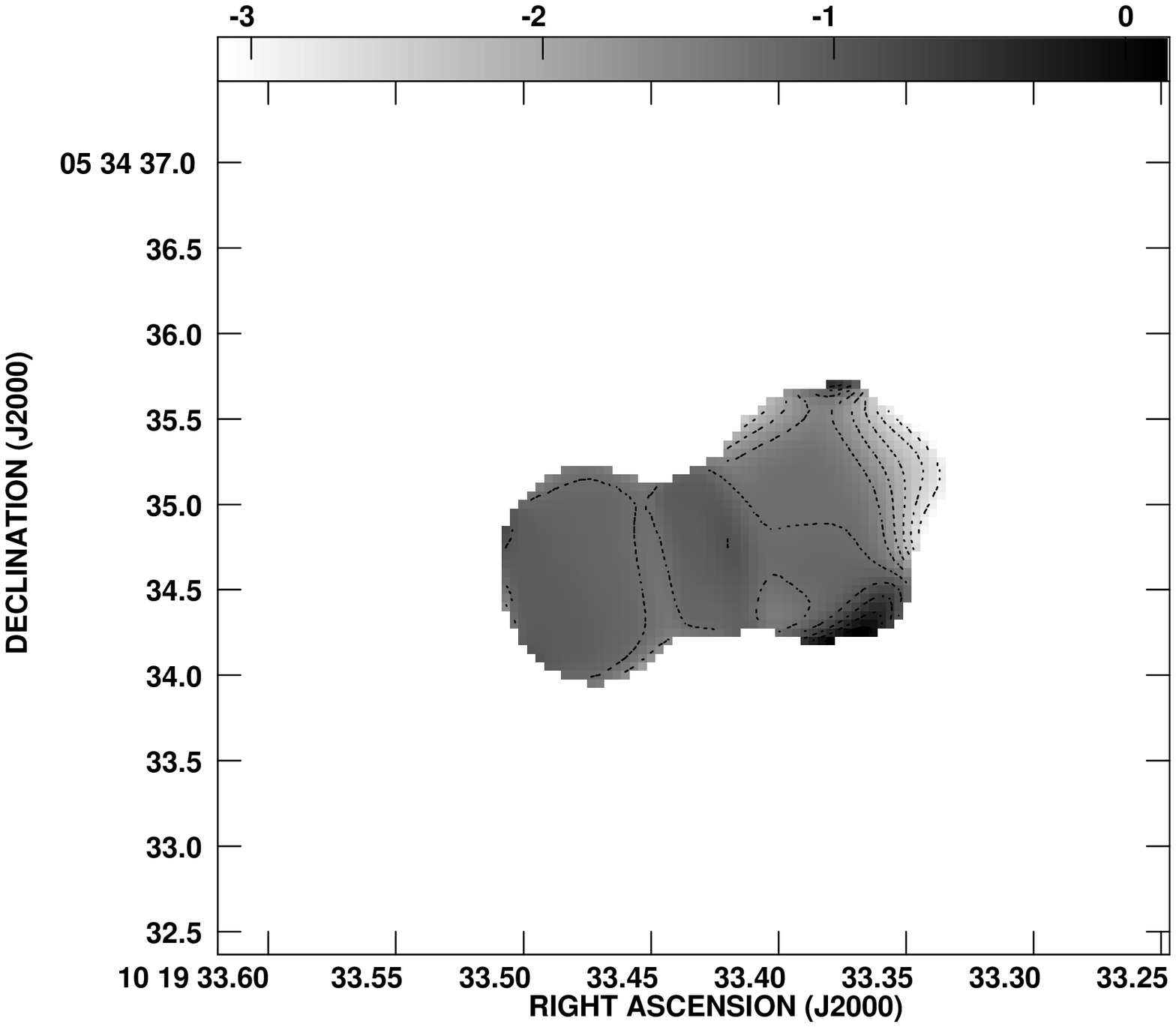,width=7.cm}}
\caption{Maps of the radio source 1019+0534 at redshift z$=$2.765.
The sequence of figures is the same as in Fig. 6. The first contour level and
the  peak surface brightness  are respectively  0.136  mJy beam$^{-1}$ and 52
 mJy beam$^{-1}$ for
the 4.5 GHz map; 0.041  mJy beam$^{-1}$ and 24 mJy beam$^{-1}$ for the 8.2
GHz map; 0.067  mJy beam$^{-1}$ and 1.6  mJy beam$^{-1}$ for the 4.7 GHz
polarized intensity map.}
\end{figure*}
\clearpage
\begin{figure*}
\centerline{
\psfig{figure=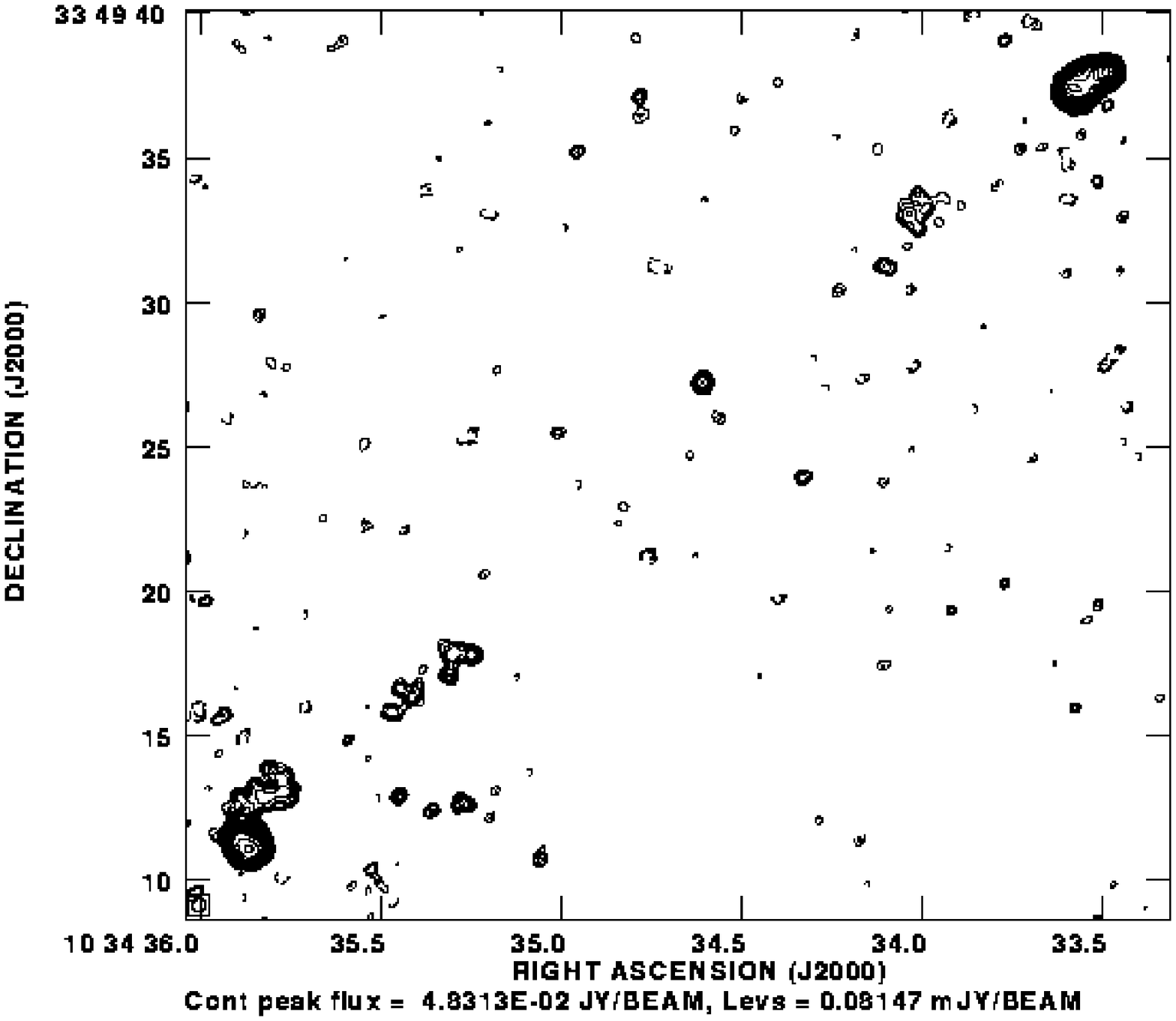,width=8cm,angle=-90}
\psfig{figure=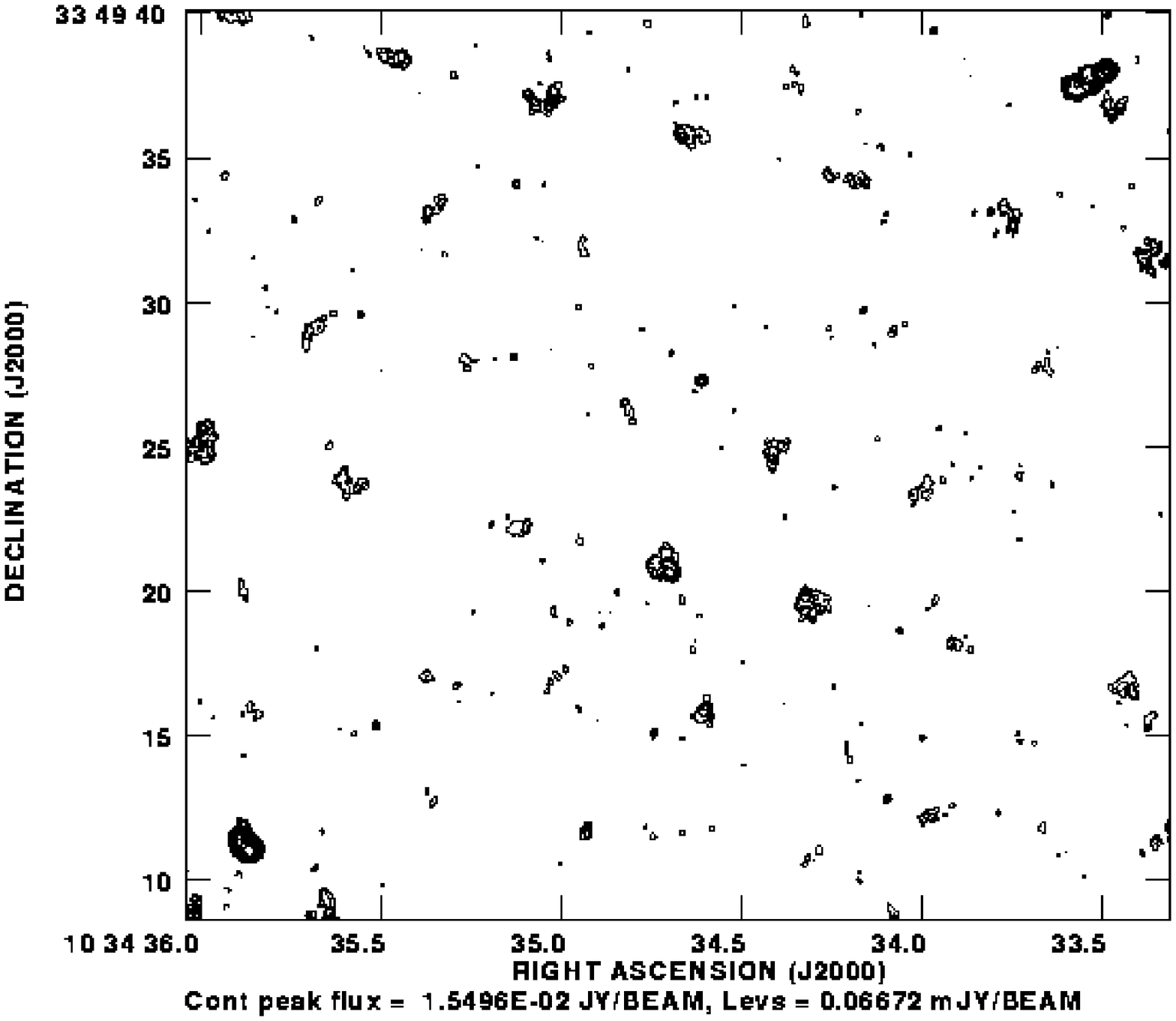,width=8cm,angle=-90}}\vskip-0.1cm\centerline{
\psfig{figure=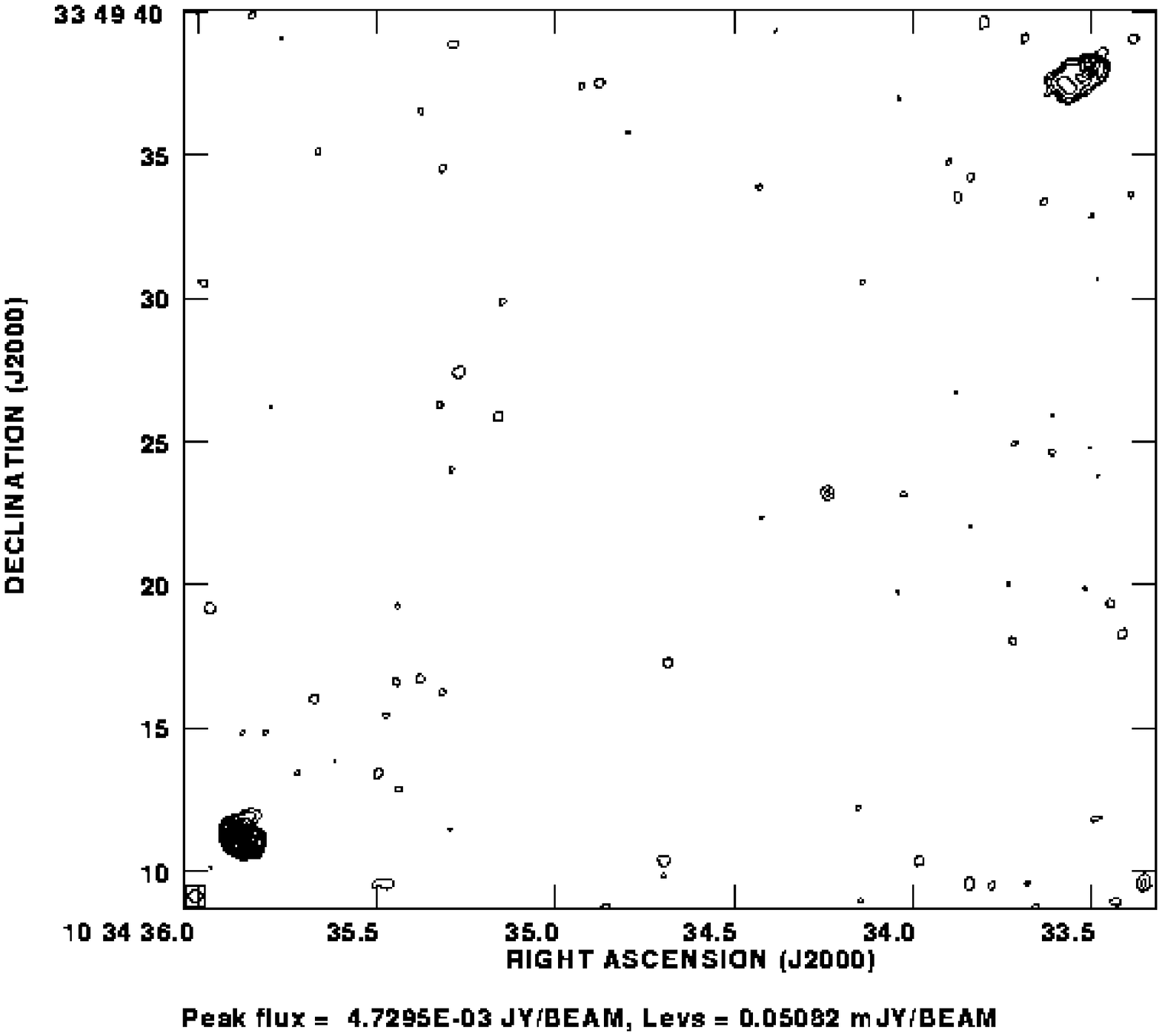,width=8cm,angle=-90}
\psfig{figure=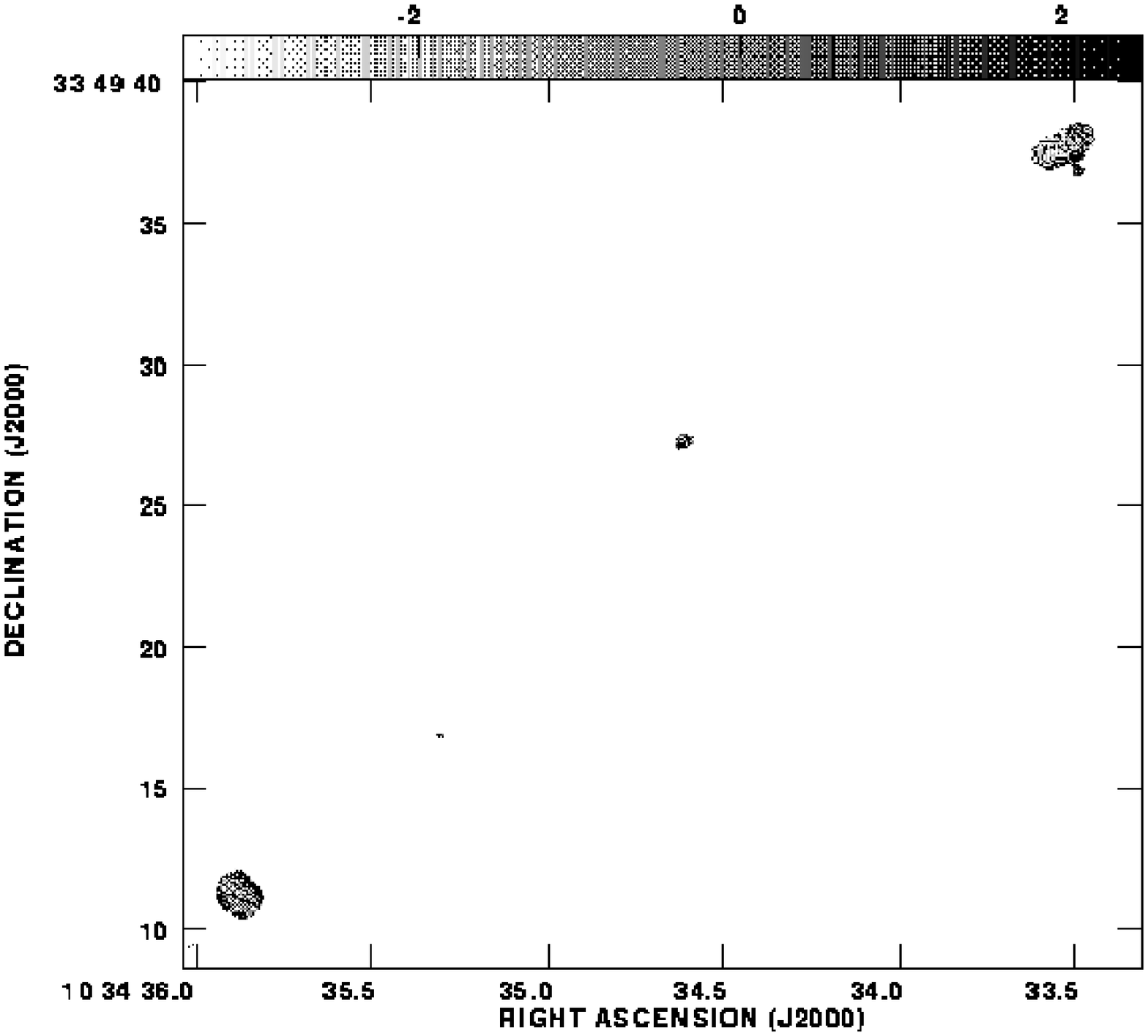,width=8cm,angle=-90}}
\caption{Maps of the radio source 1031+34 at redshift z$=$2.10
The sequence of figures is the same as in Fig. 6. The first contour level and
the  peak surface brightness  are respectively  0.081  mJy beam$^{-1}$ and 48
 mJy beam$^{-1}$ for
the 4.5 GHz map; 0.067  mJy beam$^{-1}$ and 15 mJy beam$^{-1}$ for the 8.2
GHz map; 0.051  mJy beam$^{-1}$ and 4.7  mJy beam$^{-1}$ for the 4.7 GHz
polarized intensity map.}
\end{figure*}
\clearpage
\begin{figure*}
\centerline{
\psfig{figure=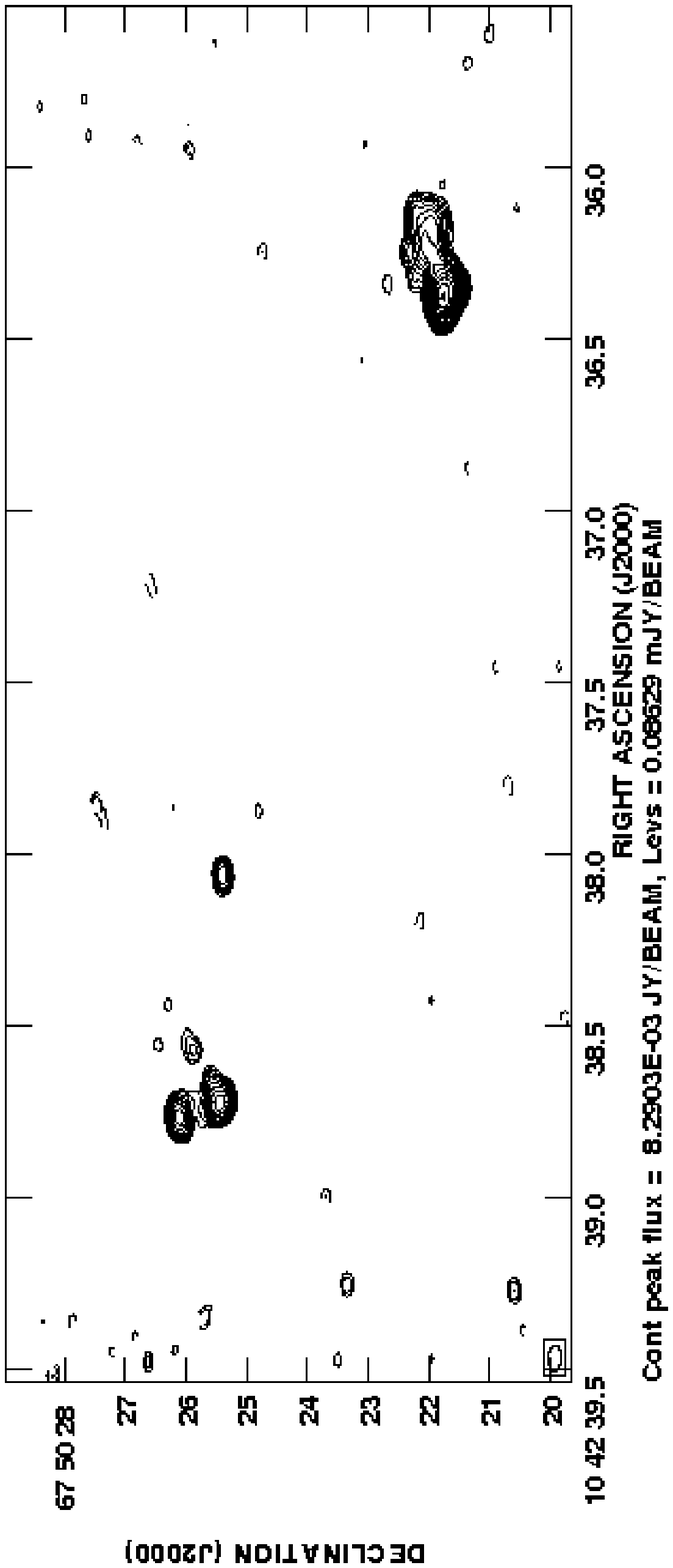,width=5.3cm}
\psfig{figure=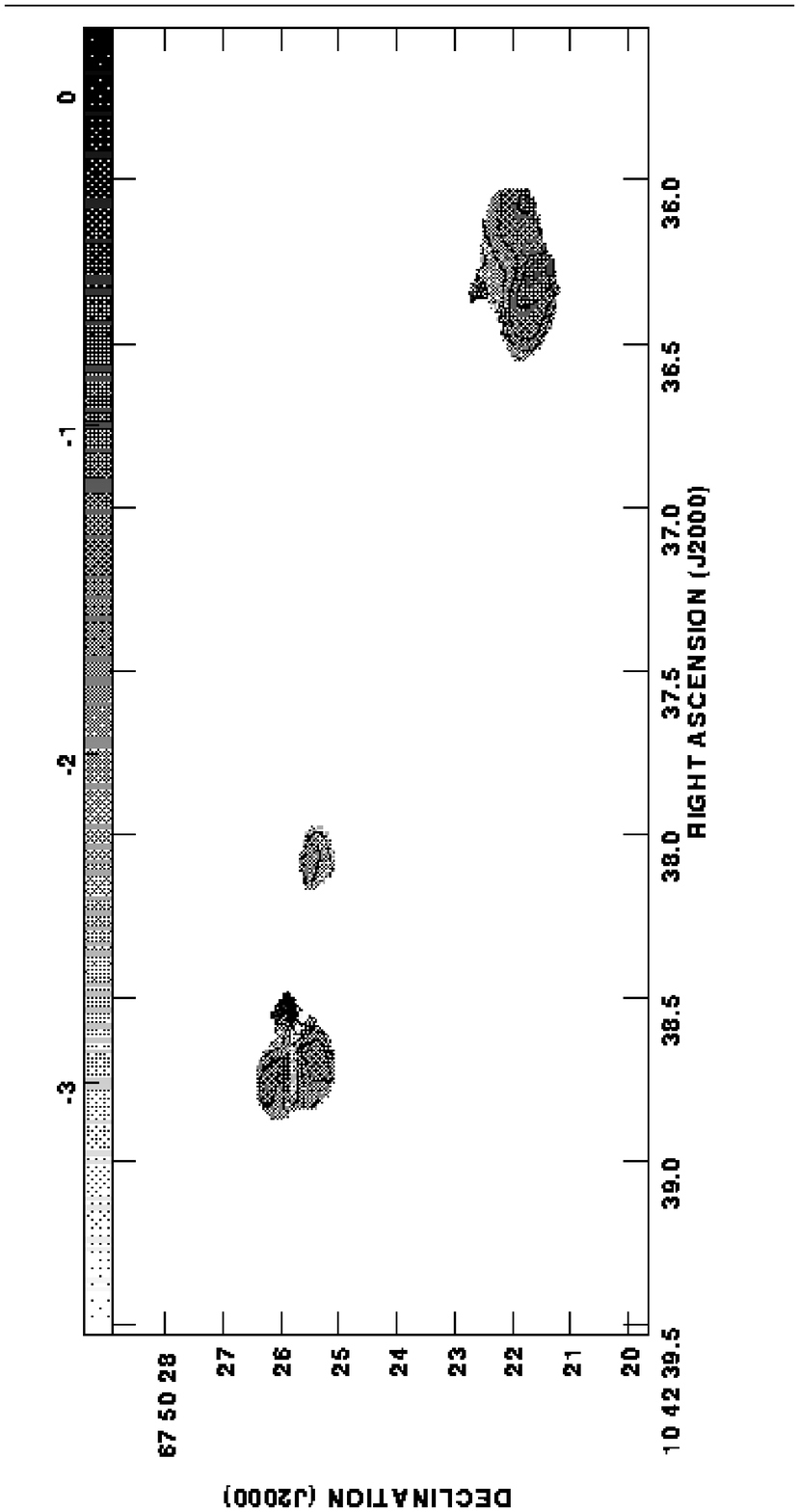,width=5.5cm}}
\centerline{
\psfig{figure=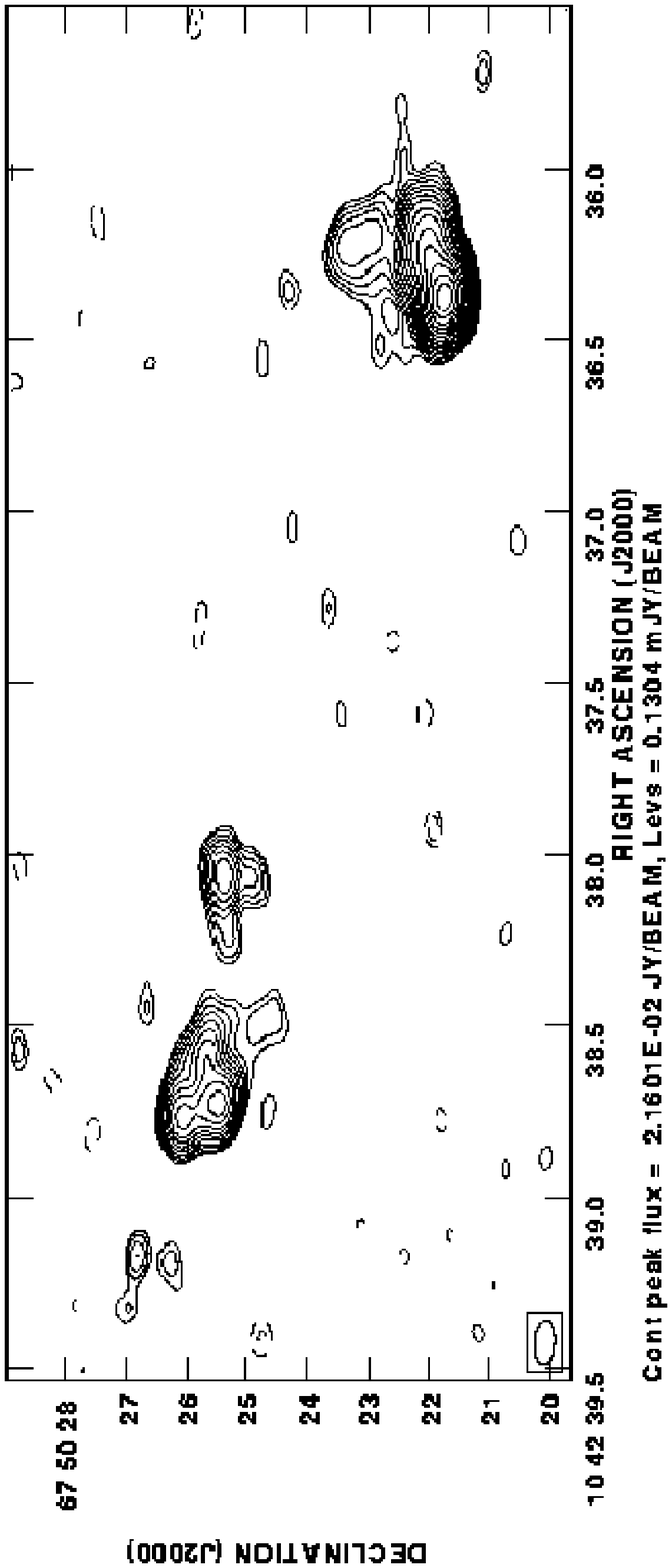,width=5.5cm}
\psfig{figure=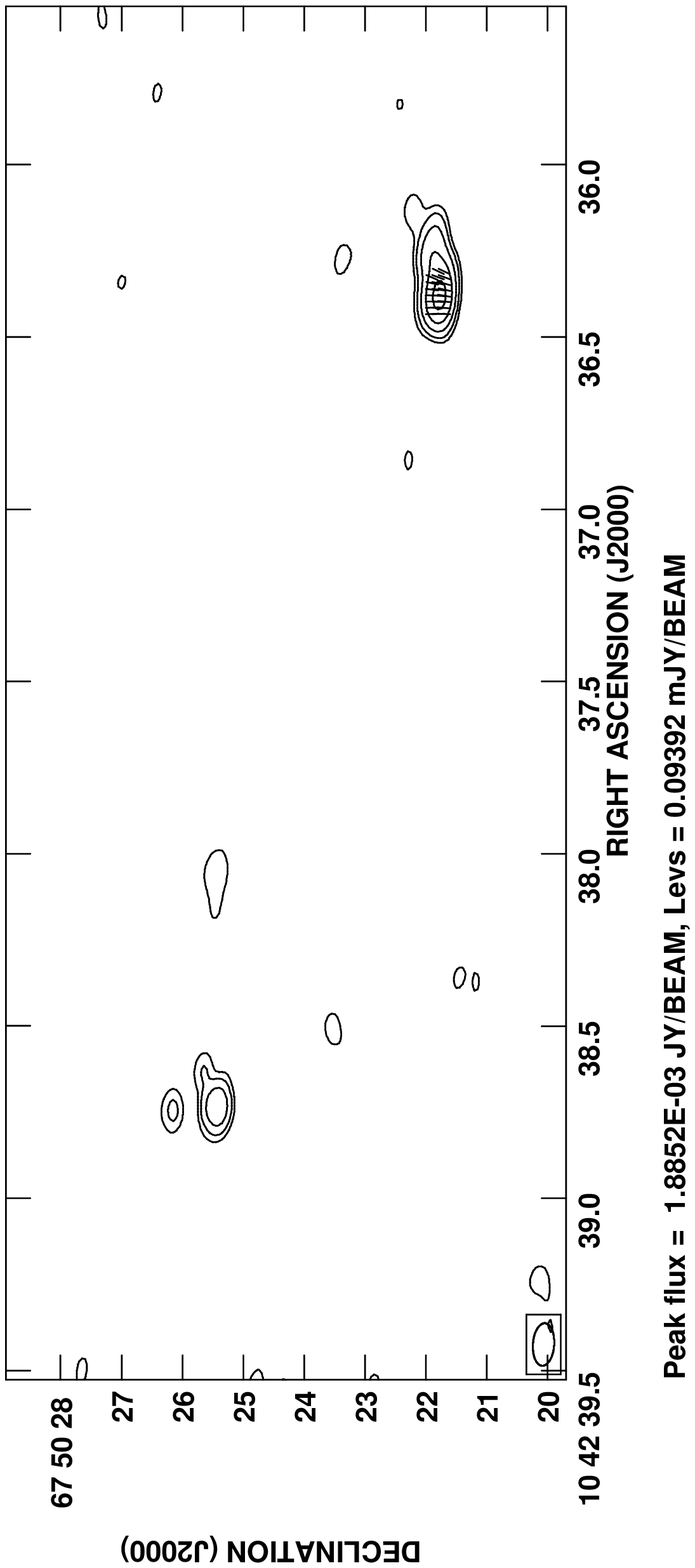,width=6cm}}
\caption{Maps of the radio source 1039+681 at redshift z$=$2.53.
The sequence of figures is the same as in Fig. 6. The first contour level and
the  peak surface brightness  are respectively  0.134  mJy beam$^{-1}$ and 22
 mJy beam$^{-1}$ for
the 4.5 GHz map; 0.086  mJy beam$^{-1}$ and 8.3 mJy beam$^{-1}$ for the 8.2
GHz map; 0.094  mJy beam$^{-1}$ and 1.9  mJy beam$^{-1}$ for the 4.7 GHz
polarized intensity map.}
\end{figure*}

\begin{figure*}
\centerline{
\psfig{figure=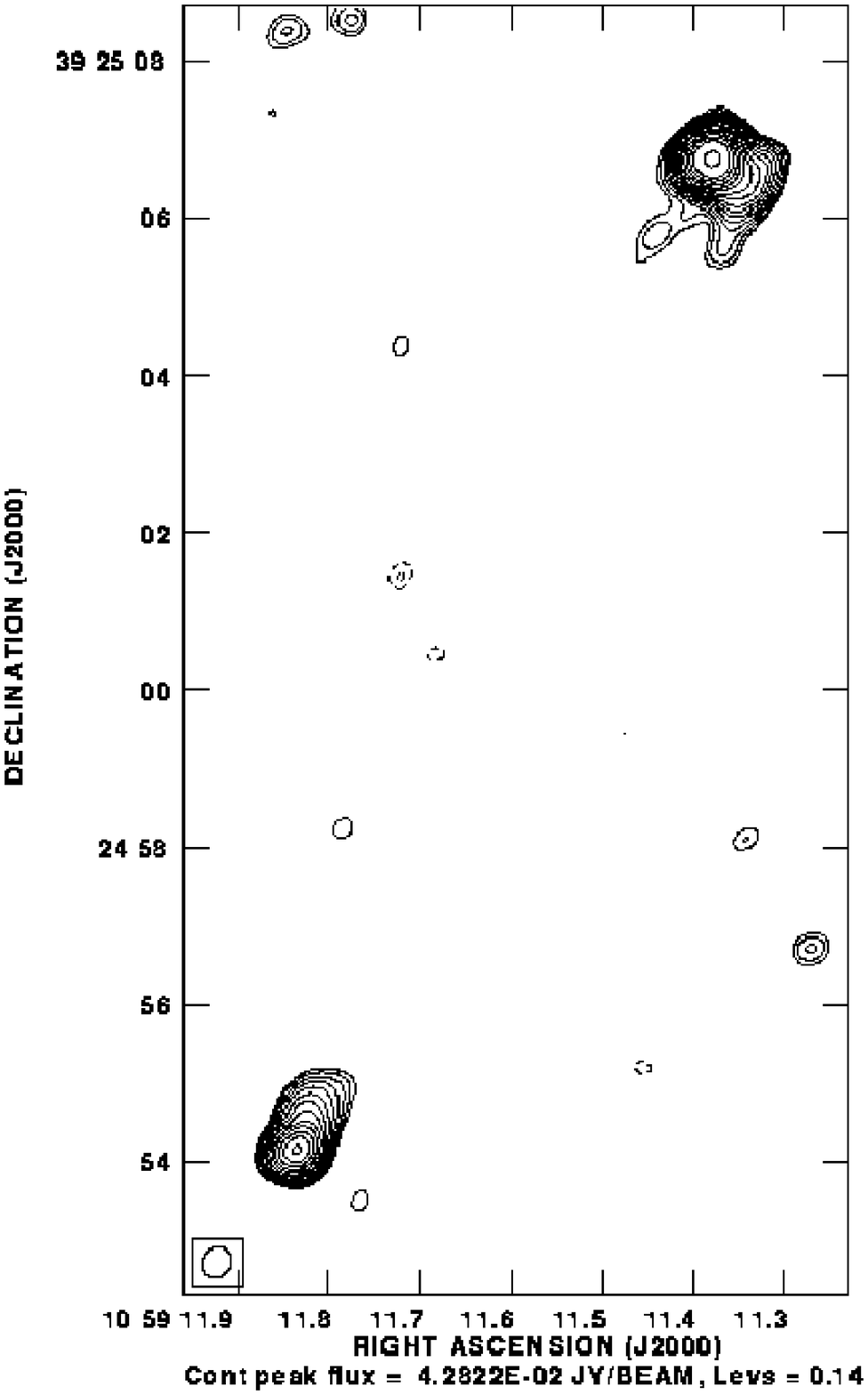,width=6cm}
\psfig{figure=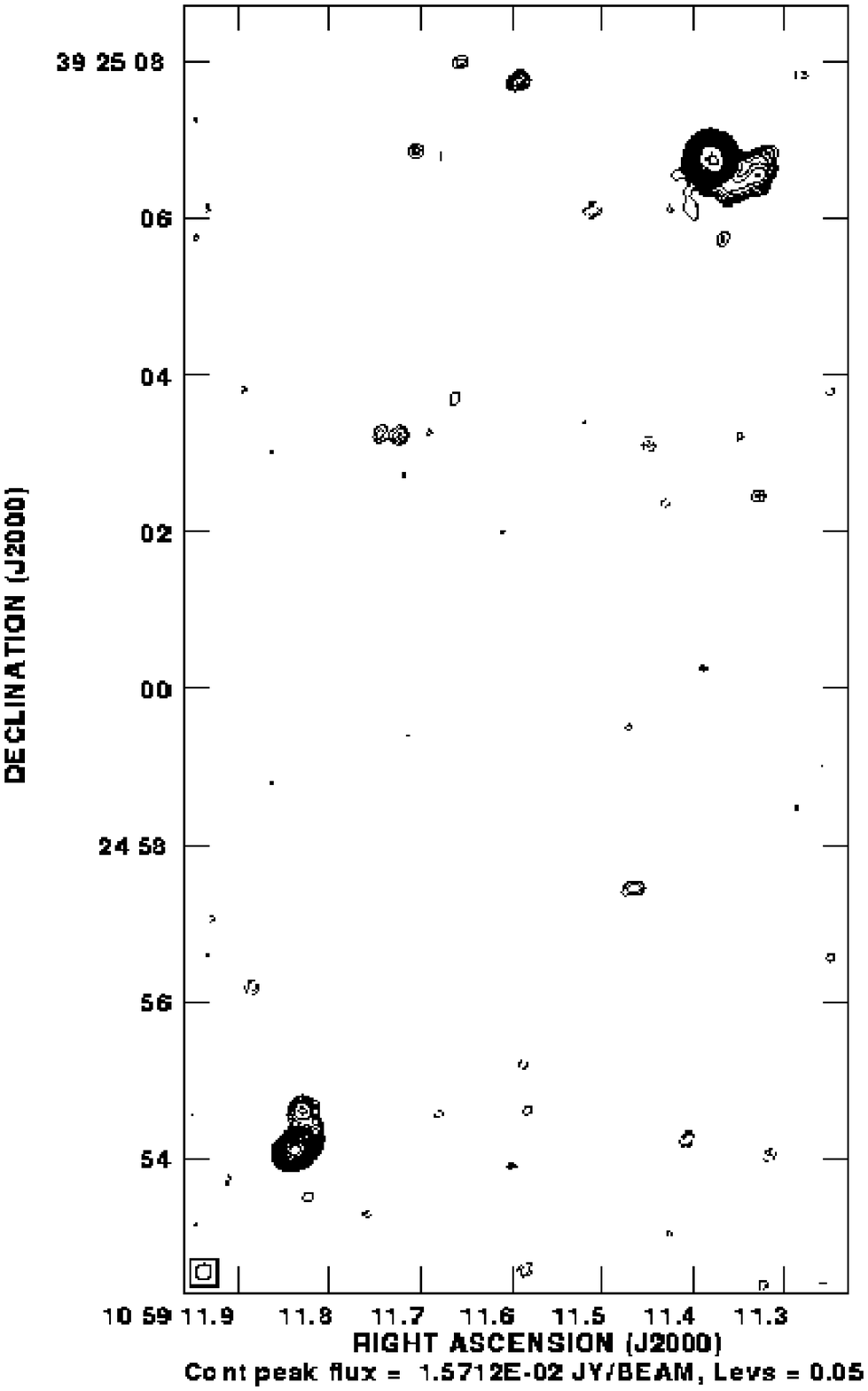,width=6.cm}}\vskip-0.5cm
\centerline{
\psfig{figure=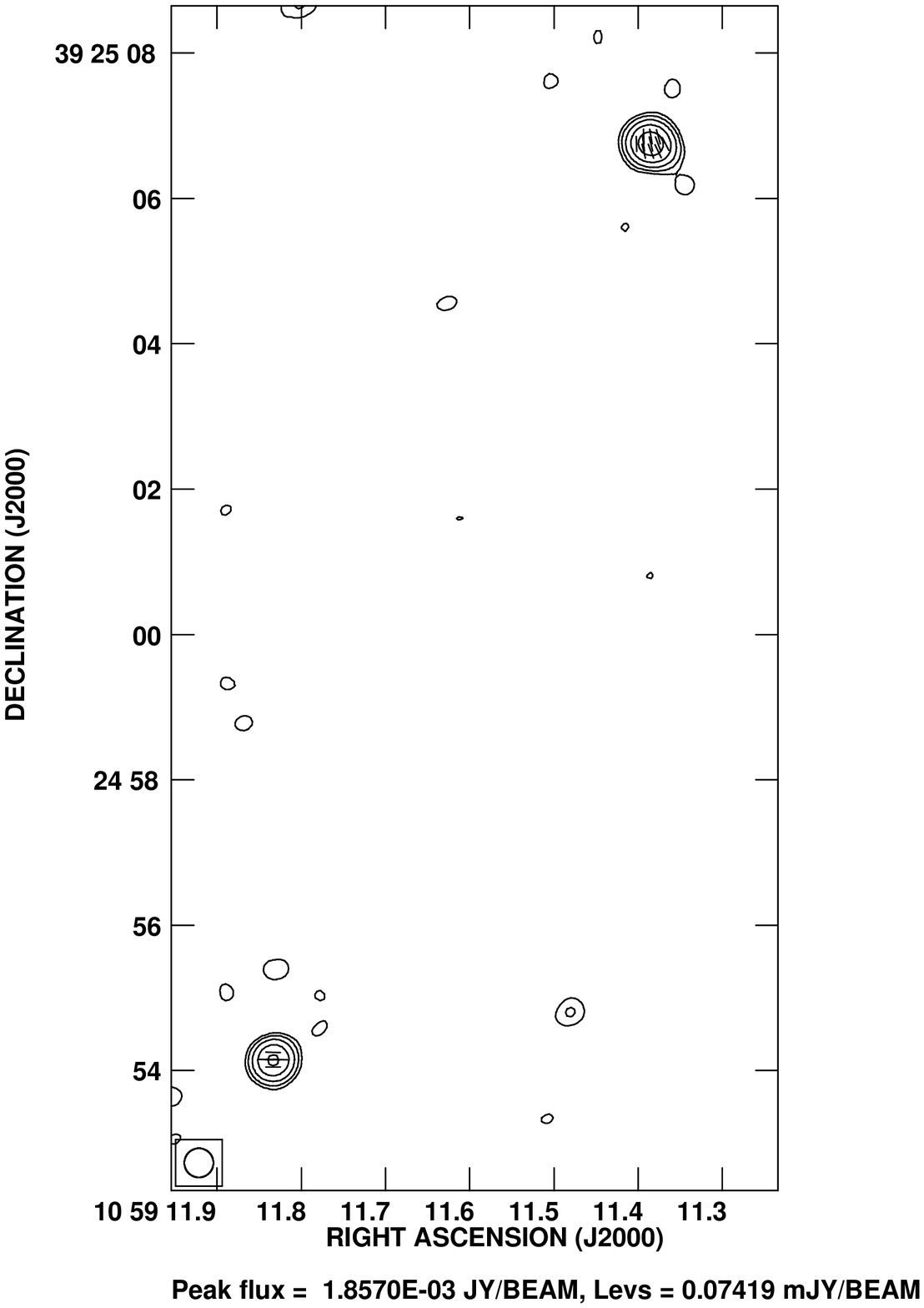,width=6.cm}
\psfig{figure=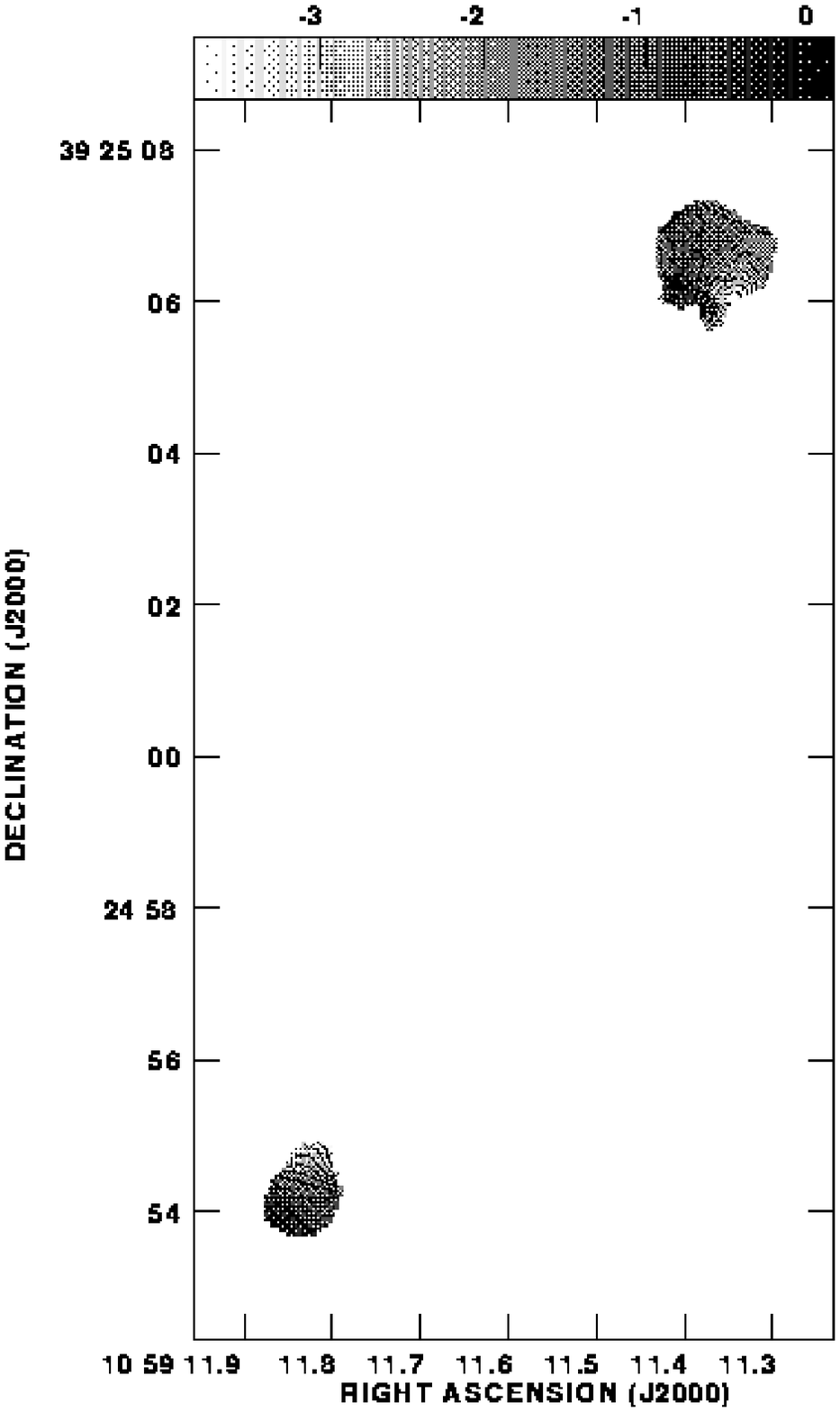,width=6.cm}}
\caption{Maps of the radio source 1056+39 at redshift z$=$2.171.
The sequence of figures is the same as in Fig. 6. The first contour level and
the  peak surface brightness  are respectively  0.143  mJy beam$^{-1}$ and 43
 mJy beam$^{-1}$ for
the 4.5 GHz map; 0.053  mJy beam$^{-1}$ and 15.7 mJy beam$^{-1}$ for the 8.2
GHz map; 0.074  mJy beam$^{-1}$ and 1.9  mJy beam$^{-1}$ for the 4.7 GHz
polarized intensity map.}
\end{figure*}

\begin{figure*}
\centerline{
\psfig{figure=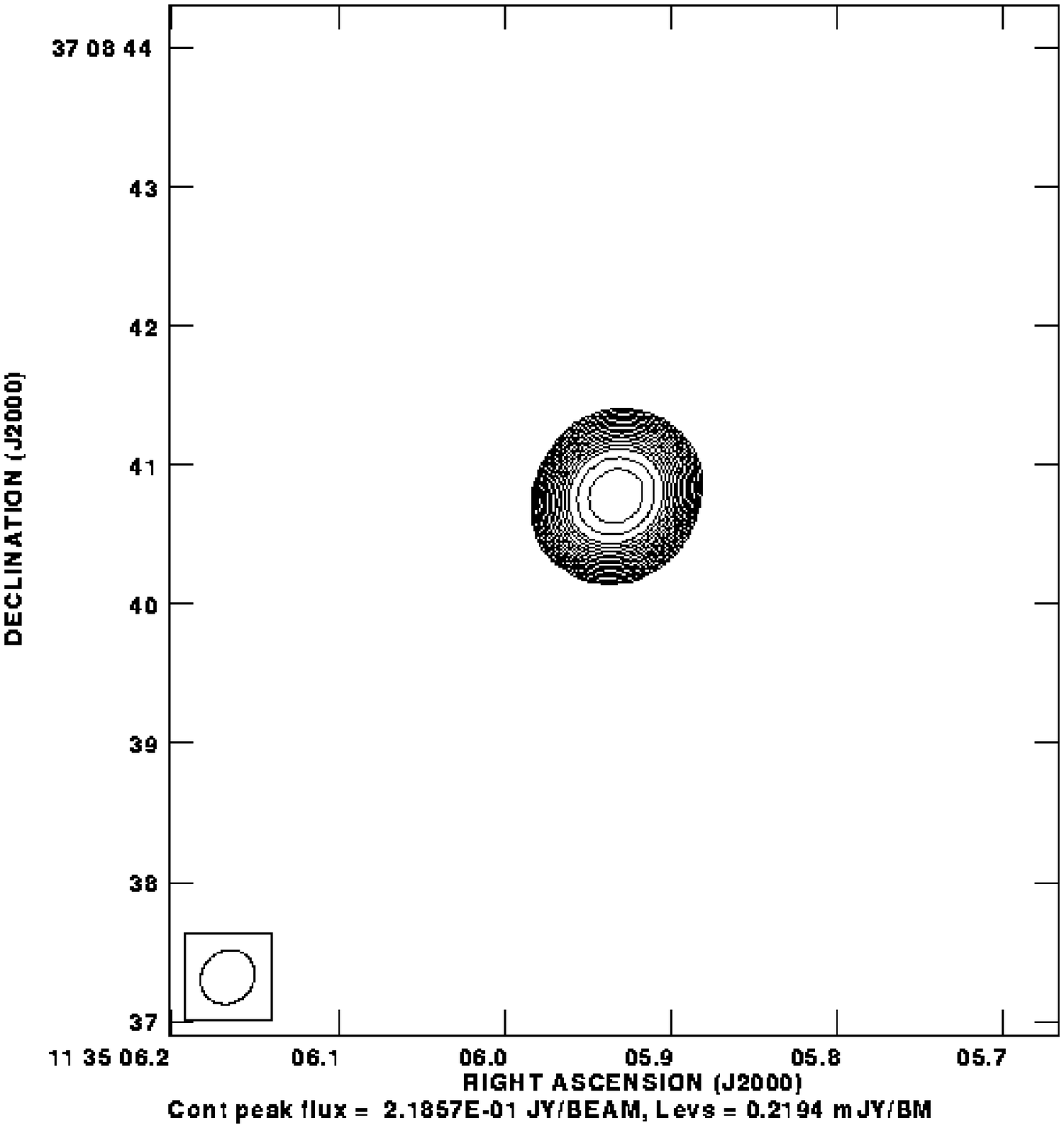,width=6.cm}
\psfig{figure=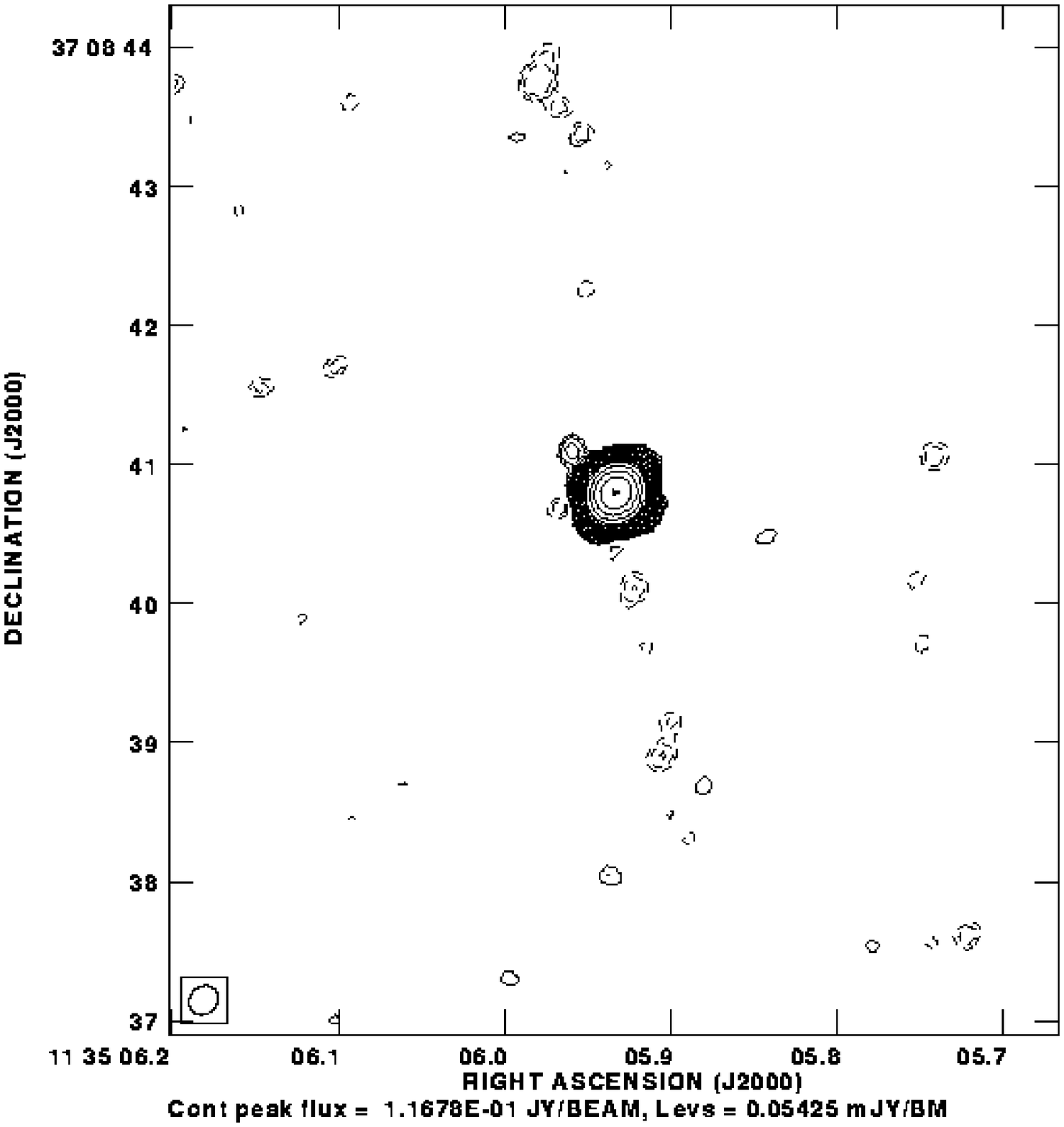,width=6.cm}}\vskip-0.3cm\centerline{
\psfig{figure=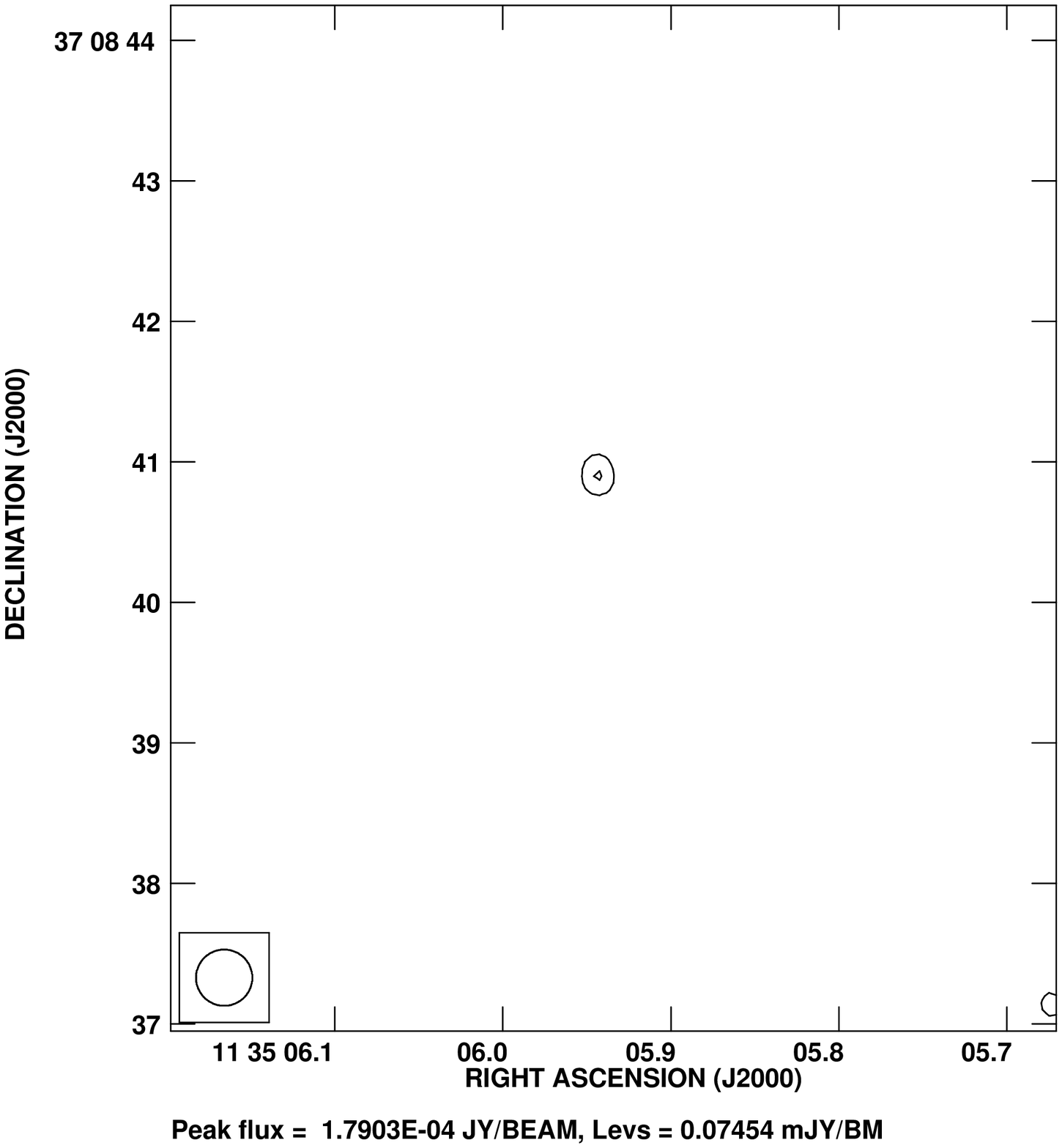,width=6.cm}
\psfig{figure=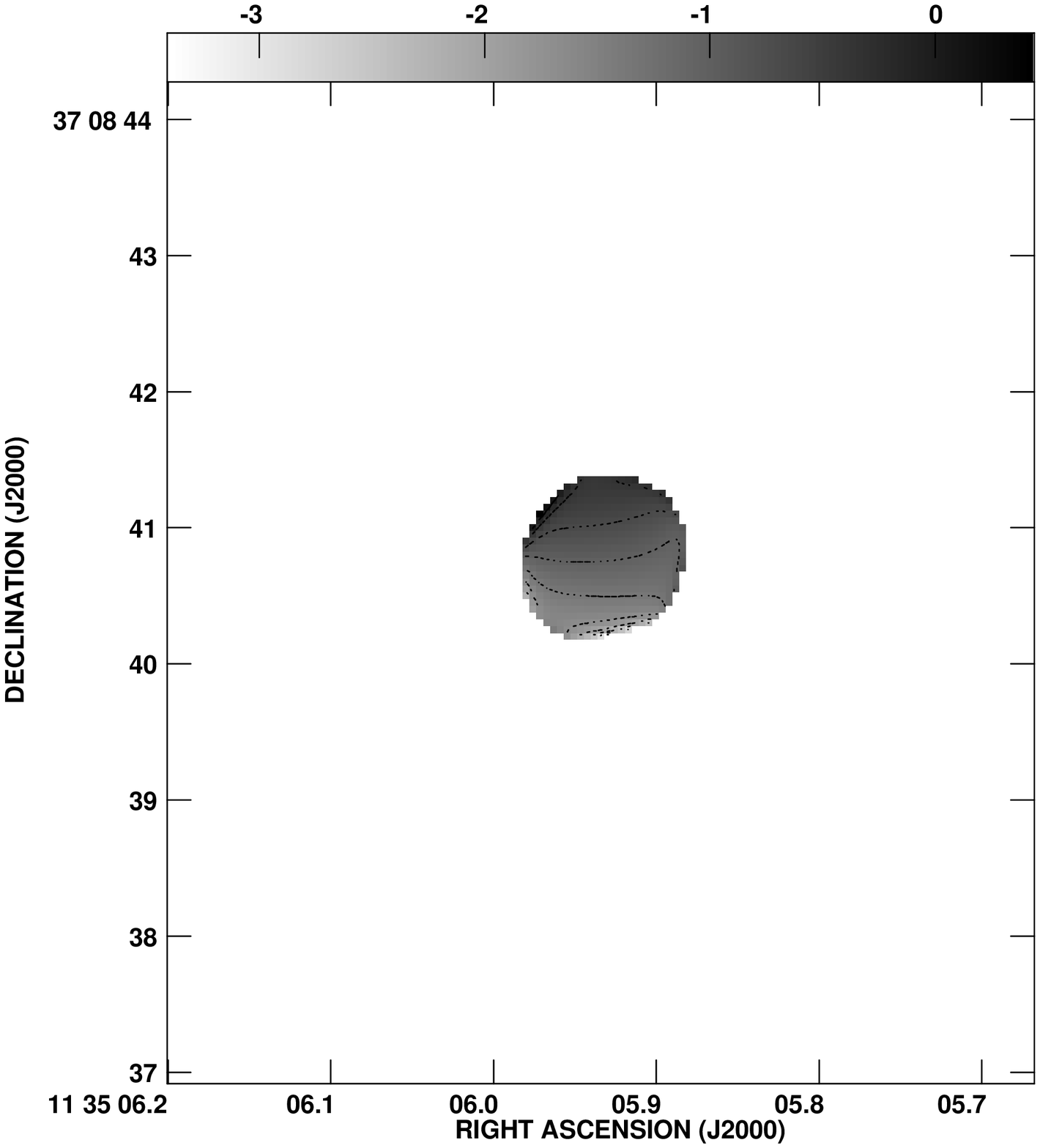,width=6.cm}}
\caption{Maps of the radio source 1132+37 at redshift z$=$2.88.
The sequence of figures is the same as in Fig. 6. The first contour level and
the  peak surface brightness  are respectively  0.22  mJy beam$^{-1}$ and 219
 mJy beam$^{-1}$ for
the 4.5 GHz map; 0.054  mJy beam$^{-1}$ and 117 mJy beam$^{-1}$ for the 8.2
GHz map; 0.074  mJy beam$^{-1}$ and 0.2  mJy beam$^{-1}$ for the 4.7 GHz
polarized intensity map.}
\end{figure*}

\begin{figure*}
\centerline{
\psfig{figure=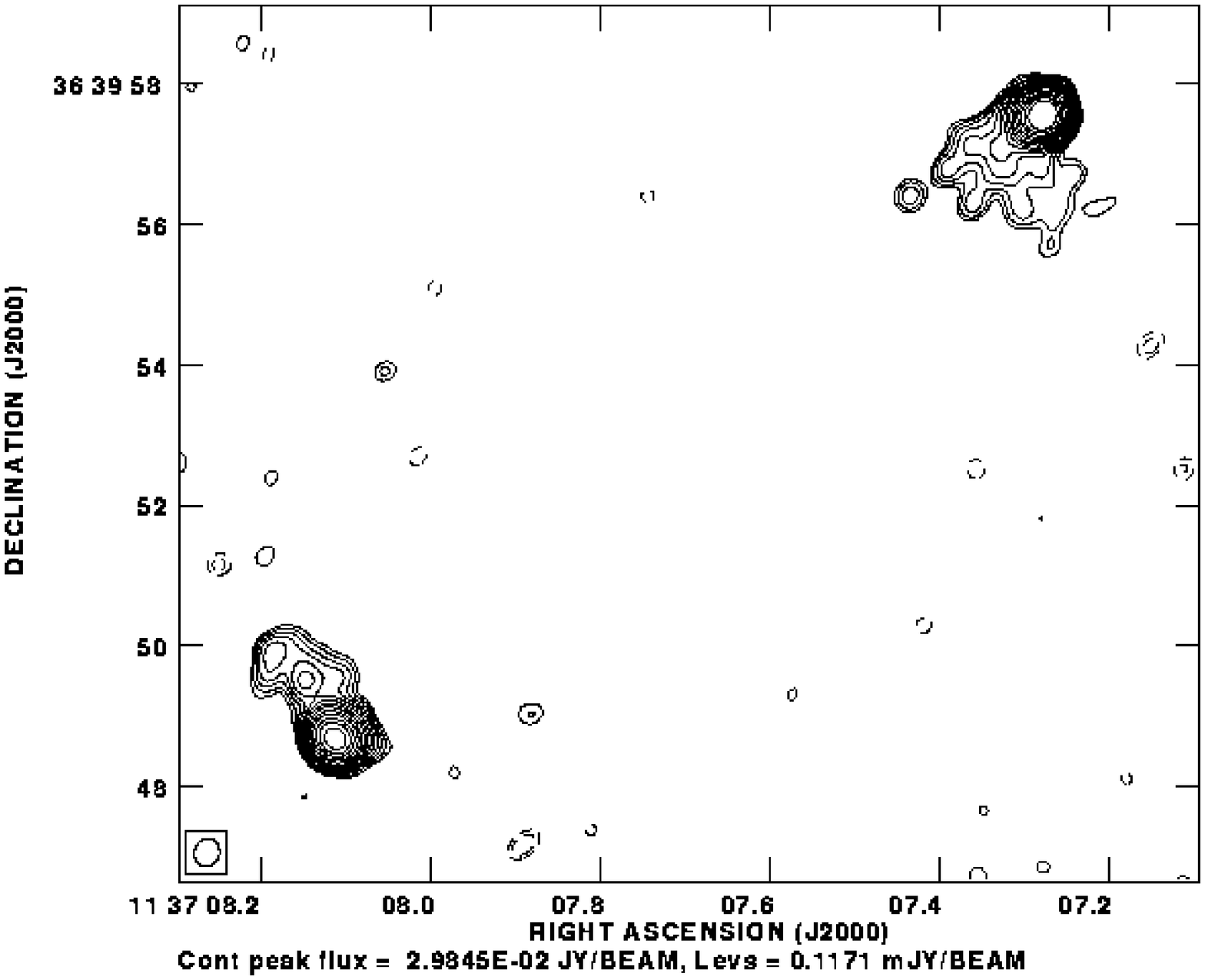,width=7.8cm,angle=-90}
\psfig{figure=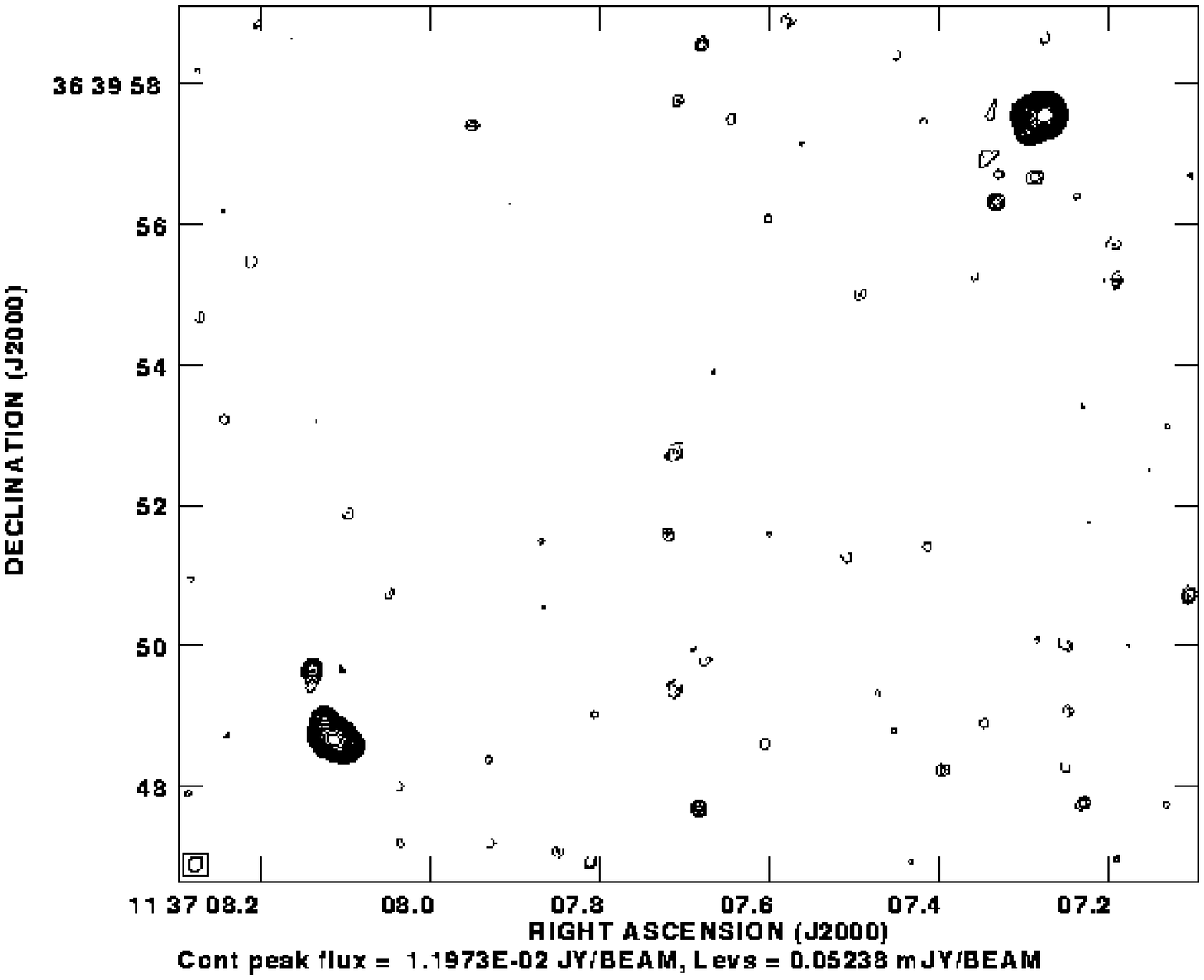,width=7.6cm,angle=-90}}\vskip-0.2cm\centerline{
\psfig{figure=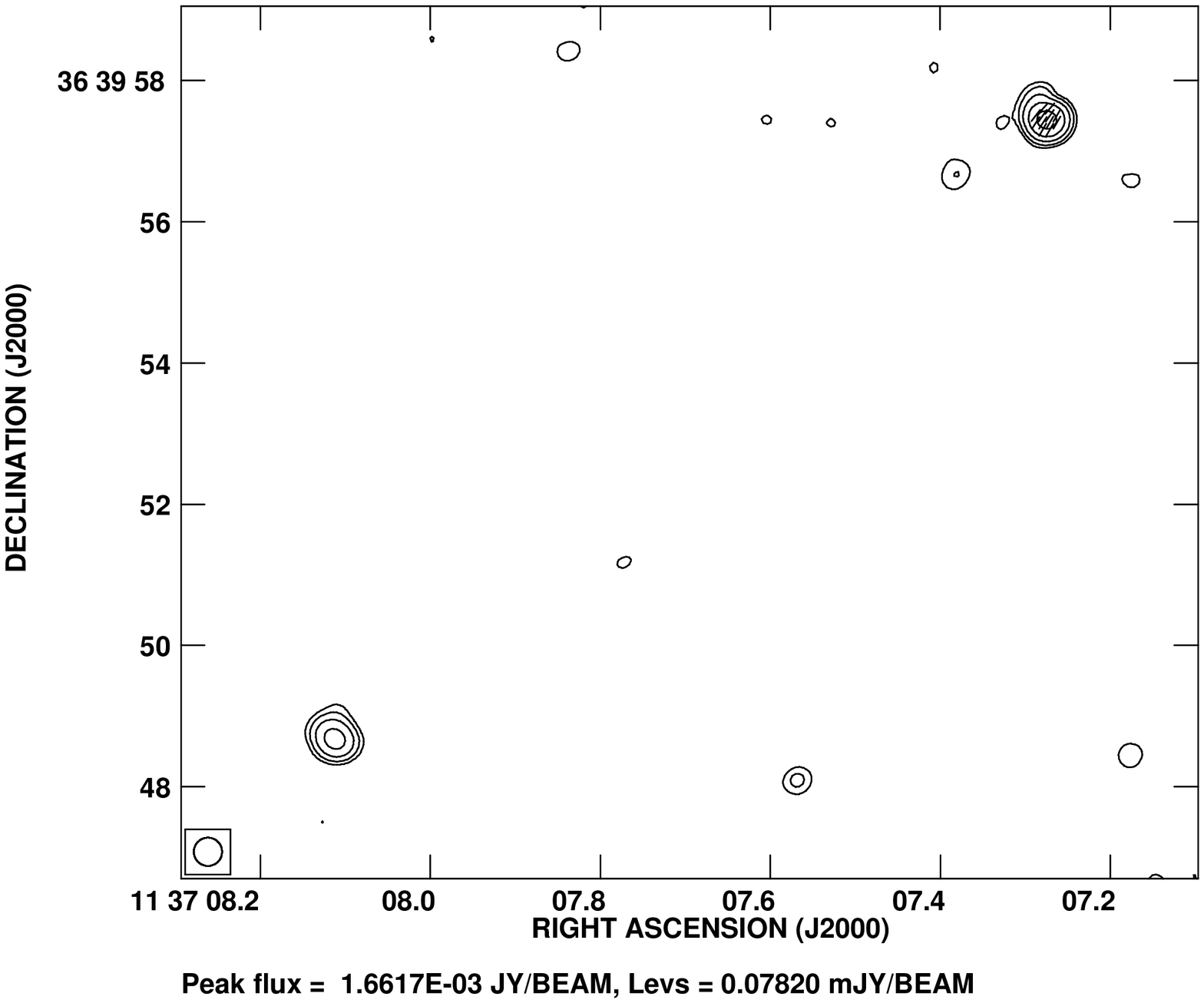,width=8.cm,angle=-90}
\psfig{figure=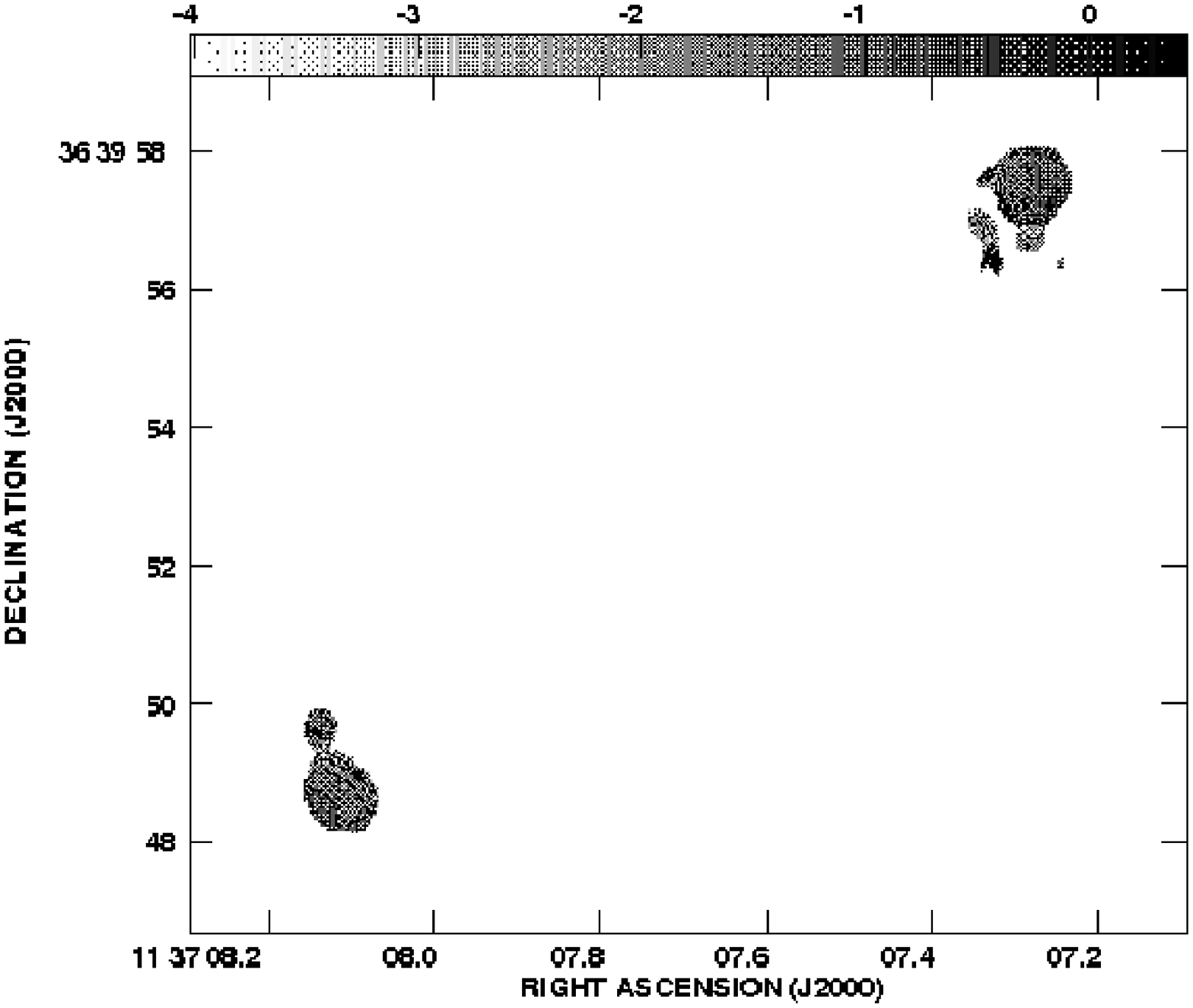,width=8.cm,angle=-90}}
\caption{Maps of the radio source 1134+369 at redshift z$=$2.12.
The sequence of figures is the same as in Fig. 6. The first contour level and
the  peak surface brightness  are respectively  0.117  mJy beam$^{-1}$ and 30
 mJy beam$^{-1}$ for
the 4.5 GHz map; 0.052  mJy beam$^{-1}$ and 12 mJy beam$^{-1}$ for the 8.2
GHz map; 0.078  mJy beam$^{-1}$ and 1.7  mJy beam$^{-1}$ for the 4.7 GHz
polarized intensity map.}
\end{figure*}
\begin{figure*}
\centerline{
\psfig{figure=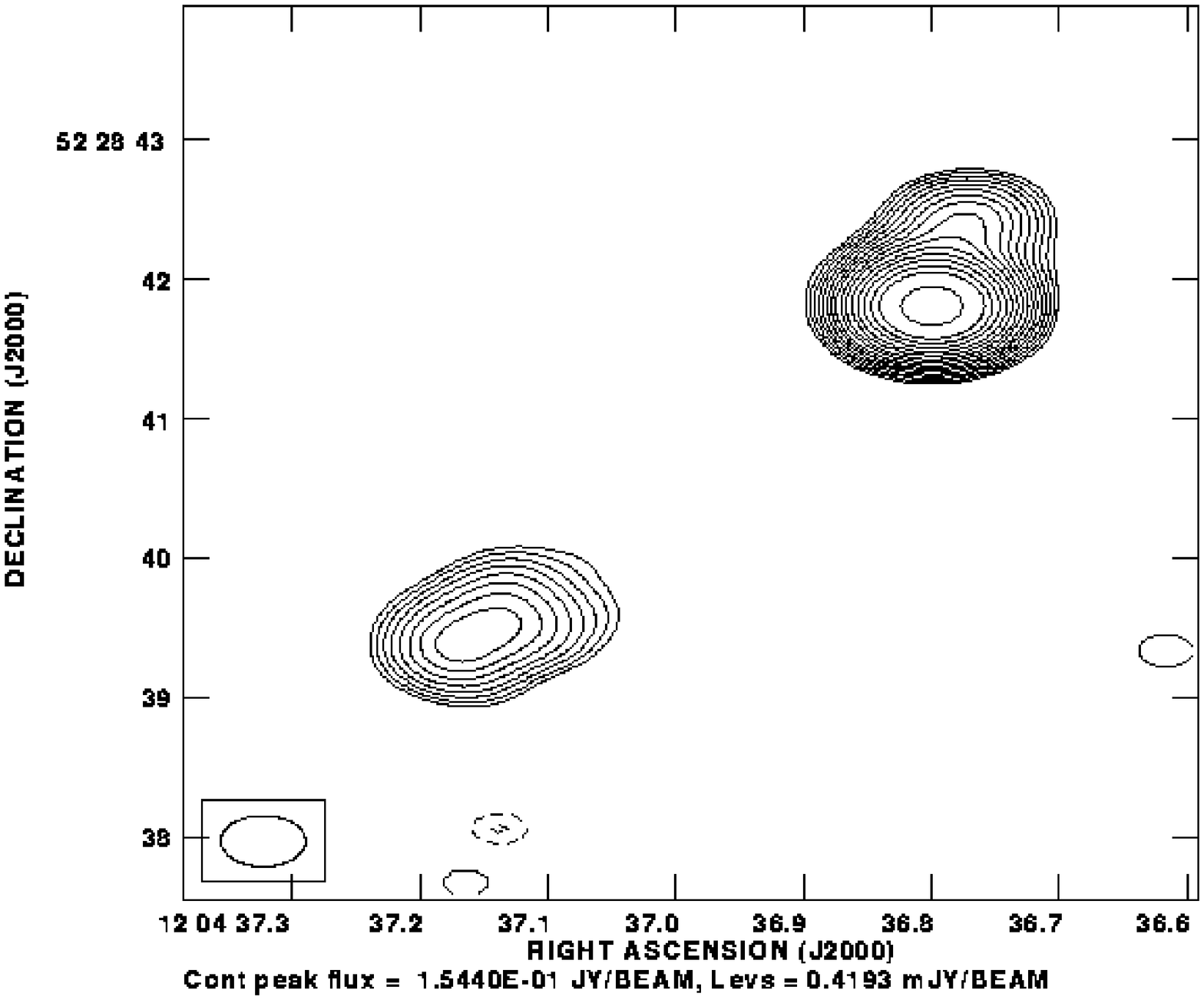,width=7.7cm,angle=-90}
\psfig{figure=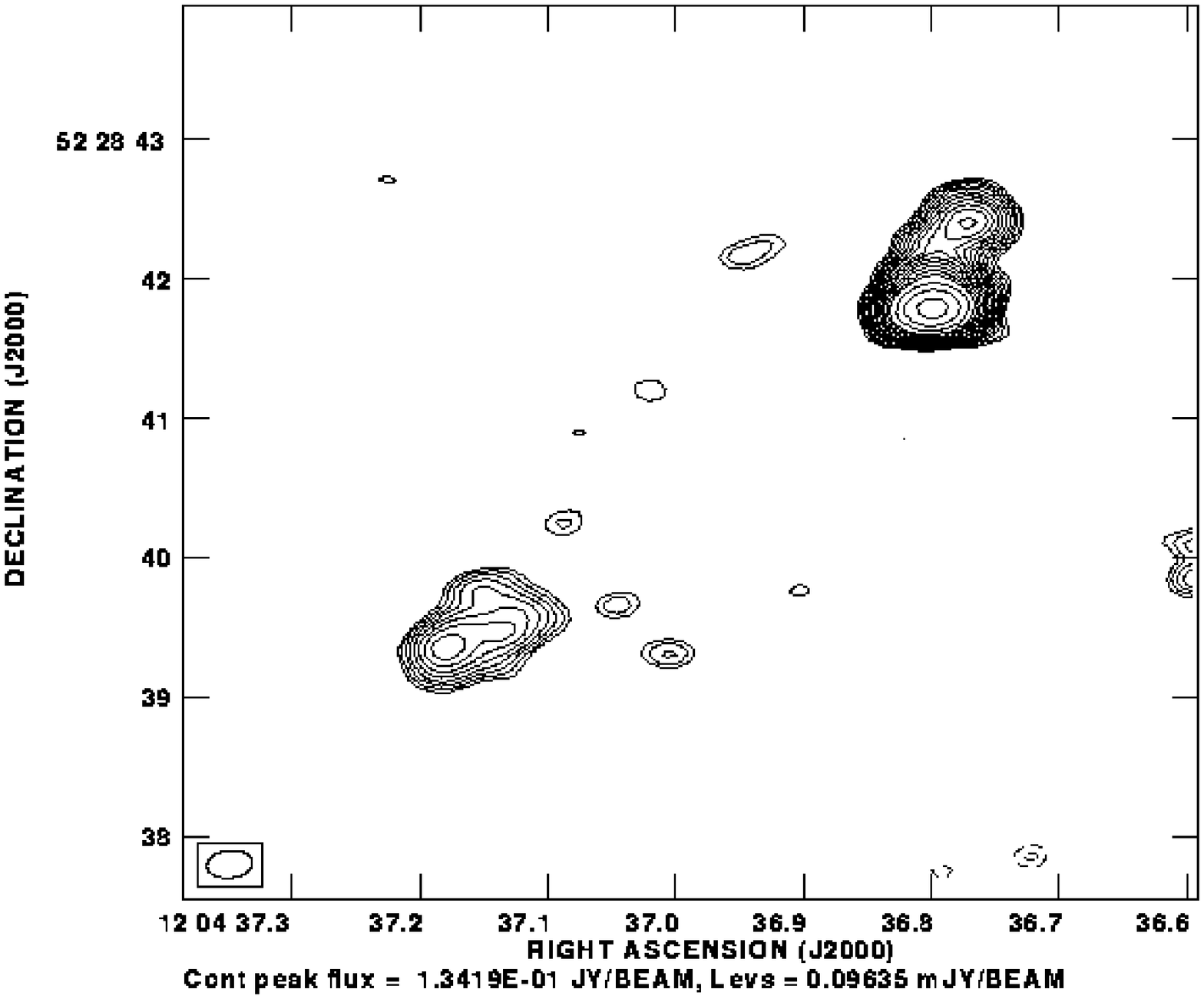,width=7.7cm,angle=-90}}
\vskip-0.2cm\centerline{
\psfig{figure=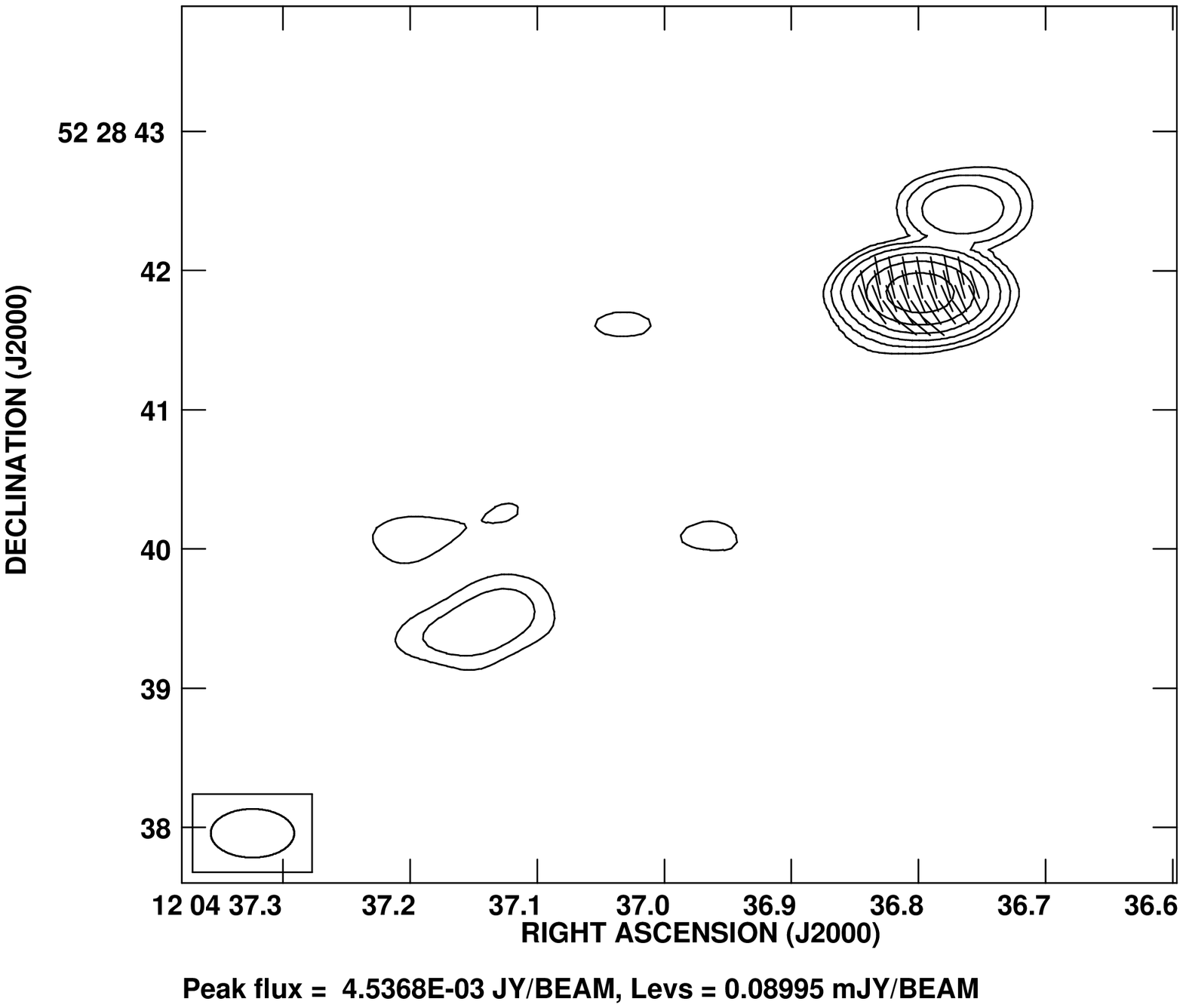,width=8.cm,angle=-90}
\psfig{figure=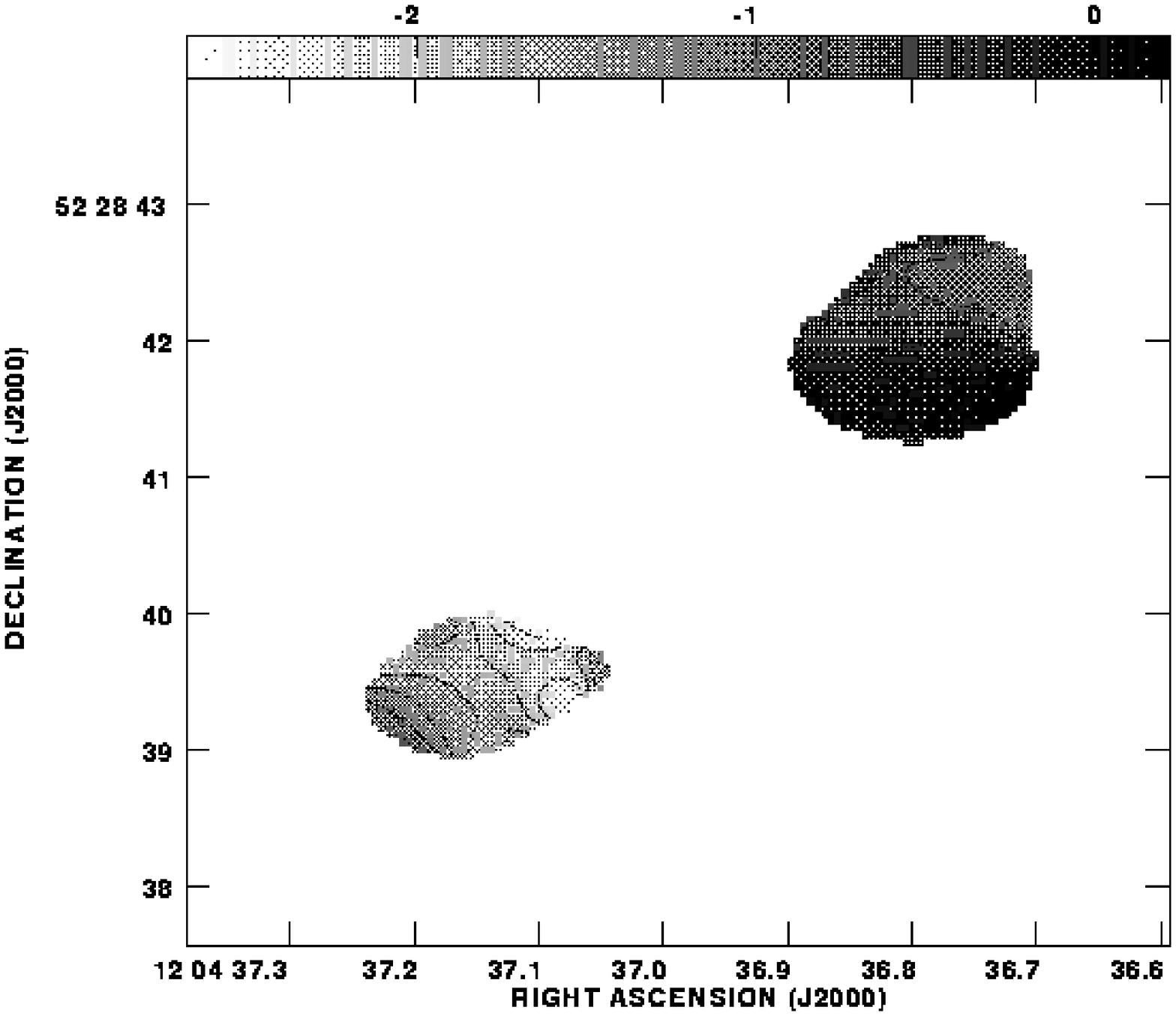,width=8.cm,angle=-90}}
\caption{Maps of the radio source 1202+527 at redshift z$=$2.73.
The sequence of figures is the same as in Fig. 6. The first contour level and
the  peak surface brightness  are respectively  0.42  mJy beam$^{-1}$ and 154
 mJy beam$^{-1}$ for
the 4.5 GHz map; 0.096  mJy beam$^{-1}$ and 134 mJy beam$^{-1}$ for the 8.2
GHz map; 0.090  mJy beam$^{-1}$ and 4.5  mJy beam$^{-1}$ for the 4.7 GHz
polarized intensity map.}
\end{figure*}
\begin{figure*}
\centerline{ 
\psfig{figure=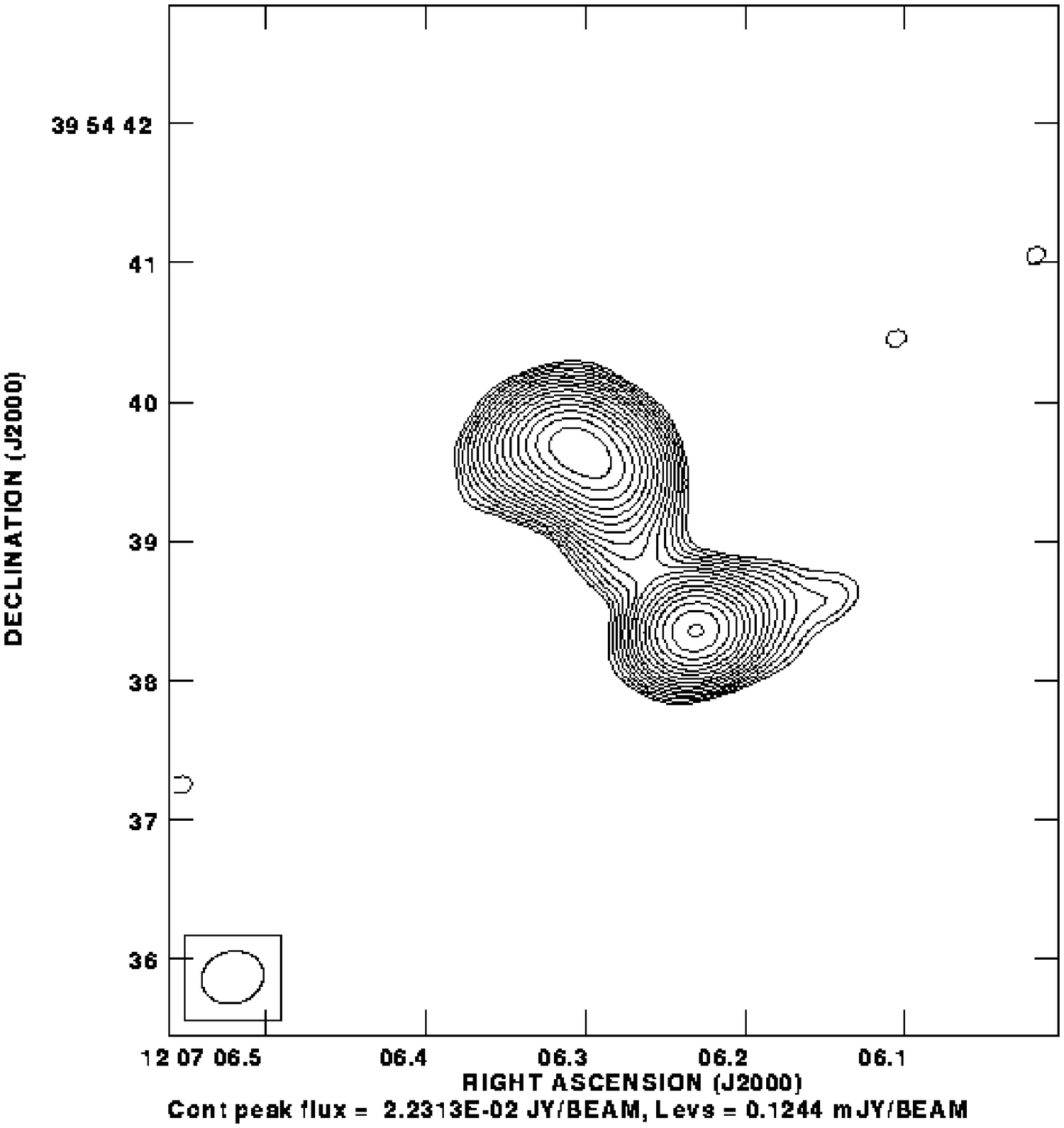,width=7.cm}
\psfig{figure=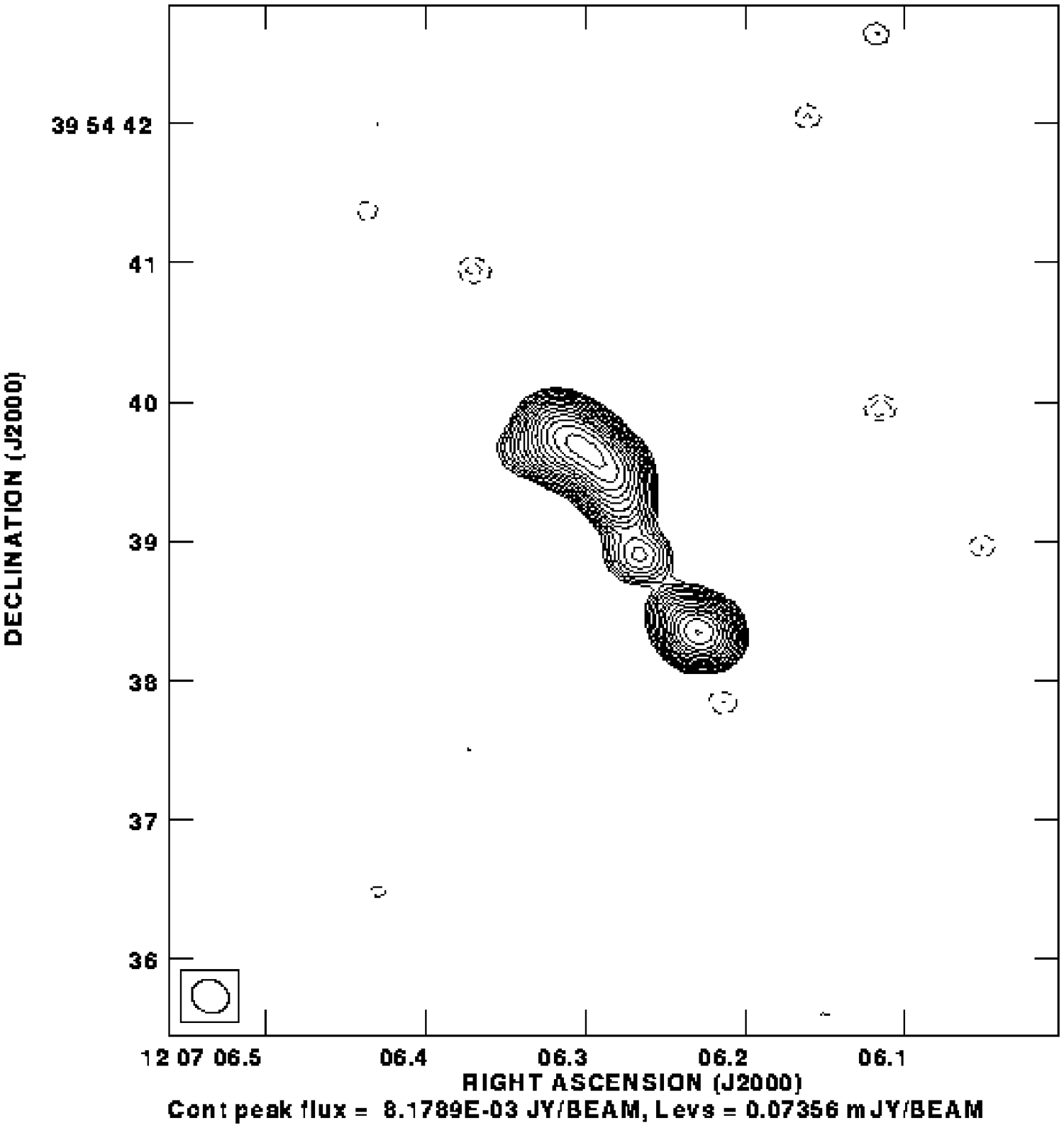,width=7.cm}}
\vskip-0.7cm\centerline{
\psfig{figure=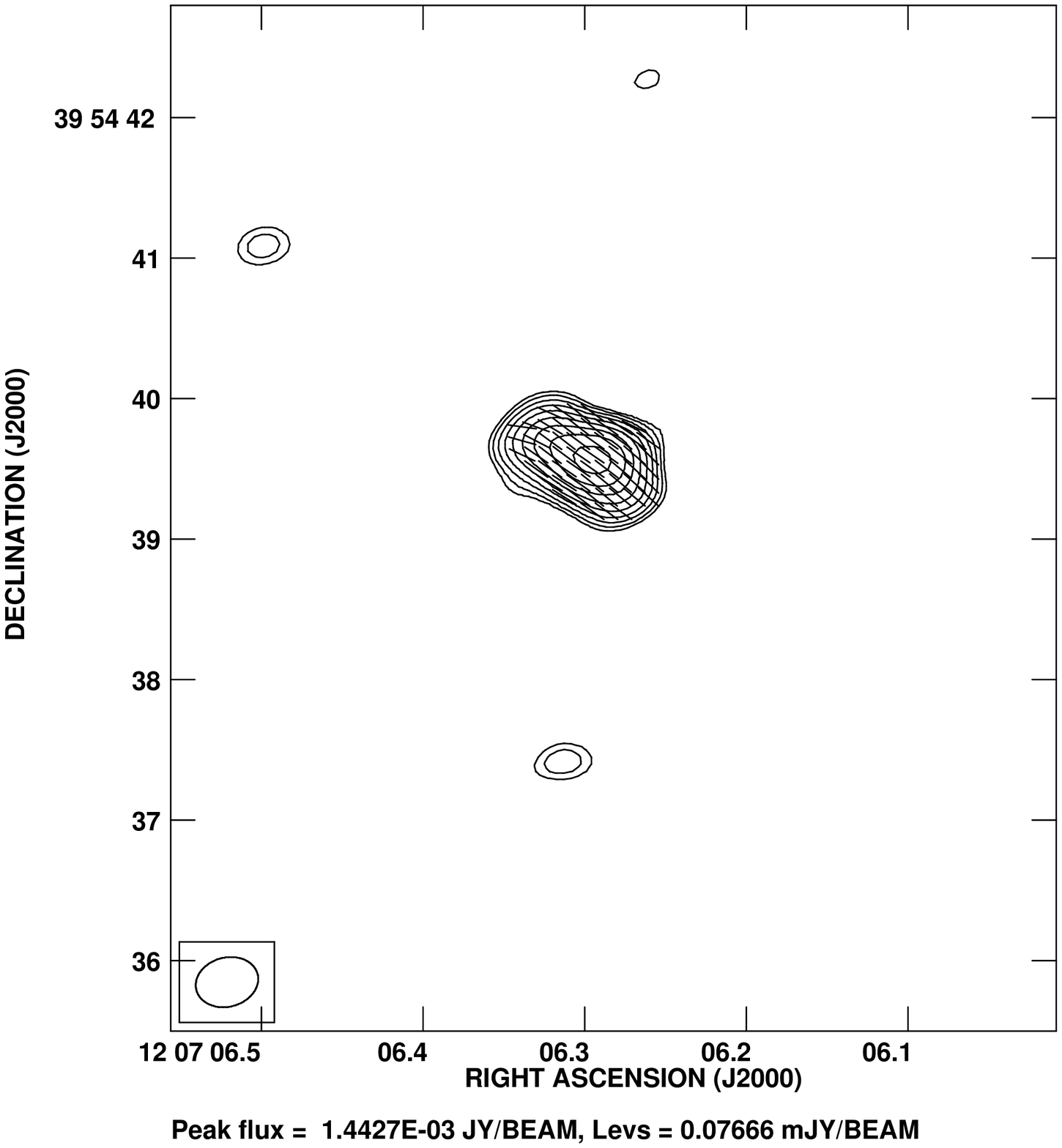,width=7.cm}
\psfig{figure=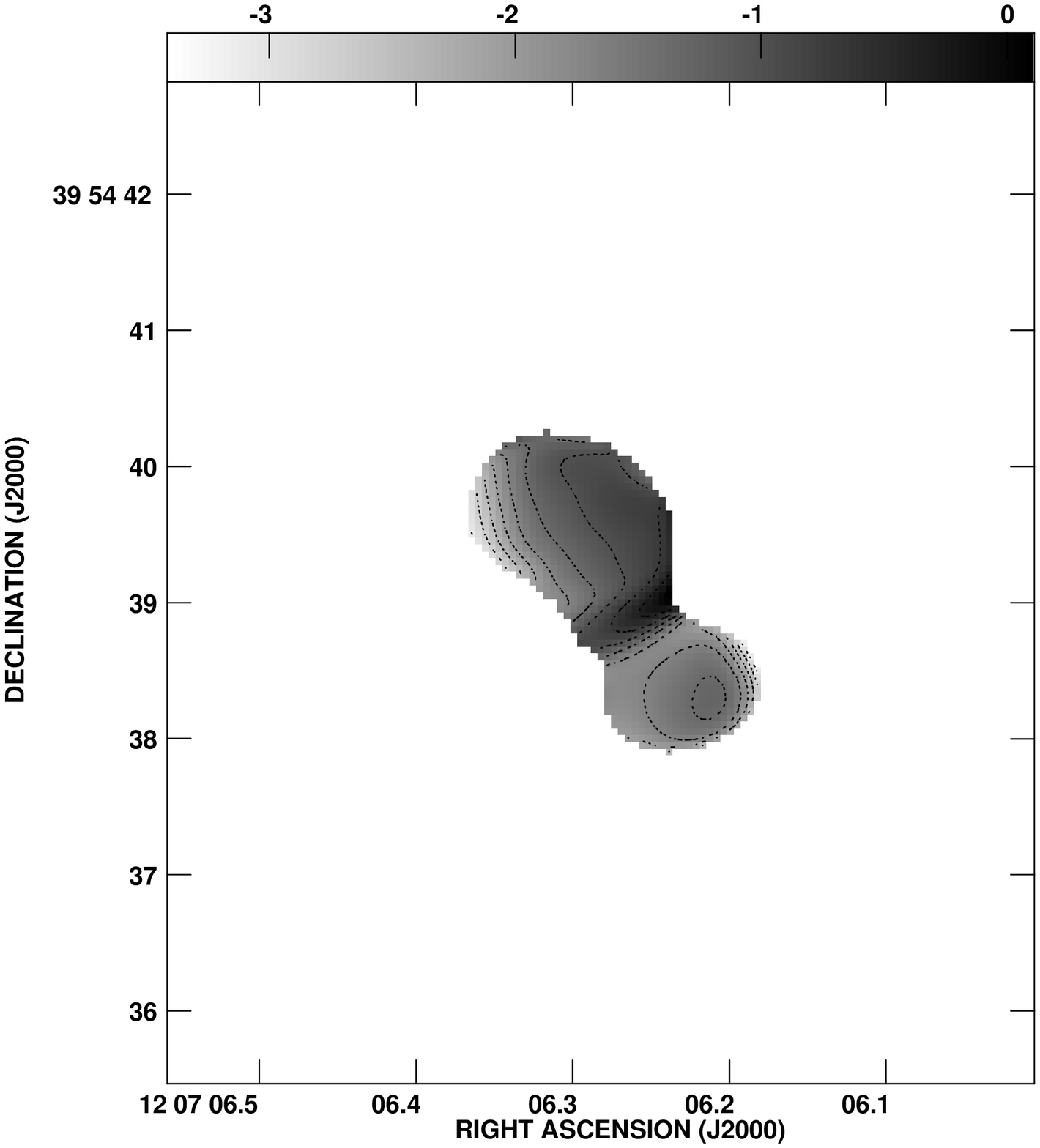,width=7.cm}}
\caption{Maps of the radio source 1204+401 at redshift z$=$2.066.
The sequence of figures is the same as in Fig. 6. The first contour level and
the  peak surface brightness  are respectively  0.124  mJy beam$^{-1}$ and 22
 mJy beam$^{-1}$ for
the 4.5 GHz map; 0.062  mJy beam$^{-1}$ and 82 mJy beam$^{-1}$ for the 8.2
GHz map; 0.077  mJy beam$^{-1}$ and 1.4  mJy beam$^{-1}$ for the 4.7 GHz
polarized intensity map.}
\end{figure*}
\begin{figure*}
\centerline{
\psfig{figure=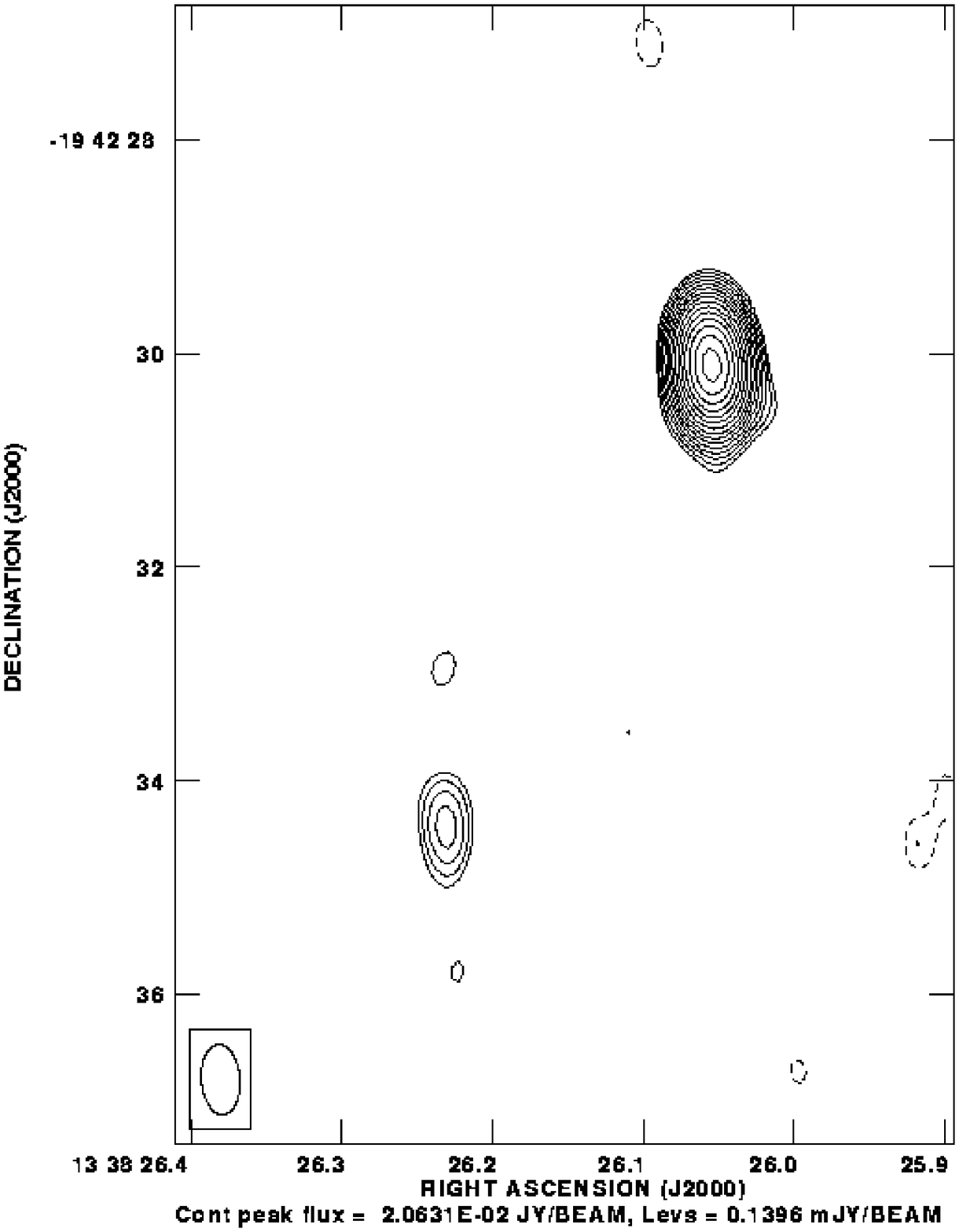,width=7.cm}
\psfig{figure=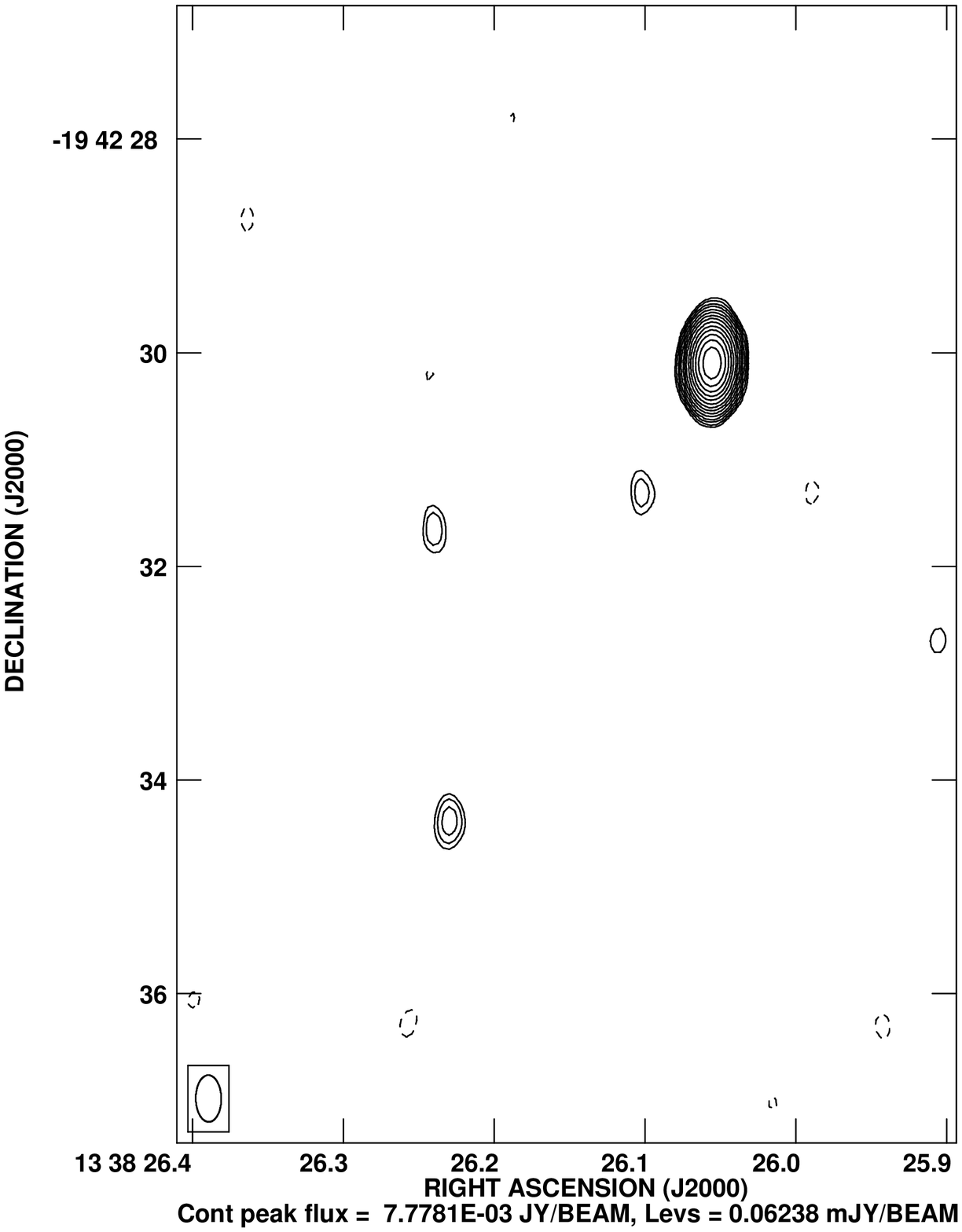,width=7.cm}}
\vskip-0.6cm\centerline{
\psfig{figure=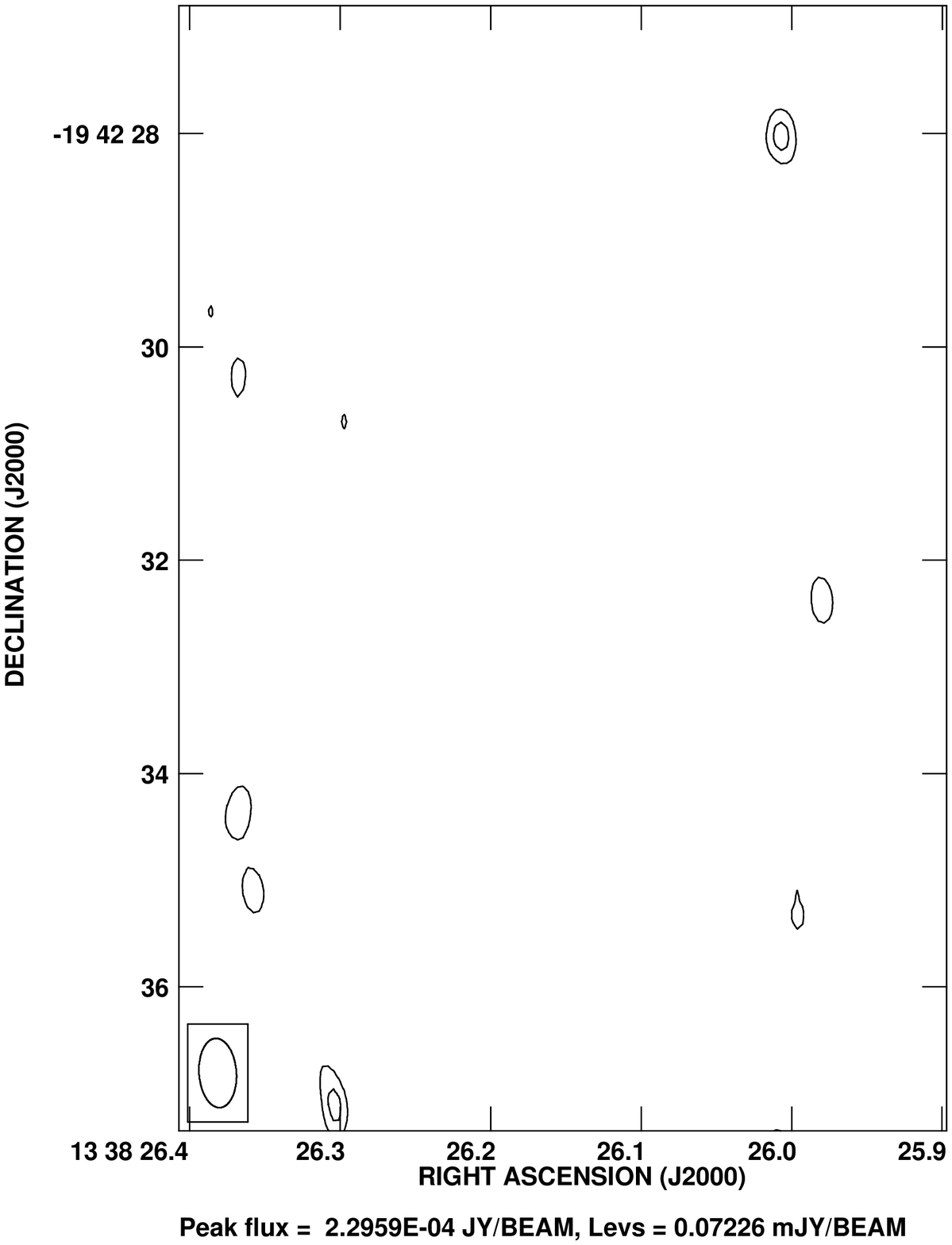,width=7.cm}
\psfig{figure=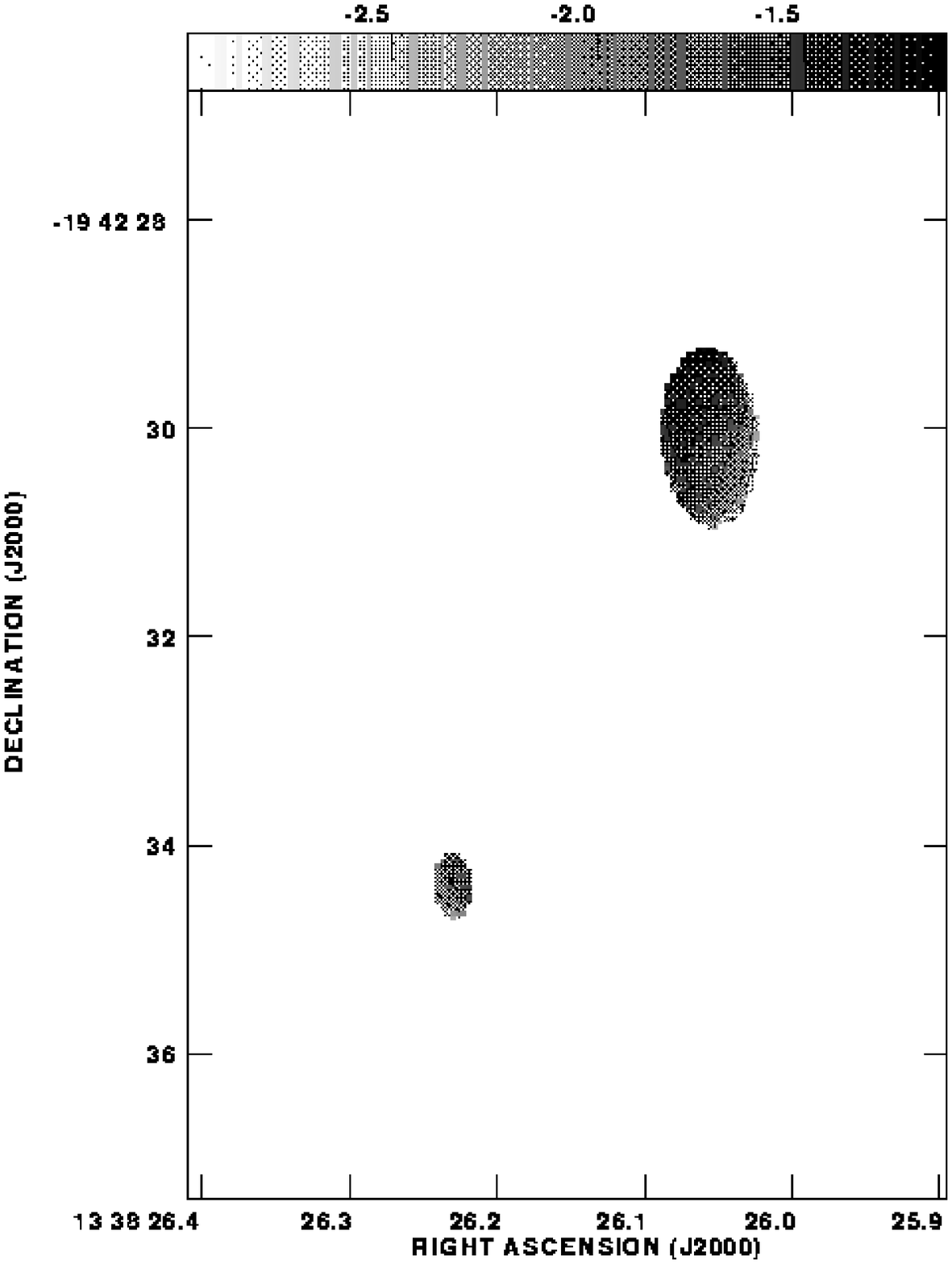,width=7.cm}}
\caption{Maps of the radio source 1338$-$19 at redshift z$=$4.11.
The sequence of figures is the same as in Fig. 6. The first contour level and
the  peak surface brightness  are respectively  0.140  mJy beam$^{-1}$ and 21
 mJy beam$^{-1}$ for
the 4.5 GHz map; 0.074  mJy beam$^{-1}$ and 8 mJy beam$^{-1}$ for the 8.2
GHz map; 0.072  mJy beam$^{-1}$  for the 4.7 GHz
polarized intensity map.}
\end{figure*}
\begin{figure*}
\centerline{
\psfig{figure=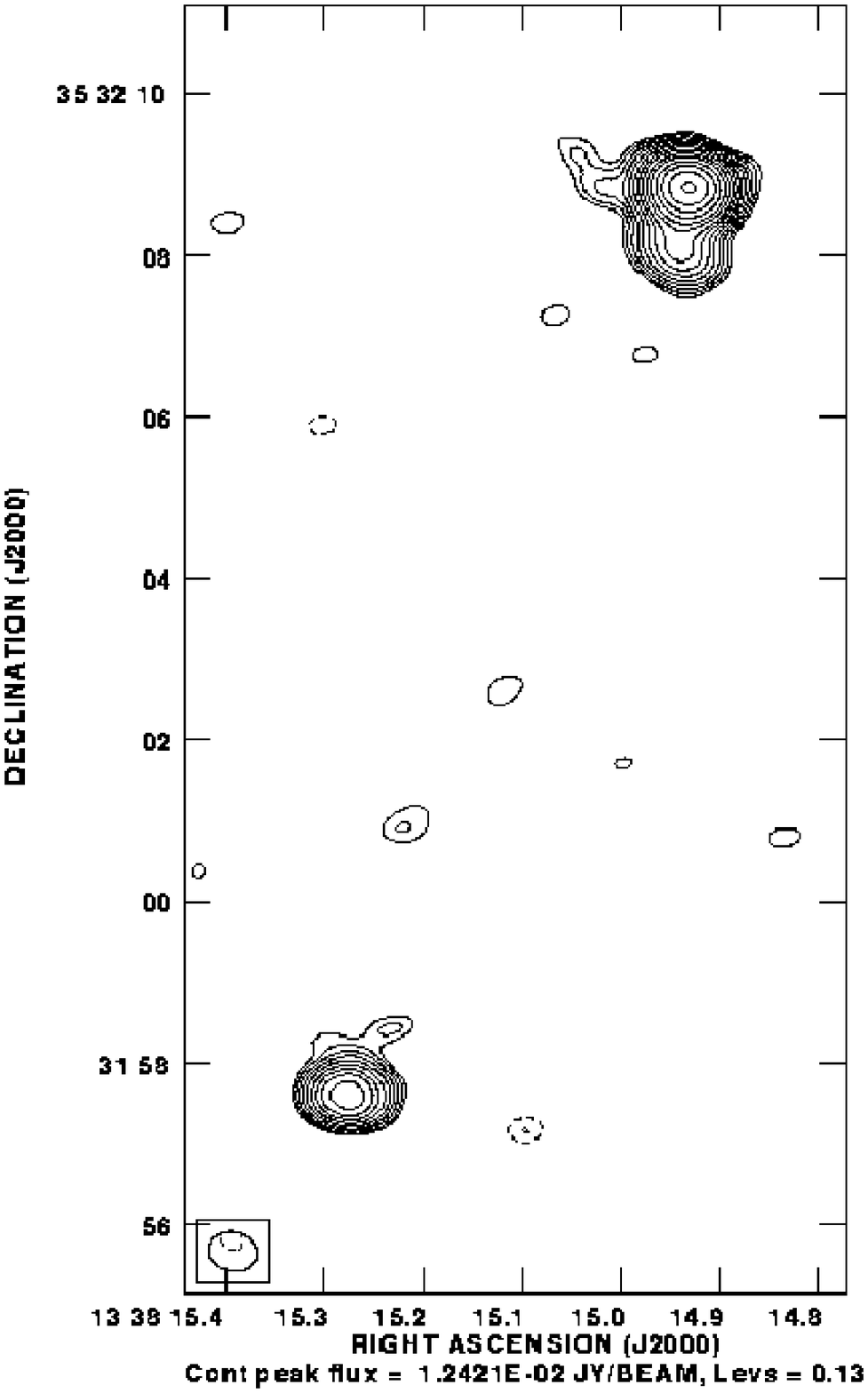,width=6.cm}
\psfig{figure=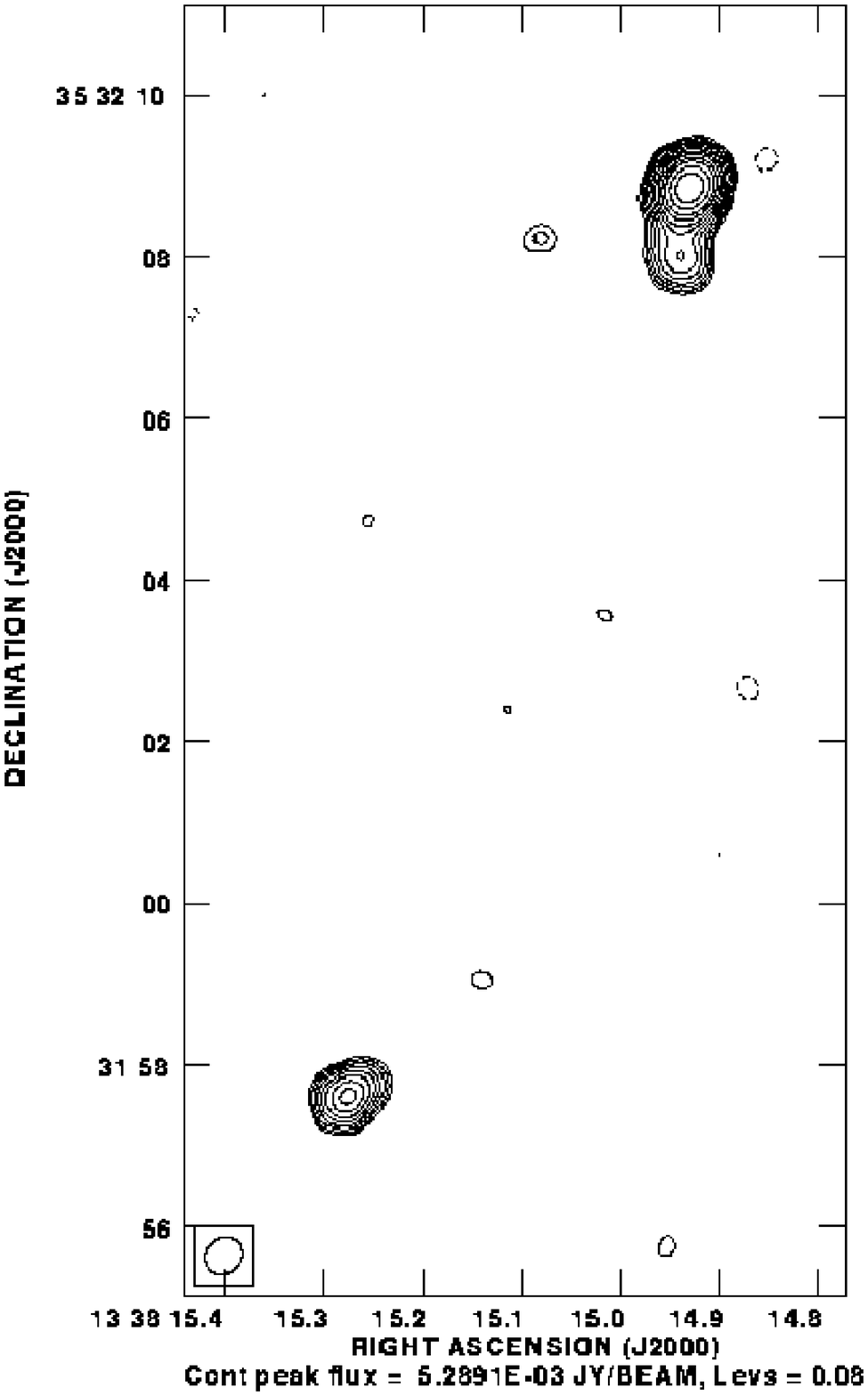,width=6.cm}}
\vskip-1cm\centerline{
\psfig{figure=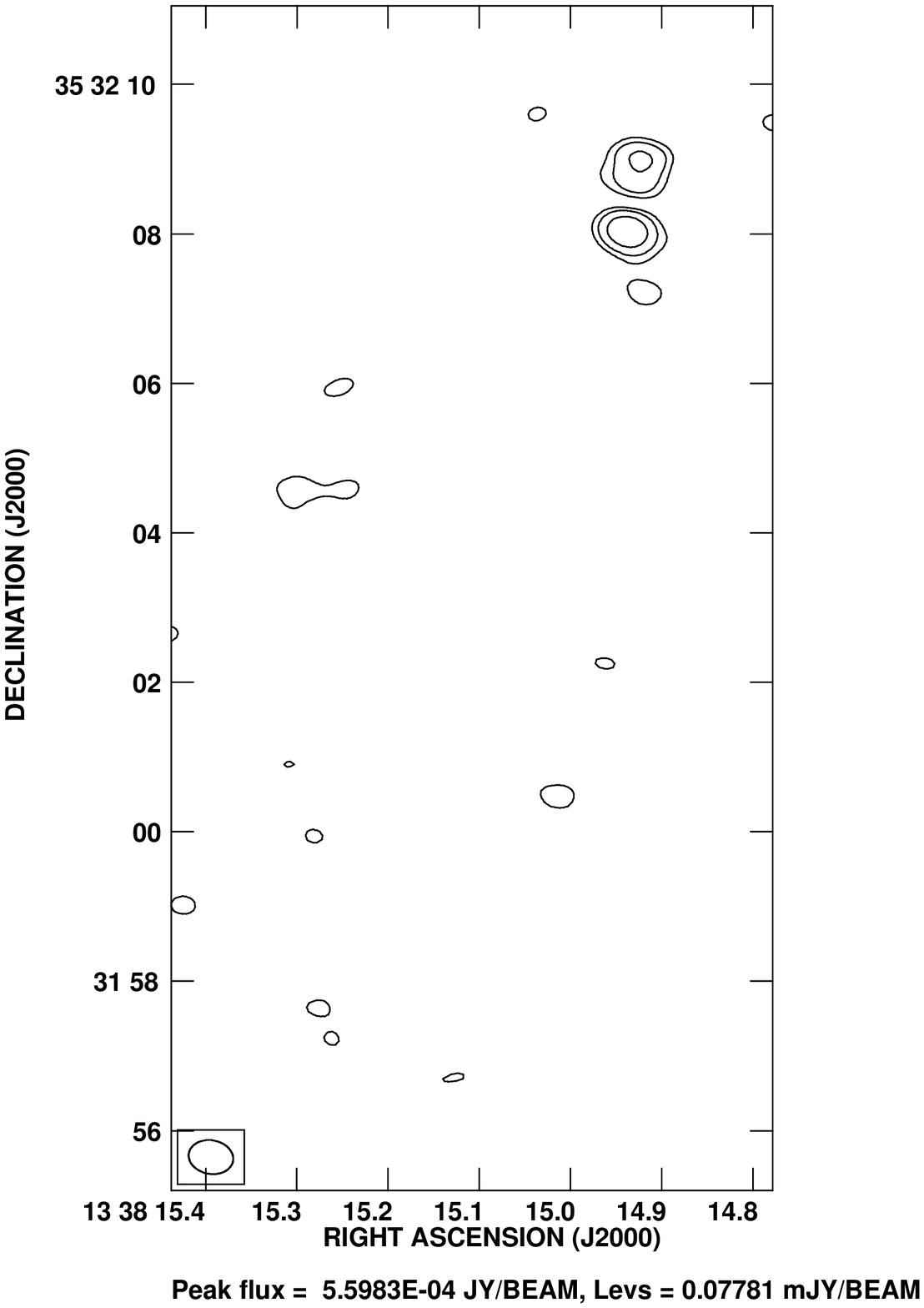,width=6.cm}
\psfig{figure=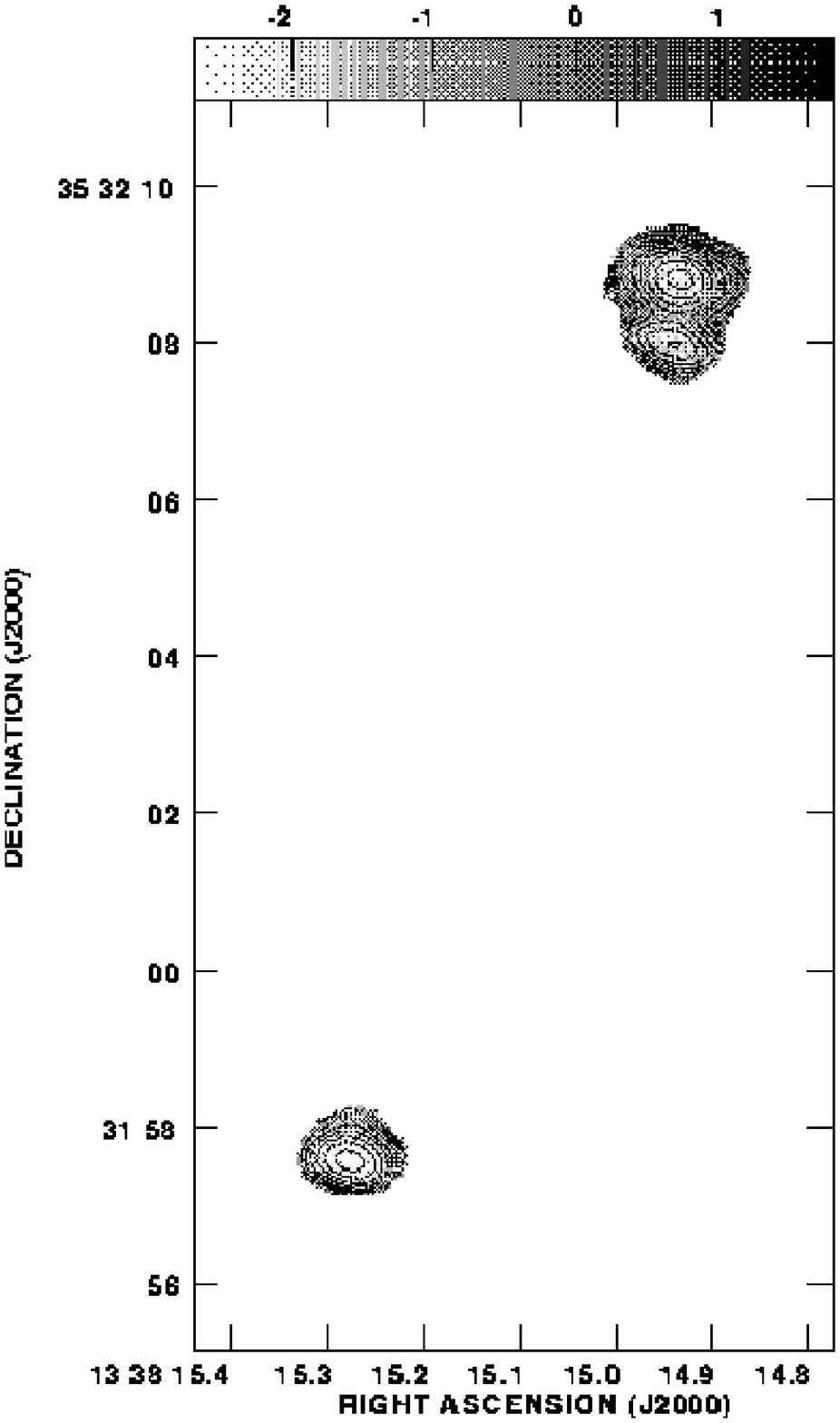,width=6cm}}
\caption{Maps of the radio source 1339+35 at redshift z$=$2.772.
The sequence of figures is the same as in Fig. 6. The first contour level and
the  peak surface brightness  are respectively  0.132  mJy beam$^{-1}$ and 12
 mJy beam$^{-1}$ for
the 4.5 GHz map; 0.084  mJy beam$^{-1}$ and 5.3 mJy beam$^{-1}$ for the 8.2
GHz map; 0.078  mJy beam$^{-1}$  and 0.6  mJy beam$^{-1}$ for the 4.7 GHz
polarized intensity map.}
\end{figure*}

\begin{figure*}
\centerline{
\psfig{figure=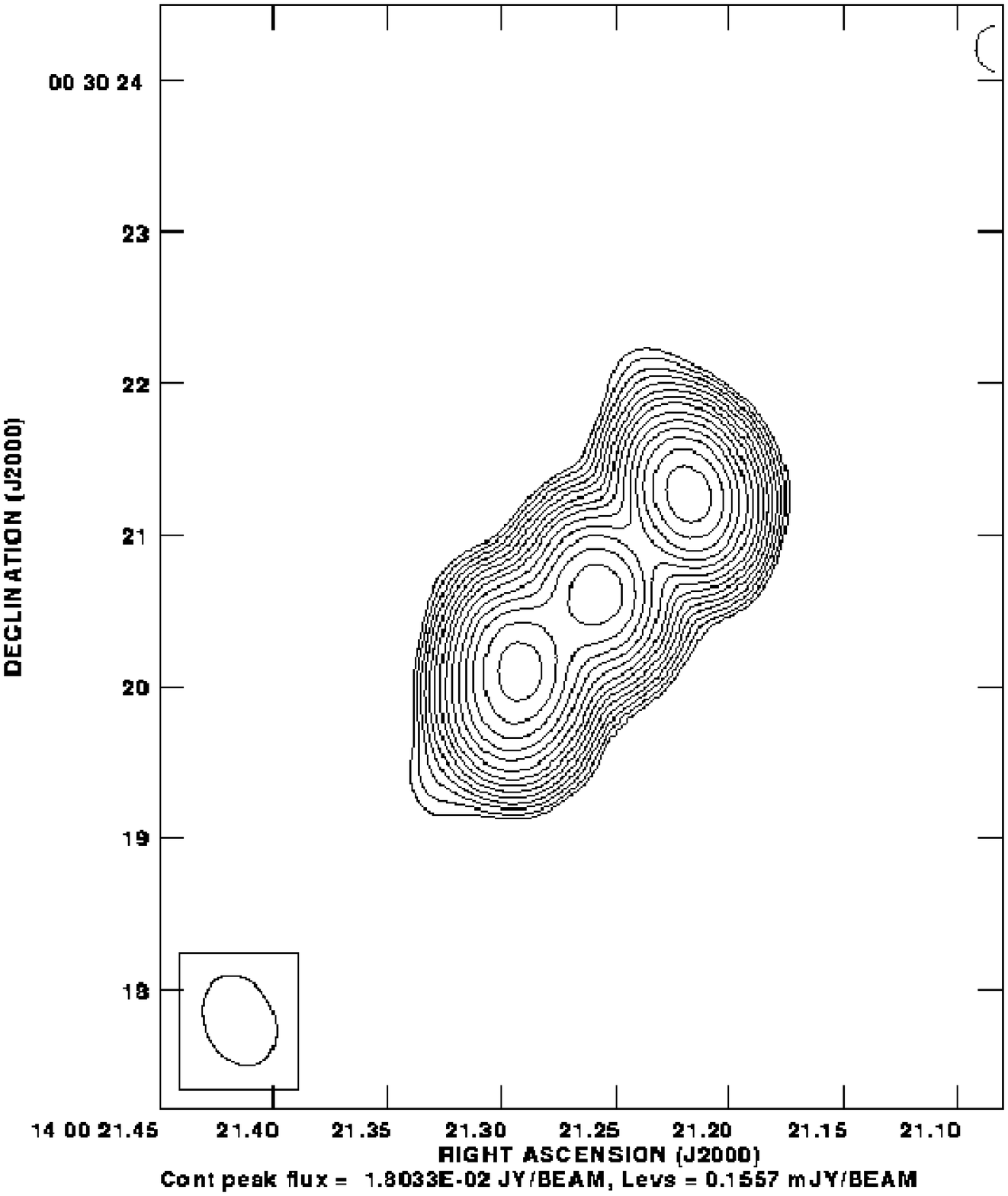,width=7.cm}
\psfig{figure=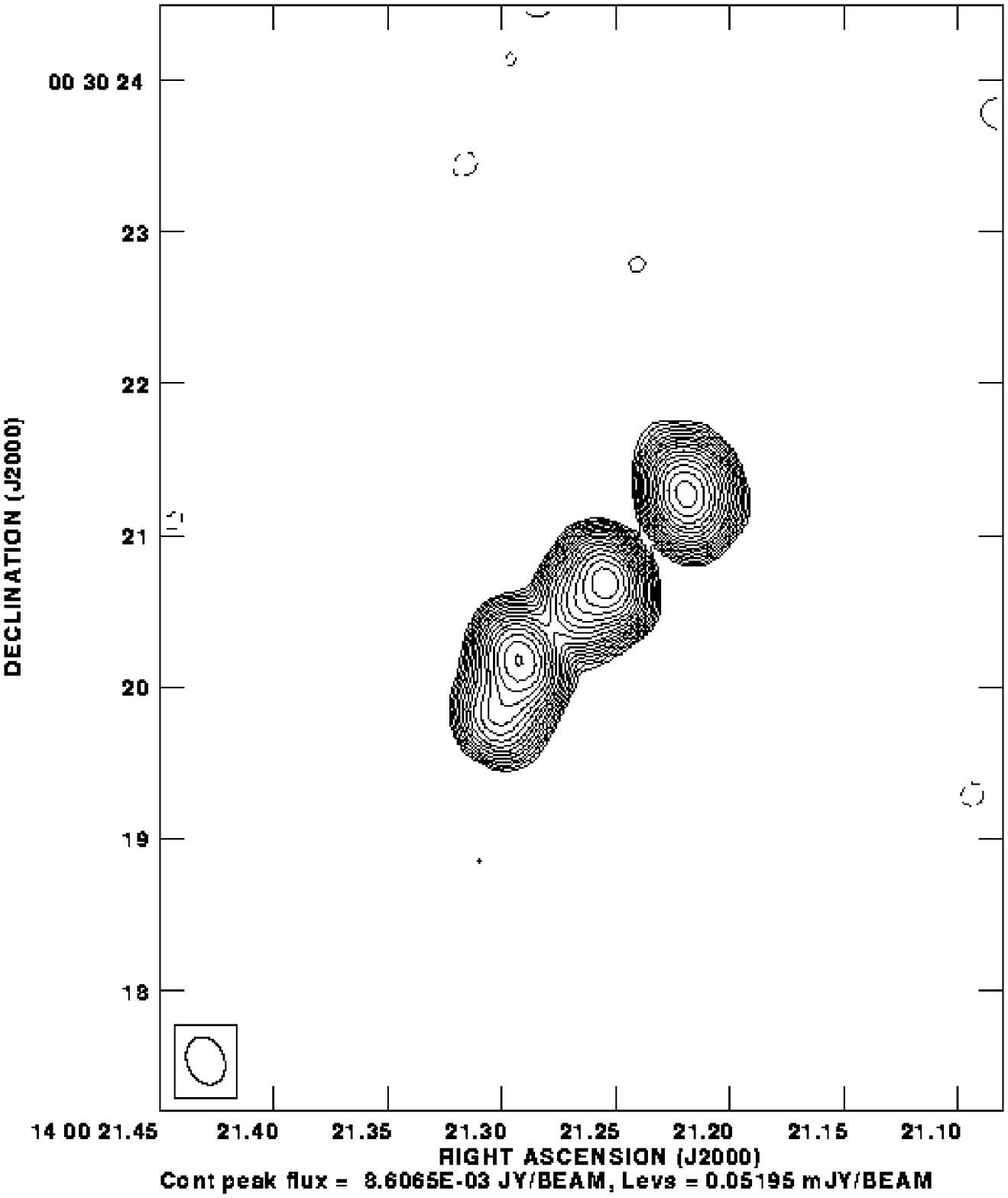,width=7.cm}}
\vskip-1cm\centerline{
\psfig{figure=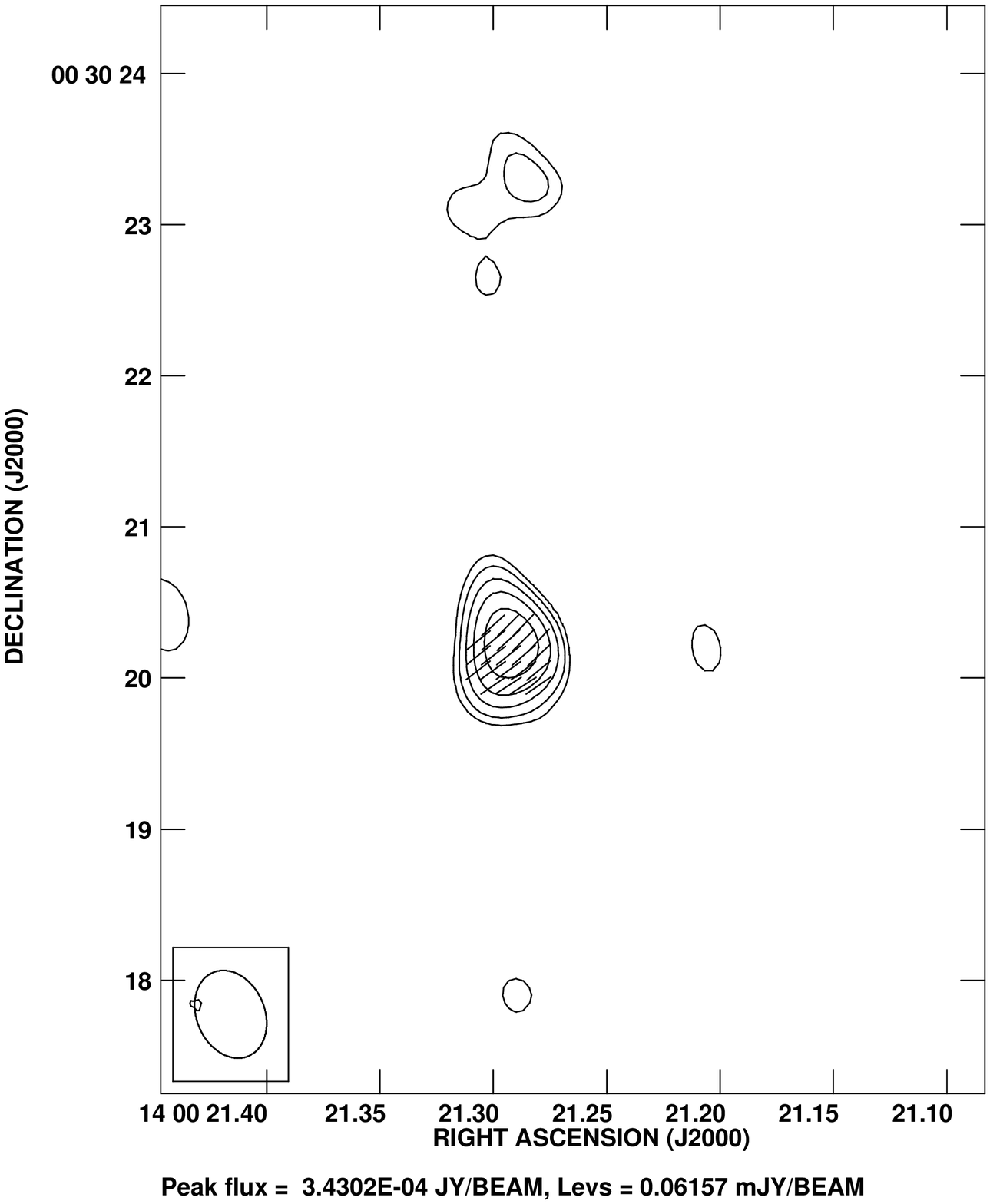,width=7.cm}
\psfig{figure=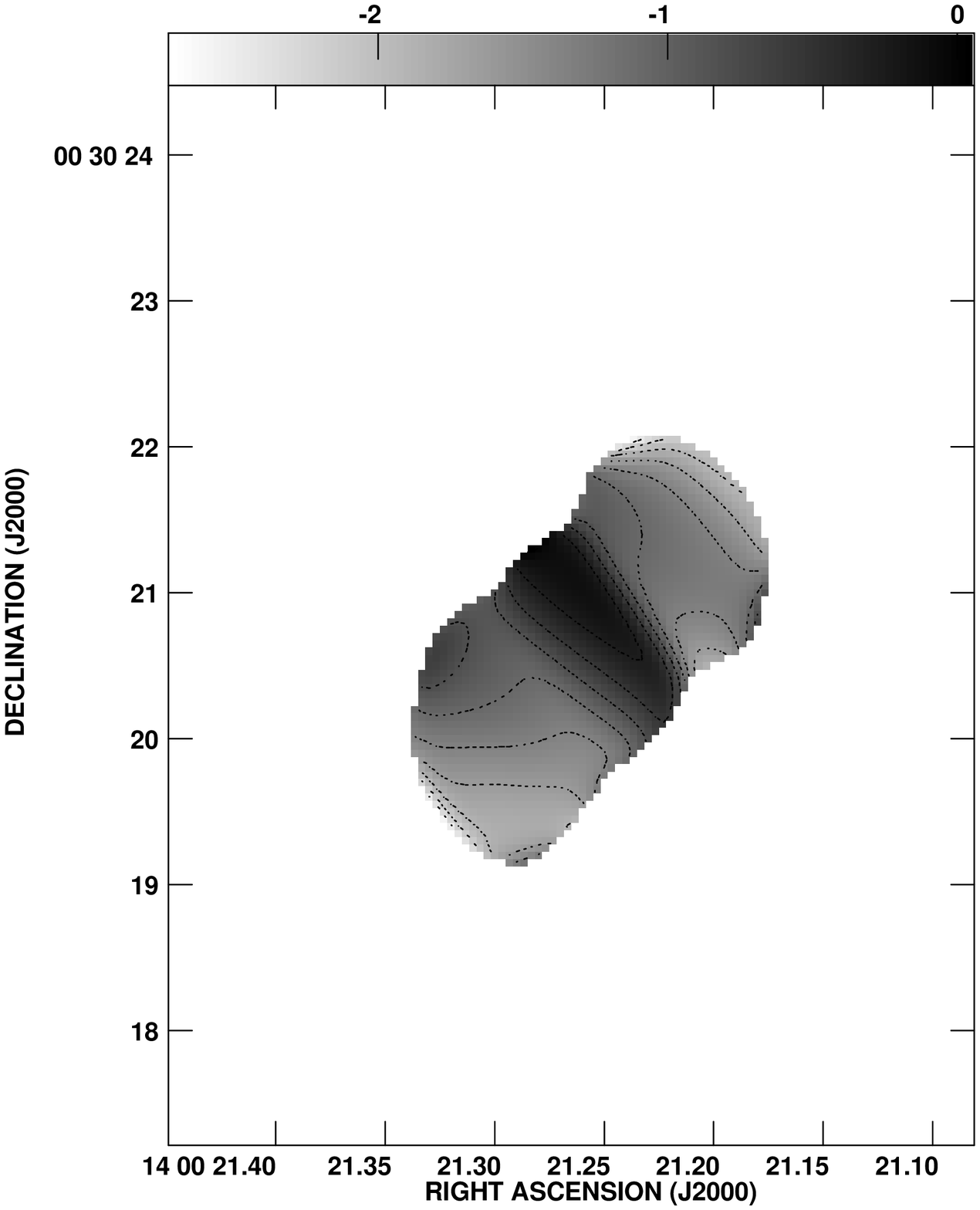,width=7.cm}}
\caption{Maps of the radio source 1357+007 at redshift z$=$2.671.
The sequence of figures is the same as in Fig. 6. The first contour level and
the  peak surface brightness  are respectively  0.156  mJy beam$^{-1}$ and 18
 mJy beam$^{-1}$ for
the 4.5 GHz map; 0.052  mJy beam$^{-1}$ and 8.6 mJy beam$^{-1}$ for the 8.2
GHz map; 0.062  mJy beam$^{-1}$  and 0.3  mJy beam$^{-1}$ for the 4.7 GHz
polarized intensity map.}
\end{figure*}
\begin{figure*}
\centerline{
\psfig{figure=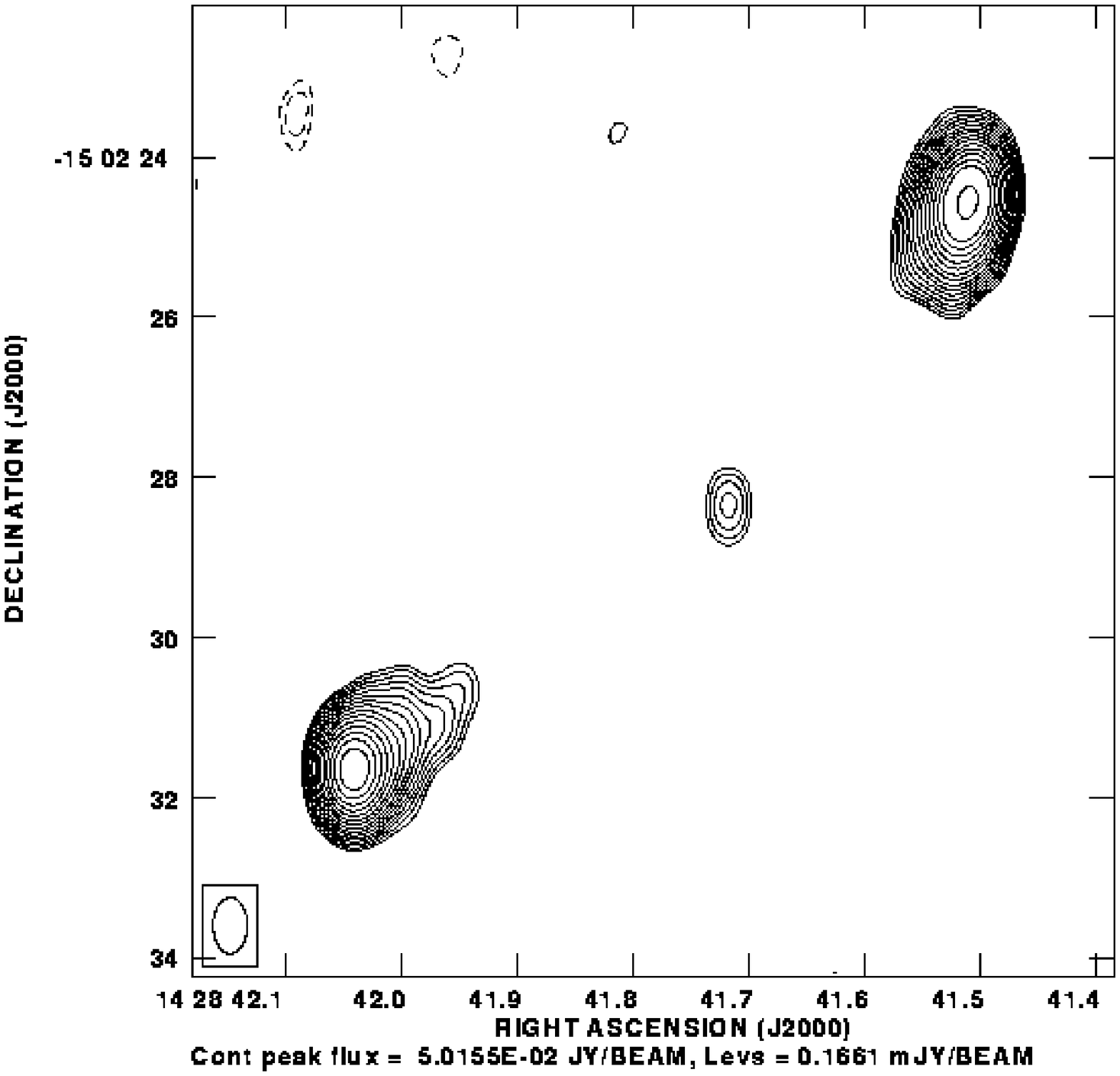,width=8.cm}
\psfig{figure=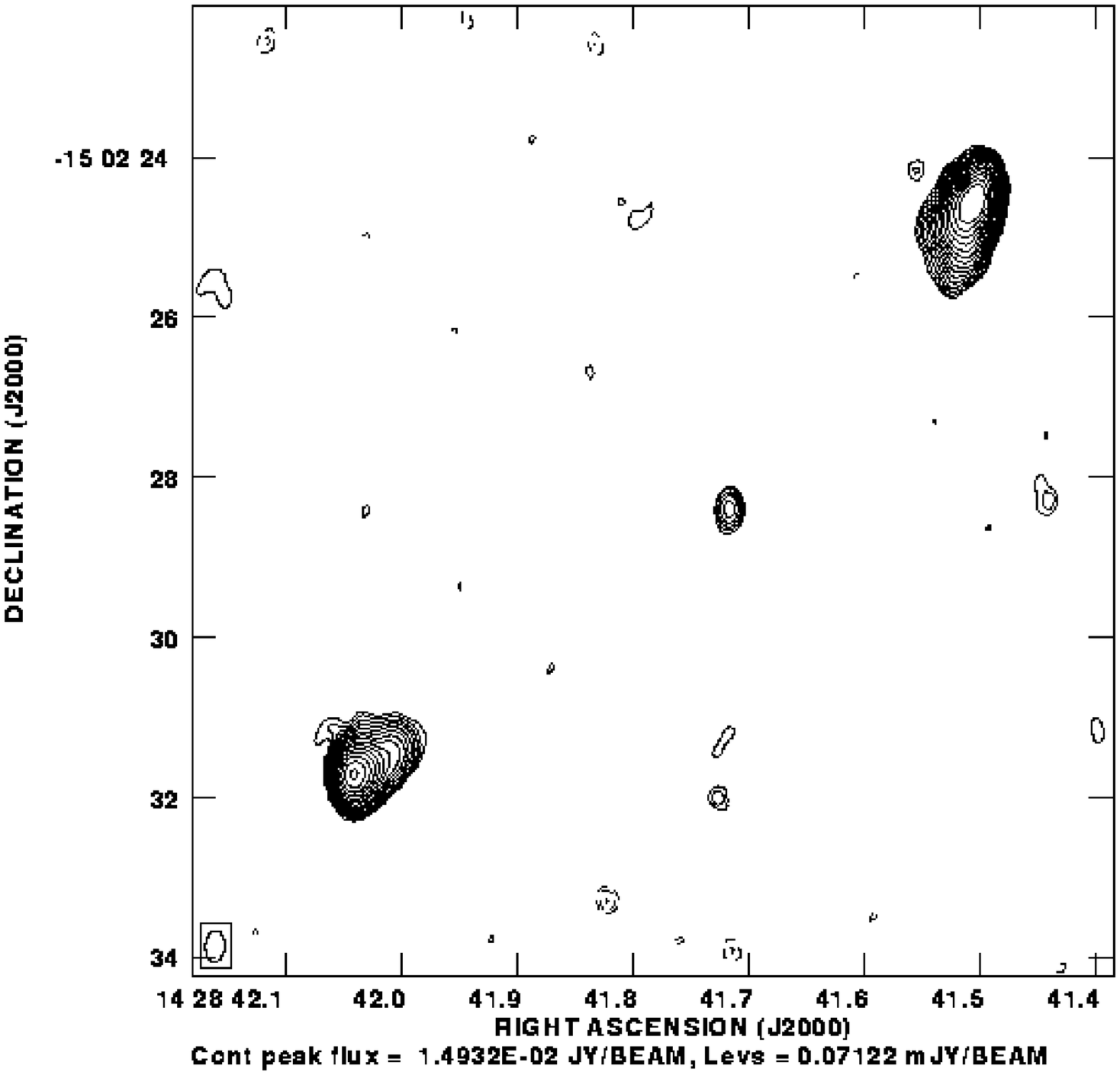,width=8.cm}}
\vskip-0.5cm\centerline{
\psfig{figure=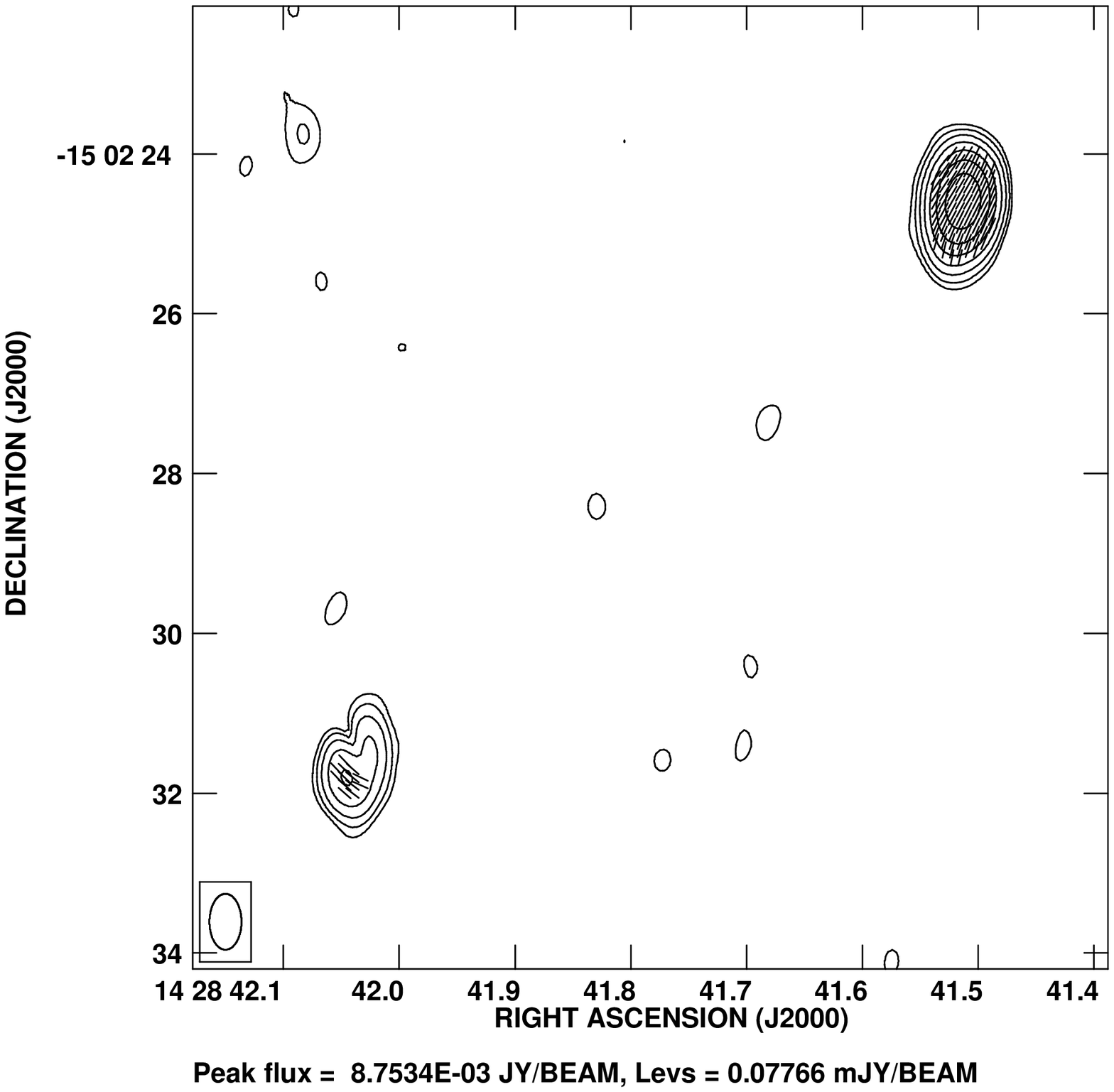,width=8.cm}
\psfig{figure=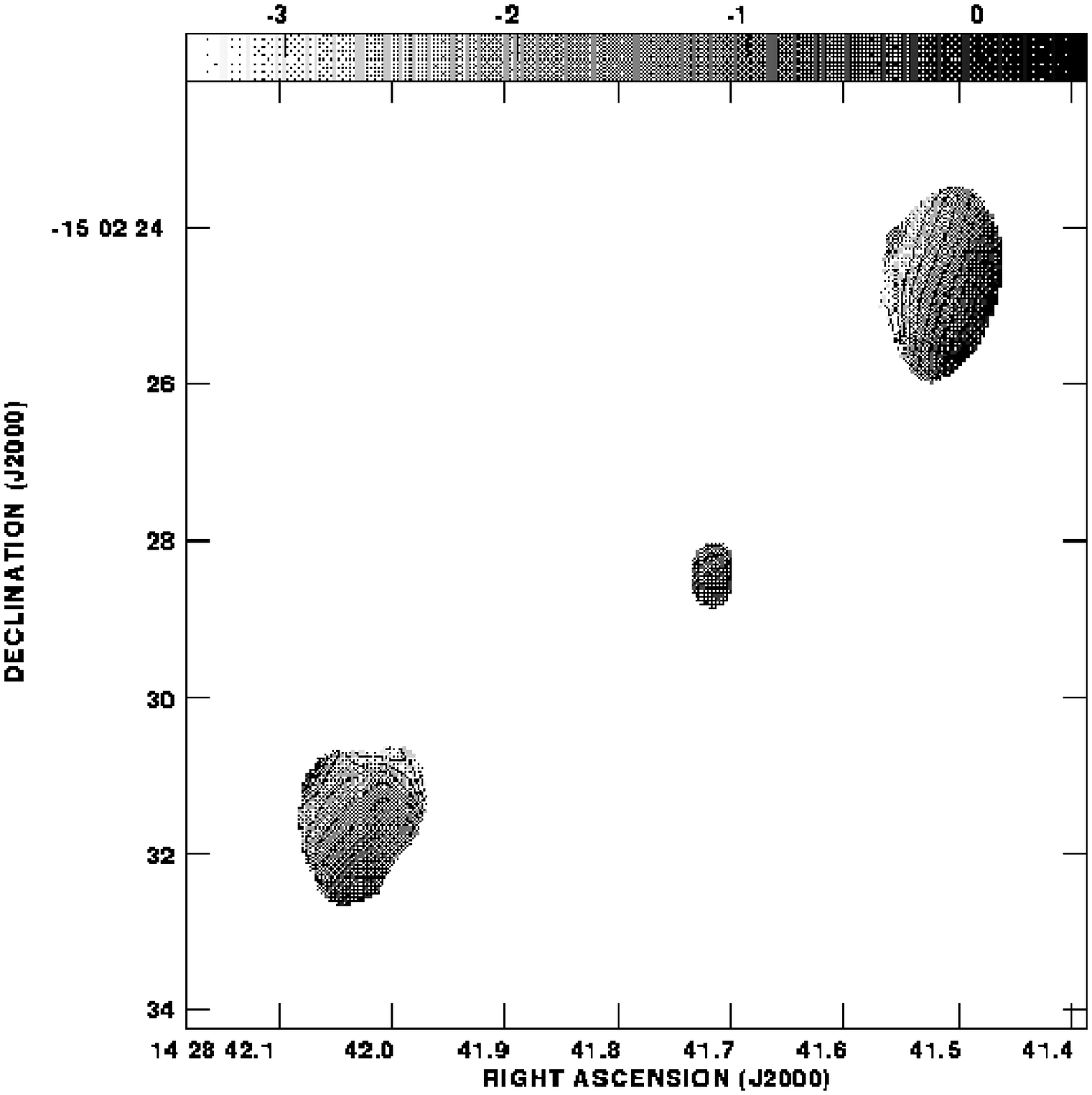,width=8.cm}}
\caption{Maps of the radio source 1425$-$148 at redshift z$=$2.355.
The sequence of figures is the same as in Fig. 6. The first contour level and
the  peak surface brightness  are respectively  0.166  mJy beam$^{-1}$ and 50
 mJy beam$^{-1}$ for
the 4.5 GHz map; 0.071  mJy beam$^{-1}$ and 15 mJy beam$^{-1}$ for the 8.2
GHz map; 0.078  mJy beam$^{-1}$  and 8.6  mJy beam$^{-1}$ for the 4.7 GHz
polarized intensity map.}
\end{figure*}

\begin{figure*}
\centerline{
\psfig{figure=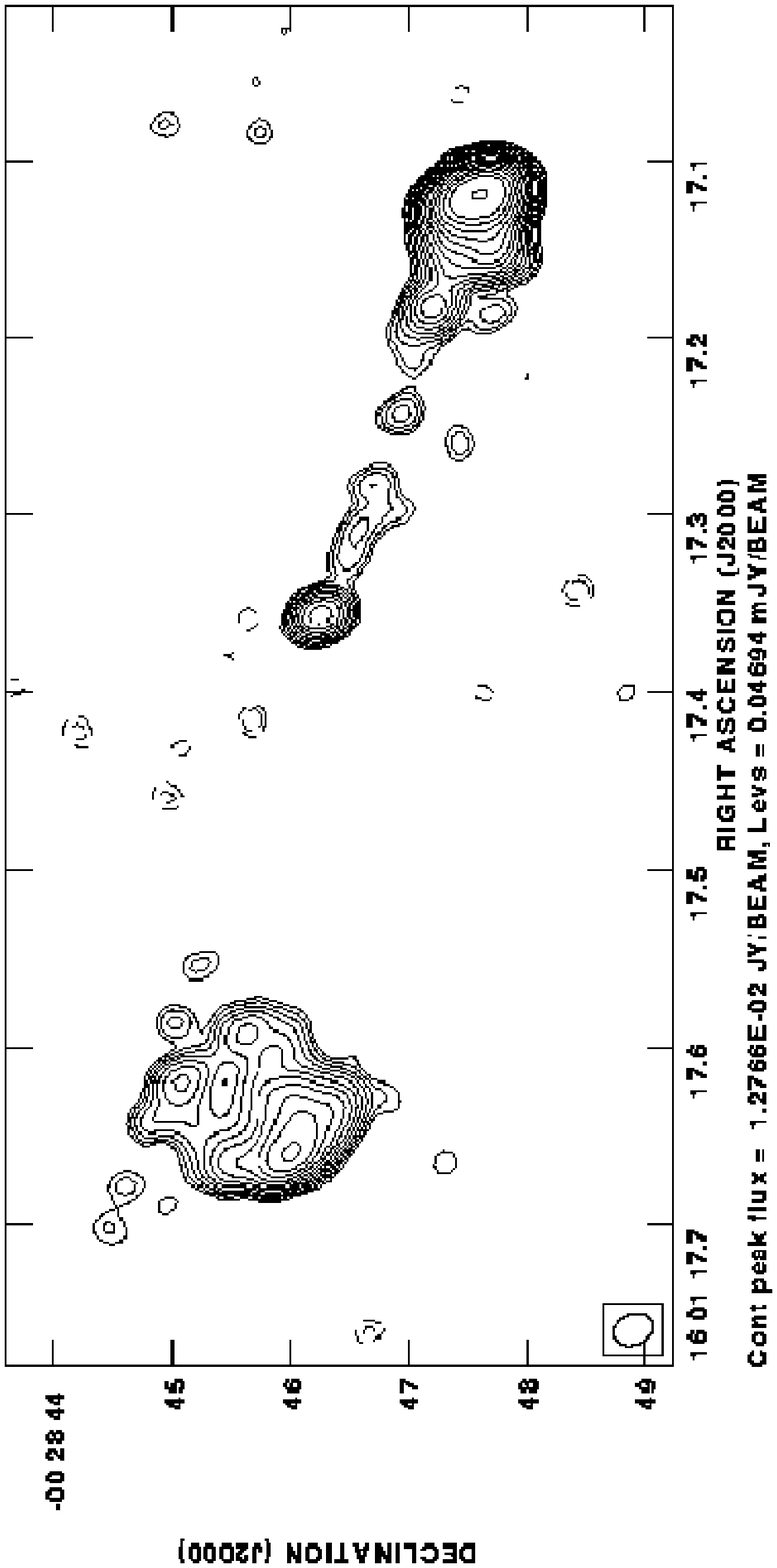,width=6.1cm}
\psfig{figure=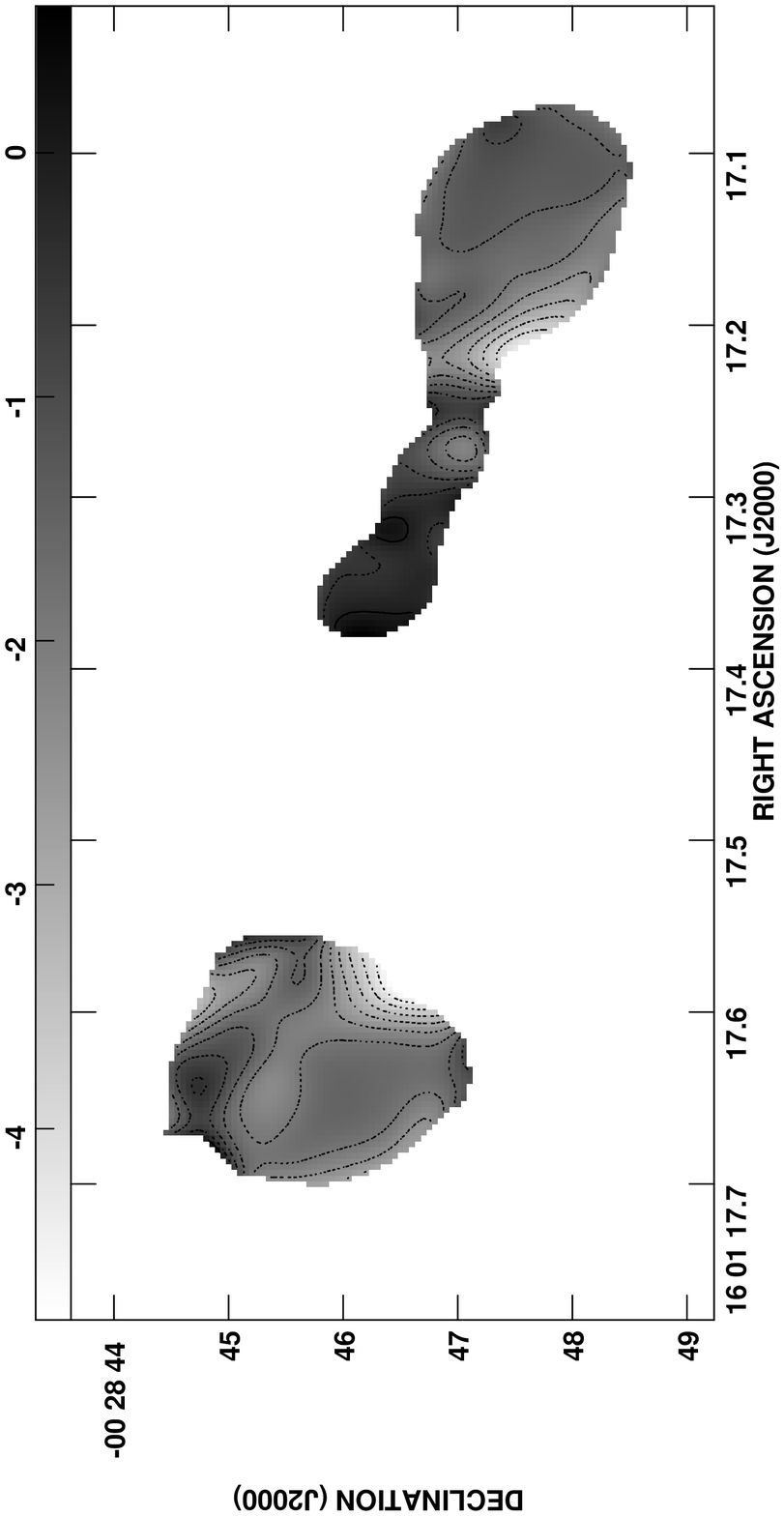,width=7.2cm}}
\vskip-0.5cm\centerline{
\psfig{figure=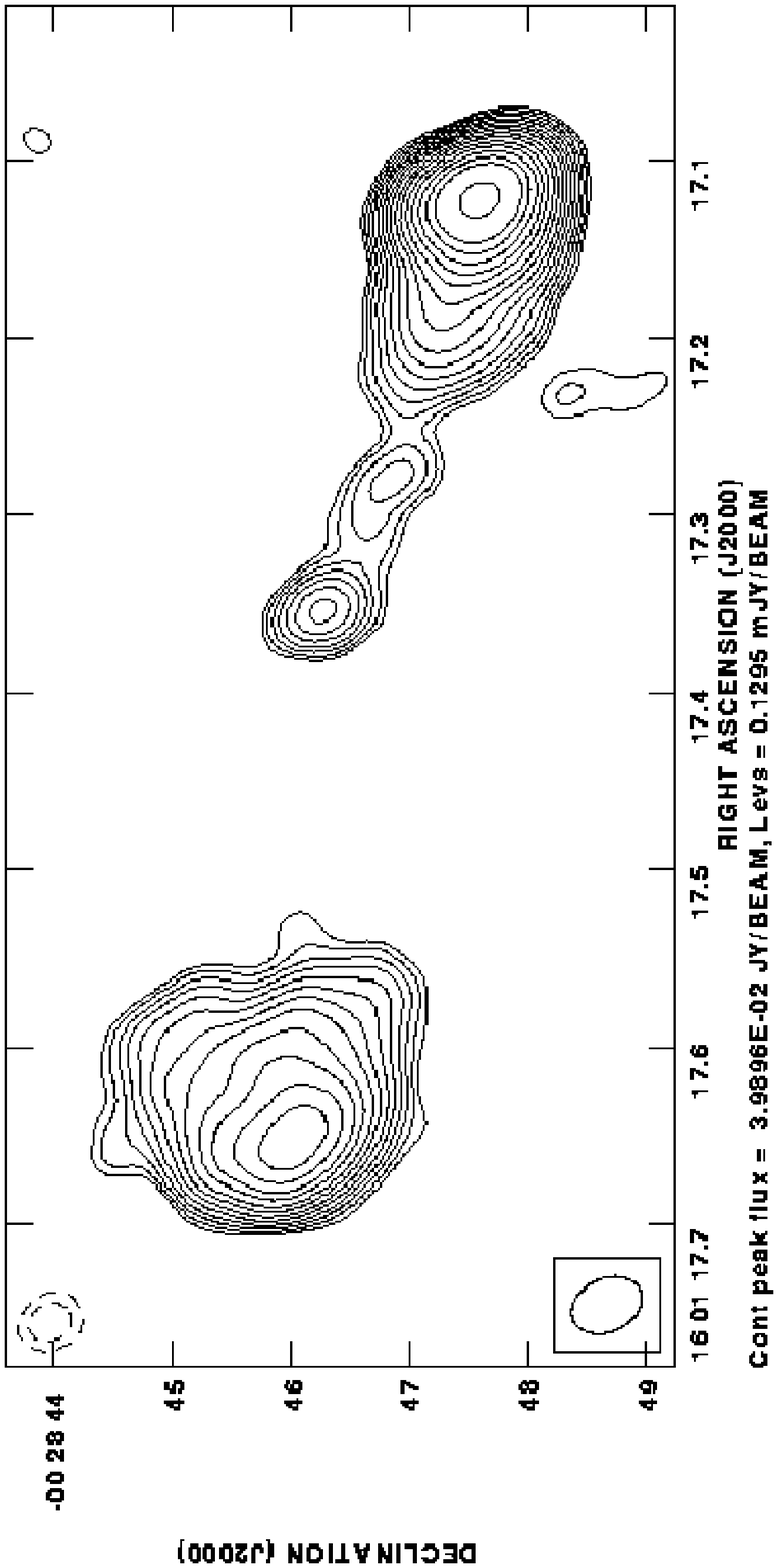,width=6.cm}
\psfig{figure=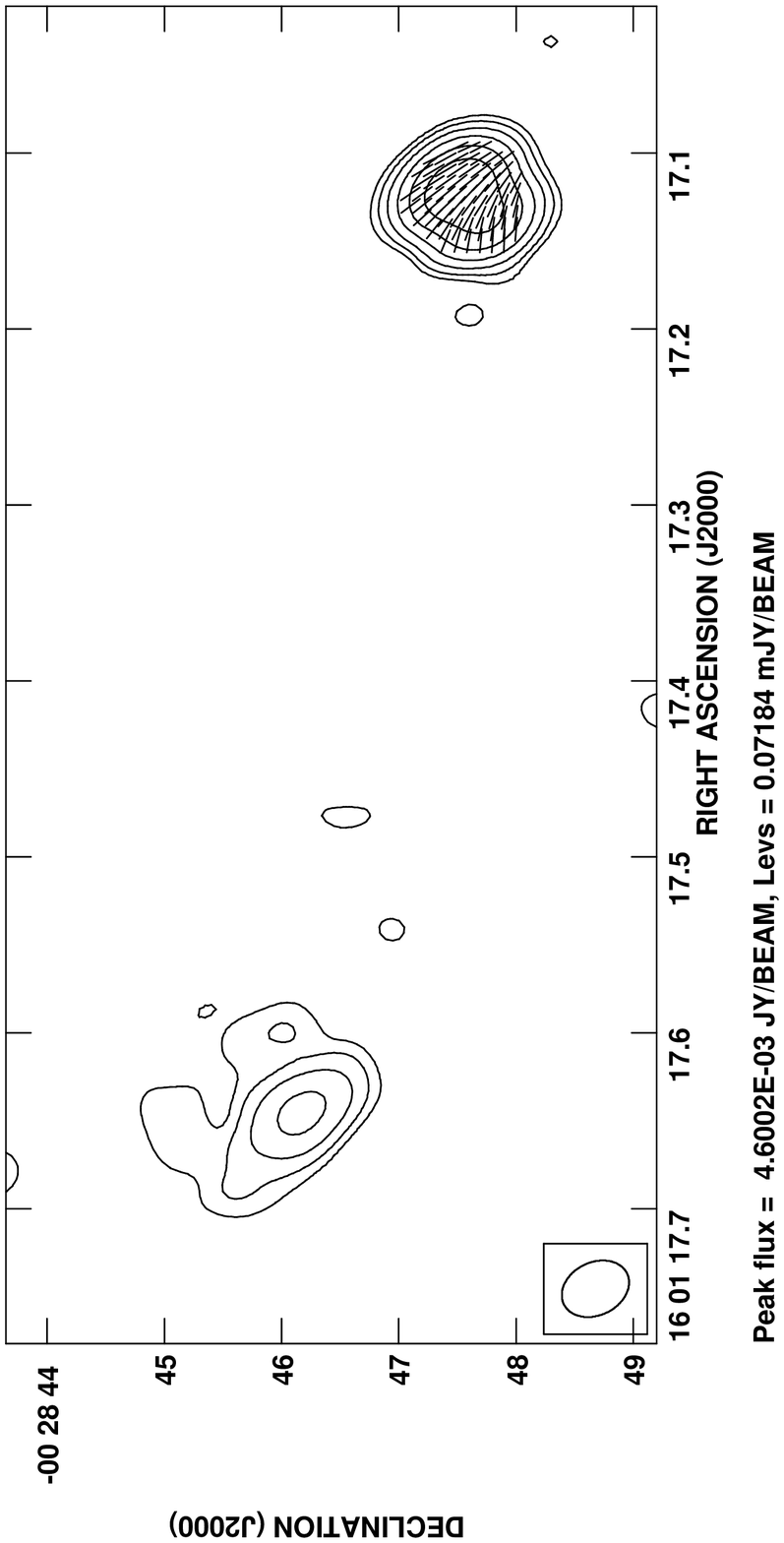,width=6.5cm}}
\caption{Maps of the radio source 1558$-$003 at redshift z$=$2.52.
The sequence of figures is the same as in Fig. 6. The first contour level and
the  peak surface brightness  are respectively  0.129  mJy beam$^{-1}$ and 40
 mJy beam$^{-1}$ for
the 4.5 GHz map; 0.047  mJy beam$^{-1}$ and 13 mJy beam$^{-1}$ for the 8.2
GHz map; 0.072  mJy beam$^{-1}$  and 4.6  mJy beam$^{-1}$ for the 4.7 GHz
polarized intensity map.}
\end{figure*}

\begin{figure*}
\centerline{
\psfig{figure=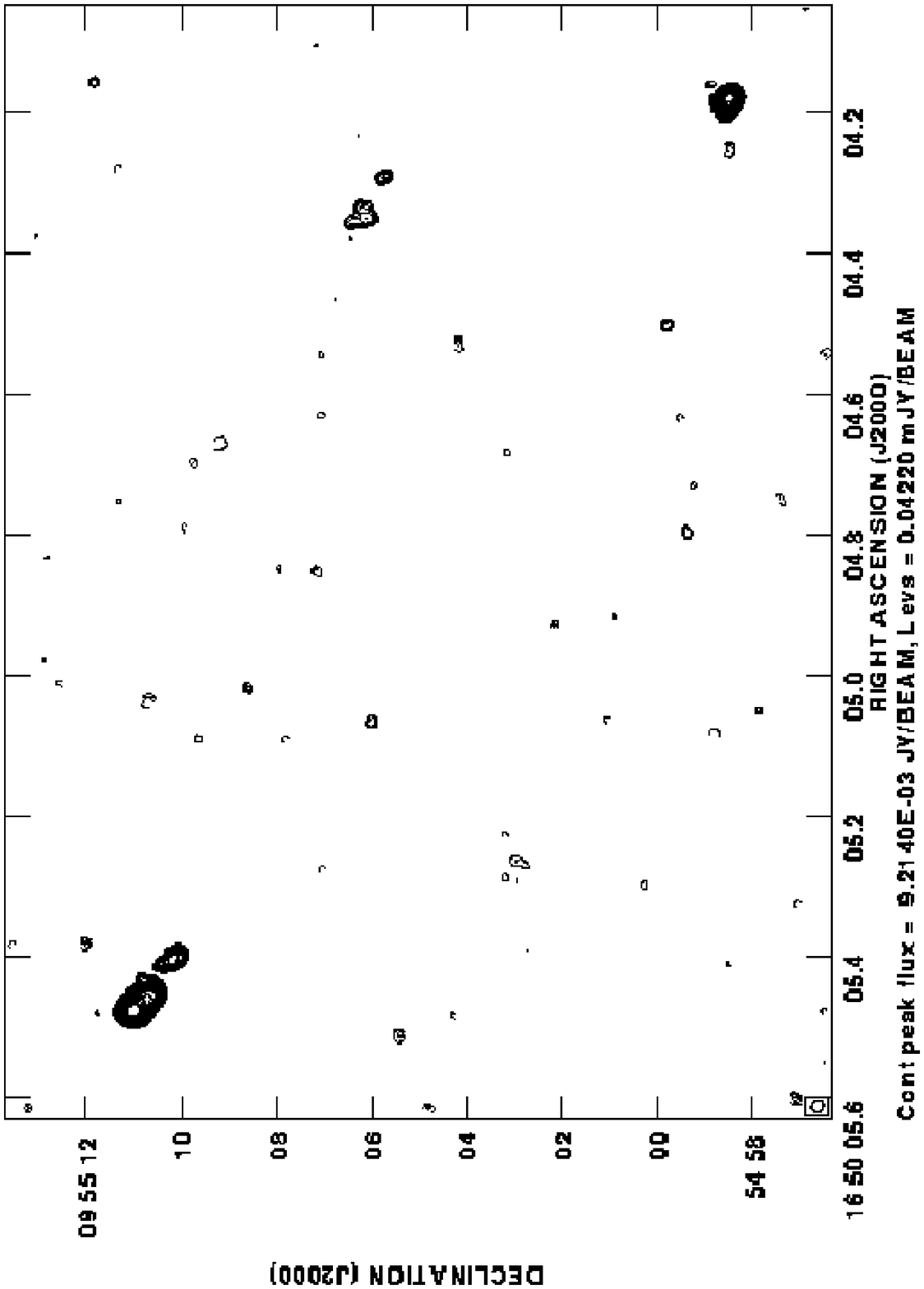,width=7.cm}
\psfig{figure=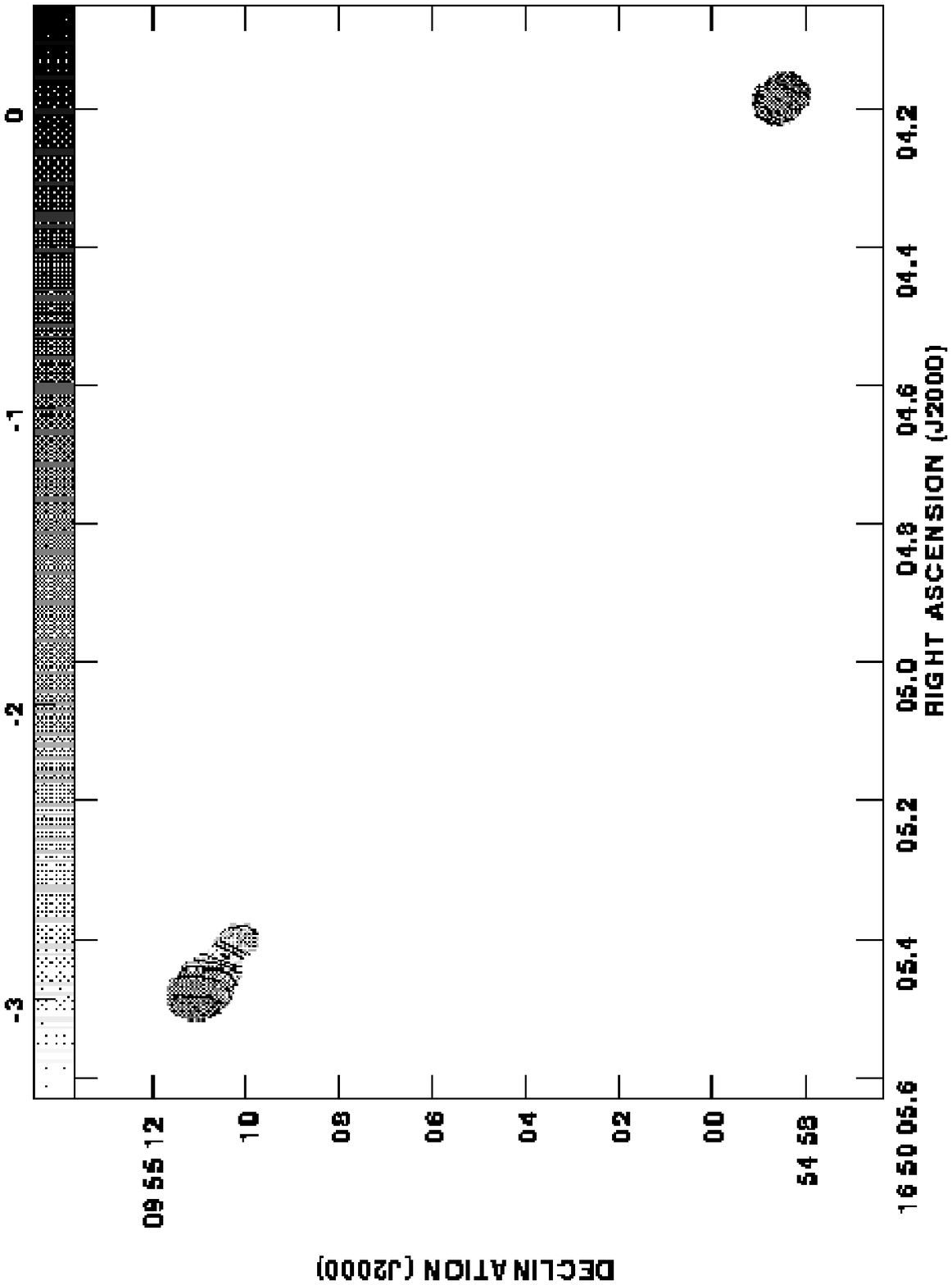,width=7.5cm}}
\centerline{
\psfig{figure=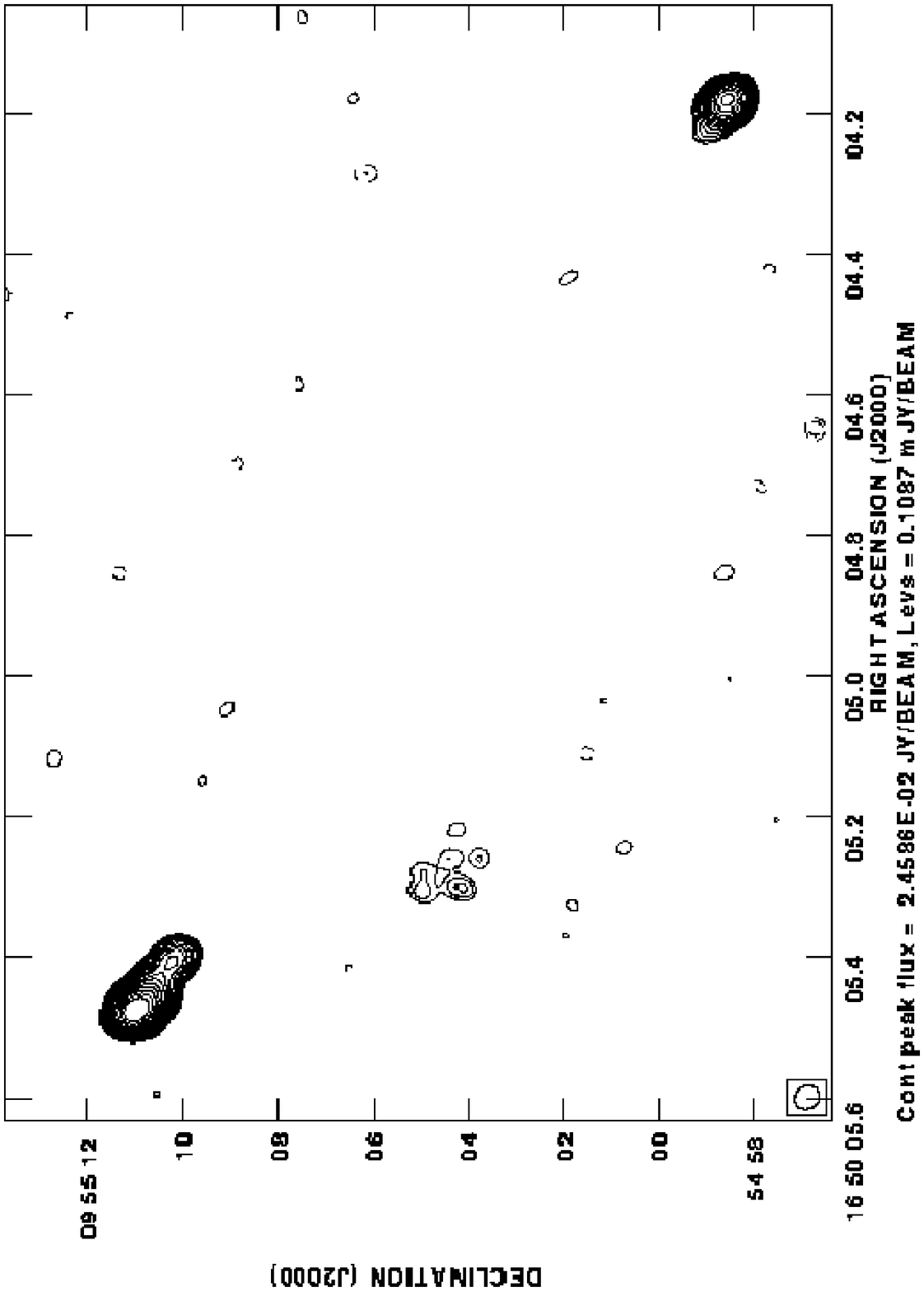,width=7.cm}
\psfig{figure=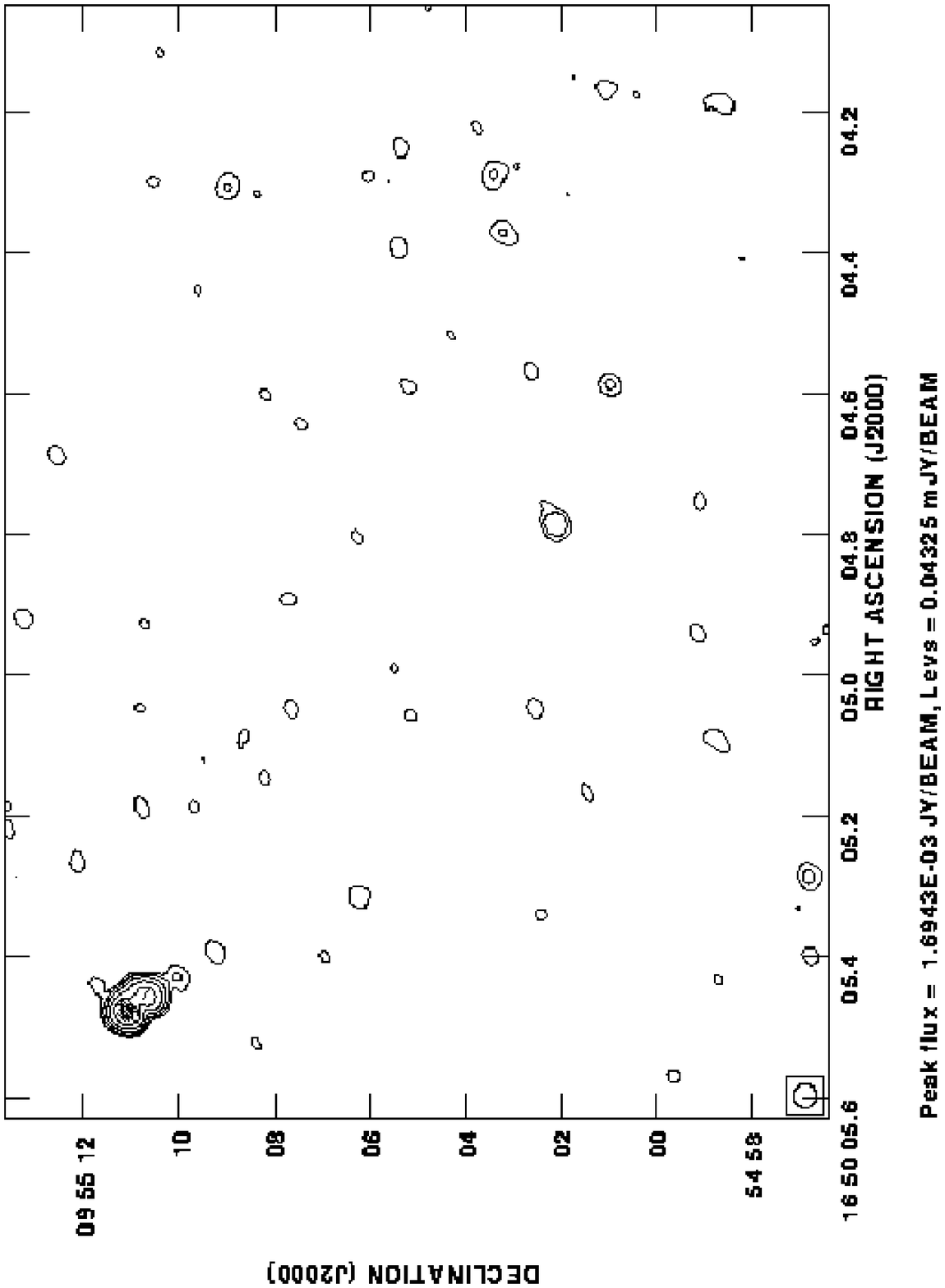,width=7.cm}}
\caption{Maps of the radio source 1647+100 at redshift z$=$2.509.
The sequence of figures is the same as in Fig. 6. The first contour level and
the  peak surface brightness  are respectively  0.109  mJy beam$^{-1}$ and 24.
 mJy beam$^{-1}$ for
the 4.5 GHz map; 0.042  mJy beam$^{-1}$ and 9.2 mJy beam$^{-1}$ for the 8.2
GHz map; 0.043  mJy beam$^{-1}$  and 1.7  mJy beam$^{-1}$ for the 4.7 GHz
polarized intensity map.}  
\end{figure*}
\clearpage
\begin{figure*}
\centerline{
\psfig{figure=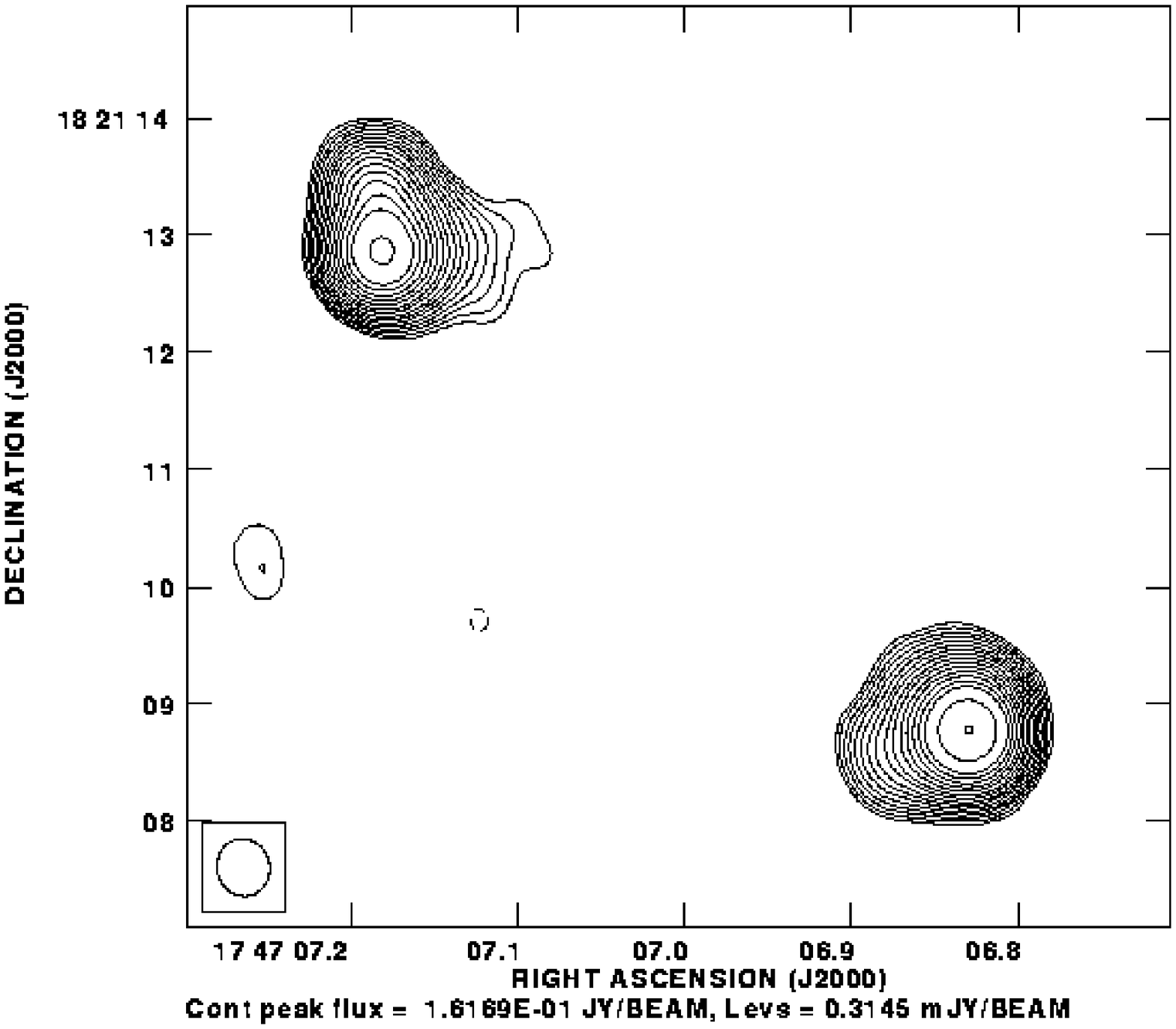,width=8.cm}
\psfig{figure=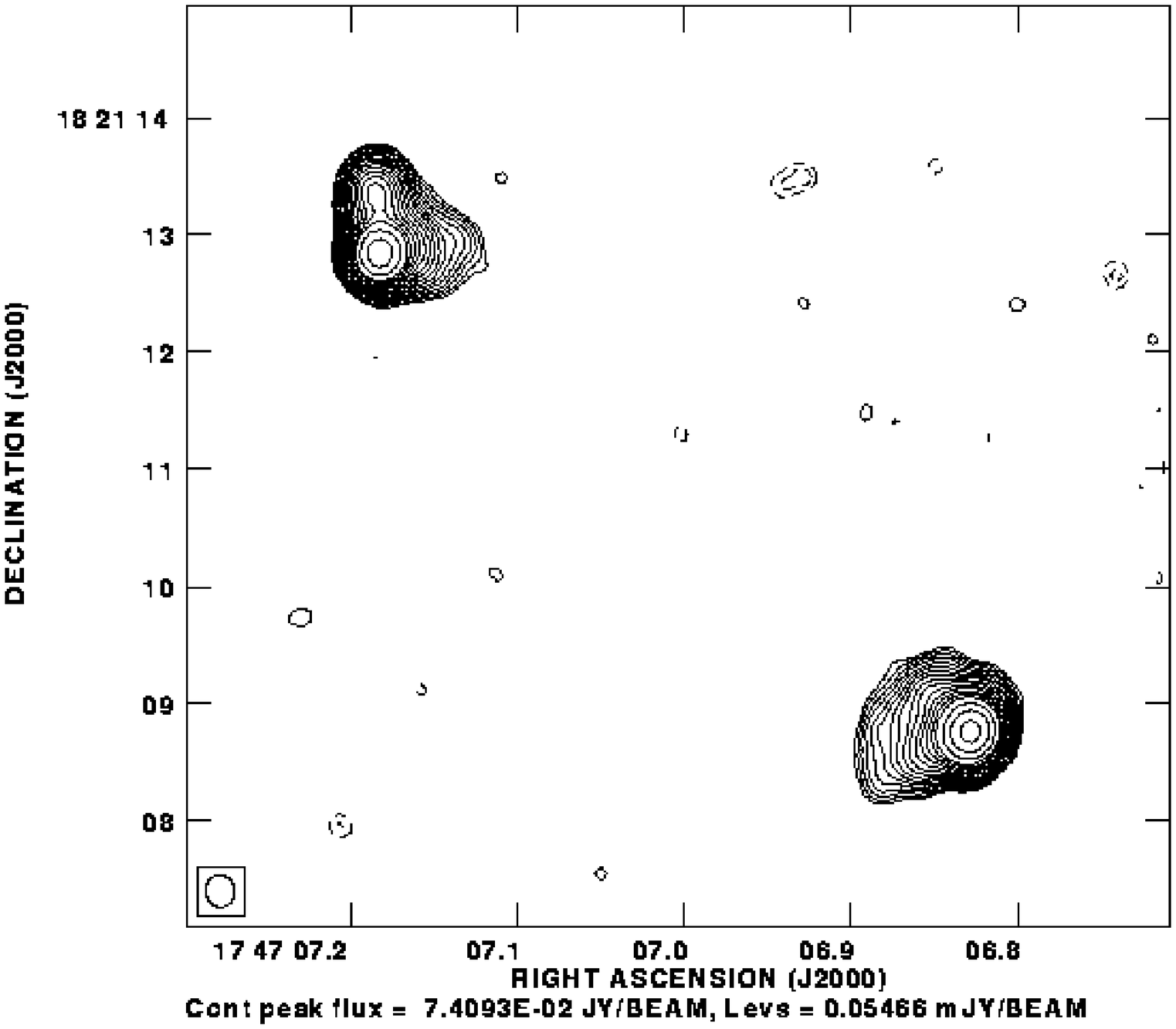,width=8.cm}}
\vskip-0.6cm\centerline{
\psfig{figure=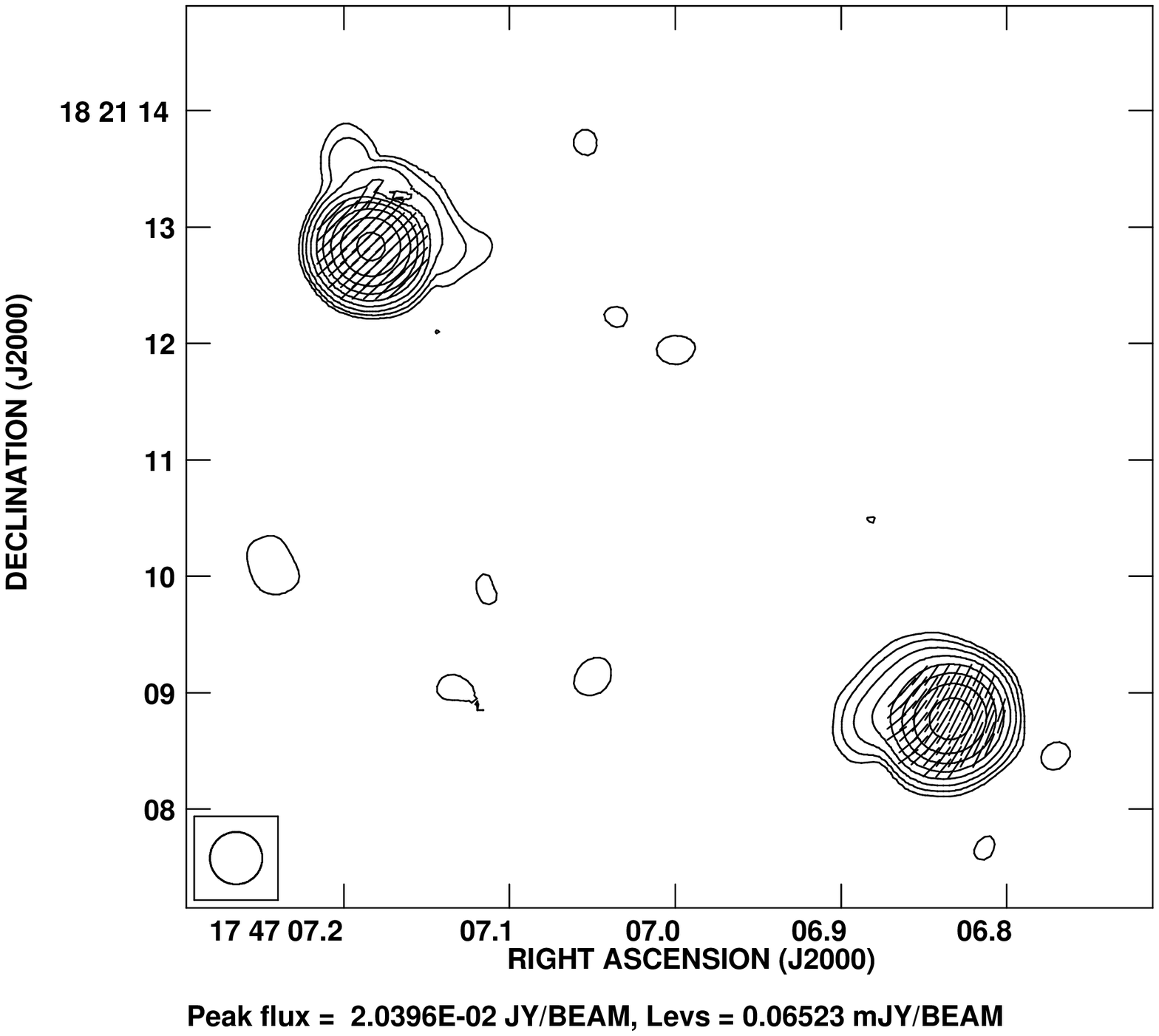,width=8.cm}
\psfig{figure=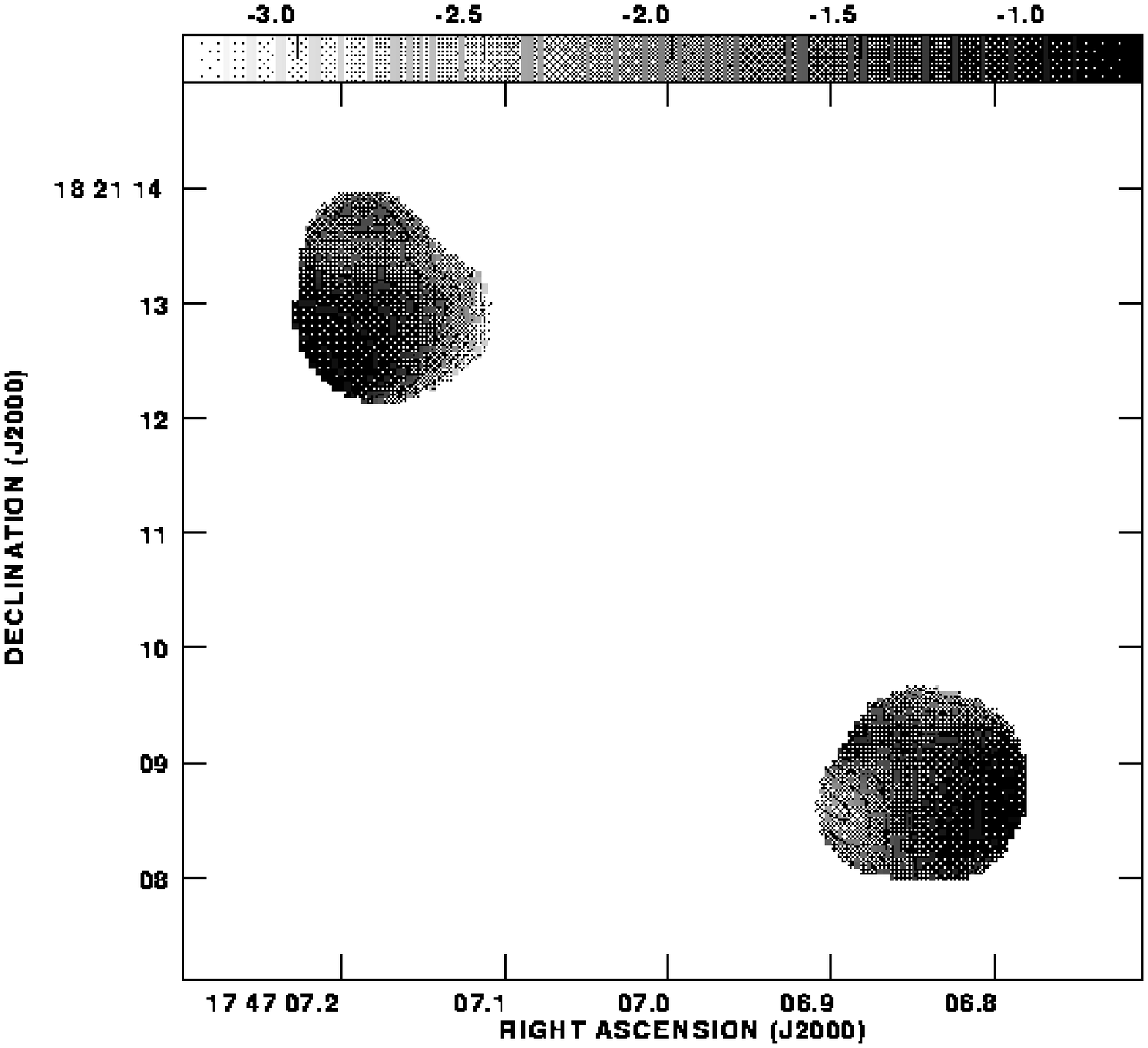,width=8.cm}}
\caption{Maps of the radio source 1747+182 at redshift z$=$2.281.
The sequence of figures is the same as in Fig. 6. The first contour level and
the  peak surface brightness  are respectively  0.314  mJy beam$^{-1}$ and 162
 mJy beam$^{-1}$ for
the 4.5 GHz map; 0.055  mJy beam$^{-1}$ and 74 mJy beam$^{-1}$ for the 8.2
GHz map; 0.065  mJy beam$^{-1}$  and 20.4  mJy beam$^{-1}$ for the 4.7 GHz
polarized intensity map.} 
\end{figure*}
\begin{figure*}
\centerline{
\psfig{figure=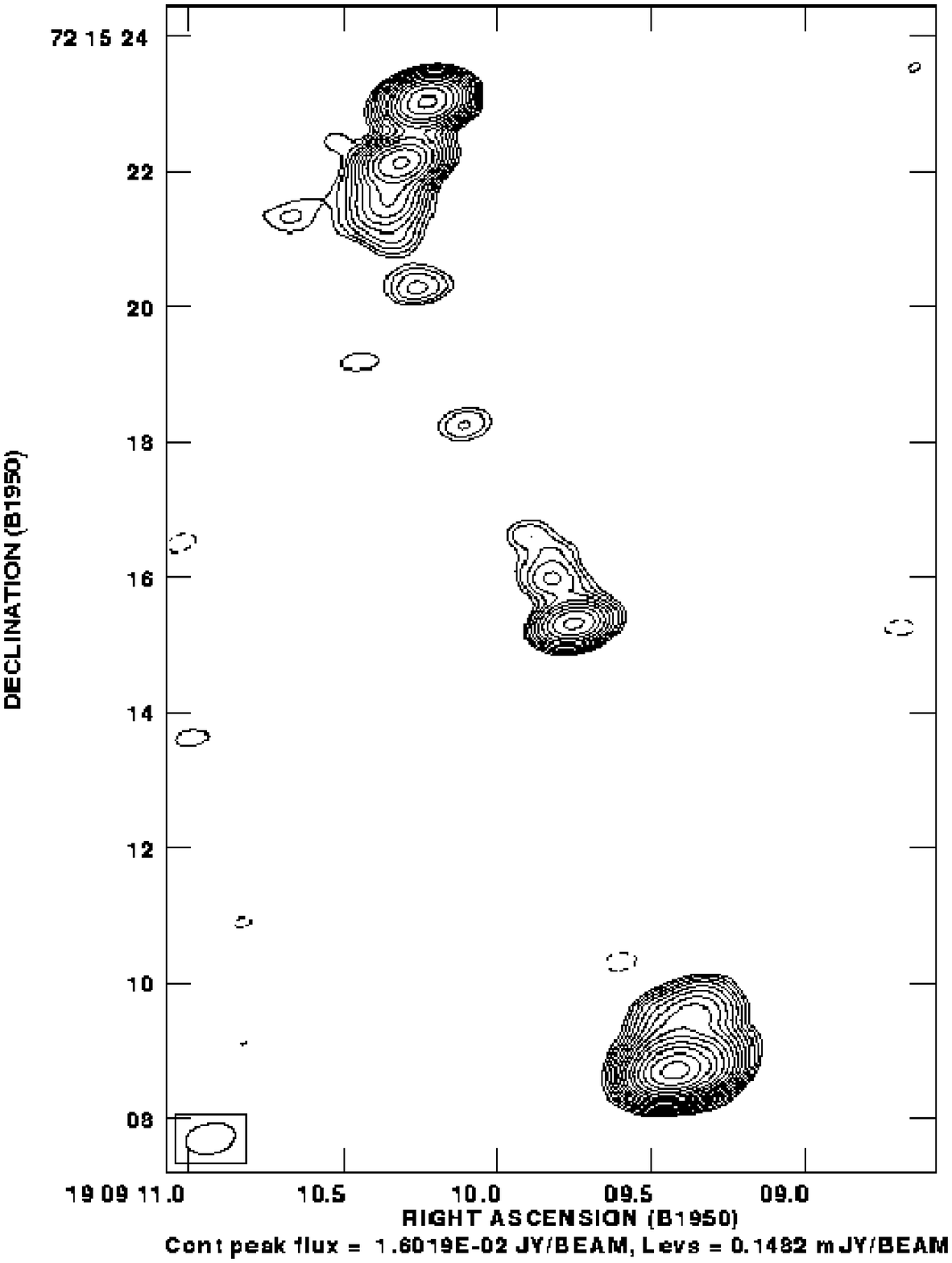,width=7.cm}
\psfig{figure=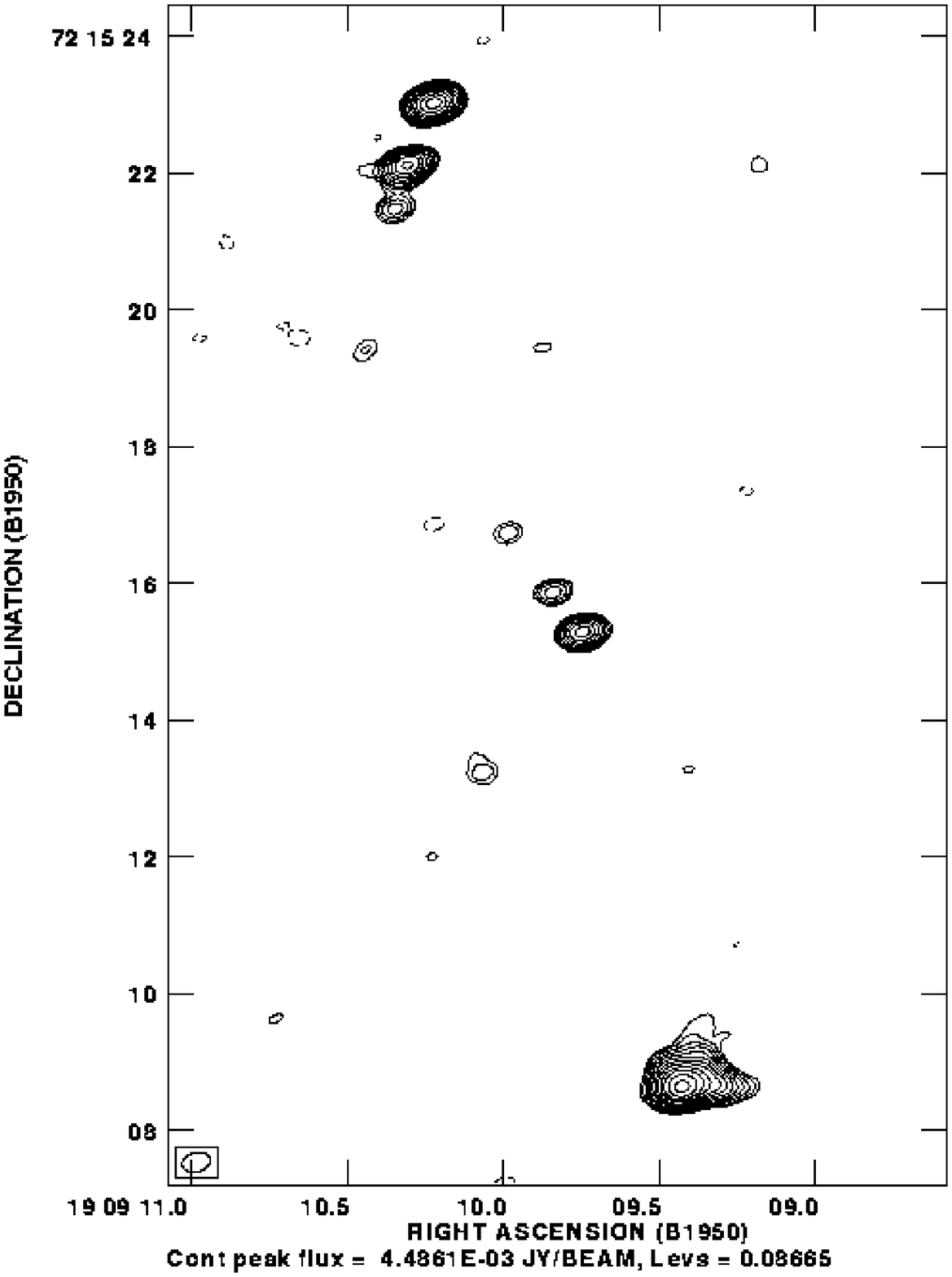,width=7.cm}}
\vskip-0.6cm\centerline{
\psfig{figure=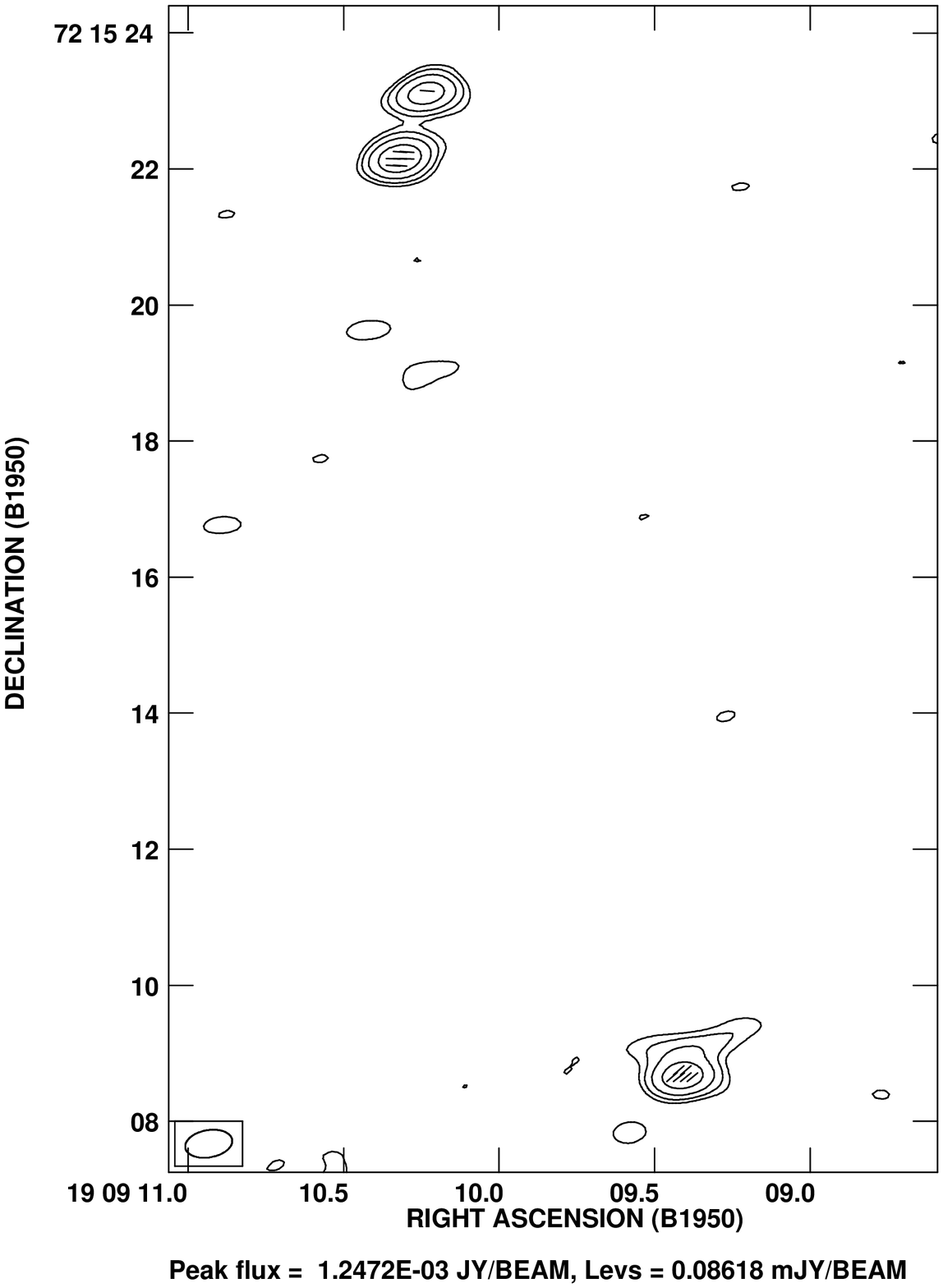,width=7.cm}
\psfig{figure=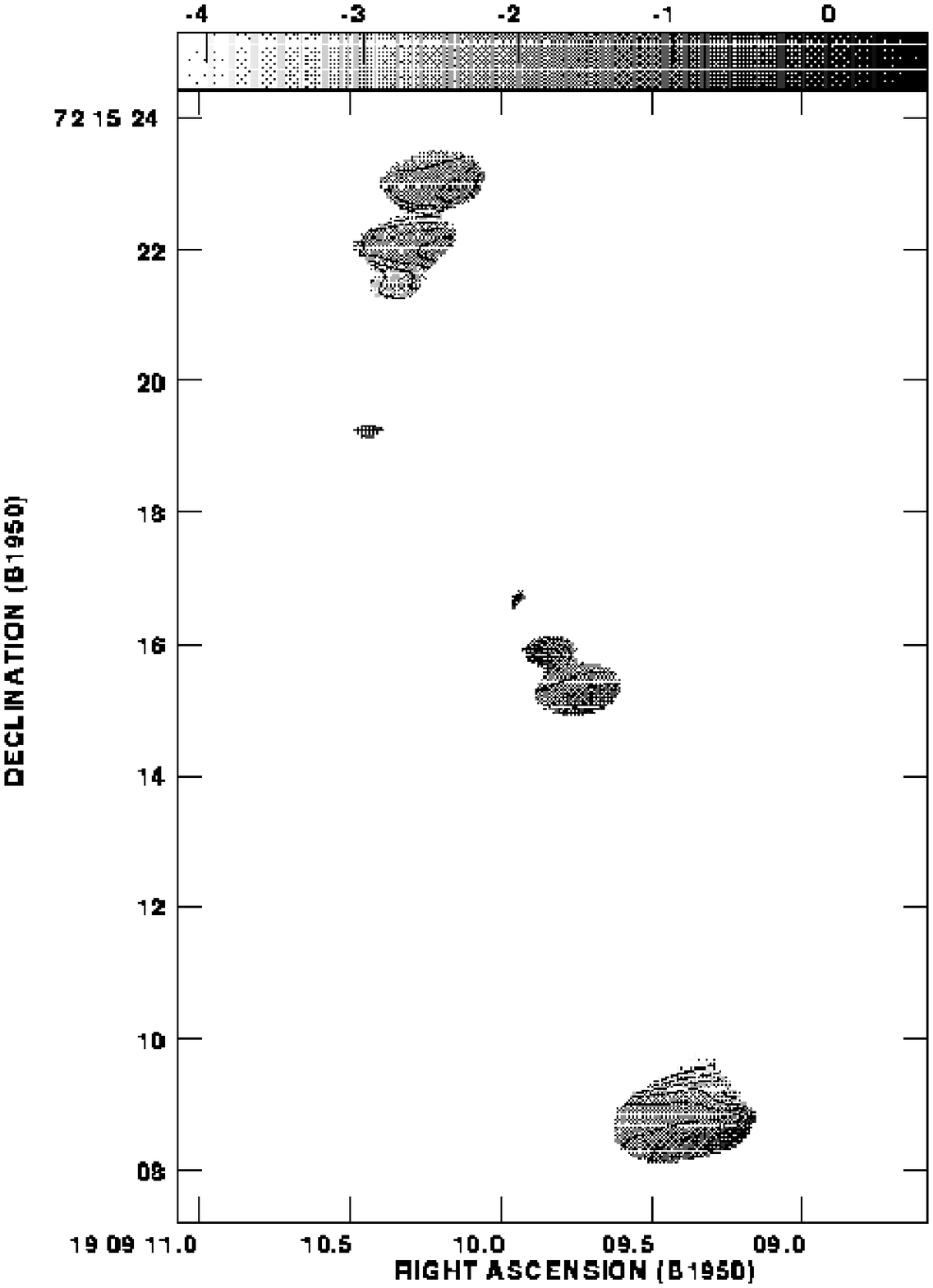,width=7.cm}}
\caption{Maps of the radio source 1908+722 at redshift z$=$3.537.
The sequence of figures is the same as in Fig. 6. The first contour level and
the  peak surface brightness  are respectively  0.148  mJy beam$^{-1}$ and 16
 mJy beam$^{-1}$ for
the 4.5 GHz map; 0.087  mJy beam$^{-1}$ and 4.5 mJy beam$^{-1}$ for the 8.2
GHz map; 0.086  mJy beam$^{-1}$  and 1.2  mJy beam$^{-1}$ for the 4.7 GHz
polarized intensity map.} 
\end{figure*}
\begin{figure*}
\centerline{
\psfig{figure=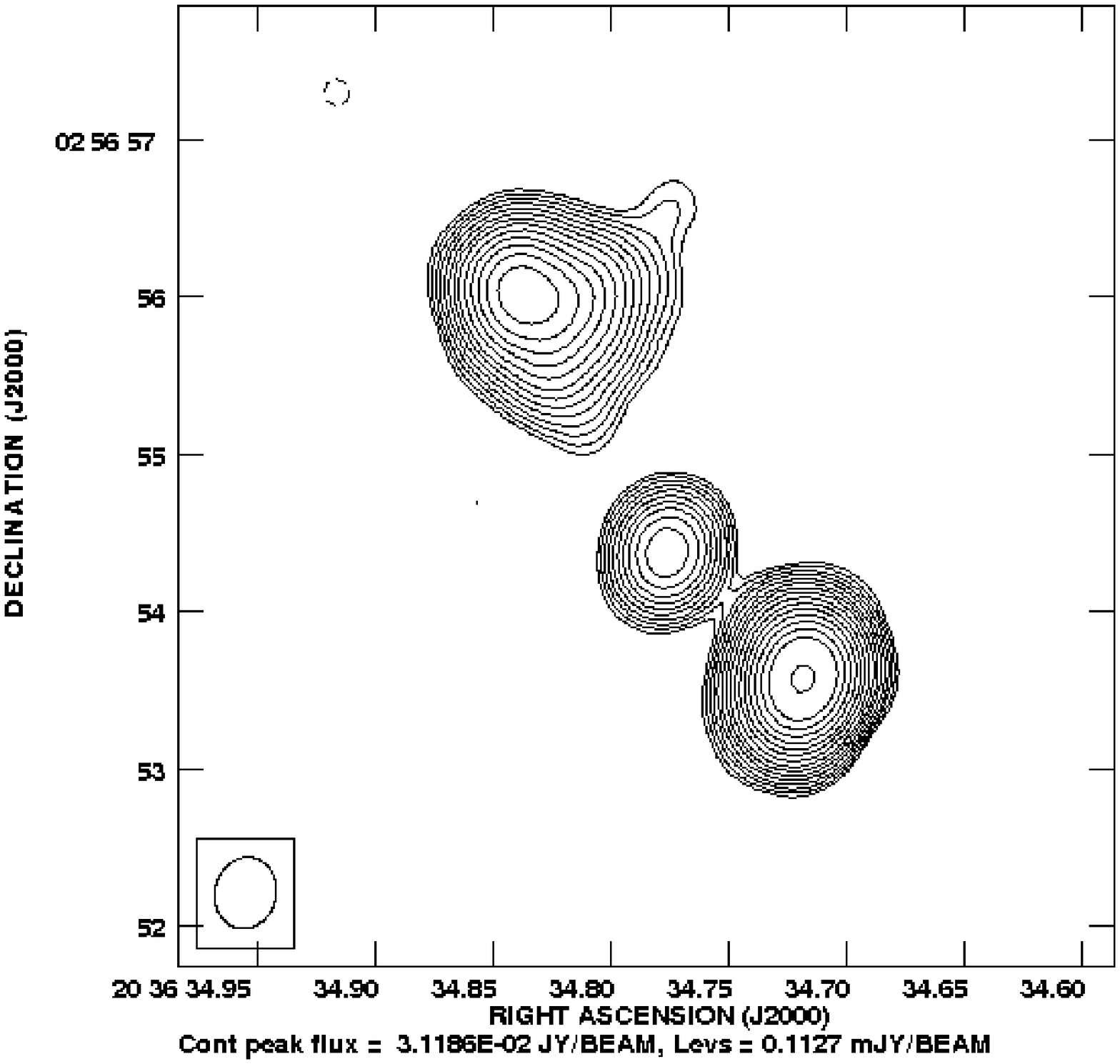,width=7.5cm}
\psfig{figure=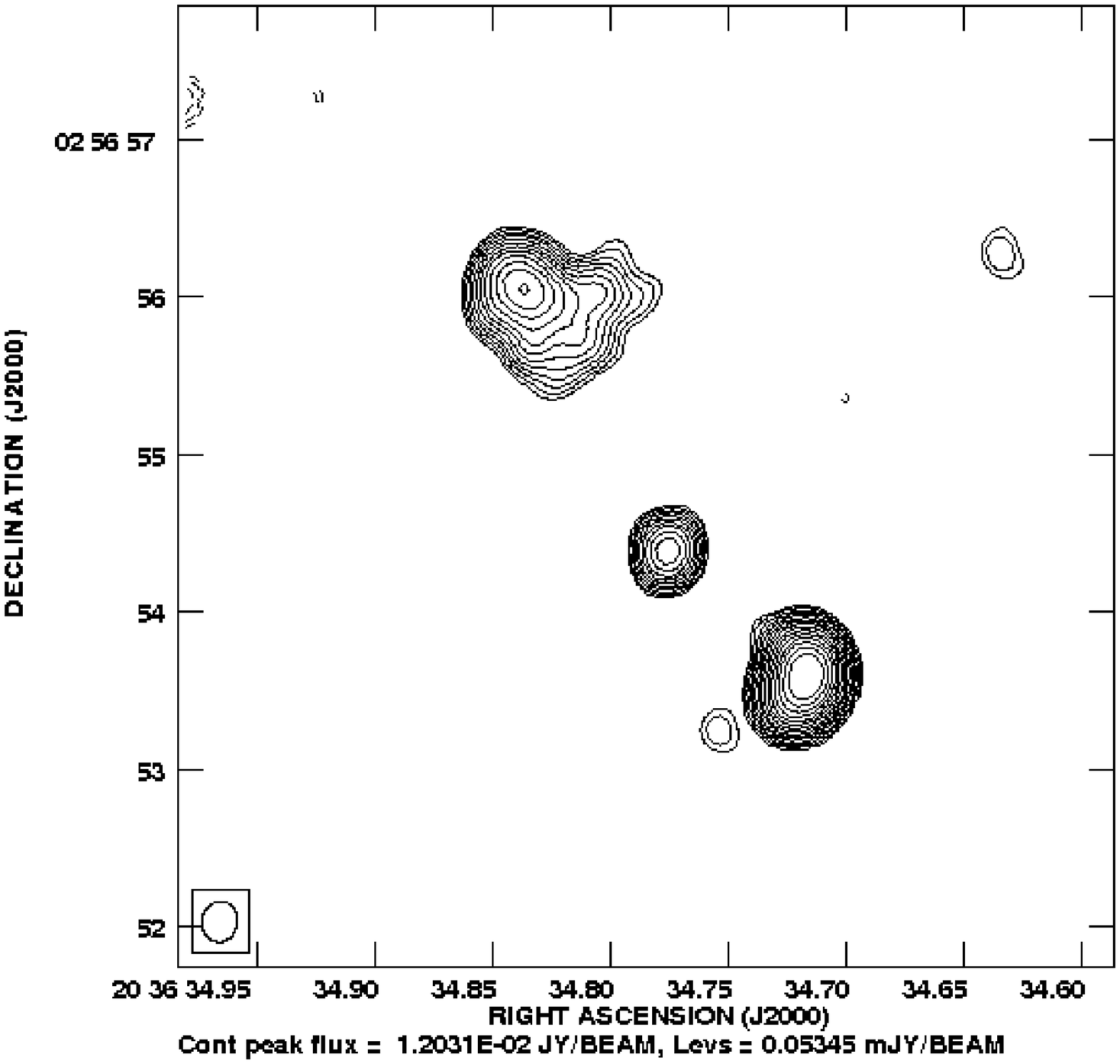,width=7.5cm}}
\vskip-1cm\centerline{
\psfig{figure=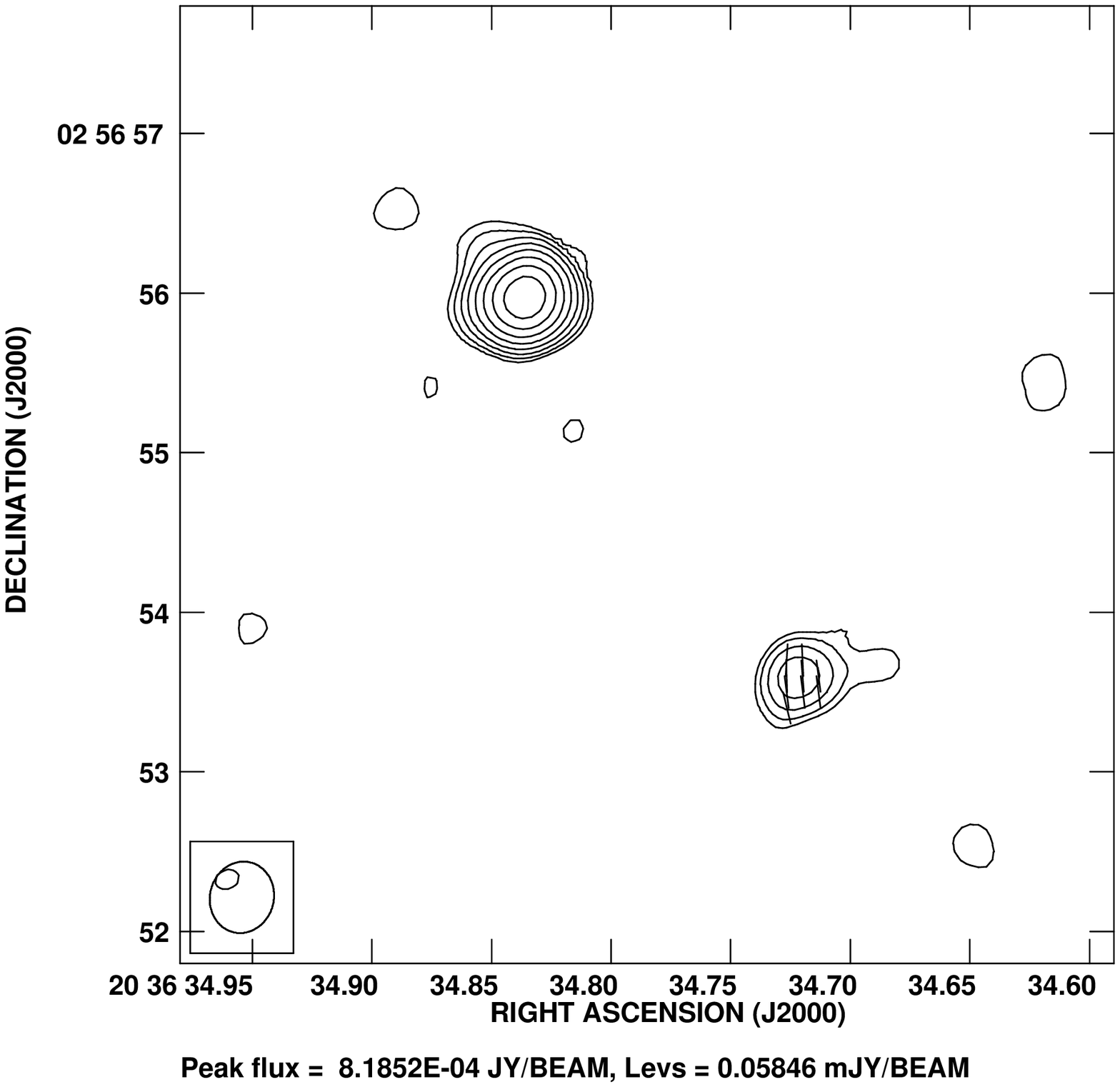,width=7.5cm}
\psfig{figure=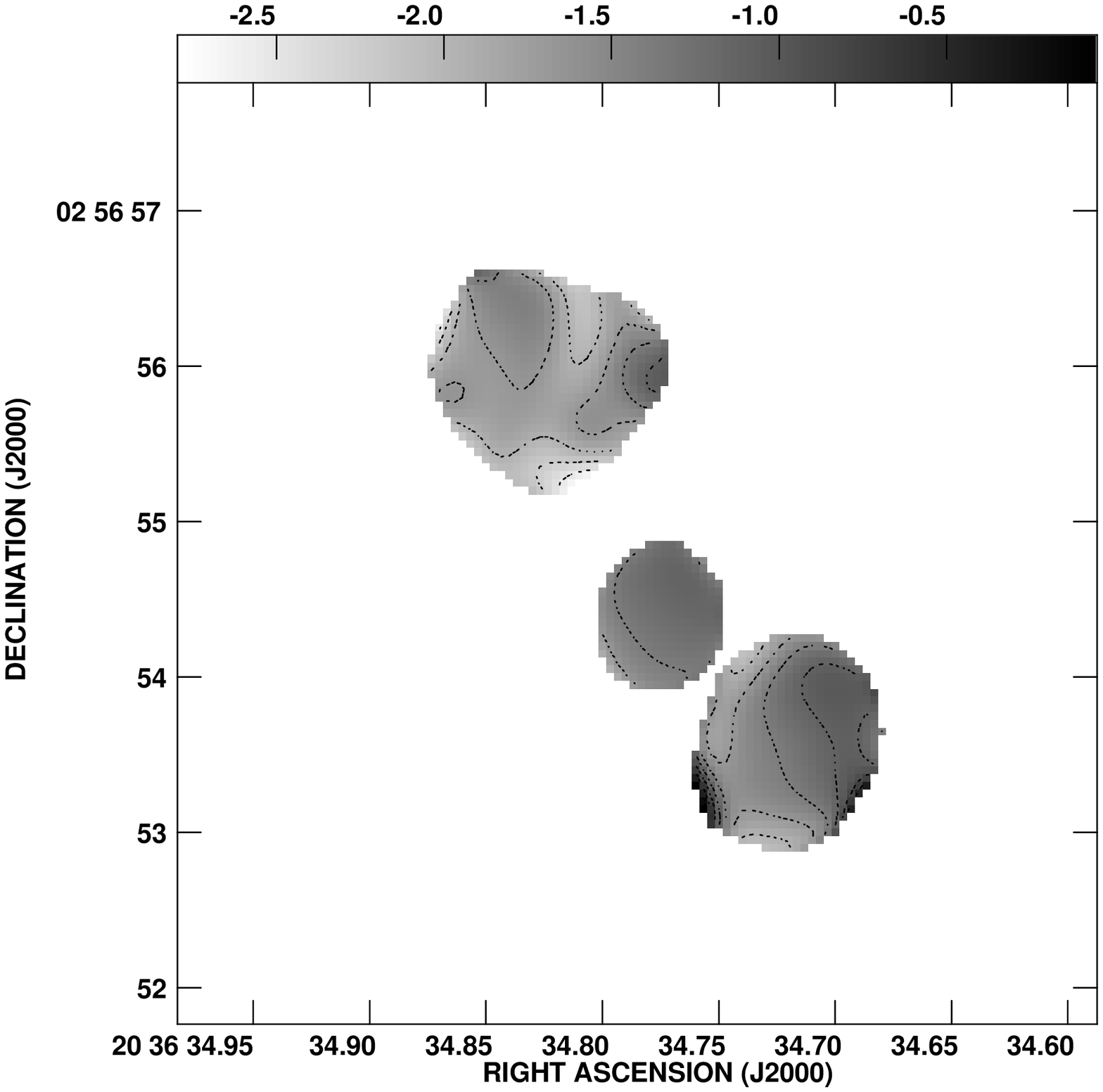,width=7.5cm}}
\caption{Maps of the radio source 2034+027 at redshift z$=$2.129.
The sequence of figures is the same as in Fig. 6. The first contour level and
the  peak surface brightness  are respectively  0.113  mJy beam$^{-1}$ and 31
 mJy beam$^{-1}$ for 
the 4.5 GHz map; 0.053  mJy beam$^{-1}$ and 12 mJy beam$^{-1}$ for the 8.2
GHz map; 0.058  mJy beam$^{-1}$  and 0.8  mJy beam$^{-1}$ for the 4.7 GHz
polarized intensity map.} 
\end{figure*}
\begin{figure*}
\centerline{
  \psfig{figure=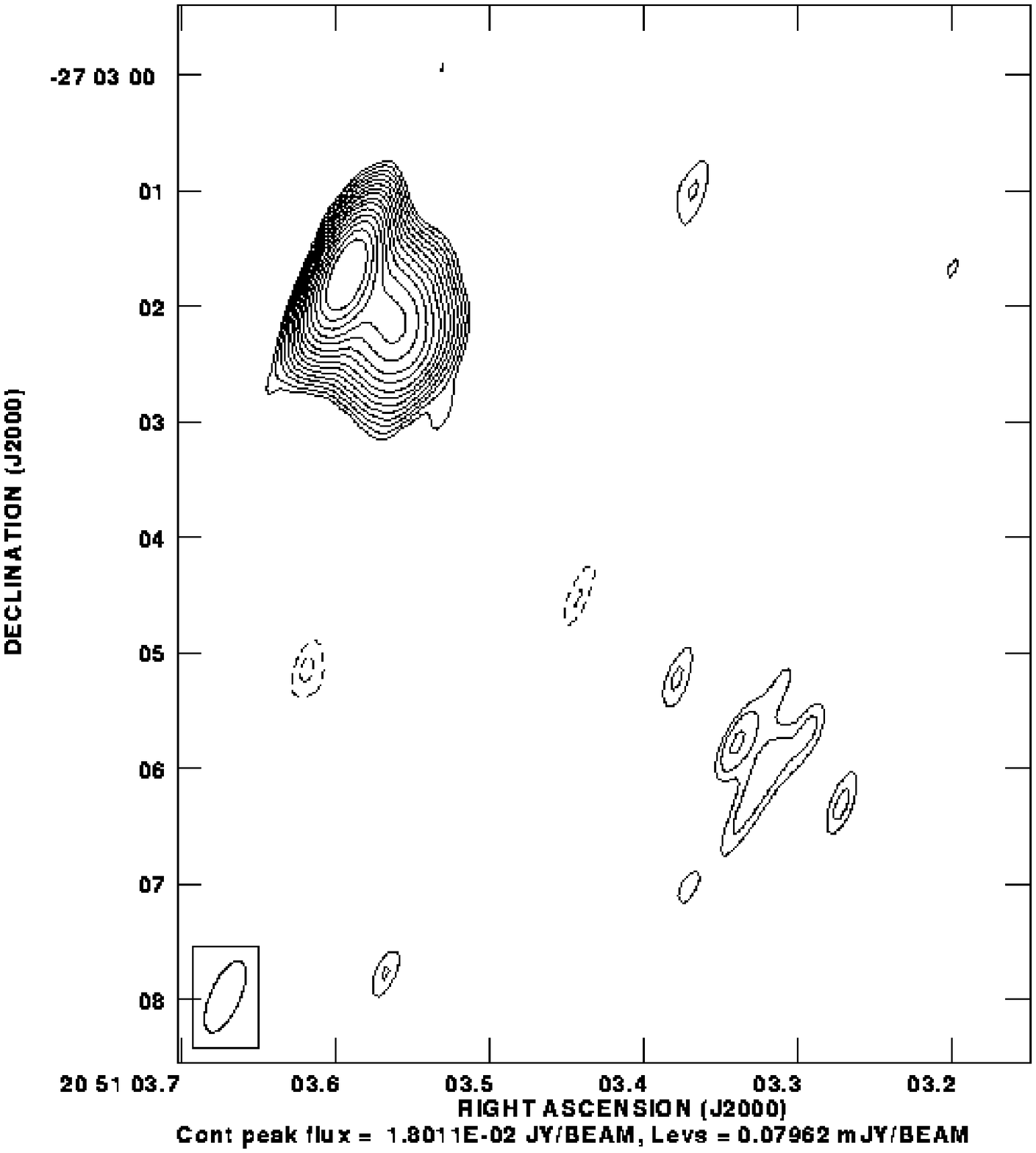,width=7.cm}
\psfig{figure=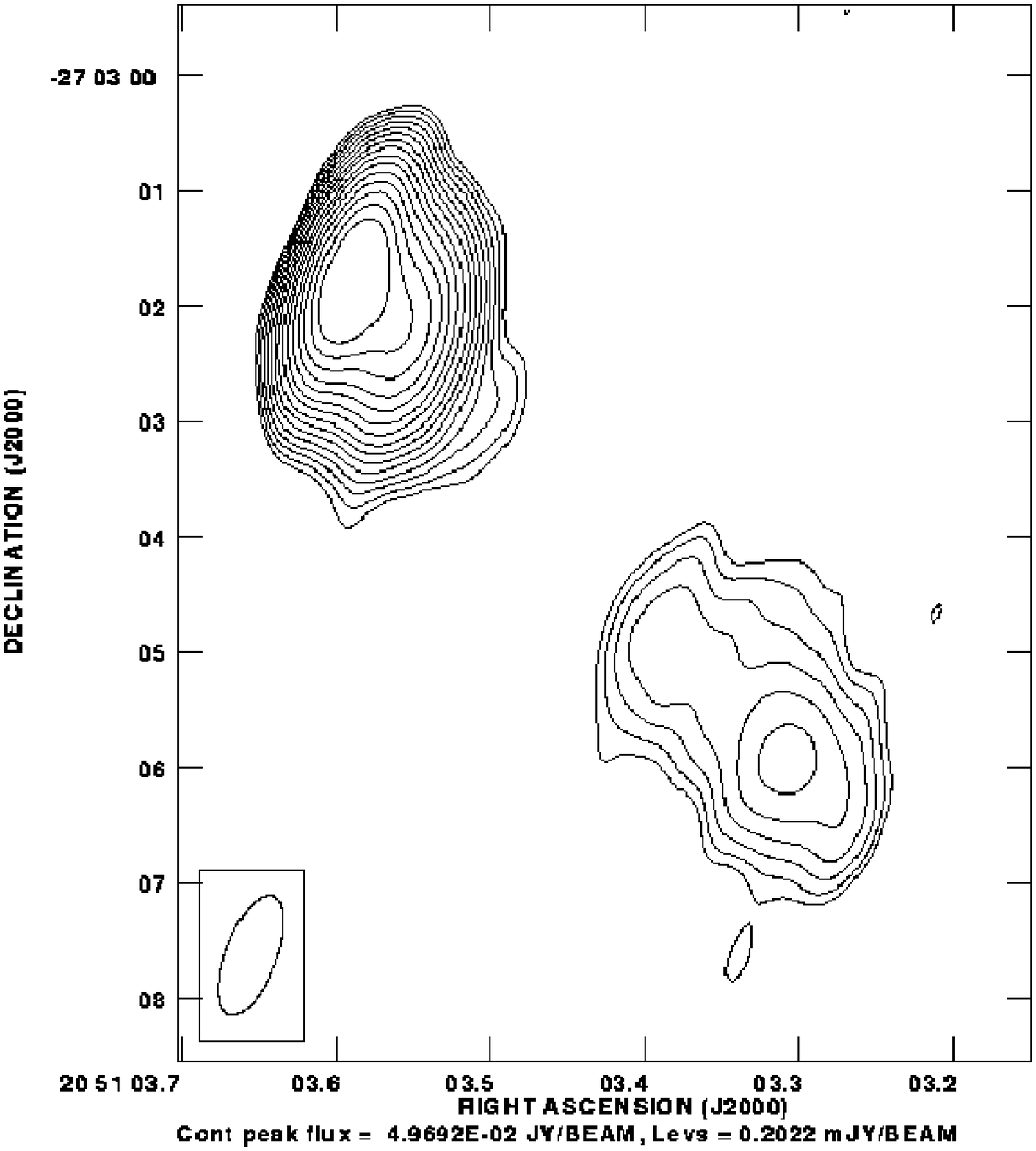,width=7.cm}}
\vskip-0.7cm\centerline{
\psfig{figure=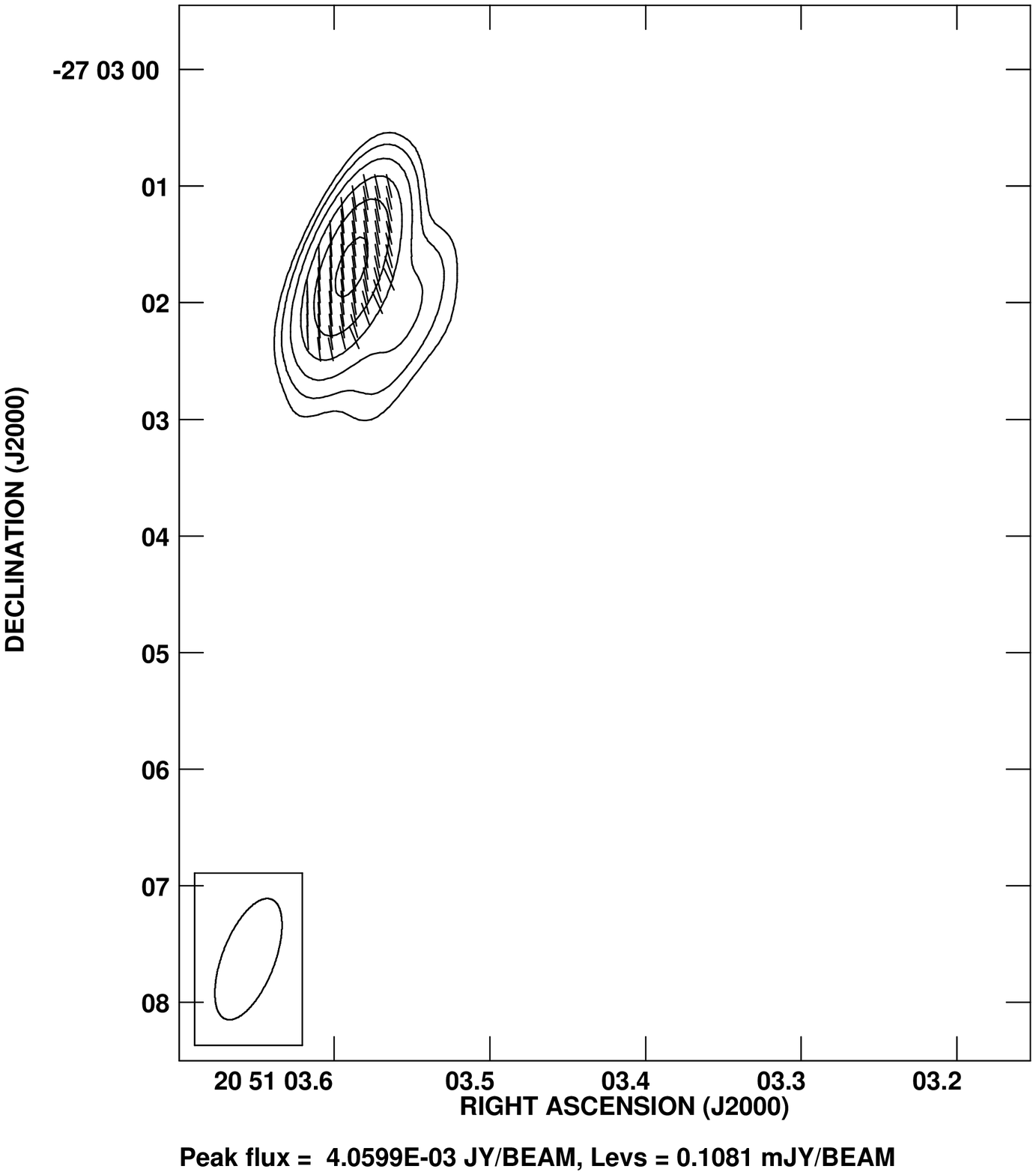,width=7.cm}
\psfig{figure=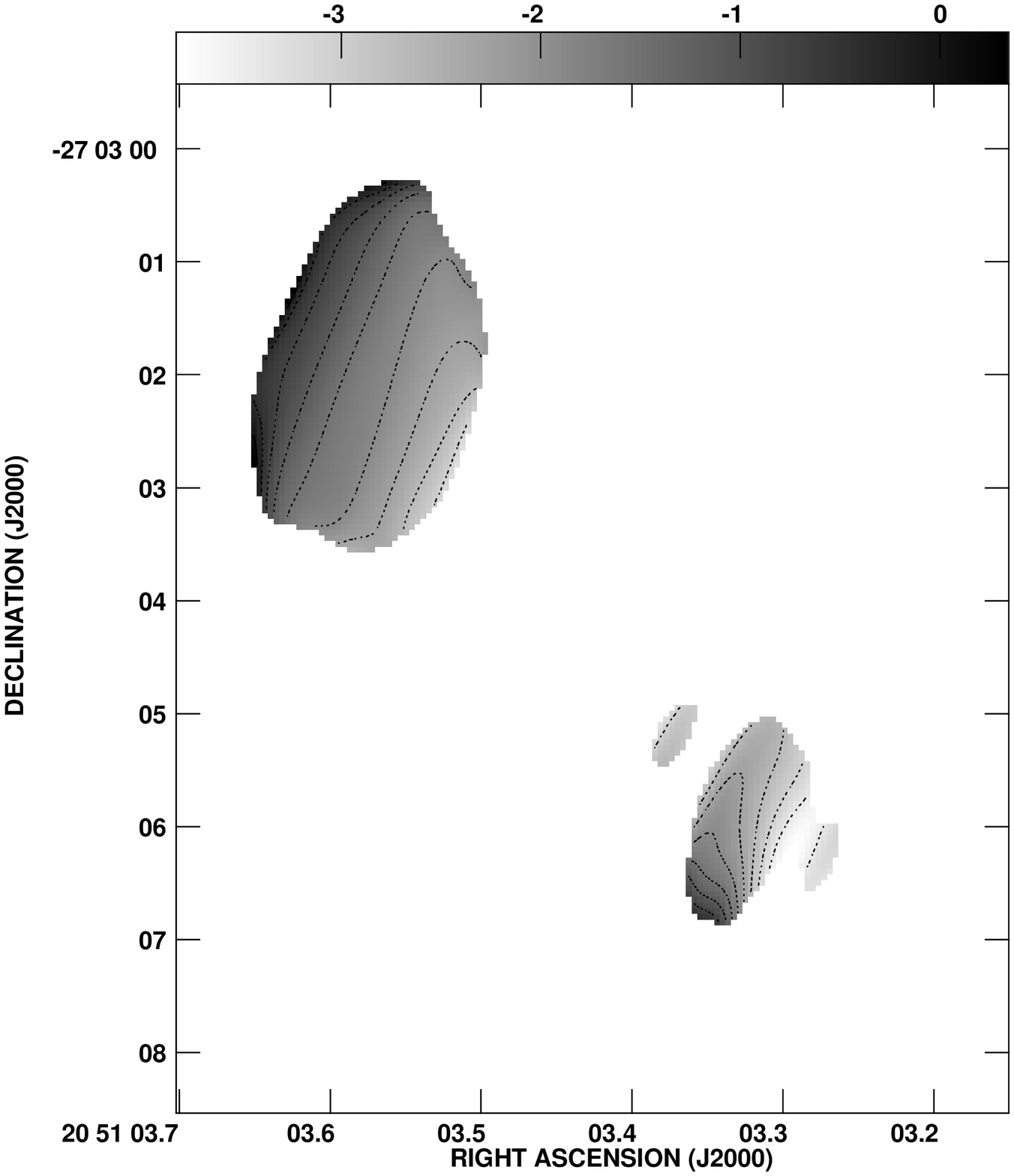,width=7.cm}}
\caption{Maps of the radio source 2048$-$272 at redshift z$=$2.06.
The sequence of figures is the same as in Fig. 6. The first contour level and
the  peak surface brightness  are respectively  0.200  mJy beam$^{-1}$ and 50
 mJy beam$^{-1}$ for 
the 4.5 GHz map; 0.079  mJy beam$^{-1}$ and 18 mJy beam$^{-1}$ for the 8.2
GHz map; 0.108  mJy beam$^{-1}$  and 4.1  mJy beam$^{-1}$ for the 4.7 GHz
polarized intensity map.} 
\end{figure*}
\begin{figure*}
\centerline{
\psfig{figure=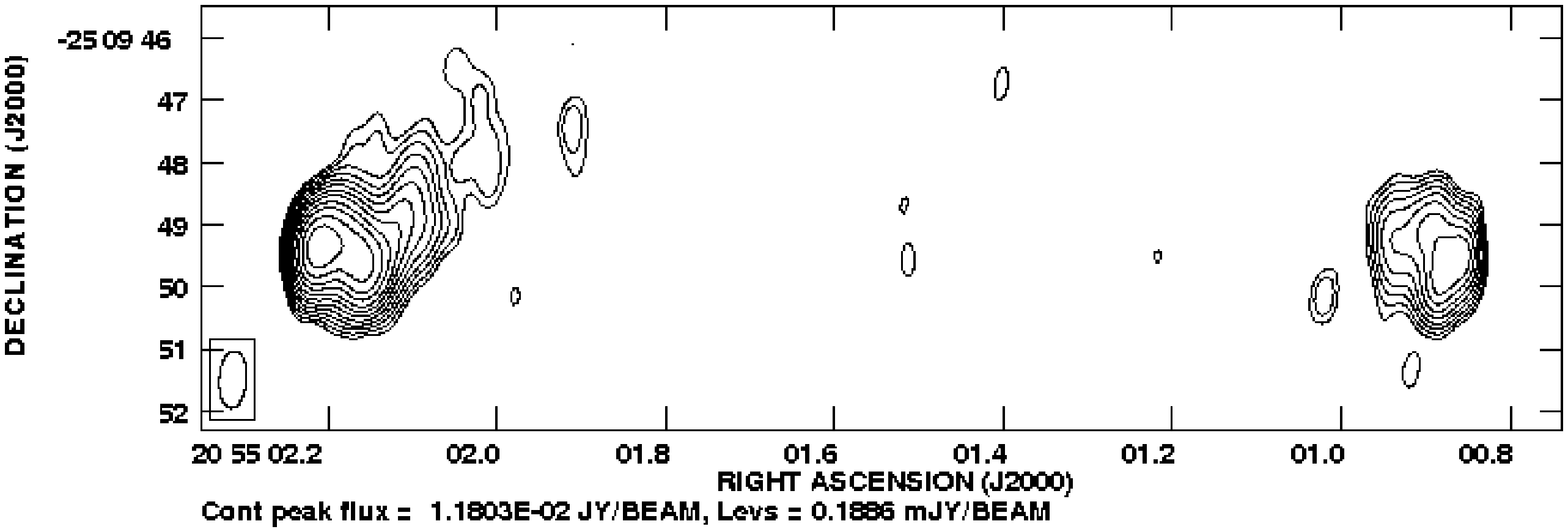,width=6.5cm,angle=-90}}\vskip-1.cm\centerline{
\psfig{figure=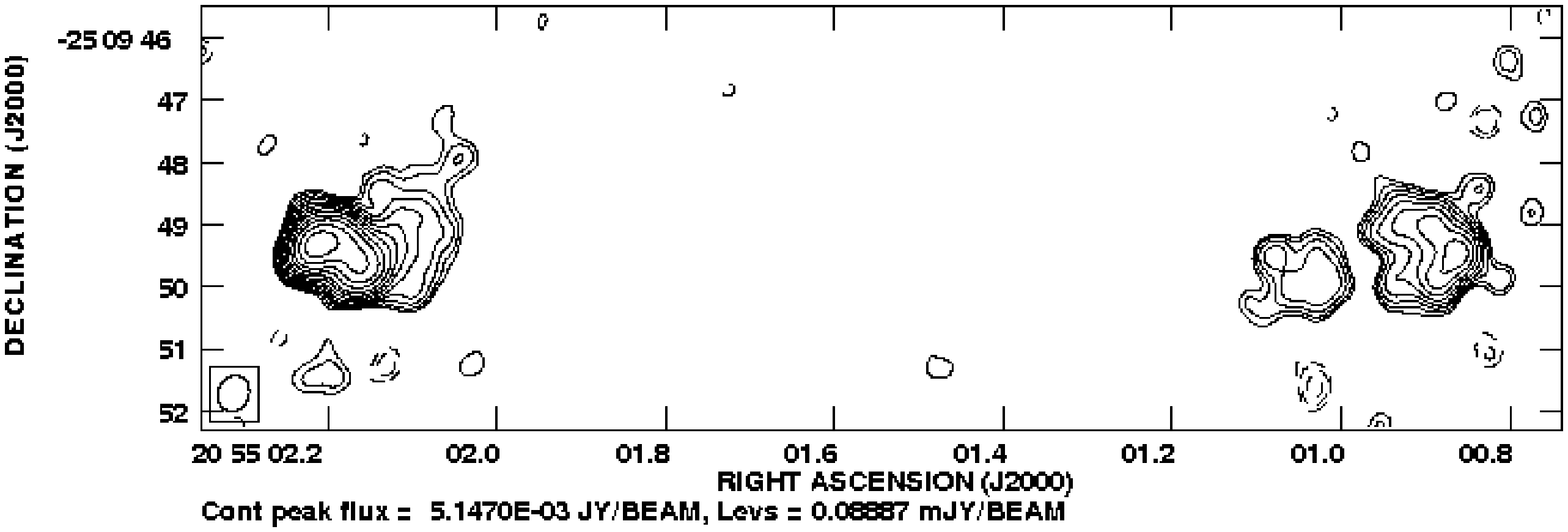,width=6.5cm,angle=-90}}\vskip-1.cm\centerline{
\psfig{figure=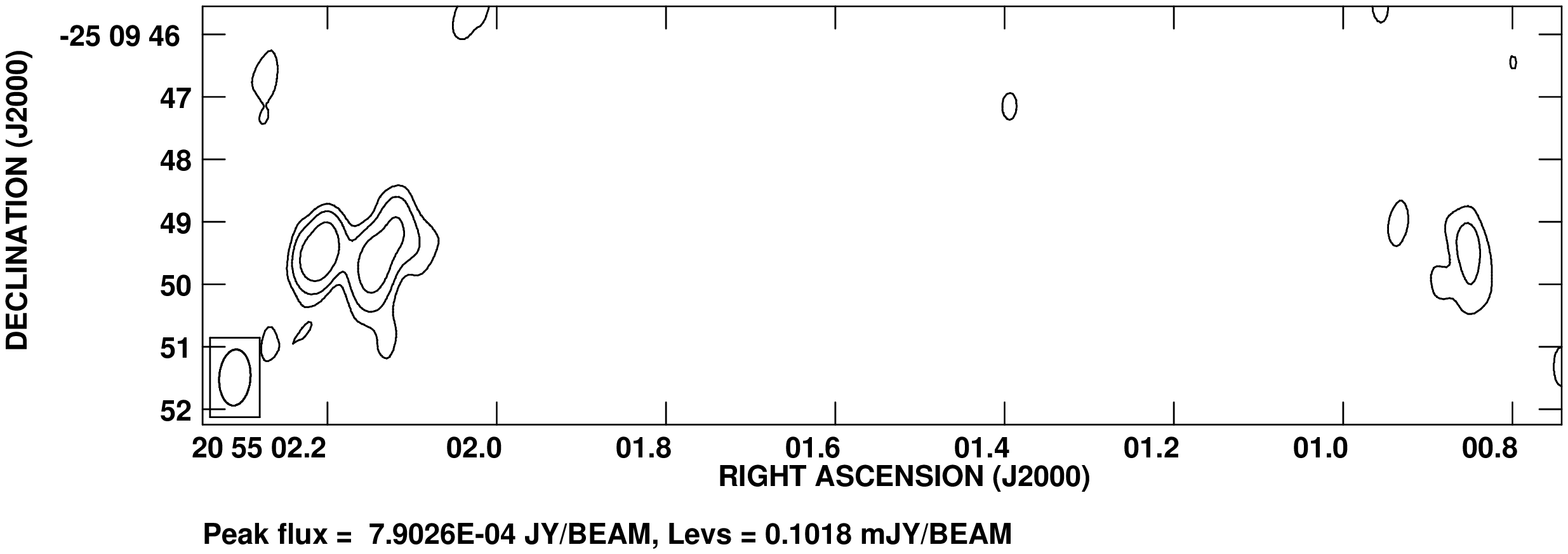,width=6.5cm,angle=-90}}\vskip-1.cm\centerline{
\psfig{figure=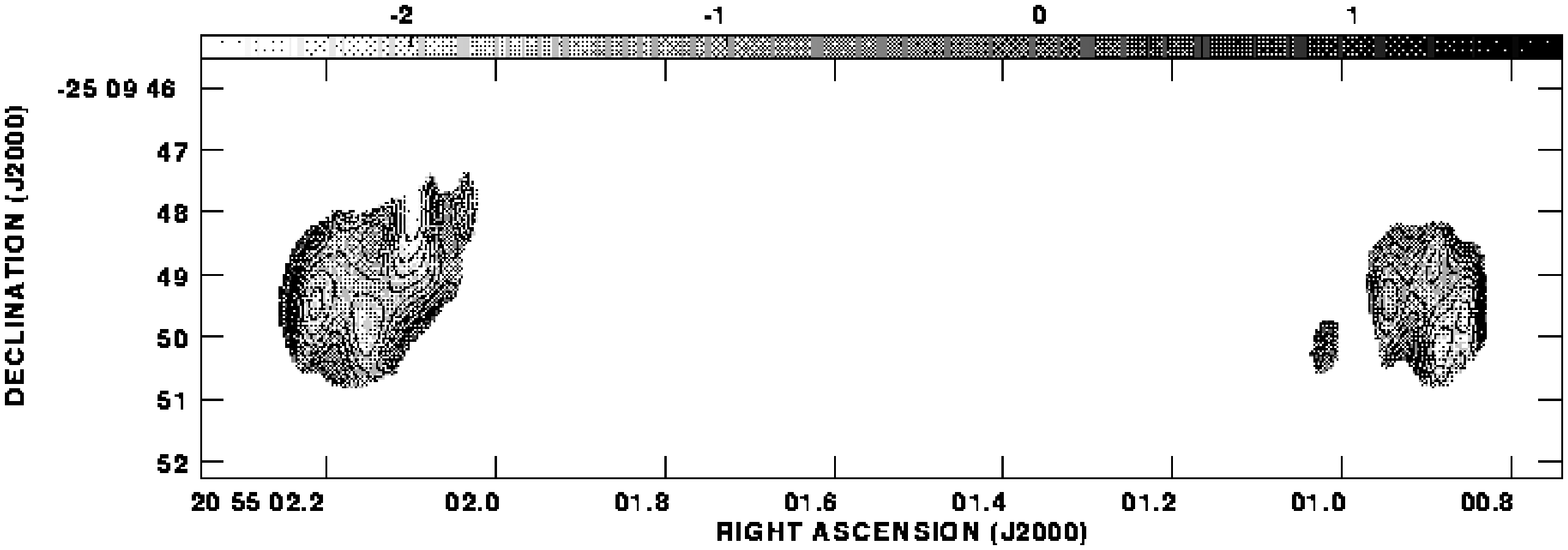,width=7.2cm,angle=-90}}\vskip-1.cm
\caption{Maps of the radio source 2052$-$253 at redshift z$=$2.63.
The sequence of figures is the same as in Fig. 6. The first contour level and
the  peak surface brightness  are respectively  0.188  mJy beam$^{-1}$ and 11.8
 mJy beam$^{-1}$ for 
the 4.5 GHz map; 0.088  mJy beam$^{-1}$ and 5.1 mJy beam$^{-1}$ for the 8.2
GHz map; 0.102  mJy beam$^{-1}$  and 0.8  mJy beam$^{-1}$ for the 4.7 GHz
polarized intensity map.} 
\end{figure*}
\begin{figure*}
\centerline{
\psfig{figure=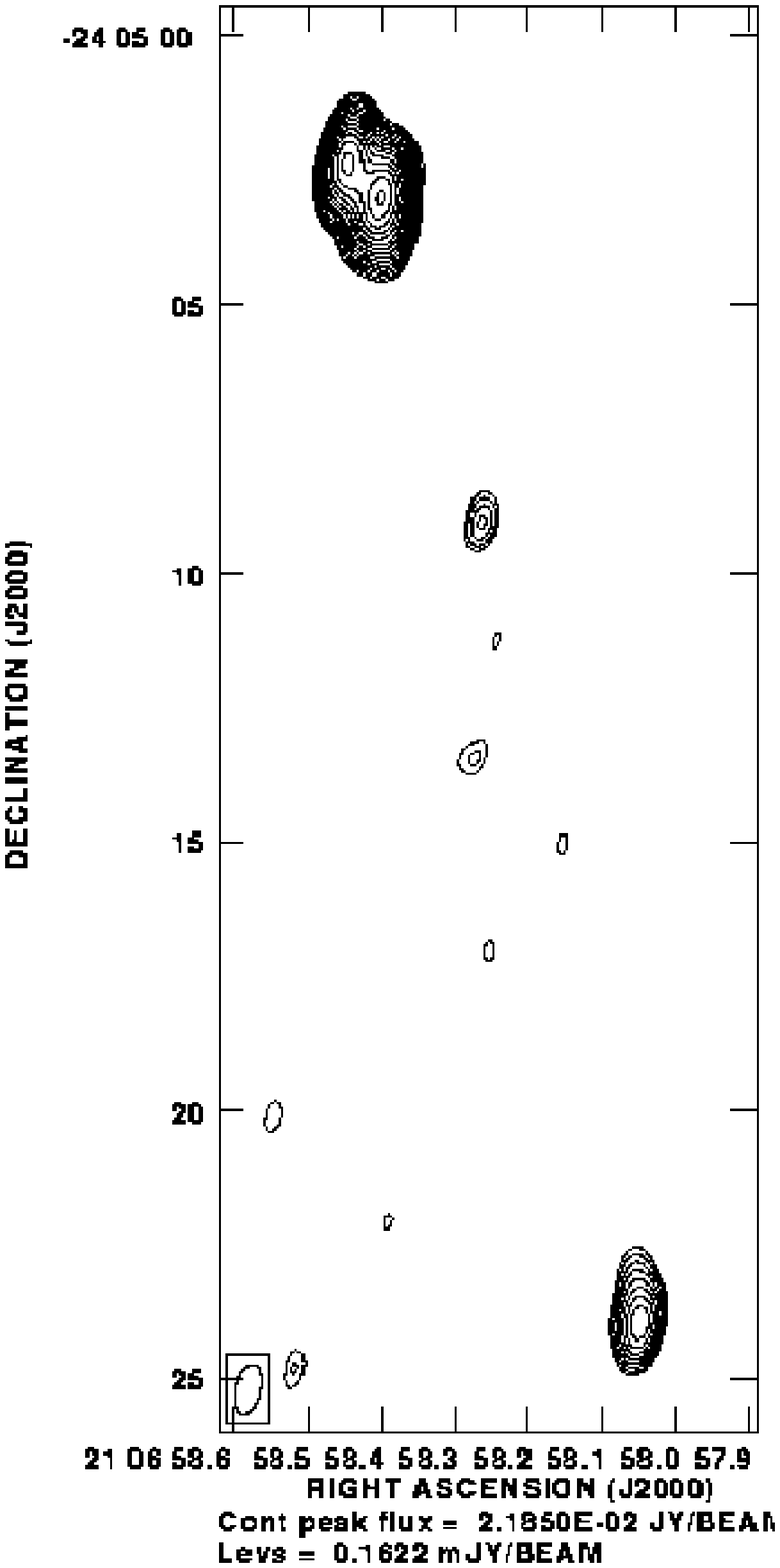,width=5.3cm}
\psfig{figure=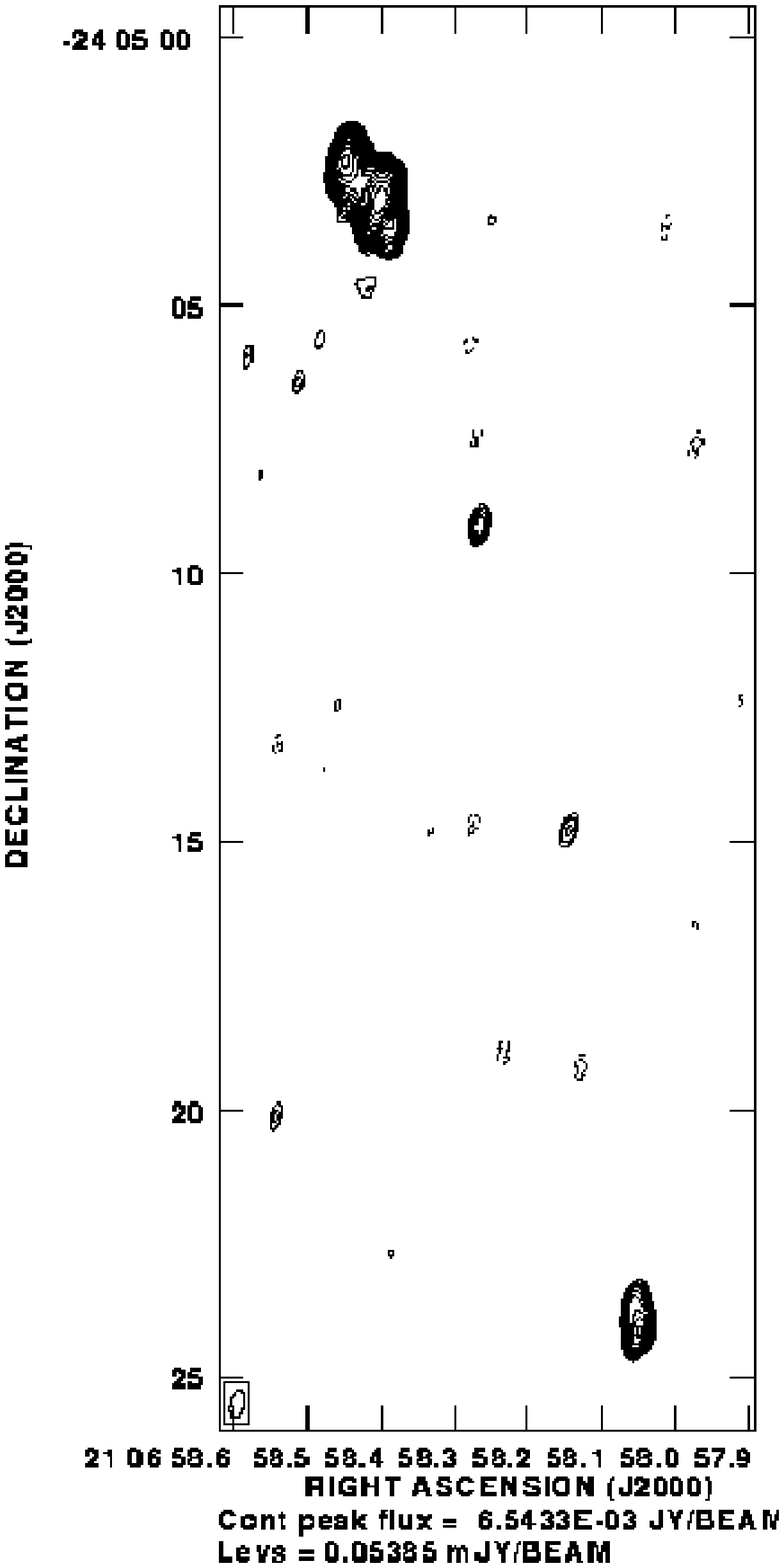,width=5.3cm}}\vskip-0.6cm
\centerline{
\psfig{figure=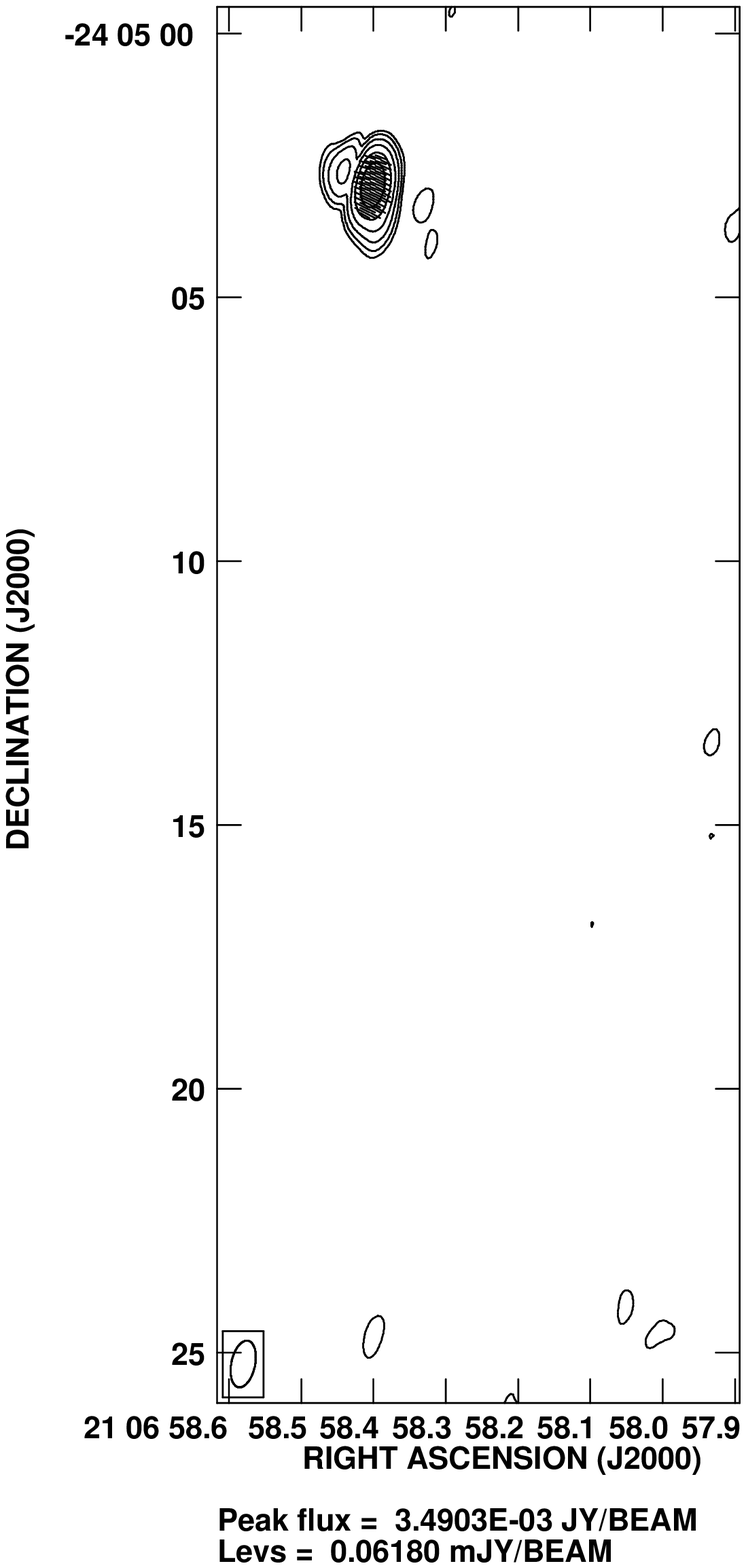,width=5.3cm}
\psfig{figure=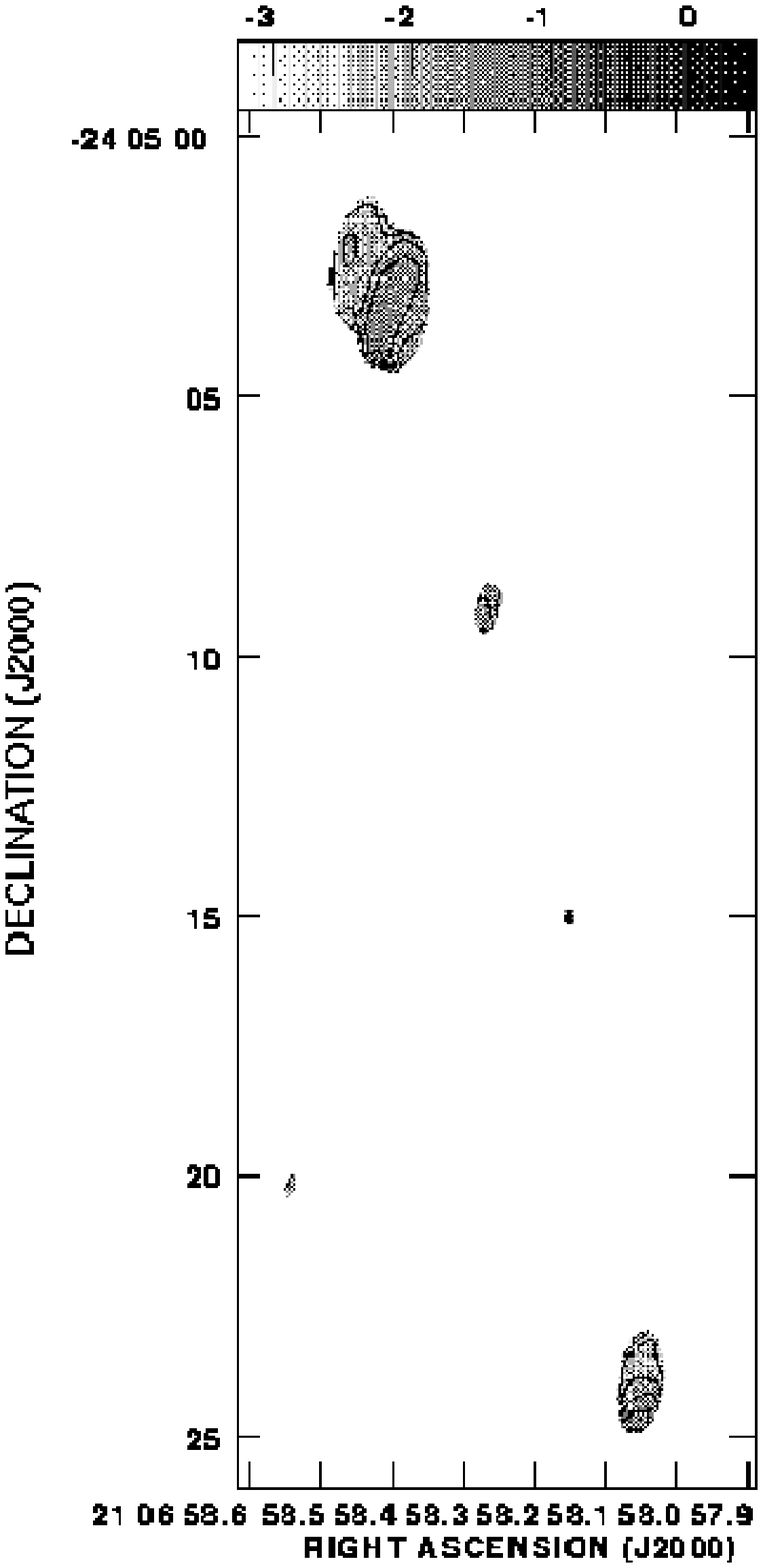,width=5.1cm}}
 \caption{Maps of the radio source 2104$-$242 at redshift z$=$2.49.
The sequence of figures is the same as in Fig. 6. The first contour level and
the  peak surface brightness  are respectively  0.162  
mJy beam$^{-1}$ and 21.8
 mJy beam$^{-1}$ for 
the 4.5 GHz map; 0.054  mJy beam$^{-1}$ and 6.5 mJy beam$^{-1}$ for the 8.2
GHz map; 0.062  mJy beam$^{-1}$  and 3.5  mJy beam$^{-1}$ for the 4.7 GHz
polarized intensity map.} 
\end{figure*}
\begin{figure*}
\centerline{
\psfig{figure=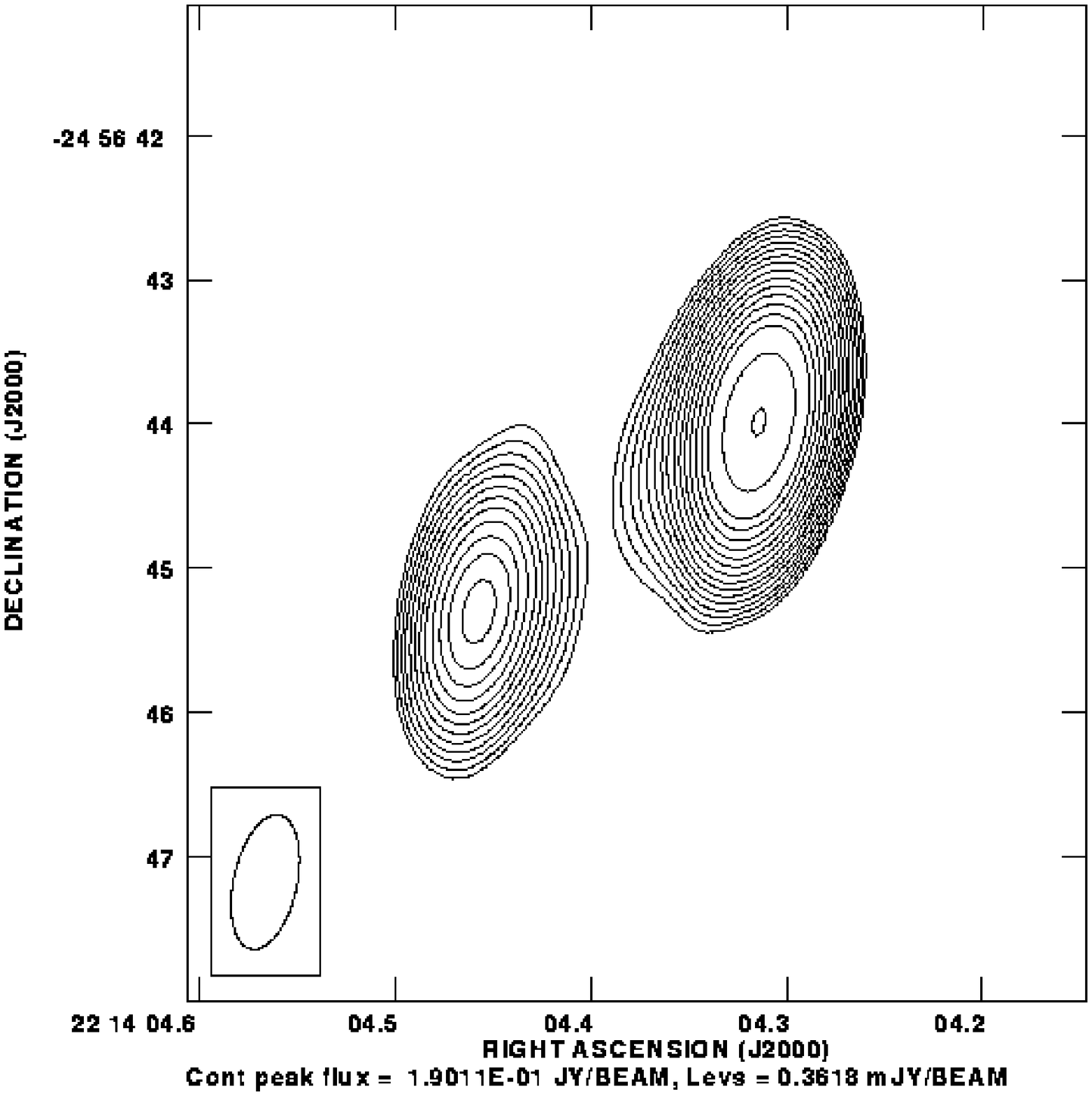,width=7.2cm}
\psfig{figure=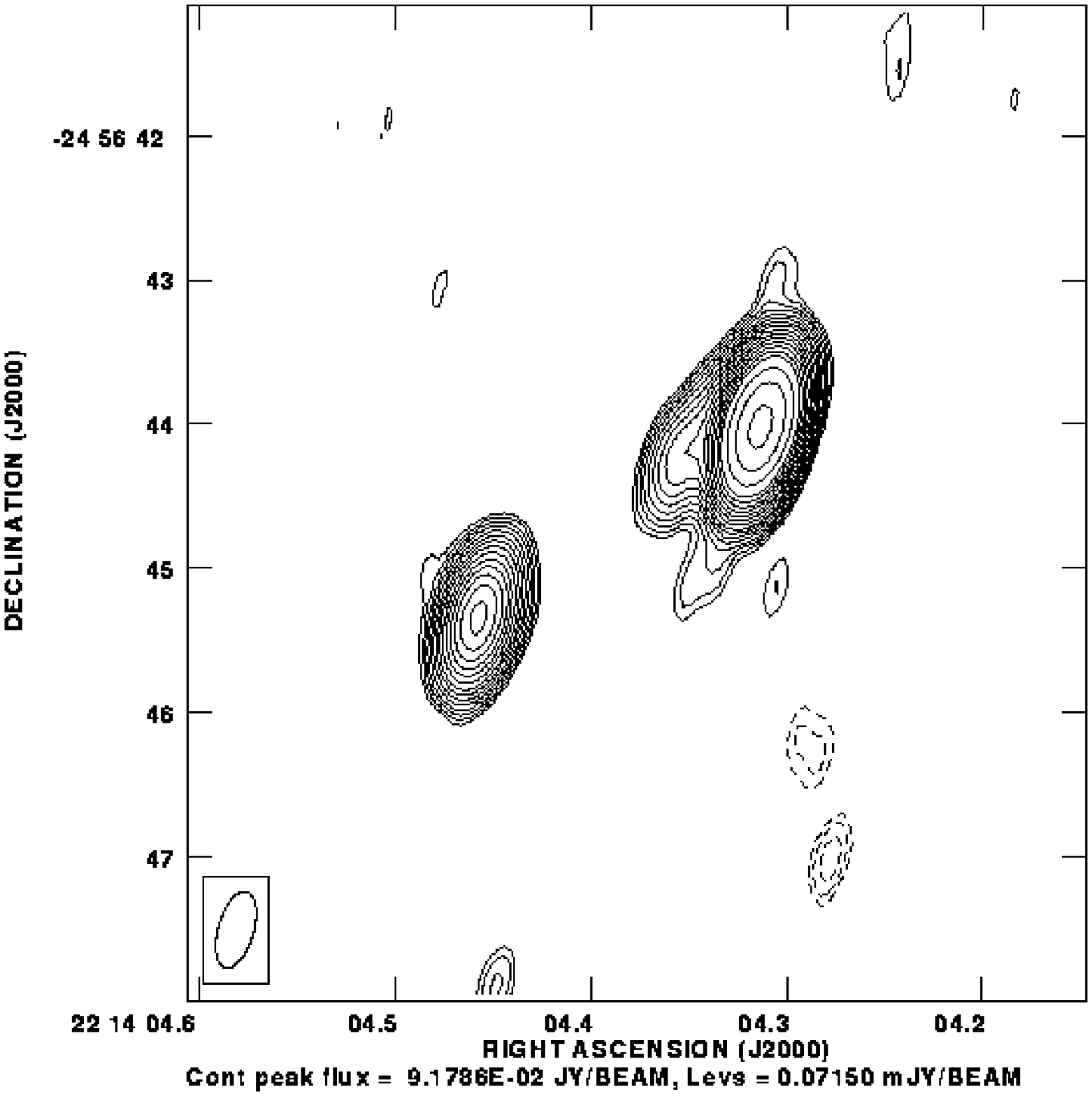,width=7.5cm}}
\vskip-0.5cm\centerline{
\psfig{figure=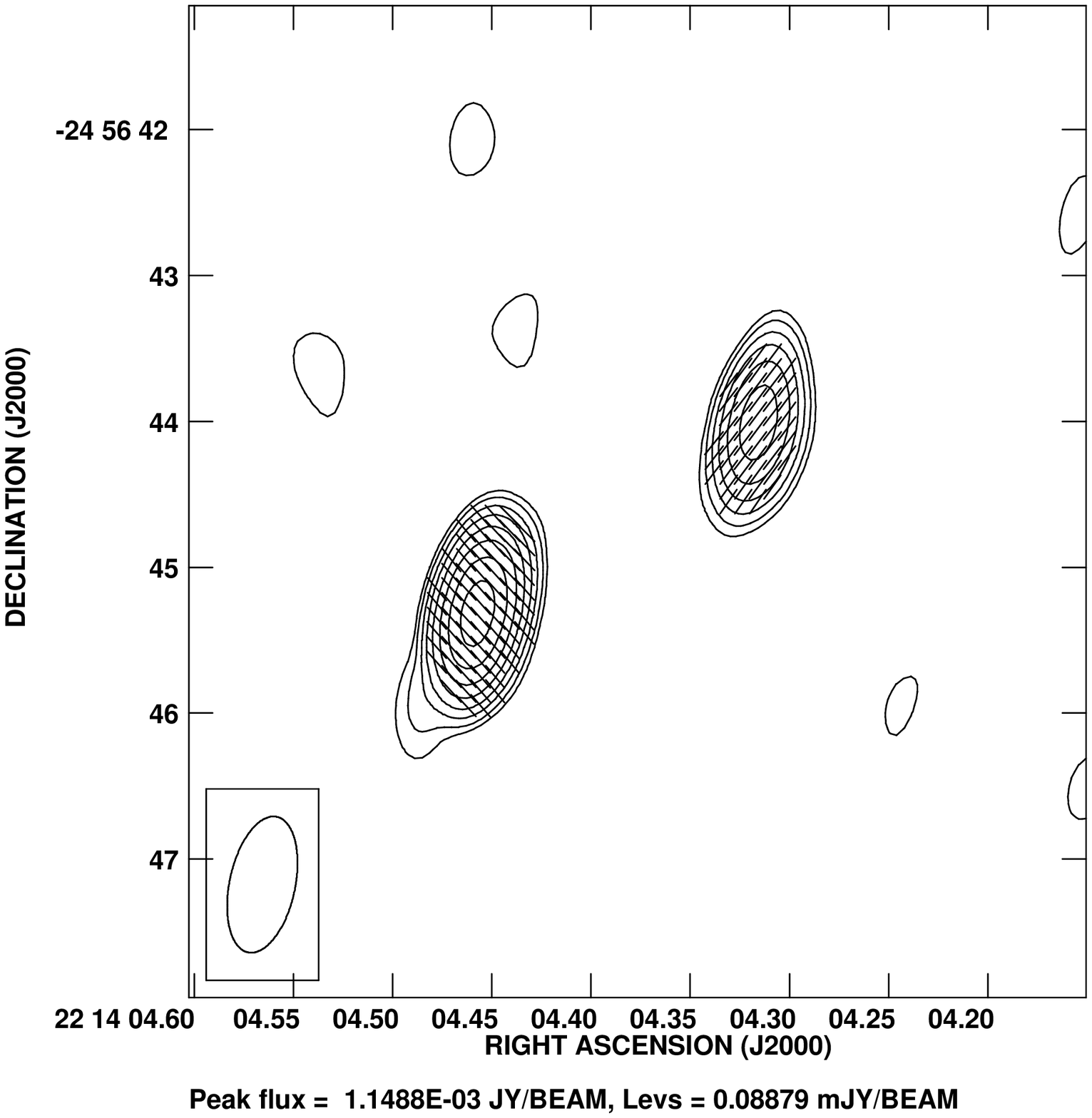,width=7.5cm}
\psfig{figure=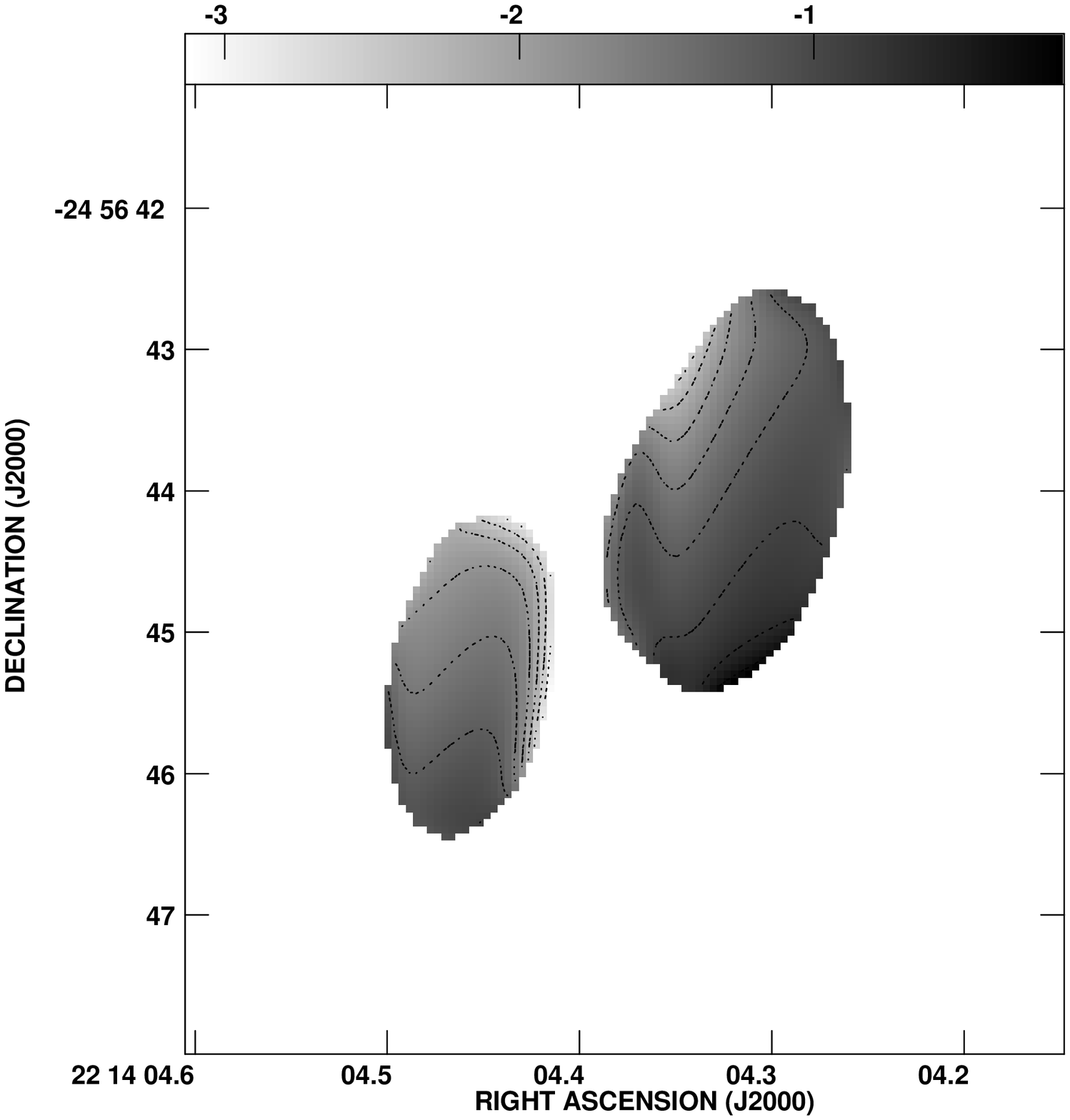,width=7.5cm}}
\caption{Maps of the radio source 2211$-$251 at redshift z$=$2.508.
The sequence of figures is the same as in Fig. 6. The first contour level and
the  peak surface brightness  are respectively  0.362  mJy beam$^{-1}$ and 190
 mJy beam$^{-1}$ for 
the 4.5 GHz map; 0.071  mJy beam$^{-1}$ and 92 mJy beam$^{-1}$ for the 8.2
GHz map; 0.089  mJy beam$^{-1}$  and 1.1  mJy beam$^{-1}$ for the 4.7 GHz
polarized intensity map.} 
\end{figure*}
\begin{figure*}
\centerline{
\psfig{figure=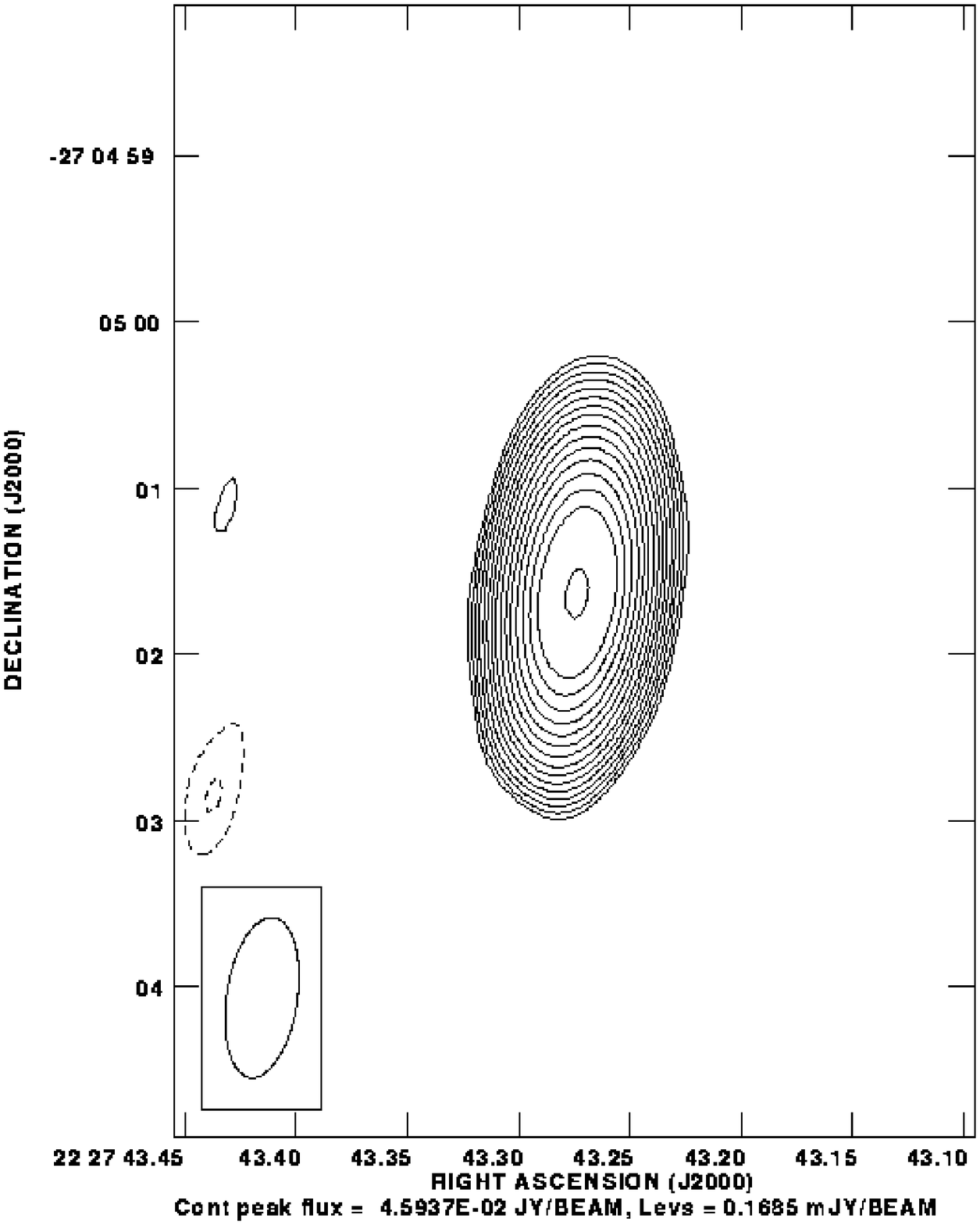,width=6.cm}
\psfig{figure=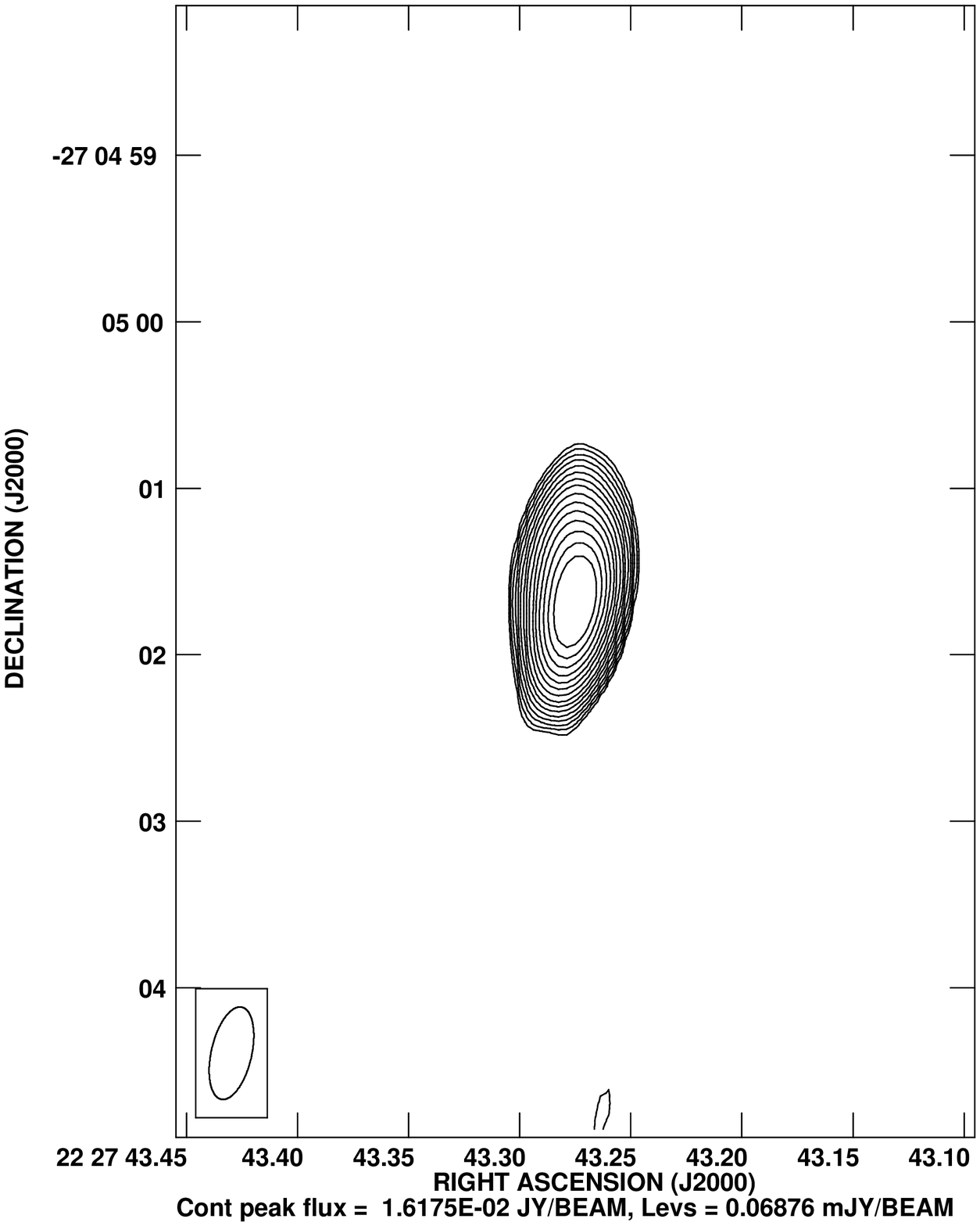,width=6.cm}}
\vskip-1cm\centerline{
\psfig{figure=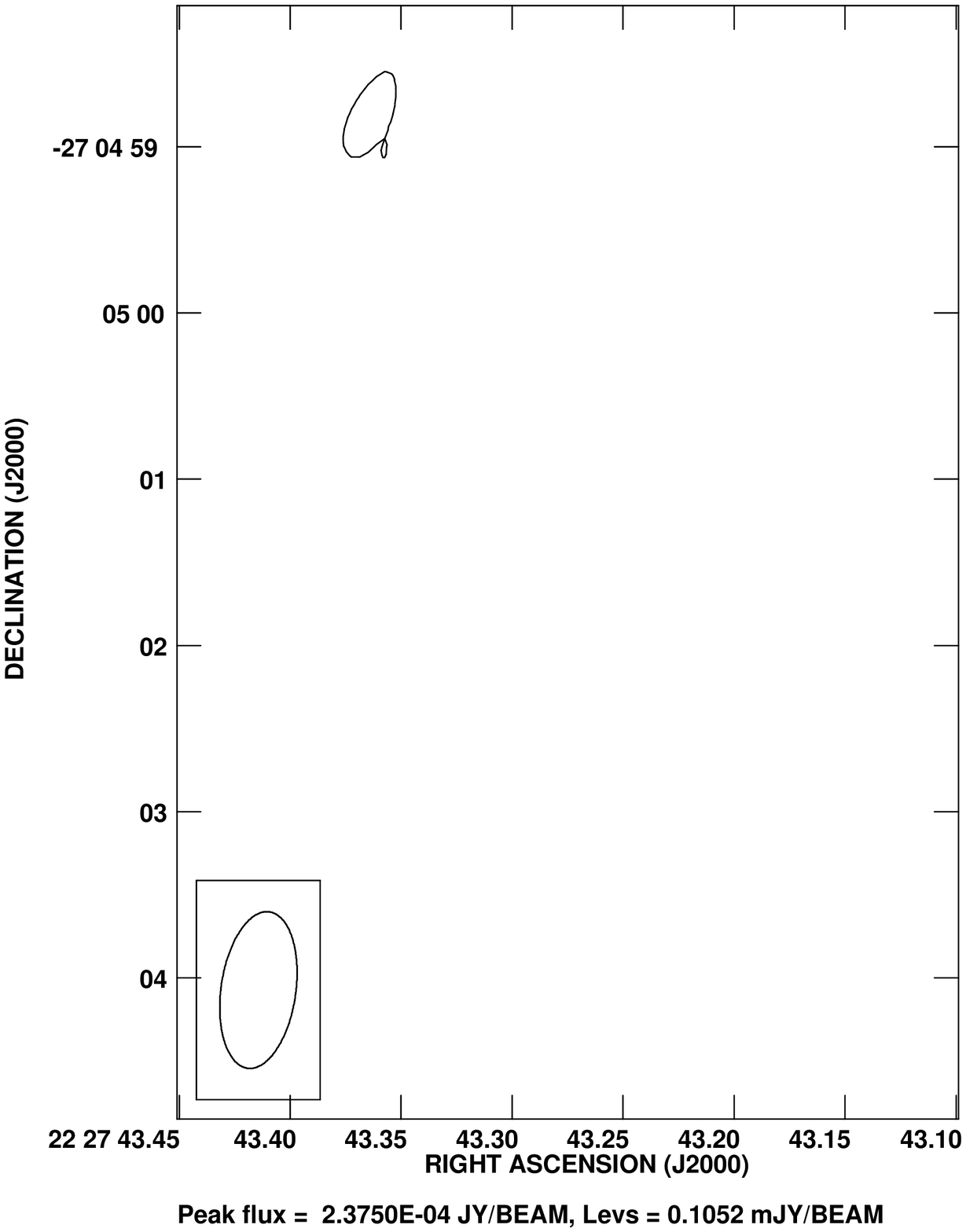,width=6.cm}
\psfig{figure=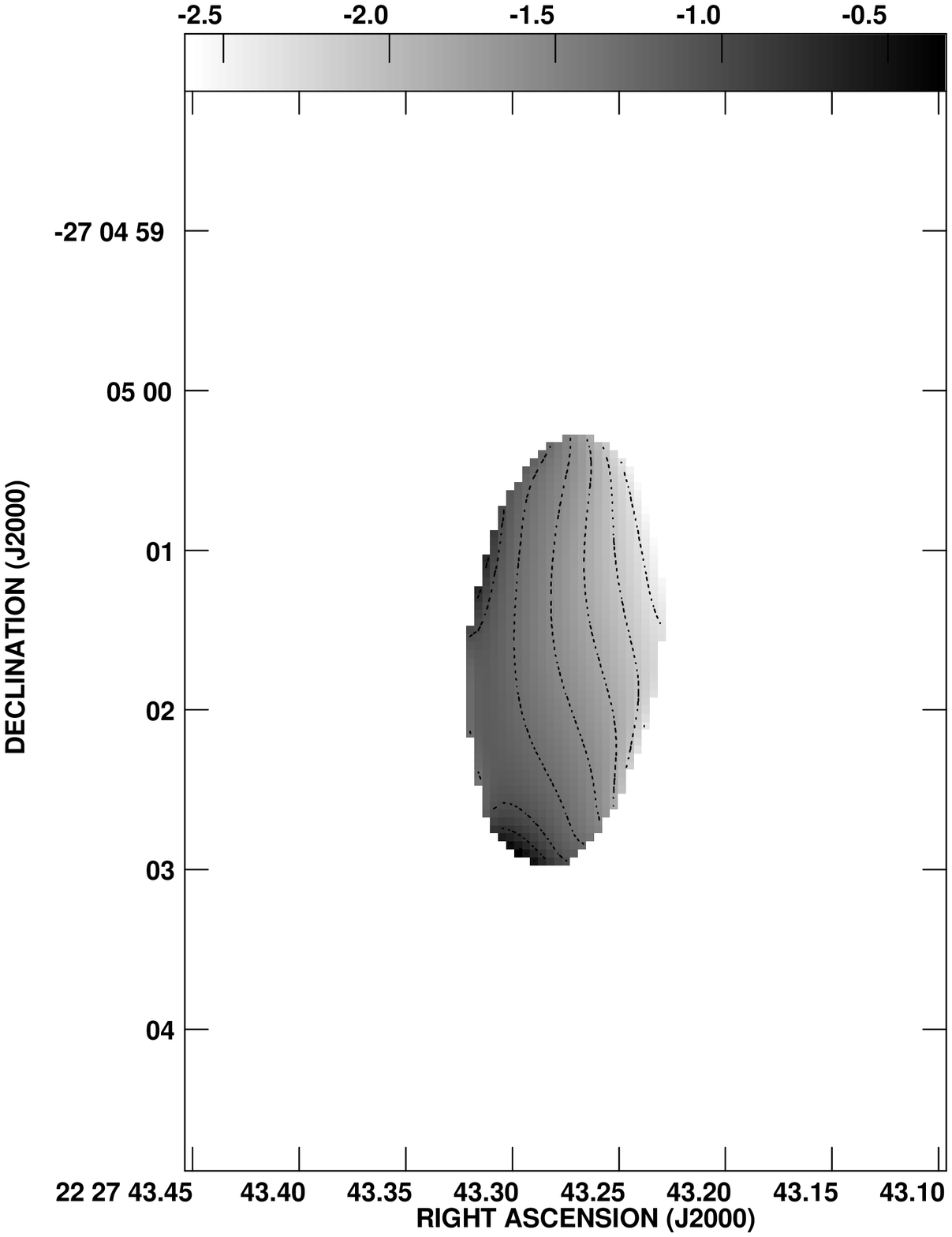,width=6.cm}}
\caption{Maps of the radio source 2224$-$273 at redshift z$=$1.68.
The sequence of figures is the same as in Fig. 6. The first contour level and
the  peak surface brightness  are respectively  0.168  mJy beam$^{-1}$ and 162
 mJy beam$^{-1}$ for 
the 4.5 GHz map; 0.068  mJy beam$^{-1}$ and 45.9 mJy beam$^{-1}$ for the 8.2
GHz map; 0.105  mJy beam$^{-1}$  for the 4.7 GHz
polarized intensity map.} 
\end{figure*}
\begin{figure*}
\centerline{
\psfig{figure=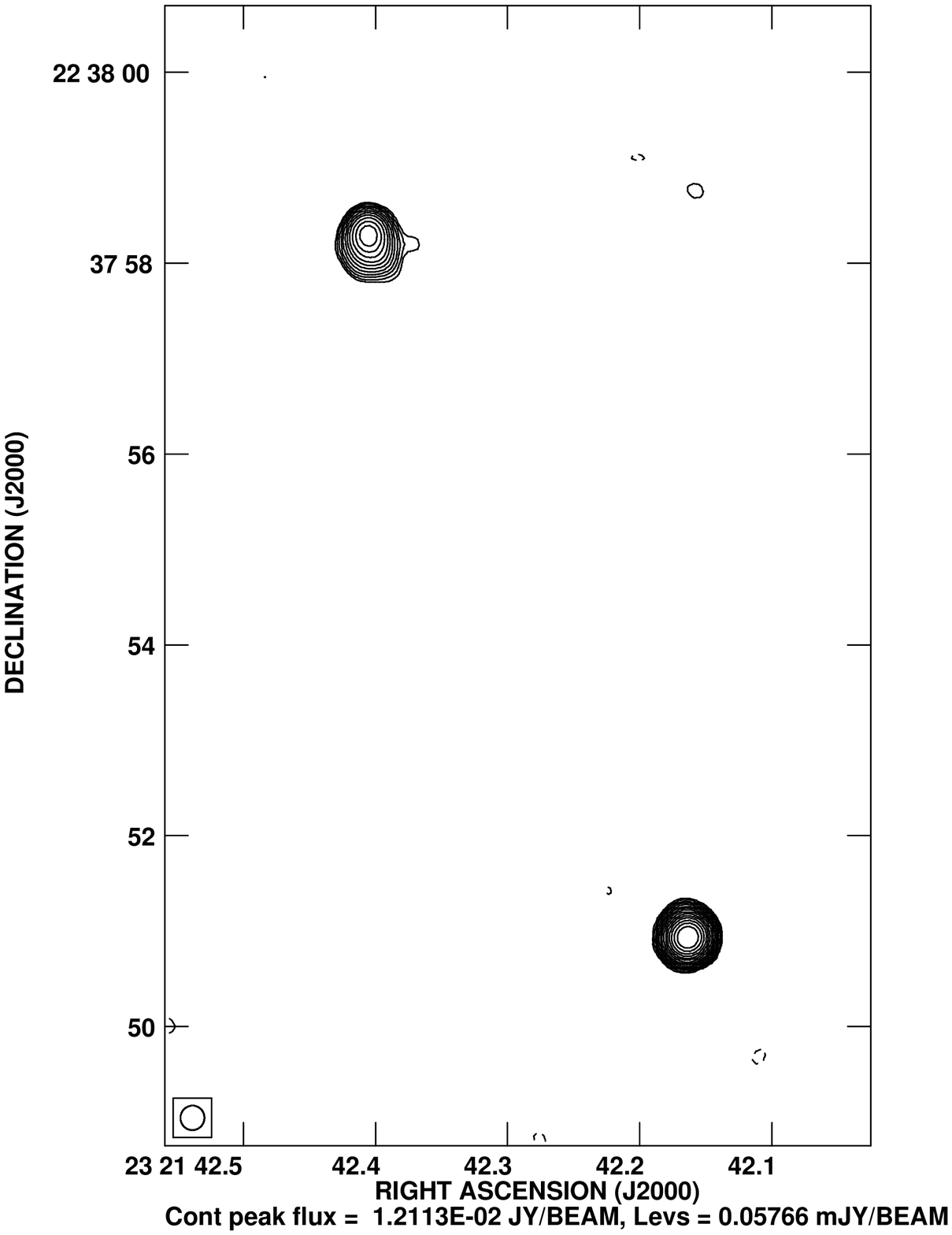,width=6.8cm}
\psfig{figure=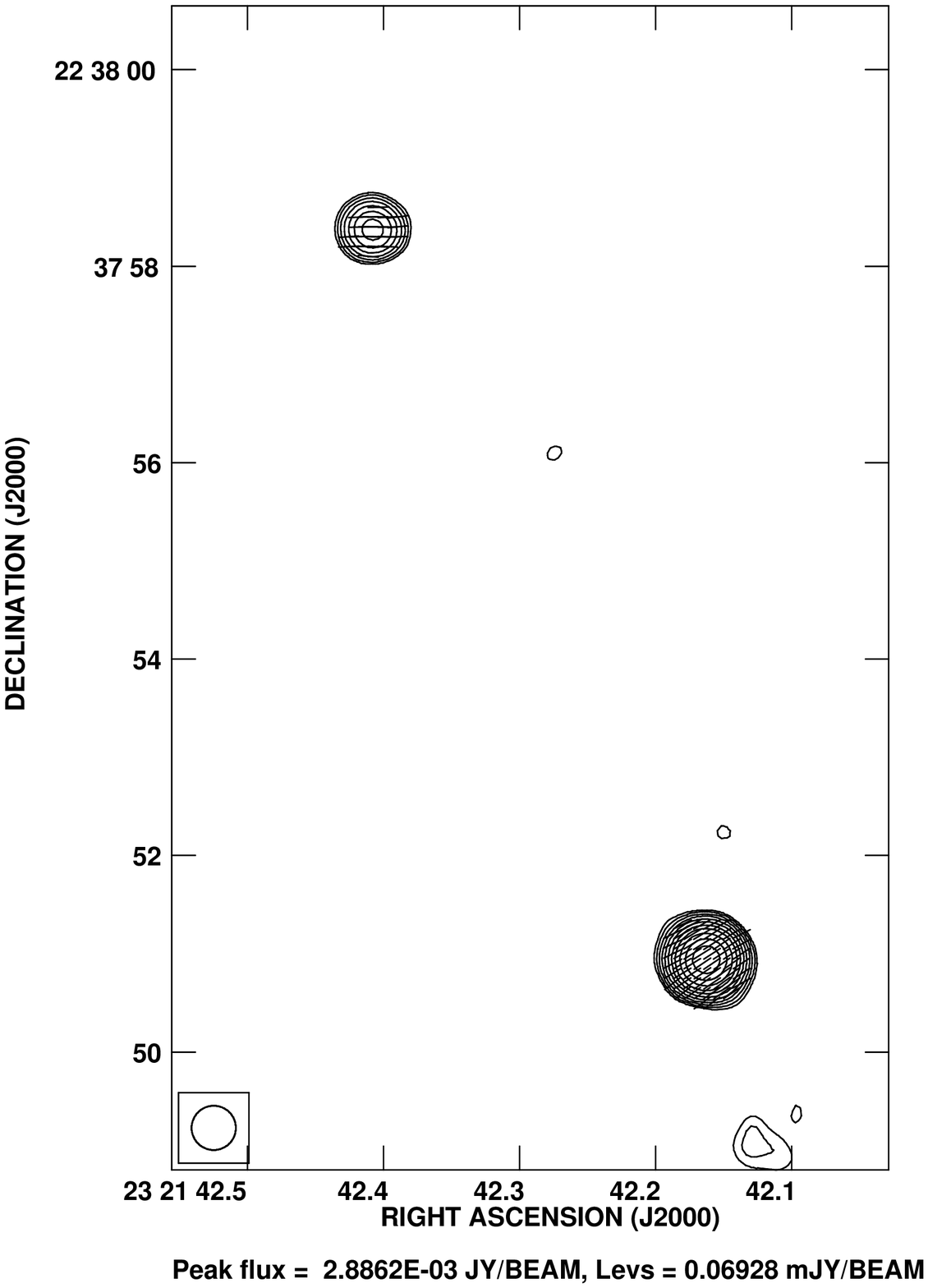,width=6.8cm}}
\vskip-0.5cm\centerline{
\psfig{figure=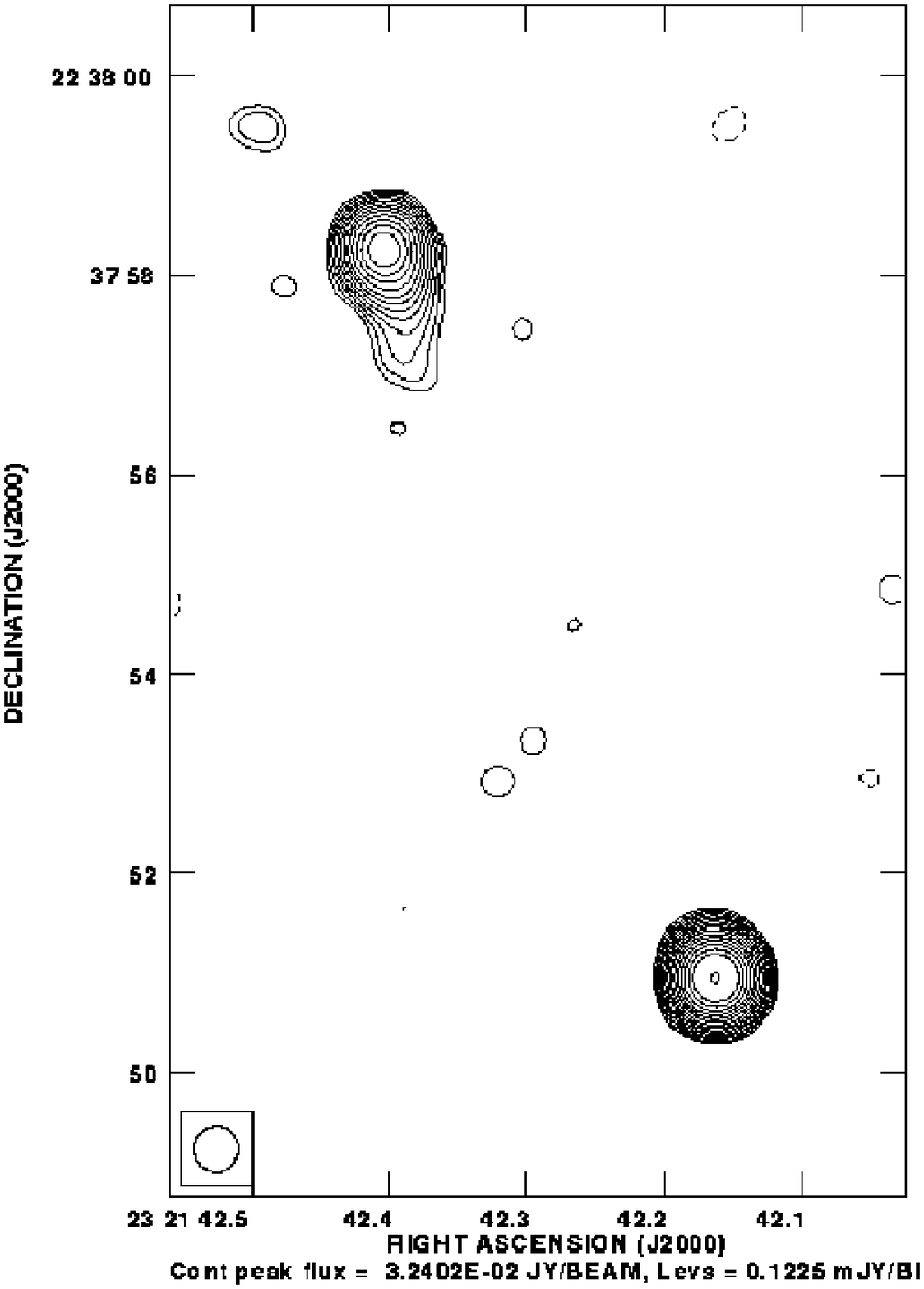,width=6.8cm}
\psfig{figure=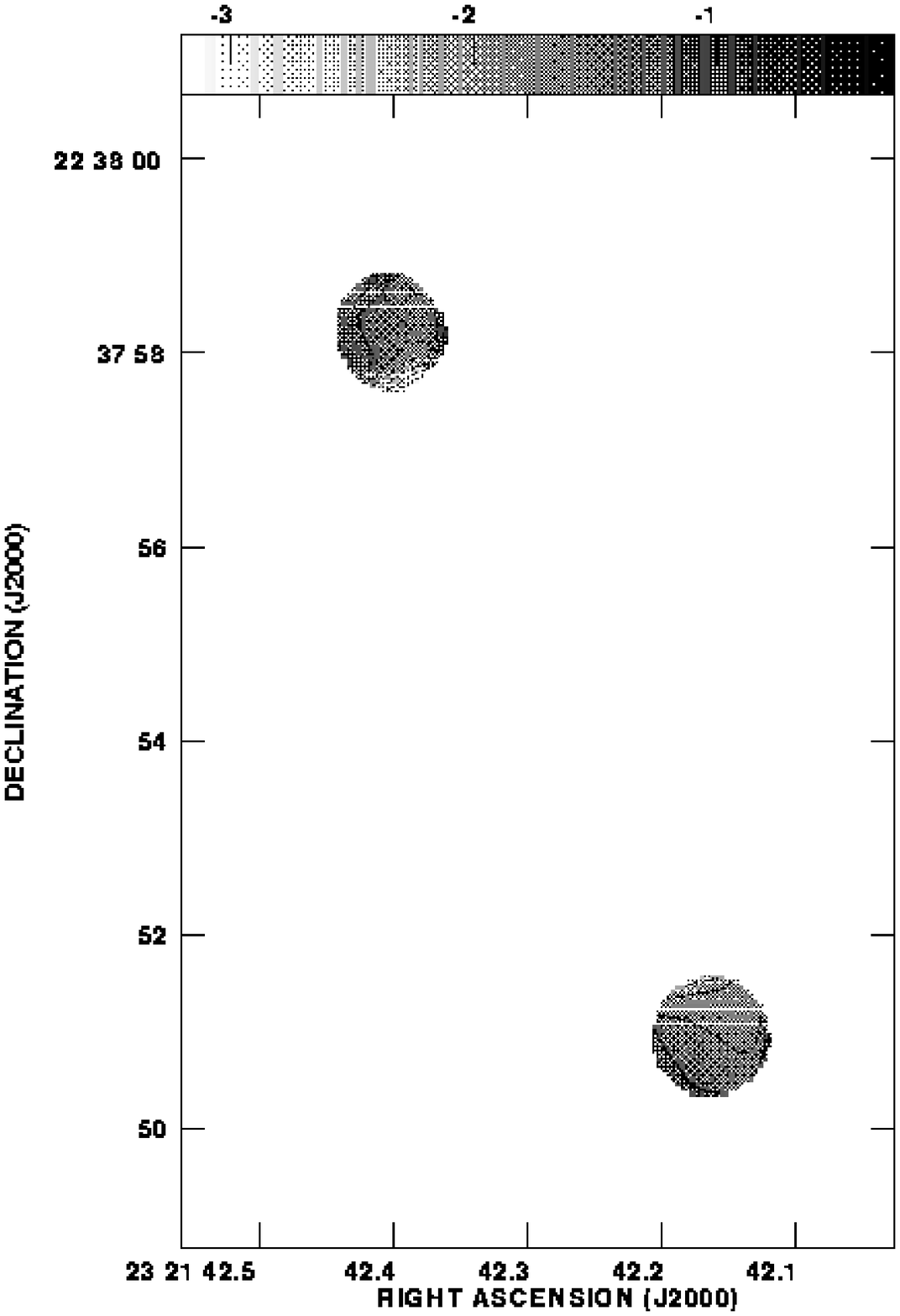,width=7.cm}}
\caption{Maps of the radio source 2319+223 at redshift z$=$2.554.
The sequence of figures is the same as in Fig. 6. The first contour level and
the  peak surface brightness  are respectively  0.122  mJy beam$^{-1}$ and 32
 mJy beam$^{-1}$ for 
the 4.5 GHz map; 0.058  mJy beam$^{-1}$ and 12.1 mJy beam$^{-1}$ for the 8.2
GHz map; 0.059  mJy beam$^{-1}$  and 2.9  mJy beam$^{-1}$ for the 4.7 GHz
polarized intensity map.} 
\end{figure*}

\end{document}